\documentclass[12pt,preprint]{aastex}

\newcommand{\Fermic}{\textit{Fermi}}
\newcommand{\Fermi}{\Fermic\ }
\newcommand{\FermiLATc}{\Fermic\ LAT}
\newcommand{\FermiLAT}{\FermiLATc\ }
\newcommand{\AFermiLATc}{\Fermic-LAT}
\newcommand{\AFermiLAT}{\AFermiLATc\ }
\newcommand{\hmsh}{\mbox{$^{\mathrm h}$}}%
\newcommand{\hmsm}{\mbox{$^{\mathrm m}$}}%
\newcommand{\hmss}{\mbox{$^{\mathrm s}$}}%

\newcommand{\grayc}{$\gamma$-ray}
\newcommand{\gray}{\grayc\ }
\newcommand{\graysc}{$\gamma$ rays}
\newcommand{\grays}{\graysc\ }

\newcommand{\ZFGLc}{\textit{0FGL}}
\newcommand{\ZFGL}{\ZFGLc\ }

\newcommand{\Icrabc}{$\phi_\mathrm{Crab}$}
\newcommand{\Icrab}{\Icrabc\ }

\newcommand{\cmsc}{cm$^{-2}$s$^{-1}$}
\newcommand{\cms}{\cmsc\ }

\newcommand{\NTeVCatExtragalactic}{30}
\newcommand{\NTeVCatAGN}{28}
\newcommand{\NTeVCatBlazar}{25}
\newcommand{\NTeVCatBLLac}{24}
\newcommand{\NTeVCatRG}{3}

\newcommand{\NObjTotal}{96}
\newcommand{\NObjTeVSrc}{\NTeVCatAGN}
\newcommand{\NObjTeVLim}{68}
\newcommand{\NDetTeVSrcZFGL}{18}
\newcommand{\NDetTeVSrcEGRET}{eight}
\newcommand{\NDetTeVSrcEGRETBLLac}{six}

\newcommand{\NDetTotal}{38}
\newcommand{\NDetTeVSrc}{21}
\newcommand{\NDetTeVLim}{17}
\newcommand{\NDetTeVLimBlazar}{16}
\newcommand{\NDetTeVLimBLLac}{15}
\newcommand{\NDetTotalNoEGRET}{29} 

\slugcomment{Accepted for The Astrophysical Journal}

\shorttitle{Fermi observations of TeV-selected AGN}
\shortauthors{Abdo et al.}

\usepackage{natbib}
\begin{document}


\title{Fermi observations of TeV-selected AGN}


\author{
A.~A.~Abdo\altaffilmark{2,3}, 
M.~Ackermann\altaffilmark{4}, 
M.~Ajello\altaffilmark{4}, 
W.~B.~Atwood\altaffilmark{5}, 
M.~Axelsson\altaffilmark{6,7}, 
L.~Baldini\altaffilmark{8}, 
J.~Ballet\altaffilmark{9}, 
G.~Barbiellini\altaffilmark{10,11}, 
M.~G.~Baring\altaffilmark{12}, 
D.~Bastieri\altaffilmark{13,14}
% ...
}
% ...
\altaffiltext{2}{Space Science Division, Naval Research Laboratory, Washington, DC 20375, USA}
\altaffiltext{3}{National Research Council Research Associate, National Academy of Sciences, Washington, DC 20001, USA}
\altaffiltext{4}{W. W. Hansen Experimental Physics Laboratory, Kavli Institute for Particle Astrophysics and Cosmology, Department of Physics and SLAC National Accelerator Laboratory, Stanford University, Stanford, CA 94305, USA}
\altaffiltext{5}{Santa Cruz Institute for Particle Physics, Department of Physics and Department of Astronomy and Astrophysics, University of California at Santa Cruz, Santa Cruz, CA 95064, USA}
\altaffiltext{6}{Department of Astronomy, Stockholm University, SE-106 91 Stockholm, Sweden}
\altaffiltext{7}{The Oskar Klein Centre for Cosmo Particle Physics, AlbaNova, SE-106 91 Stockholm, Sweden}
\altaffiltext{8}{Istituto Nazionale di Fisica Nucleare, Sezione di Pisa, I-56127 Pisa, Italy}
\altaffiltext{9}{Laboratoire AIM, CEA-IRFU/CNRS/Universit\'e Paris Diderot, Service d'Astrophysique, CEA Saclay, 91191 Gif sur Yvette, France}
\altaffiltext{10}{Istituto Nazionale di Fisica Nucleare, Sezione di Trieste, I-34127 Trieste, Italy}
\altaffiltext{11}{Dipartimento di Fisica, Universit\`a di Trieste, I-34127 Trieste, Italy}
\altaffiltext{12}{Rice University, Department of Physics and Astronomy, MS-108, P. O. Box 1892, Houston, TX 77251, USA}
\altaffiltext{13}{Istituto Nazionale di Fisica Nucleare, Sezione di Padova, I-35131 Padova, Italy}
\altaffiltext{14}{Dipartimento di Fisica ``G. Galilei", Universit\`a di Padova, I-35131 Padova, Italy}
% ...



\begin{abstract}
We report on observations of TeV-selected AGN made during the first
5.5 months of observations with the Large Area Telescope (LAT)
on-board the \textit{Fermi Gamma-ray Space Telescope} (\Fermic). In
total, \NObjTotal\ AGN were selected for study, each being either (i)
a source detected at TeV energies (\NObjTeVSrc\ sources) or (ii) an
object that has been studied with TeV instruments and for which an
upper-limit has been reported (\NObjTeVLim\ objects). The \Fermi
observations show clear detections of \NDetTotal\ of these
TeV-selected objects, of which \NDetTeVSrc\ are joint GeV--TeV sources
and \NDetTotalNoEGRET\ were not in the third EGRET catalog. For each
of the \NDetTotal\ \Fermi-detected sources, spectra and light curves
are presented. Most can be described with a power law of spectral
index harder than 2.0, with a spectral break generally required to
accommodate the TeV measurements. Based on an extrapolation of the
\Fermi spectrum, we identify sources, not previously detected at TeV
energies, which are promising targets for TeV instruments. Evidence
for systematic evolution of the \gray spectrum with redshift is
presented and discussed in the context of interaction with the EBL.
\end{abstract}


\keywords{Gamma rays: observations}


\section{Introduction}
\label{SEC::INTRO}

At energies above approximately 100\,GeV (hereafter the \textit{TeV
energy regime}), ground-based \gray observatories have detected 96
sources over the past two decades. The pace of discovery in this
energy regime has been particularly high since the inception of the
latest generation of instruments: H.E.S.S., CANGAROO, MAGIC and VERITAS
\citep[for recent review]{REF::TEV_REVIEW}. Online catalogs, such as
TeVCat\footnote{http://tevcat.uchicago.edu, see
\citet{REF::TEVCAT}}, present continuously updated views of the TeV \gray
sky. The majority of the TeV sources are galactic, however
\NTeVCatExtragalactic\ extragalactic sources have also been detected,
of which \NTeVCatAGN\ correspond to Active Galactic Nuclei (AGN), the
other two being recently detected starburst galaxies. Most
(\NTeVCatBlazar) of these TeV AGN are \textit{blazars}, an AGN
sub-category in which the jet of relativistic plasma ejected from the
core is roughly coaligned with our line-of-sight and hence appear
Doppler boosted. The majority (\NTeVCatBLLac) of the TeV blazars
belong to a further sub-category, the BL~Lac objects (from
BL~Lacertae, the prototype for the class), which do not have
significant emission or absorption features in their optical spectra,
making it difficult to measure their redshift directly.

The first blazar detected at TeV energies was Markarian 421
\citep[Mrk~421;][]{REF::TEV_MARKARIAN_421}, at a redshift of
$z=0.031$. It is seen to be highly variable, with flux varying between
$\sim0.15$ and $>10$ the flux of the Crab Nebula\footnote{The Crab
Nebula is the brightest steady TeV source and the ``standard candle''
of TeV astronomy, defining the ``Crab Unit'' in which TeV fluxes and
limits are often expressed. We adopt the value of
\Icrabc$(>E)=2.1\times10^{-11}(E/1\,\mathrm{TeV})^{-1.5}$\,\cms from
\citet{REF::CRAB_SPECTRUM_HILLAS}.} (\Icrabc). Doubling
timescales as short as 15 minutes have been observed during flares
\citep{REF::MRK_FLARE_NATURE}. Mrk~421 has a hard spectrum, with mean
photon index of $\Gamma=2.5$, and has shown clear evidence of spectral
variability during flaring episodes \citep{REF::MRK_SPECTRAL_VAR}. The
detection of such a distant object was interpreted in the context of
\gray attenuation through pair-production \citep{REF::EBL_TEV_THEORY}
to produce a limit on the power density of extragalactic background
light \citep[EBL;][]{REF::EBL_TEV_MRK}. Mrk~421 was detected by EGRET,
the predecessor of the \Fermi Large Area Telescope
\citep{REF::EGRET_CALIBRATION,REF::EGRET_INSTRUMENT}, and was reported 
in the third EGRET catalog \citep[3EG;][]{REF::3EG} with a detection
significance of $\sim10\sigma$. However, EGRET did not have sufficient
sensitivity to make detailed measurements on the short timescales
required to match the TeV observations.

Since this initial discovery, \NTeVCatAGN\ AGN sources have been
detected at TeV energies, the most distant reported being 3C~279, at a
redshift of $z=0.54$. Like Mrk~421, many of these have shown evidence
for variability, undergoing episodic flaring activity with short
doubling timescales. To date, the most extreme example of variability
has been observed from PKS~2155-304
\citep{REF::TEV_PKS_FLARE}, in which the flux was seen to reach a
maximum of $\sim15$\,\Icrab with doubling times as short as
$225$\,s. However, approximately half of the TeV blazars show no
evidence for variability, e.g.\ PKS~2005-489
\citep{REF::TEV_PKS_2005-489} and PG~1553+113
\citep{REF::TEV_PG_1553+113,REF::TEV_PG_1553+113_MAGIC}. The
detections of more distant objects lead to tighter constraints on the
the level of the EBL, suggesting that its density is close to the
minimum level required from galaxy counts
\citep{REF::TEV_1ES_1101-232}. The spectra of the majority of TeV
blazars are adequately described by a simple power law\footnote{The
exceptions being Mrk~421 and Mrk~501 which show evidence for spectral
curvature and PKS~2155-304 whose spectrum was found to have the form
of a broken power-law, during the flaring episode of June 2006.}, with
$\Gamma\geq 2$, and with those of the more distant sources being
considerably softer, up to $\Gamma\approx4$
\citep[e.g.][]{REF::TEV_3C66A_VERITAS,
REF::TEV_1ES_1011+496,REF::TEV_3C279,REF::TEV_PG_1553+113,
REF::TEV_PKS_2005-489}. The peak in the measured \gray spectrum (in
${\nu}F_\nu$ representation) of these objects lies outside of the
energy range of the TeV instruments, making it difficult to fully
constrain models of emission using TeV observations alone. The spatial
and spectral properties of the TeV-detected AGN are presented in
Tables~\ref{TAB::OBJ_DET} and \ref{TAB::OBJ_DET_SPEC}, together with
references to TeV observations of each source.

The Large Area Telescope (LAT) is a pair conversion telescope on the
\Fermi \textit{Gamma-ray Space Telescope} (formerly GLAST), launched 
in June 2008. The \Fermi LAT instrument, described in detail in
\citet{REF::LAT_INSTRUMENT}, detects \grays with energies between
20\,MeV and $>$300\,GeV (hereafter denoted the \textit{GeV energy
regime}). The bulk of the \Fermi observational program is dedicated
to a sky-survey, in which the full \gray sky is observed every 3
hours. This survey is optimized to produce a uniform exposure to the
sky on timescales of months and to facilitate the monitoring and
detection of variable and flaring \gray sources on shorter
timescales. In its first three months of operation the \FermiLAT
mapped the \gray sky with a sensitivity and precision that exceeds any
previous space mission in this energy regime. A list of the 205
brightest sources ($10\sigma$ or greater) found during that period,
and their properties, has been published by the \AFermiLAT
collaboration to guide multiwavelength studies with other instruments
\citep{REF::0FGL}. This collection of sources is henceforth referred
to as the \ZFGL list, with individual sources denoted as
0FGL~JHHMM.M$\pm$DDMM. In addition, an in-depth study of the
population of \ZFGL sources most likely associated with AGN has been
made \citep{REF::LBAS}; this population is commonly referred to as the
LAT bright AGN sample (LBAS). 

Only \NDetTeVSrcEGRET\ of the \NObjTeVSrc\ TeV-detected AGN
were detected by EGRET and included in the 3EG catalog (see
Table~\ref{TAB::OBJ_LBAS}); \NDetTeVSrcEGRETBLLac\ are BL~Lacs.  The
majority of these 3EG GeV--TeV sources were discovered as TeV emitters
only with the advent of the latest generation of TeV
observatories. With the previous generation of instruments only two
such extragalactic GeV--TeV sources were established, despite
dedicated programs to observe 3EG sources
\citep[e.g.][]{REF::TEV_SURVEY_OF_EGRET}, indicating the degree of
mismatch between the sensitivities and effective energy ranges of
these instruments. In accordance with the blazar sequence hypothesis
\citep{REF::BLASAR_SEQ}, BL~Lacs are the least luminous class of
blazars in the GeV regime, but their emission peaks at a higher energy
and they have hard photon indices in the GeV domain. In fact, above a
few 10s of GeV, BL~Lacs are often relatively brighter than other
AGN. With a rapidly falling sensitivity above $\sim 5-10$\,GeV, EGRET
preferentially detected the more luminous, lower-energy-peaked
blazars. In contrast, the \FermiLAT has a relatively flat effective
area at high energies (8000\,cm$^2$ for on-axis \grays at $E>1$\,GeV)
and an energy response that extends beyond 300\,GeV, making it much
more sensitive to the hard BL~Lac sources than was EGRET. Of the \ZFGL
sources, 14 are AGN detected at TeV energies.

One of the most powerful tools for probing the physics underlying the
emission from AGN is the dedicated multiwavelength observational
campaign, in which simultaneous observations are made across the full
spectrum. Generally, such observations of the TeV blazars reveal two
non-thermal components: one at lower energies, peaking in the UV to
X-ray regime, and showing clear evidence of polarization, and a second
peaking in the \gray regime. The low-energy component is commonly
interpreted as resulting from synchrotron emission from relativistic
electrons in the jet, while the high-energy component results from a
different process, such as inverse-Compton scattering of lower-energy
photons in the region of the jet, or the decay of $\pi^0$ particles
produced in interactions of relativistic protons. In many such
campaigns significant correlation between the X-ray and TeV
\gray emission has been detected \citep[e.g.][]{REF::MRK_MWCAMPAIGN,
REF::PKS2155_HESS_CHANDRA}, suggesting that a single population of
relativistic particles is responsible for the emission in both regimes
\citep[see][for further discussion of X-ray/TeV
correlations]{REF::XRAY_TEV_CORRELATION}. Leptonic mechanisms, such as
the synchrotron self-Compton (SSC) and external-Compton (EC)
processes, and hadronic mechanisms have been invoked to explain the
broad-band emission and correlated time variability seen between the
low-energy and high-energy components \citep[see][for further
discussion]{REF::BOTTCHER_BLAZAR_EMISSION}. Instances of isolated
\gray variability have also been detected in some cases
\citep[e.g.][]{REF::ORPHAN_FLARE}, and it is probable that no simple
mechanism will fully explain the considerable variety of the blazar
behaviour.

Despite the participation of instruments across a wide range of the
spectrum, until recently, multiwavelength campaigns have been largely
unable to probe the energy range between $\sim150$\,keV and
$\sim150$\,GeV, as no instrument with sensitivity matched to the
day-to-week timescales of typical campaigns has existed. As such,
although the synchrotron component has been well measured from radio
to X-ray, the full extent of the higher energy \gray component has
not. In the case of the TeV blazars, ground-based \gray instruments
have generally measured the falling edge of the high-energy component
(in ${\nu}F_\nu$ representation), which has usually been consistent
with a featureless power-law. The rising edge and peak of the emission
have, however, been inaccessible, and hence models of the emission
have been unconstrained in this energy range. In many cases, very
different emissions mechanisms have been invoked, and can explain the
data equally well. With the launch of \Fermic, which has the
sensitivity to measure the emission from the brighter TeV blazars on
the day to week timescales, a large part of this gap in coverage has
been closed. A recent multi-wavelength campaign on PKS~2155-304 was
the first in which the rising and falling edges of the high-energy SED
were simultaneously measured with precision
\citep{REF::PKS2155_FERMI_MWCAMPAIGN}. Measurement of the full
broadband spectrum and the pattern of correlation between the optical,
X-ray, GeV \gray and TeV \gray emission removes degeneracies in
modeling of this object which were present in the results from
previous campaigns that could not measure the high-energy component
fully.

In this paper we present the results of the first 5.5 months of
\AFermiLAT observations of the known TeV blazars and of those
AGN for which upper-limits exist at TeV energies. The motivation for
this study is two-fold: (i) to present as complete a picture of the
high-energy emission as possible by combining the GeV and TeV results
on these objects, and (ii) to help guide future TeV observations. For
each object detected by \Fermi a power-law fit to the GeV spectrum is
presented, as are light curves on monthly timescales. For the brighter
sources, lightcurves on 10-day timescale are also given. The GeV
power-law spectra are extrapolated to TeV energies assuming absorption
on the EBL, yielding predictions for TeV emission for the simplest case
where there is no curvature in the intrinsic spectra of the objects. For
those objects which are not detected by the \FermiLATc, upper limits in
the GeV range are presented and extrapolated to TeV energies.


\section{Sources}
\label{SEC::SCOPE}

The primary objects selected for this study are the \NTeVCatBlazar\
blazars and \NTeVCatRG\ radio galaxies detected at TeV energies. These
are listed in Table~\ref{TAB::OBJ_DET} with their coordinates, the AGN
sub-class of the object, redshift and references to the initial
detection at TeV energies. In summary, 19 high-frequency-peaked
BL~Lacs (HBL), 3 intermediate-frequency-peaked BL~Lacs (IBL), 2
low-frequency-peaked BL~Lacs (LBL), 1 flat-spectrum radio quasar
(FSRQ) and 3 Fanaroff-Riley radio galaxies (type FR1) have been
detected by TeV instruments, the most distant having a redshift of
$z=0.54$.  Table~\ref{TAB::OBJ_DET_SPEC} lists the parameters of a
power-law fit to the TeV spectra for these objects, where available:
the integral flux ($\phi\pm\Delta\phi$) and photon index of the fit
($\Gamma\pm\Delta\Gamma$) and the threshold energy for the observation
($E_\mathrm{thres}$), such that the differential spectrum is:
\[
\frac{dN}{dE} 
= (\Gamma-1)\frac{\phi}{E_\mathrm{thres}}
\left(\frac{E}{E_\mathrm{thres}}\right)^{-\Gamma}
= F_{200} \left(\frac{E}{200\,\mathrm{GeV}}\right)^{-\Gamma}
\]
The differential flux at $200$\,GeV (the median threshold of the
measurements), $F_{200}$, is calculated from the TeV power-law
spectrum and presented in the table to compare the TeV objects at a
single energy lying within the domain of the \AFermiLAT
observations. For some objects, multiple TeV spectra have been
measured, either by different instruments, in different epochs, or
when the object is in different flux states. Where possible, the
spectrum corresponding to a low-flux state is listed.

In addition we search for GeV emission from \NObjTeVLim\ objects for
which TeV upper-limits were published from observations with the
Whipple 10m telescope
\citep{REF::UL_WHIPPLE_HORAN,REF::UL_WHIPPLE_DLCP,
REF::UL_WHIPPLE_FALCONE}, HEGRA \citep{REF::UL_HEGRA_TLUCZYKONT},
MAGIC \citep{REF::UL_MAGIC_ALBERT} and H.E.S.S.\
\citep{REF::UL_HESS_BENBOW05, REF::UL_HESS_BENBOW08}. These targets
are listed in Table~\ref{TAB::OBJ_UL} with the lowest flux upper-limit
published.

From these \NObjTotal\ target objects, the \NDetTeVSrcZFGL\ listed in
Table~\ref{TAB::OBJ_LBAS} are identified or associated with sources in
the \ZFGL list and LBAS sample \citep{REF::0FGL,REF::LBAS}. Those
lists were limited to sources with $TS>100$ in three months of
\AFermiLAT data\footnote{The test statistic, TS, is roughly indicative 
of the significance of the LAT detection of the source squared.}; in
this study the criterion to claim a detection and derive a spectrum is
lowered to $TS>25$ and the period of observation is increased to 5.5
months. A total of \NDetTotal\ sources are detected by \Fermic, of
which \NDetTeVSrc\ are jointly detected at GeV and TeV energies. We
also give an upper-limit for TeV sources not detected by \Fermic.



\section{Analysis}
\label{SEC::ANALYSIS}

\AFermiLAT data from the 5.5 month interval from MJD 54682 to MJD
54842 are processed with the standard analysis chain
\texttt{ScienceTools} (ST; version V9R11). The latest
instrumental response functions (IRFs; version P6\_V3) are used to
characterize the PSF and effective area during the analysis. These
IRFs offer a distinct improvement over those used in the \ZFGL
analysis, which did not properly account for the presence of remnants
of non-triggering events in the tracker. This change results in a
systematic increase of $\sim15\%$ in the derived flux from \gray
sources, and a possible change in the spectral index, which is most
pronounced for softer sources. However, despite these improvements,
the P6\_V3 IRFs are based on pre-flight calibrations, and to be
conservative, only events in the energy range from 200\,MeV to
300\,GeV are retained for analysis \citep[see for
example.][]{REF::0FGL}.

The data for each of the AGN targets is analyzed in a identical
manner. Low-level processing of the spacecraft data is applied
automatically in a pipeline, reconstructing the energy, arrival
direction and particle type of the primary. Events reconstructed from
a region of interest (ROI) of $10^\circ$ around the target location
are extracted from this database and filtered such that only those
having the highest probability of being a photon (those in the
so-called ``diffuse'' event class) and having an angle of $<105^\circ$
with respect to the local zenith (to suppress the background from the
Earth albedo) are retained.

For each target, a background model is constructed, consisting of a
diffuse galactic component, predicted by the GALPROP program
\citep{REF::GALPROP,REF::GALPROP2}, a diffuse power-law component (for
the extragalactic and instrumental background) and any of the point
sources from the \Fermi three-month catalog
\citep{REF::0FGL}\footnote{We use an internal version of catalog which
is not limited to TS$>$100.} which overlap the ROI. The spectrum for
each of the point sources is modeled as a power-law. An unbinned
maximum likelihood method \citep{REF::CASH_LIKE,REF::MATOX_LIKE},
implemented as part of the ST by the \texttt{gtlike} program, is used
to optimize the parameters to best match the observations.

The validity of the optimized emission model is verified by producing
a \textit{TS map} for each region. This is done using the ST
\texttt{gttsmap} program, which adds a \textit{test source} to each
location over a prescribed region and calculates the improvement in
log-Likelihood with the inclusion of the test source. Statistically
compelling sources, not accounted for by the model, are identified
visually in the map and added to the background in another iteration
of \texttt{gtlike}. In addition, high resolution TS maps are produced
for each source of interest and the centroid of the emission and
contours defining the 68\%, 95\% and 99\% probability regions
calculated. During the construction of the \ZFGL list, a systematic
error of $\approx 1\arcmin$ in the reconstruction of the centroids of
emission from well known bright sources was identified, and this error
is folded into the emission contours displayed in
Figure~\ref{FIG::TS}. If the centroid of emission on the map is
$\vec{r}_\mathrm{c}$ and the distance from the centroid to the
(statistical) error contours are defined parametrically by the
functions $r_\mathrm{stat}^{\,i}(\theta)$, where $i=1,2,3$ for the
68\%, 95\% and 99\% probabilities ($P^{\,i}=0.68,0.95,0.99$), then the
contours which account for systematic errors are defined as,
\[ r_\mathrm{syst}^{\,i}(\theta) = 
\sqrt{r_\mathrm{stat}^{\,i}(\theta)^2 + 
(\sqrt{-2\ln(1-P^{\,i})}\times1\arcmin)^2}.\]
For well detected sources the error contours are roughly circular,
with the systematic error dominating. For weakly detected sources the
contours can be irregularly shaped, with the statistical component
dominating.

Among the outputs from the \texttt{gtlike} program are the optimized
values of the model parameters, the covariance matrix describing their
variances and correlations, and the TS of each source, which indicates
the significance of the source detection. An error contour (called
butterfly diagram) is computed from these values and plotted to
indicate the $1\sigma$ confidence range of the fitted model. If the
power-law is written as $F(E)=dN/dE=F_0(E/E_0)^{-\Gamma}$, with
normalization parameter $F_0\pm\Delta F_0$, photon index
$\Gamma\pm\Delta\Gamma$ and covariance $\mathrm{cov}(F_0,\Gamma)$, the
the contour is defined by:
\begin{equation}
\frac{\Delta F^2}{F^2} 
= \frac{\Delta F_0^2}{F_0^2}
- \frac{2\,\mathrm{cov}(F_0,\Gamma)}{F_0}\log\left(\frac{E}{E_0}\right)
+ \Delta\Gamma^2 \log^2\left(\frac{E}{E_0}\right).
\end{equation}
The narrowest point in the butterfly occurs at
$E_d=E_0\exp[\mathrm{cov}(F_0,\Gamma)/F_0\,\Delta \Gamma^2]$, the
so-called decorrelation energy. For each source, the butterfly is
drawn between the lowest energy used in the analysis (either 0.2\,GeV
or 1\,GeV) and the energy of the highest photon detected from the
source, subject to the constraint of $E<300$\,GeV.

For the sources with a detection significance of $TS<25$ ($\sim
5\sigma$), upper limits on the integral flux above 200 MeV are
computed assuming a photon index arbitrary fixed at 1.5 and 2.0. For
the very bright LAT sources, the energy range is divided into two bins
(200\,MeV--1\,GeV and 1\,GeV--300\,GeV), and spectra are fitted to
each bin separately. This analysis is limited to sources for which
$TS>100$ in \textit{each} of the energy bins, ensuring that a
sufficiently accurate spectrum can be derived in each. For all
detected sources, systematic errors on the flux and index of the
power-law fit to the full \Fermi energy range, caused by systematic
errors in the IRFs used in the analysis, are evaluated using the
``bracketing'' method of \citet{REF::FERMI_CRAB}.

In order to make predictions for the TeV energy domain and to make
comparisons between the \AFermiLAT and TeV spectra, the best-fit
spectrum and butterfly are extrapolated up to 10 TeV. This assumes
that the intrinsic spectrum of the emission is described by a single
power-law extending over the GeV and TeV energy range, the simplest
and least model-dependent assumption that can be made. Above a few
hundred GeV, the photons interact with the infrared photons from the
EBL as they propagate through the Universe, modifying the detected
spectrum from a simple power-law. \citet{REF::EBL_MODEL_FRANCESCHINI}
provide tabulated values of the optical depth as a function of the
redshift, which are used to compute the flux detectable between
200\,GeV and 10\,TeV from the extrapolated GeV spectrum. Their EBL
model is consistent with experimental measurements: the lower limits
from galaxy counts and upper limits from observations of TeV blazars,
and is widely used. However it is not necessarily the final word on
EBL density \citep[see e.g.][]{REF::KRENNRICH_DWEK_IMRAN_CONSTRAINTS},
and any errors in the model will propagate into the extrapolation of
the GeV spectra to TeV energies. In addition to absorption on the EBL,
TeV photons may undergo absorption in the neighborhood of the source
\citep[see e.g.][]{REF::INTRINSIC_ABSORPTION_AHARONIAN}, which must
also be modeled and accounted for to unfold the intrinsic accelerated
spectrum from the detected spectrum. However, such modeling is beyond the
scope of this paper.

Light curves for each source are produced with time bins of 10 and 28
days (following the lunar cycle). The light curves are produced by
binning the events by their arrival times and performing an
independent likelihood analysis for each temporal bin with the same
model (same background sources and number of free parameters) as in
the fitting of the time-averaged spectrum. The probability that the
light curve is consistent with being flat, from a $\chi^2$ fit to a
constant value, is also computed, and used to evaluate the hypothesis
that the fitted spectra, averaged over the full period, are valid.


\section{Results and discussion of individual sources} 

Of the \NObjTeVSrc\ TeV-selected sources studied here, \NDetTeVSrc\
are detected by \Fermi with $TS>25$. This degree of connection between
the TeV blazars and the GeV regime was not found by EGRET and the
previous generation of TeV instruments, and is evident now only as a
result of the improved sensitivity and greater overlap between the
effective energy ranges of \Fermi and the current generation of TeV
instruments.

The majority of the TeV blazars detected by \Fermi have a photon index
$\Gamma<2$ in the GeV regime, the median index is $\Gamma=1.9$. In
contrast, the populations of $42$ BL~Lacs and $57$ FSRQs from the LBAS
sample have median indices of $\Gamma=2.0$ and $\Gamma=2.4$
respectively. The TeV blazars are amongst the hardest extragalactic
objects detected by \Fermic.

Of the \NObjTeVLim\ extragalactic objects with TeV limits which were
considered here, a total of \NDetTeVLim\ are detected in the GeV
regime. These too have a hard median index, of $\Gamma=1.95$,
indicating that they, perhaps, are good targets for deeper observation
with TeV instruments. Of these \NDetTeVLim, only one is not a blazar
(NGC~1275, an FR1 radiogalaxy), one is an FSRQ (3C~273) and the
remainder are BL~Lacs. That the majority of these \NDetTeVLimBlazar\
blazars are BL~Lacs, rather than following the ratio of FSRQs to
BL~Lacs found in LBAS, is most likely a result of the way these
objects were selected for observation originally by the various TeV
groups, rather than anything inherently fundamental. These
\NDetTeVLimBLLac\ objects break down as follows: 3 LBL, 1 IBL and 11
HBL.

For each of the \NDetTotal\ sources detected by \Fermic, power-law
fits to the data and extrapolations into the TeV regime are presented
in Figure~\ref{FIG::SPEC_TEVDET_1} for GeV--TeV sources, and in the
online material for the GeV sources with TeV upper-limits.
Additionally, for TeV sources not detected by \Fermic, upper limits on
the spectra are presented in Figure~\ref{FIG::SPEC_LIMITS}. To justify
the validity of these ``averaged spectra'', light curves on 28-day
timescales are presented in Figure~\ref{FIG::LC28_1} for all
GeV-detected objects, and on 10-day timescales in
Figure~\ref{FIG::LC10} for the brighter GeV emitters. In the 5.5
months of data analyzed here, the majority of sources do not show
evidence of flux variation on the timescales tested. For a subset of
the sources, as discussed below, a TS map is presented in
Figure~\ref{FIG::TS}. The remainder are available in the online
material.

In addition to the figures, the parameters of the power-law spectra
and variability are given in tables. For each of the sources detected
by \Fermic, Table~\ref{TAB::OBJ_RES} lists
\begin{itemize}
\item the flux and index of the power law, the statistical and
systematic errors on these quantities, the decorrelation energy,
\item the energy of the highest and fifth highest photons detected
within $0.25^\circ$ of the source position, which corresponds to
$>99.9$\% containment according to the P6\_V3 IRFs, and
\item the probabilities that the 28-day and 10-day light curves are
consistent with a constant value.
\end{itemize}
For the brightest GeV sources, Table~\ref{TAB::OBJ_RES_HE} lists the
parameters of the power-law fits to the low-energy (0.2\,GeV--1\,GeV)
and high-energy (1\,GeV--300\,GeV) bands. For the TeV sources not
detected by \Fermic, Table~\ref{TAB::OBJ_RES_UL} gives the GeV
upper-limit over the \Fermi energy range. Finally,
Table~\ref{TAB::OBJ_PRED} presents the extrapolations of the GeV
spectra to the TeV domain, listing the differential flux extrapolated
(or measured) at 100\,GeV, the integral flux in the TeV band
(0.2\,TeV--10\,TeV) and the photon index found by fitting the
EBL-corrected spectrum between 100\,GeV and 1\,TeV with a power-law.

\subsection{TeV sources detected by the \FermiLAT}

The TeV sources detected by the \FermiLAT are discussed individually
below. For sources with a published TeV spectrum the \Fermi spectrum
is compared with the TeV. Unless otherwise noted, when more than one
TeV spectrum is available in the literature, the one corresponding to
the lowest flux state is chosen for comparison. Since references to
the initial detection of each TeV source and to the TeV spectra chosen
are given in Tables~\ref{TAB::OBJ_DET} and \ref{TAB::OBJ_DET_SPEC},
they are not repeated in the text below.  As discussed above, the TS
maps, spectra and light curves are presented in Figures~\ref{FIG::TS},
\ref{FIG::SPEC_TEVDET_1}, \ref{FIG::LC28_1} and
\ref{FIG::LC10} and in Tables~\ref{TAB::OBJ_RES},
\ref{TAB::OBJ_RES_HE} and \ref{TAB::OBJ_PRED}, and the reader is not
directed to them individually in the discussion below.

\textit{\objectname{3C 66A}/B:} 
TeV \gray emission from this region was initially reported by the
Crimea Observatory group and later, based on observations with the more
sensitive VERITAS and MAGIC instruments, with a flux that was less
than $1/100$ that claimed in the original detection. VERITAS
observations during 2007 \& 2008 lead to the detection of a flare from
3C~66A (in October 2008), an IBL, while ruling out 3C~66B, separated
by $0.12^\circ$ from 3C~66A, at a level of $4.3\sigma$. In contrast,
the MAGIC observations during 2007 are consistent with the emission
originating from 3C~66B, a radio-galaxy at a distance of $z=0.0211$;
they rule out 3C~66A at a probability of $85.4\%$. The MAGIC
observations revealed a harder source, with a significantly lower flux
than the later VERITAS observations, which were taken during a flaring
episode. The MAGIC observations showed no evidence for variability.
GeV \gray emission from the region of 3C~66A was discovered by EGRET,
although the signal was contaminated by a nearby pulsar
\citep{REF::PSR0218_EGRET}. Details of a dedicated multiwavelength
campaign on 3C~66A, involving \Fermic, during the period of the
VERITAS flare are given by \citet{REF::3C66A_FERMI_REYES_ICRC}. The
redshift of 3C~66A is assumed to be $z=0.444$. This value, however, is
based on two measurements of a single weak line in the spectrum of the
galaxy, and is considered to be uncertain
\citep{REF::3C66A_REDSHIFT}.

The positions of 3C~66A (``$\times$'') and 3C~66B (``+'') are marked
in the TS map for this field. The emission is distributed throughout a
broad region around both sources, consistent with the \FermiLAT PSF
(which is large at low energies), however it can be seen that the
centroid of the emission is constrained to a relatively small region
containing 3C~66A (at a 68\% confidence level) and excluding 3C~66B
(at $>$99\% level). The \FermiLAT emission above 1~GeV is well
described by a power law with index of $\Gamma=1.98\pm0.04$, which is
extrapolated to the TeV regime using the redshifts of both 3C~66A
(dashed line) and 3C~66B (dot-dashed line). The VERITAS and MAGIC
spectral measurements are also shown. It is clear that the spectrum
extrapolated in the distant ``3C~66A'' scenario is in better agreement
with the TeV measurements than the close-by ``3C~66B'' scenario.  The
latter case would require significant turnover in the intrinsic
spectrum above 100\,GeV to agree with the MAGIC measurements. The
28-day and 10-day lightcurves show evidence for variability, with a
factor of 5--6 between the highest and lowest fluxes. As a result of
this, and since the VERITAS measurements showed evidence for a flaring
state, and the fact that the redshift of 3C~66A is uncertain, we do
not claim that the extrapolated, averaged GeV spectrum is an exact
match to the VERITAS points, only that their superficial agreement is
suggestive of the dominance of a more distant source. Taken together
the positional and spectral information indicate that the bulk of the
GeV emission arises from 3C~66A.

\textit{\objectname{RGB J0710+591}:} 
Detected recently by the VERITAS collaboration, detailed spectral
information at TeV energies has not yet been published for this
HBL. RGB~J0710+591 is weakly detected by the \FermiLATc, with
indications of a hard spectrum. The low statistics at GeV energies
mean that the extrapolation into the TeV regime is not constraining.

\textit{\objectname{S5 0716+714}:} This recently detected MAGIC source 
was reported with a preliminary flux level of
$\phi(>400\,\mathrm{GeV})\sim 10^{-11}$\,\cmsc.
\AFermiLAT observations reveal highly significant GeV emission, 
with a falling spectrum. The extrapolated GeV spectrum gives a flux of
$\phi_\mathrm{ext}(>400\,\mathrm{GeV})\sim 0.07 \times
10^{-11}$\,\cmsc, indicating that the source was likely in a
particularly bright state during the MAGIC observations. Indeed, Swift
observations contemporaneous with the MAGIC detection revealed the
highest X-ray flux ever measured from S5~0716+714
\citep{REF::SWIFT_S5_0716+714}.

%
%

\textit{\objectname{1ES 0806+524}:} 
\Fermi detects significant emission from this object, which is consistent
with a flat spectrum of $\Gamma=2.04\pm0.14$. The extrapolation of
this power-law to TeV energies agrees well with the spectrum measured
by VERITAS. The VERITAS observations did not reveal any significant
variability on timescales of months, but the flux was too low to probe
shorter timescales. The \FermiLAT observations show only marginal
evidence for variability on 28-day timescales, and we therefore
suggest that it is reasonable to equate the time-averaged GeV and TeV
spectra.

\textit{\objectname{1ES 1011+496}:}
The spectrum of this bright \Fermi source is analyzed in two energy
bands. The high energy band ($E>1$\,GeV) is consistent with a power
law of index $\Gamma=1.96\pm0.09$. The TeV spectrum from MAGIC is
considerably softer, with the lowest spectral measurement made at
150\,GeV. The highest energy photon detected by \Fermi has an energy
of 168\,GeV, and hence is in the range covered by the MAGIC spectrum.
The flux of the two measured spectra are consistent in the overlapping
region, and the extrapolated, absorbed \Fermi spectrum agrees well
with the measured TeV points. No evidence of variability is seen in
the GeV or TeV observations.

\textit{\objectname{Markarian 421}:}
A very bright source in the GeV regime, the spectrum of Markarian~421
is measured with precision by the LAT. No indication of spectral
curvature is found; the spectrum above 1\,GeV is well described by a
simple power law of index $\Gamma=1.78\pm0.04$. A similar value is
found for the spectrum below 1\,GeV. The highest energy photon
detected by the LAT from Markarian~421 was reconstructed at
$E\approx800$\,GeV, and five photons were found with
$E>150$\,GeV. This source has the highest degree of overlap between
GeV and TeV spectra of any of the TeV blazars. There are a
considerable number of TeV spectral measurements available in the
literature; we adopt the MAGIC spectrum, which has the lowest energy
threshold, was made in a relatively low flux state (0.5\,\Icrabc). The
spectrum during an extreme flaring state, measured by Whipple in 2001,
is also shown. The differential flux measured by the \FermiLAT at 100
GeV is compatible with that found by MAGIC, nevertheless, the
extrapolation of the \Fermi spectrum leads to an overestimation of the
integral flux above 200 GeV. The extrapolated photon index is $1.9$,
clearly harder than any reported in the literature. It is impossible
to reconcile the GeV and TeV spectra on the basis of EBL absorption
alone, and we conclude there is a turnover in the intrinsic spectrum
in the neighborhood of 100\,GeV, a region of falling sensitivity for
both \Fermi and the TeV instruments.

\textit{\objectname{Markarian 180}:}
Based on detection at a level of $TS=50$, the \Fermi observations show
no evidence for variability and yield a power-law spectral index of
$\Gamma=1.91\pm0.18$. The highest energy photon associated with
Mrk~180 is 14\,GeV, a decade lower than the TeV data points reported
by MAGIC. The TeV spectrum is softer than a simple extrapolation from
the \Fermi regime, but has a larger flux at 150\,GeV. MAGIC did not
detect any variability from the object, however their observations
were triggered by a particularly high optical state, which might
indicate that the TeV spectrum is not representative of an ``average''
state.

\textit{\objectname{1ES 1218+304}:} 
This object lies close to two others considered in this study, W~Comae
and ON~325. All three occupy a single region of interest (see
section~\ref{SEC::ANALYSIS}) for \Fermic, and must hence be analyzed
together. In the GeV regime the brightest by far is ON~325, an LBL
which was detected in the \ZFGL survey but not detected by TeV
instruments. The TeV source 1ES~1218+304 lies only $\sim0.75^\circ$
from ON~325, well within the PSF of the \FermiLATc, at least at lower
energies, and W~Comae, also a TeV source, lies $\sim2^\circ$ away.

\Fermi detects significant emission from the region of
1ES~1218+304. The \Fermi spectrum is well described by a power law
with an index of $\Gamma=1.63\pm0.12$, making it one of the hardest
sources in the sample. In the TS map for this region, the centroid of
the GeV emission is offset by $\sim4$\,arcmin, with the blazar located
on the 95\% confidence contour. Since the TS map shows the residual
signal after the known sources have been accounted for, it is possible
that small errors in the modeling of ON~325 (located just beyond the
right edge of the map), possibly resulting from inaccuracies in the
IRFs, introduce a systematic shift in the centroid of the residual
emission. However, in light of the extremely hard spectrum measured by
\Fermic, we consider the GeV emission to be associated with
1ES~1218+304. No variability is detected in the \Fermi light curve.
During their original observations, MAGIC and VERITAS observed no
evidence for variability at TeV energies, and the spectra they
produced agree well. However, during observations in January and
February 2009, just after the time period considered in this study,
VERITAS detected a flare from 1ES~1218+304, during which its flux
increased by a factor of five
\citep{REF::1ES_1218+304_VERITAS_ICRC2009}. The GeV and TeV spectra
are close to overlapping at 100\,GeV, and an extrapolation from GeV
energies agrees quite well with the TeV data points. Again, the \gray
emission evidently peaks in the 50--150\,GeV range.

\textit{\objectname{W Comae}:} 
The light curves on 10 and 28-day timescales show a decline of a
factor of 2--3 in flux between the start and end of the study period.
A relatively bright \Fermi source, the averaged spectrum below 1\,GeV
is consistent with being flat or moderately increasing, while the
higher energy band is softer, with $\Gamma=2.16\pm0.10$.  The TeV
emission has also proved to be highly variable, with VERITAS reporting
a dramatic flare, during which their spectral measurements were
derived. In light of the variability in the GeV and TeV regimes, it is
not particularly surprising that the extrapolated GeV spectrum does
not match well with the TeV data.

\textit{\objectname{3C 279}:} 
The only FSRQ detected to date at TeV energies, \objectname{3C 279} is
the strongest \Fermi source in this study. The GeV spectrum shows
clear evidence for curvature, with the peak of the emission lying
below 200\,MeV. A strong flare occurred during the period of this
study, between MJD 54780 and 54840, during which the flux increased by
a factor of $\sim7$ and then declined. We show the spectra from the
pre-flare (MJD$<54780$) and peak-flaring ($54790<$MJD$<54830$) periods
separately, along with the TeV measurements from MAGIC. The spectral
indices for the $E>1$\,GeV components are similar in the two
states. The predicted TeV flux from an extrapolation of the flaring
GeV spectrum is $\phi(>100\,\mathrm{GeV})=3.5\pm1.3\times
10^{-11}$\,\cmsc, an order of magnitude below the flux reported by
MAGIC during the February 2006. The extrapolated index for both states
is $3.6\pm0.3$, in agreement with that reported by MAGIC. In contrast,
below 1\,GeV the \Fermi spectral index changed significantly between
the two flux states, becoming harder ($\Delta\Gamma=0.52\pm0.13$)
during the flaring period, suggesting that the peak of the high energy
component increased in energy.

\textit{\objectname{PKS 1424+240}:} Prompted by an initial detection in 
the \ZFGL list by \Fermic, this object was subsequently detected at TeV
energies by VERITAS and confirmed by MAGIC. The GeV spectrum is hard
($\Gamma=1.85\pm0.05$). No redshift measurement has been made for this
object, and, therefore, we do not extrapolate the GeV spectrum into
the TeV regime. At the present time, no TeV spectrum has been
published by VERITAS or MAGIC.

\textit{\objectname{H 1426+428}:} 
\Fermi detects weak emission from the region of this HBL, with a
spectral index of $\Gamma=1.47\pm0.30$. No evidence for variability is
seen. A powerful, distant FSRQ (B3~1428+422, $z=4.72$) lies
$\approx40$ arc seconds from H~1426+428, too close to be resolved
separately by \Fermi \citep{REF::FABIAN_1428,REF::COSTAMENTE_1426}.
However, the hard GeV spectral index strongly suggests that H~1426+428
is the source of the bulk of the emission detected by \Fermic. At TeV
energies, H~1426+428 was detected during active periods by Whipple,
HEGRA, CAT and others, but no detections have been reported with the
more sensitive third-generation IACTs. Extrapolated to TeV energies,
the GeV spectrum would result in an integral flux of $\approx
0.08$\,\Icrabc, compared with $0.17$\,\Icrab ($E>350$\,GeV) measured
by Whipple and $0.08$ \& $0.03$\,\Icrab ($E>1$\,TeV) by HEGRA over
two different periods.

\textit{\objectname{PG 1553+113}:} 
Detected by H.E.S.S., PG~1553+113 is one of the softest TeV sources,
with $\Gamma_\mathrm{TeV}=4.0\pm0.6$. In contrast, in the GeV regime,
\Fermi detects a bright, hard source, with a spectral index of
$\Gamma=1.69\pm0.04$. The spectra measured by \Fermi overlaps with
those measured by H.E.S.S. and MAGIC, and are in good agreement at
around 150\,GeV.  This source therefore has a strong spectral break of
$\Delta\Gamma\approx 2.3$, which must be explained either through
absorption with the EBL or through some mechanism intrinsic to the
blazar.

Due to the almost complete lack of measurable spectral lines, the
redshift of this HBL has not been established. Several indirect
methods place it in the range $0.09\le z\le 0.78$. Therefore we
extrapolate the flux from the GeV to TeV energy ranges by assuming
both the lower and upper limit on redshift. In the case of $z=0.78$,
the EBL absorption is sufficient that the extrapolated GeV spectrum is
in good agreement with the TeV measurements, leading credence to the
hypothesis that the redshift of PG~1553+113 is significantly larger
than the lower bound of the allowable redshift range. Indeed,
PG~1553+113 might be the most distant TeV object detected to
date. This source is the subject of an independent \AFermiLAT paper
\citep{REF::FERMI_PG_1553+113}.

\textit{\objectname{Markarian 501}:} 
Many episodes of flaring have been detected from Mrk~501 with previous
generation TeV instruments. To date, however, the spectrum for a
low-level flux state has not been published by third-generation
instruments. In the GeV regime, EGRET detected emission from Mrk~501
after the initial discovery at TeV energies, and significant emission
is observed by the LAT. The GeV spectrum of Mrk~501 is well fitted
with a simple power law with $\Gamma=1.79\pm0.06$. In contrast to the
historical TeV emission, the GeV flux shows no evidence of
variability. With $\Gamma_\mathrm{ext}=1.86\pm0.05$, the extrapolated
GeV spectrum is harder than all spectral measurements made with TeV
instruments, indicating the presence of curvature in the intrinsic
spectrum of the source. TeV instruments have long detected evidence
for curvature in the spectrum of this object at 1\,TeV. The GeV
spectrum is shown with TeV measurements made during a moderate flare
(0.4\,\Icrabc, $E>150$\,GeV) and during a very high state
(1.8\,\Icrabc, $E>500$\,GeV).

\textit{\objectname{1ES 1959+650}:}
\Fermi detects emission from this object with a flat spectrum, finding
no evidence of variability over the period of this study. At TeV
energies, 1ES~1959+650 had long been detected only while flaring.
However, during a dedicated multiwavelength campaign in 2006, MAGIC
measured its spectrum in a low flux state. An extrapolation of the GeV
butterfly overlaps these MAGIC measurements indicating that the
underlying spectrum is largely compatible with a single power law over
the full \gray regime. The difference between the measured and
extrapolated spectral indices, however, is
$\Gamma_\mathrm{TeV}-\Gamma_\mathrm{ext}=0.41\pm0.19$, indicating that
there is some evidence for curvature between the two bands (at the
$2\sigma$ level).

\textit{\objectname{PKS 2005-489}:} 
A southern hemisphere HBL detected in the TeV domain by H.E.S.S., this
source is one of the softer TeV blazars with
$\Gamma_\mathrm{TeV}=4.0\pm0.4$. \Fermi detects significant, hard GeV
emission with an index of $\Gamma=1.91\pm0.09$. No evidence of
variability is seen by \Fermi on timescales of months while
H.E.S.S. observes variability only on timescales longer than a
year. The difference between the indices of the extrapolated GeV and
H.E.S.S.\ spectra is $\Delta\Gamma=1.8\pm 0.4$, indicating a clear
break at a few hundred GeV.

\textit{\objectname{PKS 2155-304}:} 
The results of a dedicated multiwavelength campaign on this object,
including GeV and TeV observations with \Fermi and H.E.S.S.\ and
simple SSC modeling, are reported by
\citet{REF::PKS2155_FERMI_MWCAMPAIGN}. Since the publication of these
results, an improved set of IRFs have become available, which correct
for an overestimate of the effective area at lower energies that
results from on-orbit ``pile-up'' of events in the detector. A
reanalysis of the \Fermi data with these IRFs results in an increase
in flux of $\approx 15\%$ with almost no change in the spectral index.
During the 5.5 month period of this study, the GeV flux of
PKS~2155-304 varied by a factor of $\approx 3$, with a maximum of
$145\times10^{-9}$\,\cmsc, $\approx 1.5$ times higher than the average
flux. The change in spectral index between the GeV and TeV
measurements can only partly be explained by EBL absorption, the
remainder presumably resulting from some process intrinsic to the
source.

\textit{\objectname{BL Lacertae}:} 
The spectral index measured by \Fermi from this LBL is relatively
soft, with $\Gamma=2.43\pm0.1$. No evidence for variability is seen
over the period of the study. The extrapolated flux is approximately
one third that measured by MAGIC during a flaring episode. Since this
flux represents an estimate of the TeV flux in the optimistic case
that there is no intrinsic curvature, we conclude that the low flux
state from this object will likely be difficult to measure at TeV
energies without a significant investment of observing time.

\textit{\objectname{1ES 2344+514}:} 
One of the fainter sources in this study, 1ES~2344+514 is detected
with a TS of only 37 and the highest energy photon collected has
$E=53$\,GeV. The \Fermi spectrum is not consistent with the
MAGIC spectrum obtained in a low state, but the poor statistics are
insufficient to make a reliable prediction of the TeV flux.

\textit{\objectname{M 87} \& \objectname{Centaurus A}:}
These nearby sources are two of the three radio galaxies thus far
detected at TeV energies. Both are classified as FR1, thought to be
the parent class of BL~Lacs. While Cen~A has not shown evidence of
variability on time scales of days or months, M~87 has undergone
several flaring episodes on time scales as short as a day. In the GeV
domain, \Fermi detects both of these sources. The flux of M~87 between
0.2 GeV and 300 GeV, is ten times lower than the flux of Cen~A and is
too faint to make strong predictions for the TeV emission. In contrast,
the detection of Cen~A with $TS=308$ yields a strongly constraining
extrapolation to the TeV domain. The \Fermi butterfly underestimates
the TeV measurements from Cen~A by a factor 10. M~87 and Cen~A are the
subject of dedicated \Fermi papers \citep{REF::FERMI_M87,
REF::FERMI_CENA}.

\subsection{TeV sources not detected by the \FermiLAT}

Of the \NObjTeVSrc\ TeV-detected AGN, only the six listed in
Table~\ref{TAB::OBJ_RES_UL} (and 3C~66B, for which we cannot produce
upper limits due to contamination by 3C~66A) have not been detected at
GeV energies in 5.5 months of data taking with \Fermic. The
differential fluxes of these objects at 200\,GeV, calculated from
measured TeV spectra (column $F_{200}$ in
Table~\ref{TAB::OBJ_DET_SPEC}), are all below the median flux,
$\mathrm{med}(F_{200})=0.171$, from the full sample. In fact, they are
six of the eight TeV objects with the smallest $F_{200}$ fluxes.

Upper limits are calculated from the GeV observations assuming two
scenarios for the spectral index, $\Gamma_\mathrm{hard}=1.5$ and
$\Gamma_\mathrm{soft}=2.0$, and are given in Table~\ref{TAB::OBJ_RES_UL}.
These limits are extrapolated into the TeV regime in the usual way,
and are presented in Figure~\ref{FIG::SPEC_LIMITS}. In general, these
two sets of extrapolated limits bracket the TeV measurements.

\subsection{GeV sources with upper limits in the TeV regime}

In addition to TeV-detected sources, \Fermi detects emission from
\NDetTeVLim\ AGN for which only upper limits exist at TeV
energies. These GeV detections can be extrapolated to predict the flux
that might be observable by TeV observatories, with the caveat that
assuming the intrinsic spectrum is described by a single power-law up
to the TeV regime can lead to overly optimistic predictions for
detection.  For most of the TeV sources, the spectra in the TeV regime
are well described by a power law with an index no harder than
$\Gamma=2$, and it is reasonable to think that this will be the case
for future detections. In this context, sources with a predicted
photon index harder than $\Gamma=2$ should exhibit an intrinsic
spectral break.

\textit{\objectname{1ES 0033+595}:} 
Observed by Whipple for 12 hours without detection, this HBL, visible
only from the northern hemisphere, has an uncertain redshift. Perlman
tentatively measured $z=0.086$ \citep[see comment
in][]{REF::FALOMO_BLLACS}, while others have argued that it is more
distant \citep[e.g.][who claim $z>0.24$]{REF::SBARUFATTI_REDSHIFTS}.
We adopt the tentative, direct measurement. The \Fermi spectrum is
consistent with a flat power law with $\Gamma=2.00\pm0.13$) and
photons detected up to 150 GeV, resulting in an extrapolated index of
$\Gamma_\mathrm{ext}=2.35\pm0.08$. The Whipple upper limit,
$\phi_\mathrm{TeV}<$0.166\,\Icrabc, is in agreement with the
prediction of 0.053\,\Icrab obtained here. To reach this flux level
with the sensitivity of VERITAS \citep[see][]{REF::VERITAS_STATUS}
would require an observation of approximately 2.5 hours.

\textit{\objectname{PKS~B0521-365}, \objectname{PKS~0829+046} \&
\objectname{3C~273}:}
The objects have an extrapolated flux less than 0.01\,\Icrabc. In the
absence of significant flaring, it will likely be difficult to detect
TeV emission from them unless a large amount of time is dedicated to their
observation.

\textit{\objectname{1ES 0647+250}:}
The \FermiLAT detects the emission of this object with a very hard
spectral index of $\Gamma=1.66\pm0.15 $ and very high maximum photon
energy of 257\,GeV. The source is visible in the northern hemisphere
and was observed by HEGRA for 4 hours. The predicted flux from
the \Fermi data is 0.0543\,\Icrabc, consistent with the HEGRA upper
limit and corresponding to 2.5 hours of observations with VERITAS.

\textit{\objectname{1ES 1028+511}, \objectname{1ES 1118+424} \&
\objectname{I Zw 187}}:
These AGN lie outside the 95\% probability contour for the origin of
the emission on the TS maps. Given that there are \NDetTeVLim\ AGN
detected in this category, the chance probability of discovering three
outside the 95\% contours is $P_\mathrm{Bin}(\ge3,17,1-0.95)=0.050$,
equivalent to a $2.0\sigma$ Gaussian event. The possibility that the
emission detected from one or more of these three regions is not
associated with the particular AGN under study cannot be
discounted. For completeness, the chance probability based on the
\NDetTotal\ AGN detected in this study is
$P_\mathrm{Bin}(\ge3,38,1-0.95)=0.30$, not unreasonable, but begging
the question as to why all such objects belong to this category 
(TeV non-detected sources in Table~\ref{TAB::OBJ_RES}).

\textit{\objectname{1ES 1440+122}:}
This \AFermiLAT source is formally the hardest detected in this study.
However, since the LAT detected only 10 photons with energy
$E>1\,GeV$, more observations are required before firm statements can
be made about the spectral index. With a redshift of $z=0.162$, the
extrapolated index between 200 GeV and 1 TeV is 1.68, which is harder
than any TeV source yet detected. The predicted flux of 0.231\,\Icrab
is likely overestimated by a large factor and a turnover, not
accounted for by EBL absorption only, is required to explain the
discrepancy between the predicted flux and the H.E.S.S.\ upper limits
of 0.03\,\Icrabc.


\section{Evolution of detected GeV--TeV spectra with redshift}
\label{SEC::EBL}

In the LBAS study, the dependence of GeV spectral index of the FSRQ
and BL~Lac populations on redshift was presented \citep[Figure 11
of][]{REF::LBAS}. Although it was found that the two populations have
different spectral properties, no significant relation between the
gamma-ray photon index and redshift was found within each source
class. The GeV--TeV sources provide a population in which the effects
of spectral evolution with redshift can be studied across a much wider
energy range than LBAS (although, admittedly in a much smaller
redshift range). The presence of a redshift-dependent spectral break
in these sources could be indicative of the effects of absorption on
the EBL, and provide experimental evidence for this absorption in a
manner independent of any specific EBL-density
model. Appendix~\ref{APP::DELTA_GAMMA} discusses the relationship
between EBL absorption and the redshift dependent change in spectral
index between the GeV and TeV ranges,
\begin{equation}
\Delta\Gamma = \Gamma_\mathrm{TeV} - \Gamma_\mathrm{GeV}
\ge \Gamma_\mathrm{TeV} - \Gamma_\mathrm{Int} = \delta(z,E^*)
\approx \left.\frac{\mathrm{d}\tau(E,z)}{\mathrm{d}\log E}\right|_{E=E^*},
\end{equation}
i.e.\ in the presence of EBL absorption, and assuming that the
intrinsic spectra do not get harder with energy, measured spectral
breaks must lie in a region of the $(\Delta\Gamma,z)$ plane defined by
$\Delta\Gamma \ge \delta(z,E^*)$, with $\delta(z,E^*)$ defined by the
EBL optical depth. 

Figure~\ref{FIG::DELTAG_VS_Z} depicts the difference in the measured
TeV and GeV spectral indices, $\Delta\Gamma$, as a function of the
redshift, $z$, for 15 of the \NDetTeVSrc\ extragalactic GeV--TeV
sources\footnote{We exclude 3C~66A and PG~1553+113 as $z$ is not
known, RGB J0710+59, S5~0716+714 and PKS~1424+240 as
$\Gamma_\mathrm{TeV}$ has not been published and 3C~279 as the GeV and
TeV states are badly mismatched.}. It is evident that the difference
between the GeV and TeV spectral indices increases with redshift. At
low redshifts the radio galaxies M~87 and Cen~A have
$\Delta\Gamma\approx0$, as do the near-by BL~Lacs. At redshifts
greater than $0.1$, all of the BL~Lacs are consistent with
$\Delta\Gamma\ge1.5$.

Pearson's correlation factor, $r$, is widely used to quantify the
correlation between two variables. However, this correlation factor
does not take measurement errors into account. Thus, to evaluate the
significance of the correlation, a series of Gaussian random variables
centered on the values of the measured spectral changes,
$\Delta\Gamma_i$, and with width equal to their measurement errors are
used to generate multiple simulated data sets. The full width at
half maximum of the distribution obtained for $r$ gives an estimate of
the error\footnote{The analytic expression for the error gives
$r=0.76\pm0.16$, consistent with the Monte Carlo method.}. The value
obtained is $r=0.76\pm0.14$, which indicates a clear correlation. To
check the robustness of this result, the Kendall rank, defined by
\begin{displaymath}
\tau_K=\frac{2S}{N\cdot(N-1)},
\end{displaymath}
where N is the size of the data set ($N=15$), is been calculated. For
each pair of points from the dataset $(\Delta\Gamma,z)$, those in the
same order are assigned a value of $+1$, and the others assigned
$-1$. The sum, $S$, of these $N\cdot(N-1)/2$ combinations is then
constructed \citep[][and references therein]{REF::KENDAL}, giving a
value of $\tau_K$, ranging from $-1\le\tau_K\le+1$, indicating the
degree of correlation. The error on $\tau_K$ is calculated in the same
way as for $r$. The same conclusion is established with this test,
$\tau_K = 0.68\pm 0.15$.

The effects of systematics present in Figure~\ref{FIG::DELTAG_VS_Z}
must be considered when determining the validity of concluding that it
reflects the effects of EBL absorption. Such systematics are difficult
to evaluate in a quantitative manner, appendix~\ref{APP::EBL_SYST}
discusses possible contributions in more detail. In light of these
difficulties it cannot be claimed with 100\% certainty that the
observed deficit of sources at $\Delta\Gamma\approx0$ for large
redshift is a real effect. We anticipate that the TeV AGN not detected
by the LAT in this study will ultimately be detected and that this
figure will then be limited only by the selection bias at TeV
energies. Finally, EBL absorption is not the only effect that could
cause a redshift dependent spectral break, evolution of the intrinsic
spectra with redshift also cannot be excluded, although this may be
difficult to reconcile with the LBAS study, which did not detect any
evolution in the larger LAT BL~Lac sample.


\section{Conclusions}

In 5.5 months of observation the \FermiLAT has detected GeV emission
from \NDetTeVSrc\ TeV-detected AGN, and from \NDetTeVLim\ AGN
previously observed by TeV groups and for which upper limits have been
published at TeV energies. Whereas EGRET detected only a small number
of such TeV sources, and detected those only with integration times of
months and years, \Fermi has detected the majority of the TeV blazars
in its first few months of operation. \Fermi observations will help
TeV observatories optimize the limited observation time available each
year, and will be useful in evaluating possible targets for
observation. \Fermi has sufficient sensitivity to participate
meaningfully in simultaneous multi-wavelength campaigns and to measure
spectra and light curves from the brighter blazars with a resolution
of days to months. In future campaigns, it can be expected that the
high-energy emission will be as well covered by
\Fermi and the TeV observatories as the low-energy peak is by
instruments in the radio to X-ray bands.

Many of the TeV sources exhibit an increasing spectrum ($\Gamma<2$) in
the GeV range confirming the presence of a high energy peak in $\nu
F_\nu$ representation. This is the first large-scale characterization
of the full \gray emission component for this class of energetic
AGN. More detailed modeling of these \gray sources, which is beyond
the scope of this paper, will become possible as more data are acquired
by \Fermi and the flight-calibrated IRFs become available (extending
the effective energy below 100\,MeV). The MAGIC-II and H.E.S.S.\ II
instruments will increase the range over which the sensitives of TeV
instruments overlap with \Fermi at lower energies, producing better
coverage at energies between 50 and 200\,GeV where a number of TeV
sources seem to have turnovers in their measured spectra.

The intrinsic spectrum for some of the TeV sources can be well
described by a single power-law across the energy range spanned by the
\FermiLAT and the TeV observatories, with any breaks in the measured
\gray spectra between the two regimes being consistent with the effects of
absorption with a model of minimal EBL density. For other objects,
however, a softening of the intrinsic spectrum is required to match
the TeV measurements. This could be due to softening intrinsic to the
IC component -- itself reflecting curvature in the relativistic
electron or seed photon distributions.

Based on an extrapolation of the GeV spectra measured by \Fermi to TeV
energies, a number of previously observed TeV candidates are good
candidates for re-observation with TeV instruments: 1ES~0033+595 and
1ES 0647+250 are two such targets.

Redshift-dependent evolution is detected in the spectra of objects
detected at GeV and TeV energies. The most reasonable explanation for
this is absorption on the EBL, and as such, it would represent the
first model-independent evidence for absorption of \grays on the
EBL. Future observations with \Fermi and TeV instruments have the
potential to probe $\tau(E,z)$ in a more quantitative manner.


\acknowledgments

The \AFermiLAT Collaboration acknowledges the generous support of a
number of agencies and institutes that have supported the \AFermiLAT
Collaboration. These include the National Aeronautics and Space
Administration and the Department of Energy in the United States, the
Commissariat \`a l'Energie Atomique and the Centre National de la
Recherche Scientifique / Institut National de Physique Nucl\'eaire et
de Physique des Particules in France, the Agenzia Spaziale Italiana
and the Istituto Nazionale di Fisica Nucleare in Italy, the Ministry
of Education, Culture, Sports, Science and Technology (MEXT), High
Energy Accelerator Research Organization (KEK) and Japan Aerospace
Exploration Agency (JAXA) in Japan, and the K.\ A.\ Wallenberg
Foundation, the Swedish Research Council and the Swedish National
Space Board in Sweden.

Additional support for science analysis during the operations phase is
gratefully acknowledged from the Istituto Nazionale di Astrofisica in
Italy.

This research has made use of NASA's Astrophysics Data System
Bibliographic Services, the NASA/IPAC Extragalactic Database, operated
by JPL, Caltech, under contract from NASA, and the SIMBAD database,
operated at CDS, Strasbourg, France.



{\it Facilities:} \facility{Fermi LAT}

\appendix


\section{Relationship between GeV--TeV break index and EBL}
\label{APP::DELTA_GAMMA}

Writing the measured TeV spectrum as $F_\mathrm{TeV}(E)$, the
unabsorbed (intrinsic) spectrum as $F_\mathrm{Int}(E)$ and the
redshift-dependent optical depth due to EBL absorption as $\tau(E,z)$,
the effects of the EBL on the measured spectrum is given by,
\begin{equation}\label{EQN::ABSORPTION}
F_\mathrm{TeV}(E) = e^{-\tau(E,z)} F_\mathrm{Int}(E).
\end{equation}
If both the measured and intrinsic spectra can be approximated as
power laws over the range of the TeV observations, $F_\mathrm{TeV}(E)
\approx C_\mathrm{TeV} (E/E_0)^{-\Gamma_\mathrm{TeV}}$ and
$F_\mathrm{Int}(E) \approx C_\mathrm{Int}
(E/E_0)^{-\Gamma_\mathrm{Int}}$, then the EBL effects must also be
given by a power law, say $e^{-\tau(E,z)} \approx C_\tau
(E/E_0)^{-\delta(z,E^*)}$.  The power-law index of the EBL absorption,
$\delta(z,E^*)$, is a function of the redshift and the energy range
over which the TeV observations are made (denoted for convenience as a
dependence on some energy, $E^*$, at which the measured TeV data most
constrains the fitted spectrum). Equating the two expressions for the
absorption to first order in $x=\log(E/E^*)$ gives
$\delta(z,E^*)\approx\mathrm{d}\tau(x,z)/\mathrm{d}x|_{x=0}$
\citep[see][for further discussion of linear and polynomial expansions of
the EBL optical depth]{REF::EBL_VASSILIEV,
REF::EBL_STECKER_SCULLY_LINEAR}. Equation~\ref{EQN::ABSORPTION} relates
$\delta(z,E^*)$ to the intrinsic and measured spectral indices,
\begin{displaymath}
C_\mathrm{TeV} (E/E_0)^{-\Gamma_\mathrm{TeV}} =
C_\tau C_\mathrm{Int} (E/E_0)^{-\delta(z,E^*) - 
\Gamma_\mathrm{Int}},\ \mathrm{or}
\end{displaymath}
\begin{equation}\label{EQN::DELTA}
\delta(z,E^*) = 
\Gamma_\mathrm{TeV} - \Gamma_\mathrm{Int} \sim
\Gamma_\mathrm{TeV} - \Gamma_\mathrm{GeV},	
\end{equation}
where it is further assumed that the measured GeV spectral index can
be used as a proxy for the intrinsic index in the TeV regime. This is
equivalent to assuming that there is no curvature in the intrinsic
spectrum between the GeV and TeV energy regimes and that absorption on
the EBL does not affect the spectrum in the LAT energy range, which is
true for all reasonable EBL models for sources with $z<0.5$. For
GeV--TeV detected sources, this suggests that the variable
$\Delta\Gamma=\Gamma_\mathrm{TeV}-\Gamma_\mathrm{GeV}$ probes the
effects of EBL absorption in a manner independent of any specific
model of EBL density.

For real sources there is curvature in the intrinsic spectrum between
the GeV and TeV regimes, and the measured $\Gamma_\mathrm{GeV}$ is not
a perfect estimator for $\Gamma_\mathrm{Int}$ in the TeV regime. In
general it has been found that the curvature in the differential
spectra is concave, and the intrinsic spectrum at TeV energies is
expected to be softer than the measured GeV spectrum. Therefore it
is expected that for real sources:
\begin{equation}
\Delta\Gamma = \Gamma_\mathrm{TeV} - \Gamma_\mathrm{GeV}
\ge \Gamma_\mathrm{TeV} - \Gamma_\mathrm{Int} = \delta(z,E^*),
\end{equation}
so that sources would occupy a region of the space of
$(\Delta\Gamma,z)$ above some curve defined by the EBL.


\section{Discussion of systematics in evolution of spectra with redshift}
\label{APP::EBL_SYST}

A number of components contribute to the systematics in
Figure~\ref{FIG::DELTAG_VS_Z} and are addressed below: (i) systematic
errors on the points themselves and (ii) the effects of the criteria
used to select targets for the study.

The systematic errors on the measurements of $\Gamma_\mathrm{GeV}$ for
the individual sources are presented in Table~\ref{TAB::OBJ_RES}, and
are generally smaller than statistical errors, with the largest being
$\approx0.25$. Similarly, the TeV groups estimate and report
systematic errors on $\Gamma_\mathrm{TeV}$ for each of the detected
TeV sources. See, for example, \citet{REF::HESS_CRAB} for a discussion
of systematic error estimation with H.E.S.S. In general, the GeV and
TeV systematic errors are too small to explain the trend in
Figure~\ref{FIG::DELTAG_VS_Z}, in which $\Delta\Gamma$ changes by
$>2.0$ over the range of redshift in question.

The sources in this study are subject to a two-stage selection
process, which may lead to regions of the phase-space of
$(\Delta\Gamma,z)$ being inaccessible due to limitations imposed by of
the sensitivities of \Fermi and the TeV instruments. This could, in
turn, lead to a false correlation being evident in the data -- for
example, see Figure~7 from the LBAS study \citep{REF::LBAS} for a
correlation which might incorrectly be claimed based on a
sensitivity-limited sample. Targets were originally selected by TeV
astronomers for observation with TeV instruments. Given the
sensitivities of those instruments and the amount of time dedicated to
each target, some fraction were detected, leading to \NObjTeVSrc\
sources used in this study. The \Fermi sensitivity further restricts
the sample to the 15 GeV--TeV sources displayed in
Figure~\ref{FIG::DELTAG_VS_Z}. Given the complexity (and randomness)
of the selection process, it is almost impossible to quantify where
its ``sensitivity'' limits are in the space of $(\Delta\Gamma,z)$. In
the null hypothesis, that there is no EBL effect present in the
results of Figure~\ref{FIG::DELTAG_VS_Z}, we attempt to evaluate in a
qualitative manner whether the lack of sources at
$\Delta\Gamma\approx0$ for larger redshifts could arise from a
selection effect. The primary selection of targets is based on
detection at TeV energies. Sources in this region of the plot would
have harder TeV spectra than those actually detected at the larger
redshifts. Since, it is unlikely that TeV astronomers are deliberately
biasing the sample toward softer distant AGN, the major effect
producing this bias would have to result from the sensitivity limit of
the instrument. The instrumental sensitivity directly limits the space
of $(\phi,\Gamma)$, requiring $\phi>\phi_\mathrm{Lim}(\Gamma)$ for
detection (as in the LBAS figure cited above), and these limits
transfer to the space of $(\Delta\Gamma,z)$ only through convolution
with the source function
$f(\phi,\Gamma;z)=\mathrm{d}^2P/\mathrm{d}\phi\,\mathrm{d}\Gamma$.
Therefore, a sharp cut-off in this space should not be expected,
rather a slower decrease in source counts into the ``forbidden''
region. In the no-EBL hypothesis, and further assuming that evolution
in the source function, the primary consideration is whether the
decrease in measured flux with distance coupled with the sensitivity
limit would lead to an evolution in the population of detectable
sources with redshift. There is very little published material
addressing the flux sensitivity of current TeV instruments as a
function of spectral index, however it is reasonable to presume that
TeV instruments are more sensitive to sources with harder spectra,
than to those with softer: they have better background rejection and
an improved PSF at higher energies. Indeed, TeV instruments often use
``hard cuts'' to improve the sensitivity for hard sources. In this
case, it would be expected that TeV instruments would preferentially
detect \textit{harder} sources (with lower fluxes) at larger
redshifts, whereas this is exactly the opposite of what is actually
observed, i.e.\ that the majority of the distant AGN are softer -- no
hard, weak TeV AGN has been detected to date.





\bibliography{references}

\begin{thebibliography}{96}
\expandafter\ifx\csname natexlab\endcsname\relax\def\natexlab#1{#1}\fi

\bibitem[{{Abdo} {et~al.}(2009{\natexlab{a}})}]{REF::LBAS}
{Abdo}, A.~A., {et~al.} 2009{\natexlab{a}}, \apj, 700, 597

\bibitem[{{Abdo} {et~al.}(2009{\natexlab{b}})}]{REF::0FGL}
---. 2009{\natexlab{b}}, \apjs, 183, 46

\bibitem[{{Abdo} {et~al.}(2009{\natexlab{c}})}]{REF::FERMI_M87}
---. 2009{\natexlab{c}}, \apj, submitted

\bibitem[{{Abdo} {et~al.}(2009{\natexlab{d}})}]{REF::FERMI_CRAB}
---. 2009{\natexlab{d}}, \apj, submitted

\bibitem[{{Abdo} {et~al.}(2009{\natexlab{e}})}]{REF::FERMI_EXTRAGALACTIC}
---. 2009{\natexlab{e}}, in preparation

\bibitem[{{Abdo} {et~al.}(2009{\natexlab{f}})}]{REF::FERMI_CENA}
---. 2009{\natexlab{f}}, in preparation

\bibitem[{{Abdo} {et~al.}(2009{\natexlab{g}})}]{REF::FERMI_PG_1553+113}
---. 2009{\natexlab{g}}, \apj, submitted

\bibitem[{{Acciari} {et~al.}(2008){Acciari}, {Aliu}, {Beilicke}, {Benbow},
  {B{\"o}ttcher}, {Bradbury}, {Buckley}, {Bugaev}, {Butt}, {Celik}, {Cesarini},
  {Ciupik}, {Chow}, {Cogan}, {Colin}, {Cui}, {Daniel}, {Ergin}, {Falcone},
  {Fegan}, {Finley}, {Finnegan}, {Fortin}, {Fortson}, {Furniss}, {Gall},
  {Gillanders}, {Grube}, {Guenette}, {Gyuk}, {Hanna}, {Hays}, {Holder},
  {Horan}, {Hui}, {Humensky}, {Imran}, {Kaaret}, {Karlsson}, {Kertzman},
  {Kieda}, {Konopelko}, {Krawczynski}, {Krennrich}, {Lang}, {LeBohec}, {Lee},
  {Maier}, {McCann}, {McCutcheon}, {Moriarty}, {Mukherjee}, {Nagai}, {Niemiec},
  {Ong}, {Pandel}, {Perkins}, {Petry}, {Pohl}, {Quinn}, {Ragan}, {Reyes},
  {Reynolds}, {Roache}, {Rose}, {Schroedter}, {Sembroski}, {Smith}, {Steele},
  {Swordy}, {Toner}, {Vassiliev}, {Wagner}, {Wakely}, {Ward}, {Weekes},
  {Weinstein}, {White}, {Williams}, {Wissel}, {Wood}, \&
  {Zitzer}}]{REF::TEV_W_COMAE}
{Acciari}, V.~A., {et~al.} 2008, \apjl, 684, L73

\bibitem[{{Acciari} {et~al.}(2009{\natexlab{a}}){Acciari}, {Aliu}, {Arlen},
  {Bautista}, {Beilicke}, {Benbow}, {B{\"o}ttcher}, {Bradbury}, {Buckley},
  {Bugaev}, {Butt}, {Byrum}, {Cannon}, {Celik}, {Cesarini}, {Chow}, {Ciupik},
  {Cogan}, {Colin}, {Cui}, {Dickherber}, {Duke}, {Ergin}, {Falcone}, {Fegan},
  {Finley}, {Finnegan}, {Fortin}, {Fortson}, {Furniss}, {Gall}, {Gibbs},
  {Gillanders}, {Grube}, {Guenette}, {Gyuk}, {Hanna}, {Hays}, {Holder},
  {Horan}, {Hui}, {Humensky}, {Imran}, {Kaaret}, {Karlsson}, {Kertzman},
  {Kieda}, {Kildea}, {Konopelko}, {Krawczynski}, {Krennrich}, {Lang},
  {LeBohec}, {Maier}, {McCann}, {McCutcheon}, {Millis}, {Moriarty},
  {Mukherjee}, {Nagai}, {Ong}, {Otte}, {Pandel}, {Perkins}, {Petry}, {Pohl},
  {Quinn}, {Ragan}, {Reyes}, {Reynolds}, {Roache}, {Rose}, {Schroedter},
  {Sembroski}, {Smith}, {Steele}, {Swordy}, {Theiling}, {Toner}, {Valcarcel},
  {Varlotta}, {Vassiliev}, {Wagner}, {Wakely}, {Ward}, {Weekes}, {Weinstein},
  {White}, {Williams}, {Wissel}, {Wood}, \& {Zitzer}}]{REF::TEV_1ES_0806+524}
---. 2009{\natexlab{a}}, \apjl, 690, L126

\bibitem[{{Acciari} {et~al.}(2009{\natexlab{b}}){Acciari}, {Aliu}, {Arlen},
  {Beilicke}, {Benbow}, {B{\"o}ttcher}, {Bradbury}, {Buckley}, {Bugaev},
  {Butt}, {Byrum}, {Cannon}, {Celik}, {Cesarini}, {Chow}, {Ciupik}, {Cogan},
  {Cui}, {Daniel}, {Dickherber}, {Ergin}, {Falcone}, {Fegan}, {Finley},
  {Fortin}, {Fortson}, {Furniss}, {Gall}, {Gibbs}, {Gillanders}, {Godambe},
  {Grube}, {Guenette}, {Gyuk}, {Hanna}, {Hays}, {Holder}, {Horan}, {Hui},
  {Humensky}, {Imran}, {Kaaret}, {Karlsson}, {Kertzman}, {Kieda}, {Kildea},
  {Konopelko}, {Krawczynski}, {Krennrich}, {Lang}, {LeBohec}, {Maier},
  {McCann}, {McCutcheon}, {Millis}, {Moriarty}, {Mukherjee}, {Nagai}, {Ong},
  {Otte}, {Pandel}, {Perkins}, {Petry}, {Pizlo}, {Pohl}, {Quinn}, {Ragan},
  {Reyes}, {Reynolds}, {Roache}, {Rose}, {Schroedter}, {Sembroski}, {Smith},
  {Steele}, {Swordy}, {Theiling}, {Toner}, {Varlotta}, {Vassiliev}, {Wagner},
  {Wakely}, {Ward}, {Weekes}, {Weinstein}, {Williams}, {Wissel}, {Wood}, \&
  {Zitzer}}]{REF::TEV_3C66A_VERITAS}
---. 2009{\natexlab{b}}, \apjl, 693, L104

\bibitem[{{Acciari} {et~al.}(2009{\natexlab{c}}){Acciari}, {Aliu}, {Arlen},
  {Beilicke}, {Benbow}, {Bradbury}, {Buckley}, {Bugaev}, {Butt}, {Byrum},
  {Celik}, {Cesarini}, {Ciupik}, {Chow}, {Cogan}, {Colin}, {Cui}, {Daniel},
  {Ergin}, {Falcone}, {Fegan}, {Finley}, {Fortin}, {Fortson}, {Furniss},
  {Gillanders}, {Grube}, {Guenette}, {Gyuk}, {Hanna}, {Hays}, {Holder},
  {Horan}, {Hui}, {Humensky}, {Imran}, {Kaaret}, {Karlsson}, {Kertzman},
  {Kieda}, {Kildea}, {Konopelko}, {Krawczynski}, {Krennrich}, {Lang},
  {LeBohec}, {Maier}, {McCann}, {McCutcheon}, {Moriarty}, {Mukherjee}, {Nagai},
  {Niemiec}, {Ong}, {Pandel}, {Perkins}, {Pohl}, {Quinn}, {Ragan}, {Reyes},
  {Reynolds}, {Rose}, {Schroedter}, {Sembroski}, {Smith}, {Steele}, {Swordy},
  {Toner}, {Valcarcel}, {Vassiliev}, {Wagner}, {Wakely}, {Ward}, {Weekes},
  {Weinstein}, {White}, {Williams}, {Wissel}, {Wood}, \&
  {Zitzer}}]{REF::TEV_1ES_1218+304_VERITAS}
---. 2009{\natexlab{c}}, \apj, 695, 1370

\bibitem[{{Aharonian} {et~al.}(2003{\natexlab{a}}){Aharonian}, {Akhperjanian},
  {Beilicke}, {Bernl{\"o}hr}, {B{\"o}rst}, {Bojahr}, {Bolz}, {Coarasa},
  {Contreras}, {Cortina}, {Denninghoff}, {Fonseca}, {Girma}, {G{\"o}tting},
  {Heinzelmann}, {Hermann}, {Heusler}, {Hofmann}, {Horns}, {Jung}, {Kankanyan},
  {Kestel}, {Kohnle}, {Konopelko}, {Kornmeyer}, {Kranich}, {Lampeitl}, {Lopez},
  {Lorenz}, {Lucarelli}, {Mang}, {Meyer}, {Mirzoyan}, {Moralejo},
  {Ona-Wilhelmi}, {Panter}, {Plyasheshnikov}, {P{\"u}hlhofer}, {de los Reyes},
  {Rhode}, {Ripken}, {Rowell}, {Sahakian}, {Samorski}, {Schilling}, {Siems},
  {Sobzynska}, {Stamm}, {Tluczykont}, {Vitale}, {V{\"o}lk}, {Wiedner}, \&
  {Wittek}}]{REF::TEV_M_87}
{Aharonian}, F., {et~al.} 2003{\natexlab{a}}, \aap, 403, L1

\bibitem[{{Aharonian} {et~al.}(2003{\natexlab{b}}){Aharonian}, {Akhperjanian},
  {Beilicke}, {Bernl{\"o}hr}, {B{\"o}rst}, {Bojahr}, {Bolz}, {Coarasa},
  {Contreras}, {Cortina}, {Costamante}, {Denninghoff}, {Fonseca}, {Girma},
  {G{\"o}tting}, {Heinzelmann}, {Hermann}, {Heusler}, {Hofmann}, {Horns},
  {Jung}, {Kankanyan}, {Kestel}, {Kohnle}, {Konopelko}, {Kornmeyer}, {Kranich},
  {Lampeitl}, {Lopez}, {Lorenz}, {Lucarelli}, {Mang}, {Mazine}, {Meyer},
  {Mirzoyan}, {Moralejo}, {Ona-Wilhelmi}, {Panter}, {Plyasheshnikov}, {Prahl},
  {P{\"u}hlhofer}, {de los Reyes}, {Rhode}, {Ripken}, {Rowell}, {Sahakian},
  {Samorski}, {Schilling}, {Siems}, {Sobzynska}, {Stamm}, {Tluczykont},
  {Vitale}, {V{\"o}lk}, {Wiedner}, \& {Wittek}}]{REF::TEV_H_1426+428_HEGRA}
---. 2003{\natexlab{b}}, \aap, 403, 523

\bibitem[{{Aharonian} {et~al.}(2004){Aharonian}, {Akhperjanian}, {Beilicke},
  {Bernl{\"o}hr}, {B{\"o}rst}, {Bojahr}, {Bolz}, {Coarasa}, {Contreras},
  {Cortina}, {Denninghoff}, {Fonseca}, {Girma}, {G{\"o}tting}, {Heinzelmann},
  {Hermann}, {Heusler}, {Hofmann}, {Horns}, {Jung}, {Kankanyan}, {Kestel},
  {Konopelko}, {Kornmeyer}, {Kranich}, {Lampeitl}, {Lopez}, {Lorenz},
  {Lucarelli}, {Mang}, {Mazin}, {Meyer}, {Mirzoyan}, {Moralejo},
  {Ona-Wilhelmi}, {Panter}, {Plyasheshnikov}, {P{\"u}hlhofer}, {de los Reyes},
  {Rhode}, {Ripken}, {Rowell}, {Sahakian}, {Samorski}, {Schilling}, {Siems},
  {Sobzynska}, {Stamm}, {Tluczykont}, {Vitale}, {V{\"o}lk}, {Wiedner}, \&
  {Wittek}}]{REF::UL_HEGRA_TLUCZYKONT}
---. 2004, \aap, 421, 529

\bibitem[{{Aharonian} {et~al.}(2005{\natexlab{a}}){Aharonian}, {Akhperjanian},
  {Aye}, {Bazer-Bachi}, {Beilicke}, {Benbow}, {Berge}, {Berghaus},
  {Bernl{\"o}hr}, {Boisson}, {Bolz}, {Braun}, {Breitling}, {Brown}, {Bussons
  Gordo}, {Chadwick}, {Chounet}, {Cornils}, {Costamante}, {Degrange},
  {Djannati-Ata{\"i}}, {O'C.~Drury}, {Dubus}, {Emmanoulopoulos}, {Espigat},
  {Feinstein}, {Fleury}, {Fontaine}, {Fuchs}, {Funk}, {Gallant}, {Giebels},
  {Gillessen}, {Glicenstein}, {Goret}, {Hadjichristidis}, {Hauser},
  {Heinzelmann}, {Henri}, {Hermann}, {Hinton}, {Hofmann}, {Holleran}, {Horns},
  {de Jager}, {Kh{\'e}lifi}, {Komin}, {Konopelko}, {Latham}, {Le Gallou},
  {Lemi{\`e}re}, {Lemoine-Goumard}, {Leroy}, {Lohse}, {Martineau-Huynh},
  {Marcowith}, {Masterson}, {McComb}, {de Naurois}, {Nolan}, {Noutsos},
  {Orford}, {Osborne}, {Ouchrif}, {Panter}, {Pelletier}, {Pita},
  {P{\"u}hlhofer}, {Punch}, {Raubenheimer}, {Raue}, {Raux}, {Rayner},
  {Redondo}, {Reimer}, {Reimer}, {Ripken}, {Rob}, {Rolland}, {Rowell},
  {Sahakian}, {Saug{\'e}}, {Schlenker}, {Schlickeiser}, {Schuster}, {Schwanke},
  {Siewert}, {Sol}, {Steenkamp}, {Stegmann}, {Tavernet}, {Terrier},
  {Th{\'e}oret}, {Tluczykont}, {Vasileiadis}, {Venter}, {Vincent}, {V{\"o}lk},
  \& {Wagner}}]{REF::TEV_PKS_2005-489}
---. 2005{\natexlab{a}}, \aap, 436, L17

\bibitem[{{Aharonian} {et~al.}(2005{\natexlab{b}}){Aharonian}, {Akhperjanian},
  {Aye}, {Bazer-Bachi}, {Beilicke}, {Benbow}, {Berge}, {Berghaus},
  {Bernl{\"o}hr}, {Bolz}, {Boisson}, {Borgmeier}, {Breitling}, {Brown},
  {Bussons Gordo}, {Chadwick}, {Chitnis}, {Chounet}, {Cornils}, {Costamante},
  {Degrange}, {Djannati-Ata{\"i}}, {Drury}, {Ergin}, {Espigat}, {Feinstein},
  {Fleury}, {Fontaine}, {Funk}, {Gallant}, {Giebels}, {Gillessen}, {Goret},
  {Guy}, {Hadjichristidis}, {Hauser}, {Heinzelmann}, {Henri}, {Hermann},
  {Hinton}, {Hofmann}, {Holleran}, {Horns}, {de Jager}, {Jung I.},
  {Kh{\'e}lifi}, {Komin}, {Konopelko}, {Latham}, {Le Gallou}, {Lemoine},
  {Lemi{\`e}re}, {Leroy}, {Lohse}, {Marcowith}, {Masterson}, {McComb}, {de
  Naurois}, {Nolan}, {Noutsos}, {Orford}, {Osborne}, {Ouchrif}, {Panter},
  {Pelletier}, {Pita}, {Pohl}, {P{\"u}hlhofer}, {Punch}, {Raubenheimer},
  {Raue}, {Raux}, {Rayner}, {Redondo}, {Reimer}, {Reimer}, {Ripken}, {Rivoal},
  {Rob}, {Rolland}, {Rowell}, {Sahakian}, {Saug{\'e}}, {Schlenker},
  {Schlickeiser}, {Schuster}, {Schwanke}, {Siewert}, {Sol}, {Steenkamp},
  {Stegmann}, {Tavernet}, {Th{\'e}oret}, {Tluczykont}, {van der Walt},
  {Vasileiadis}, {Vincent}, {Visser}, {V{\"o}lk}, \&
  {Wagner}}]{REF::TEV_PKS_2155-304_HESS_2003}
---. 2005{\natexlab{b}}, \aap, 430, 865

\bibitem[{{Aharonian} {et~al.}(2005{\natexlab{c}}){Aharonian}, {Akhperjanian},
  {Bazer-Bachi}, {Beilicke}, {Benbow}, {Berge}, {Bernl{\"o}hr}, {Boisson},
  {Bolz}, {Borrel}, {Braun}, {Breitling}, {Brown}, {Chadwick}, {Chounet},
  {Cornils}, {Costamante}, {Degrange}, {Dickinson}, {Djannati-Ata{\"i}},
  {O'C.~Drury}, {Dubus}, {Emmanoulopoulos}, {Espigat}, {Feinstein}, {Fontaine},
  {Fuchs}, {Funk}, {Gallant}, {Giebels}, {Gillessen}, {Glicenstein}, {Goret},
  {Hadjichristidis}, {Hauser}, {Heinzelmann}, {Henri}, {Hermann}, {Hinton},
  {Hofmann}, {Holleran}, {Horns}, {Jacholkowska}, {de Jager}, {Kh{\'e}lifi},
  {Komin}, {Konopelko}, {Latham}, {Le Gallou}, {Lemi{\`e}re},
  {Lemoine-Goumard}, {Leroy}, {Lohse}, {Martin}, {Martineau-Huynh},
  {Marcowith}, {Masterson}, {McComb}, {de Naurois}, {Nolan}, {Noutsos},
  {Orford}, {Osborne}, {Ouchrif}, {Panter}, {Pelletier}, {Pita},
  {P{\"u}hlhofer}, {Punch}, {Raubenheimer}, {Raue}, {Raux}, {Rayner}, {Reimer},
  {Reimer}, {Ripken}, {Rob}, {Rolland}, {Rowell}, {Sahakian}, {Saug{\'e}},
  {Schlenker}, {Schlickeiser}, {Schuster}, {Schwanke}, {Siewert}, {Sol},
  {Spangler}, {Steenkamp}, {Stegmann}, {Tavernet}, {Terrier}, {Th{\'e}oret},
  {Tluczykont}, {Vasileiadis}, {Venter}, {Vincent}, {V{\"o}lk}, \&
  {Wagner}}]{REF::UL_HESS_BENBOW05}
---. 2005{\natexlab{c}}, \aap, 441, 465

\bibitem[{{Aharonian} {et~al.}(2006{\natexlab{a}}){Aharonian}, {Akhperjanian},
  {Bazer-Bachi}, {Beilicke}, {Benbow}, {Berge}, {Bernl{\"o}hr}, {Boisson},
  {Bolz}, {Borrel}, {Braun}, {Breitling}, {Brown}, {Chadwick}, {Chounet},
  {Cornils}, {Costamante}, {Degrange}, {Dickinson}, {Djannati-Ata{\"i}},
  {Drury}, {Dubus}, {Emmanoulopoulos}, {Espigat}, {Feinstein}, {Fontaine},
  {Fuchs}, {Funk}, {Gallant}, {Giebels}, {Gillessen}, {Glicenstein}, {Goret},
  {Hadjichristidis}, {Hauser}, {Hauser}, {Heinzelmann}, {Henri}, {Hermann},
  {Hinton}, {Hofmann}, {Holleran}, {Horns}, {Jacholkowska}, {de Jager},
  {Kh{\'e}lifi}, {Klages}, {Komin}, {Konopelko}, {Latham}, {Le Gallou},
  {Lemi{\`e}re}, {Lemoine-Goumard}, {Leroy}, {Lohse}, {Martin},
  {Martineau-Huynh}, {Marcowith}, {Masterson}, {McComb}, {de Naurois}, {Nolan},
  {Noutsos}, {Orford}, {Osborne}, {Ouchrif}, {Panter}, {Pelletier}, {Pita},
  {P{\"u}hlhofer}, {Punch}, {Raubenheimer}, {Raue}, {Raux}, {Rayner}, {Reimer},
  {Reimer}, {Ripken}, {Rob}, {Rolland}, {Rowell}, {Sahakian}, {Saug{\'e}},
  {Schlenker}, {Schlickeiser}, {Schuster}, {Schwanke}, {Siewert}, {Sol},
  {Spangler}, {Steenkamp}, {Stegmann}, {Tavernet}, {Terrier}, {Th{\'e}oret},
  {Tluczykont}, {van Eldik}, {Vasileiadis}, {Venter}, {Vincent}, {V{\"o}lk}, \&
  {Wagner}}]{REF::TEV_1ES_1101-232}
---. 2006{\natexlab{a}}, \nat, 440, 1018

\bibitem[{{Aharonian} {et~al.}(2006{\natexlab{b}}){Aharonian}, {Akhperjanian},
  {Bazer-Bachi}, {Beilicke}, {Benbow}, {Berge}, {Bernl{\"o}hr}, {Boisson},
  {Bolz}, {Borrel}, {Braun}, {Breitling}, {Brown}, {B{\"u}hler},
  {B{\"u}sching}, {Carrigan}, {Chadwick}, {Chounet}, {Cornils}, {Costamante},
  {Degrange}, {Dickinson}, {Djannati-Ata{\"i}}, {O'C.~Drury}, {Dubus},
  {Egberts}, {Emmanoulopoulos}, {Espigat}, {Feinstein}, {Ferrero}, {Fontaine},
  {Funk}, {Funk}, {Gallant}, {Giebels}, {Glicenstein}, {Goret},
  {Hadjichristidis}, {Hauser}, {Hauser}, {Heinzelmann}, {Henri}, {Hermann},
  {Hinton}, {Hofmann}, {Holleran}, {Horns}, {Jacholkowska}, {de Jager},
  {Kh{\'e}lifi}, {Komin}, {Konopelko}, {Latham}, {Le Gallou}, {Lemi{\`e}re},
  {Lemoine-Goumard}, {Lohse}, {Martin}, {Martineau-Huynh}, {Marcowith},
  {Masterson}, {McComb}, {de Naurois}, {Nedbal}, {Nolan}, {Noutsos}, {Orford},
  {Osborne}, {Ouchrif}, {Panter}, {Pelletier}, {Pita}, {P{\"u}hlhofer},
  {Punch}, {Raubenheimer}, {Raue}, {Rayner}, {Reimer}, {Reimer}, {Ripken},
  {Rob}, {Rolland}, {Rowell}, {Sahakian}, {Saug{\'e}}, {Schlenker},
  {Schlickeiser}, {Schwanke}, {Sol}, {Spangler}, {Spanier}, {Steenkamp},
  {Stegmann}, {Superina}, {Tavernet}, {Terrier}, {Th{\'e}oret}, {Tluczykont},
  {van Eldik}, {Vasileiadis}, {Venter}, {Vincent}, {V{\"o}lk}, {Wagner}, \&
  {Ward}}]{REF::TEV_H_2356-309}
---. 2006{\natexlab{b}}, \aap, 455, 461

\bibitem[{{Aharonian} {et~al.}(2006{\natexlab{c}}){Aharonian}, {Akhperjanian},
  {Bazer-Bachi}, {Beilicke}, {Benbow}, {Berge}, {Bernl{\"o}hr}, {Boisson},
  {Bolz}, {Borrel}, {Braun}, {Breitling}, {Brown}, {B{\"u}hler}, {Carrigan},
  {Chadwick}, {Chounet}, {Cornils}, {Costamante}, {Degrange}, {Dickinson},
  {Djannati-Ata{\"i}}, {O'C.~Drury}, {Dubus}, {Egberts}, {Emmanoulopoulos},
  {Espigat}, {Feinstein}, {Fontaine}, {Funk}, {Gallant}, {Giebels},
  {Glicenstein}, {Goret}, {Hadjichristidis}, {Hauser}, {Hauser}, {Heinzelmann},
  {Henri}, {Hermann}, {Hinton}, {Hofmann}, {Holleran}, {Horns}, {Jacholkowska},
  {de Jager}, {Kh{\'e}lifi}, {Komin}, {Konopelko}, {Latham}, {Le Gallou},
  {Lemi{\`e}re}, {Lemoine-Goumard}, {Lohse}, {Martin}, {Martineau-Huynh},
  {Marcowith}, {Masterson}, {McComb}, {de Naurois}, {Nedbal}, {Nolan},
  {Noutsos}, {Orford}, {Osborne}, {Ouchrif}, {Panter}, {Pelletier}, {Pita},
  {P{\"u}hlhofer}, {Punch}, {Raubenheimer}, {Raue}, {Rayner}, {Reimer},
  {Reimer}, {Ripken}, {Rob}, {Rolland}, {Rowell}, {Sahakian}, {Saug{\'e}},
  {Schlenker}, {Schlickeiser}, {Schuster}, {Schwanke}, {Siewert}, {Sol},
  {Spangler}, {Steenkamp}, {Stegmann}, {Superina}, {Tavernet}, {Terrier},
  {Th{\'e}oret}, {Tluczykont}, {van Eldik}, {Vasileiadis}, {Venter}, {Vincent},
  {V{\"o}lk}, {Wagner}, \& {Ward}}]{REF::TEV_PG_1553+113}
---. 2006{\natexlab{c}}, \aap, 448, L19

\bibitem[{{Aharonian} {et~al.}(2006{\natexlab{d}}){Aharonian}, {Akhperjanian},
  {Bazer-Bachi}, {Beilicke}, {Benbow}, {Berge}, {Bernl{\"o}hr}, {Boisson},
  {Bolz}, {Borrel}, {Braun}, {Brown}, {B{\"u}hler}, {B{\"u}sching}, {Carrigan},
  {Chadwick}, {Chounet}, {Coignet}, {Cornils}, {Costamante}, {Degrange},
  {Dickinson}, {Djannati-Ata{\"i}}, {O'C.~Drury}, {Dubus}, {Egberts},
  {Emmanoulopoulos}, {Espigat}, {Feinstein}, {Ferrero}, {Fiasson}, {Fontaine},
  {Funk}, {Funk}, {F{\"u}{\ss}ling}, {Gallant}, {Giebels}, {Glicenstein},
  {Goret}, {Hadjichristidis}, {Hauser}, {Hauser}, {Heinzelmann}, {Henri},
  {Hermann}, {Hinton}, {Hoffmann}, {Hofmann}, {Holleran}, {Hoppe}, {Horns},
  {Jacholkowska}, {de Jager}, {Kendziorra}, {Kerschhaggl}, {Kh{\'e}lifi},
  {Komin}, {Konopelko}, {Kosack}, {Lamanna}, {Latham}, {Le Gallou},
  {Lemi{\`e}re}, {Lemoine-Goumard}, {Lenain}, {Lohse}, {Martin},
  {Martineau-Huynh}, {Marcowith}, {Masterson}, {Maurin}, {McComb}, {Moulin},
  {de Naurois}, {Nedbal}, {Nolan}, {Noutsos}, {Orford}, {Osborne}, {Ouchrif},
  {Panter}, {Pelletier}, {Pita}, {P{\"u}hlhofer}, {Punch}, {Ranchon},
  {Raubenheimer}, {Raue}, {Rayner}, {Reimer}, {Ripken}, {Rob}, {Rolland},
  {Rosier-Lees}, {Rowell}, {Sahakian}, {Santangelo}, {Saug{\'e}}, {Schlenker},
  {Schlickeiser}, {Schr{\"o}der}, {Schwanke}, {Schwarzburg}, {Schwemmer},
  {Shalchi}, {Sol}, {Spangler}, {Spanier}, {Steenkamp}, {Stegmann}, {Superina},
  {Tam}, {Tavernet}, {Terrier}, {Tluczykont}, {van Eldik}, {Vasileiadis},
  {Venter}, {Vialle}, {Vincent}, {V{\"o}lk}, {Wagner}, \&
  {Ward}}]{REF::TEV_M_87_HESS_2006}
---. 2006{\natexlab{d}}, Science, 314, 1424

\bibitem[{{Aharonian} {et~al.}(2006{\natexlab{e}}){Aharonian}, {Akhperjanian},
  {Bazer-Bachi}, {Beilicke}, {Benbow}, {Berge}, {Bernl{\"o}hr}, {Boisson},
  {Bolz}, {Borrel}, {Braun}, {Breitling}, {Brown}, {B{\"u}hler},
  {B{\"u}sching}, {Carrigan}, {Chadwick}, {Chounet}, {Cornils}, {Costamante},
  {Degrange}, {Dickinson}, {Djannati-Ata{\"i}}, {O'C.~Drury}, {Dubus},
  {Egberts}, {Emmanoulopoulos}, {Espigat}, {Feinstein}, {Ferrero}, {Fiasson},
  {Fontaine}, {Funk}, {Funk}, {Gallant}, {Giebels}, {Glicenstein}, {Goret},
  {Hadjichristidis}, {Hauser}, {Hauser}, {Heinzelmann}, {Henri}, {Hermann},
  {Hinton}, {Hofmann}, {Holleran}, {Horns}, {Jacholkowska}, {de Jager},
  {Kh{\'e}lifi}, {Komin}, {Konopelko}, {Kosack}, {Latham}, {Le Gallou},
  {Lemi{\`e}re}, {Lemoine-Goumard}, {Lohse}, {Martin}, {Martineau-Huynh},
  {Marcowith}, {Masterson}, {McComb}, {de Naurois}, {Nedbal}, {Nolan},
  {Noutsos}, {Orford}, {Osborne}, {Ouchrif}, {Panter}, {Pelletier}, {Pita},
  {P{\"u}hlhofer}, {Punch}, {Raubenheimer}, {Raue}, {Rayner}, {Reimer},
  {Reimer}, {Ripken}, {Rob}, {Rolland}, {Rowell}, {Sahakian}, {Saug{\'e}},
  {Schlenker}, {Schlickeiser}, {Schwanke}, {Sol}, {Spangler}, {Spanier},
  {Steenkamp}, {Stegmann}, {Superina}, {Tavernet}, {Terrier}, {Th{\'e}oret},
  {Tluczykont}, {van Eldik}, {Vasileiadis}, {Venter}, {Vincent}, {V{\"o}lk},
  {Wagner}, \& {Ward}}]{REF::HESS_CRAB}
---. 2006{\natexlab{e}}, \aap, 457, 899

\bibitem[{{Aharonian} {et~al.}(2007{\natexlab{a}}){Aharonian}, {Akhperjanian},
  {Bazer-Bachi}, {Behera}, {Beilicke}, {Benbow}, {Berge}, {Bernl{\"o}hr},
  {Boisson}, {Bolz}, {Borrel}, {Boutelier}, {Braun}, {Brion}, {Brown},
  {B{\"u}hler}, {B{\"u}sching}, {Bulik}, {Carrigan}, {Chadwick}, {Clapson},
  {Chounet}, {Coignet}, {Cornils}, {Costamante}, {Degrange}, {Dickinson},
  {Djannati-Ata{\"i}}, {Domainko}, {Drury}, {Dubus}, {Dyks}, {Egberts},
  {Emmanoulopoulos}, {Espigat}, {Farnier}, {Feinstein}, {Fiasson},
  {F{\"o}rster}, {Fontaine}, {Funk}, {Funk}, {F{\"u}{\ss}ling}, {Gallant},
  {Giebels}, {Glicenstein}, {Gl{\"u}ck}, {Goret}, {Hadjichristidis}, {Hauser},
  {Hauser}, {Heinzelmann}, {Henri}, {Hermann}, {Hinton}, {Hoffmann}, {Hofmann},
  {Holleran}, {Hoppe}, {Horns}, {Jacholkowska}, {de Jager}, {Kendziorra},
  {Kerschhaggl}, {Kh{\'e}lifi}, {Komin}, {Kosack}, {Lamanna}, {Latham}, {Le
  Gallou}, {Lemi{\`e}re}, {Lemoine-Goumard}, {Lenain}, {Lohse}, {Martin},
  {Martineau-Huynh}, {Marcowith}, {Masterson}, {Maurin}, {McComb}, {Moderski},
  {Moulin}, {de Naurois}, {Nedbal}, {Nolan}, {Olive}, {Orford}, {Osborne},
  {Ostrowski}, {Panter}, {Pedaletti}, {Pelletier}, {Petrucci}, {Pita},
  {P{\"u}hlhofer}, {Punch}, {Ranchon}, {Raubenheimer}, {Raue}, {Rayner},
  {Renaud}, {Ripken}, {Rob}, {Rolland}, {Rosier-Lees}, {Rowell}, {Rudak},
  {Ruppel}, {Sahakian}, {Santangelo}, {Saug{\'e}}, {Schlenker}, {Schlickeiser},
  {Schr{\"o}der}, {Schwanke}, {Schwarzburg}, {Schwemmer}, {Shalchi}, {Sol},
  {Spangler}, {Stawarz}, {Steenkamp}, {Stegmann}, {Superina}, {Tam},
  {Tavernet}, {Terrier}, {van Eldik}, {Vasileiadis}, {Venter}, {Vialle},
  {Vincent}, {Vivier}, {V{\"o}lk}, {Volpe}, {Wagner}, {Ward}, \&
  {Zdziarski}}]{REF::TEV_PKS_FLARE}
---. 2007{\natexlab{a}}, \apjl, 664, L71

\bibitem[{{Aharonian} {et~al.}(2007{\natexlab{b}}){Aharonian}, {Akhperjanian},
  {Barres de Almeida}, {Bazer-Bachi}, {Behera}, {Beilicke}, {Benbow},
  {Bernl{\"o}hr}, {Boisson}, {Bolz}, {Borrel}, {Braun}, {Brion}, {Brown},
  {B{\"u}hler}, {Bulik}, {B{\"u}sching}, {Boutelier}, {Carrigan}, {Chadwick},
  {Chounet}, {Clapson}, {Coignet}, {Cornils}, {Costamante}, {Dalton},
  {Degrange}, {Dickinson}, {Djannati-Ata{\"i}}, {Domainko}, {O'C.~Drury},
  {Dubois}, {Dubus}, {Dyks}, {Egberts}, {Emmanoulopoulos}, {Espigat},
  {Farnier}, {Feinstein}, {Fiasson}, {F{\"o}rster}, {Fontaine}, {Funk},
  {F{\"u}{\ss}ling}, {Gallant}, {Giebels}, {Glicenstein}, {Gl{\"u}ck}, {Goret},
  {Hadjichristidis}, {Hauser}, {Hauser}, {Heinzelmann}, {Henri}, {Hermann},
  {Hinton}, {Hoffmann}, {Hofmann}, {Holleran}, {Hoppe}, {Horns},
  {Jacholkowska}, {de Jager}, {Jung}, {Katarzy{\'n}ski}, {Kendziorra},
  {Kerschhaggl}, {Kh{\'e}lifi}, {Keogh}, {Komin}, {Kosack}, {Lamanna},
  {Latham}, {Lemi{\`e}re}, {Lemoine-Goumard}, {Lenain}, {Lohse}, {Martin},
  {Martineau-Huynh}, {Marcowith}, {Masterson}, {Maurin}, {Maurin}, {McComb},
  {Moderski}, {Moulin}, {de Naurois}, {Nedbal}, {Nolan}, {Ohm}, {Olive}, {de
  O{\~n}a Wilhelmi}, {Orford}, {Osborne}, {Ostrowski}, {Panter}, {Pedaletti},
  {Pelletier}, {Petrucci}, {Pita}, {P{\"u}hlhofer}, {Punch}, {Ranchon},
  {Raubenheimer}, {Raue}, {Rayner}, {Renaud}, {Ripken}, {Rob}, {Rolland},
  {Rosier-Lees}, {Rowell}, {Rudak}, {Ruppel}, {Sahakian}, {Santangelo},
  {Schlickeiser}, {Sch{\"o}ck}, {Schr{\"o}der}, {Schwanke}, {Schwarzburg},
  {Schwemmer}, {Shalchi}, {Sol}, {Spangler}, {Stawarz}, {Steenkamp},
  {Stegmann}, {Superina}, {Tam}, {Tavernet}, {Terrier}, {van Eldik},
  {Vasileiadis}, {Venter}, {Vialle}, {Vincent}, {Vivier}, {V{\"o}lk}, {Volpe},
  {Wagner}, {Ward}, {Zdziarski}, \& {Zech}}]{REF::TEV_1ES_0347-121}
---. 2007{\natexlab{b}}, \aap, 473, L25

\bibitem[{{Aharonian} {et~al.}(2007{\natexlab{c}}){Aharonian}, {Akhperjanian},
  {Barres de Almeida}, {Bazer-Bachi}, {Behera}, {Beilicke}, {Benbow},
  {Bernl{\"o}hr}, {Boisson}, {Bolz}, {Borrel}, {Braun}, {Brion}, {Brown},
  {B{\"u}hler}, {Bulik}, {B{\"u}sching}, {Boutelier}, {Carrigan}, {Chadwick},
  {Chounet}, {Clapson}, {Coignet}, {Cornils}, {Costamante}, {Dalton},
  {Degrange}, {Dickinson}, {Djannati-Ata{\"i}}, {Domainko}, {O'C.~Drury},
  {Dubois}, {Dubus}, {Dyks}, {Egberts}, {Emmanoulopoulos}, {Espigat},
  {Farnier}, {Feinstein}, {Fiasson}, {F{\"o}rster}, {Fontaine}, {Funk},
  {F{\"u}{\ss}ling}, {Gallant}, {Giebels}, {Glicenstein}, {Gl{\"u}ck}, {Goret},
  {Hadjichristidis}, {Hauser}, {Hauser}, {Heinzelmann}, {Henri}, {Hermann},
  {Hinton}, {Hoffmann}, {Hofmann}, {Holleran}, {Hoppe}, {Horns},
  {Jacholkowska}, {de Jager}, {Jung}, {Katarzy{\'n}ski}, {Kendziorra},
  {Kerschhaggl}, {Kh{\'e}lifi}, {Keogh}, {Komin}, {Kosack}, {Lamanna},
  {Latham}, {Lemi{\`e}re}, {Lemoine-Goumard}, {Lenain}, {Lohse}, {Martin},
  {Martineau-Huynh}, {Marcowith}, {Masterson}, {Maurin}, {Maurin}, {McComb},
  {Moderski}, {Moulin}, {de Naurois}, {Nedbal}, {Nolan}, {Ohm}, {Olive}, {de
  O{\~n}a Wilhelmi}, {Orford}, {Osborne}, {Ostrowski}, {Panter}, {Pedaletti},
  {Pelletier}, {Petrucci}, {Pita}, {P{\"u}hlhofer}, {Punch}, {Ranchon},
  {Raubenheimer}, {Raue}, {Rayner}, {Renaud}, {Ripken}, {Rob}, {Rolland},
  {Rosier-Lees}, {Rowell}, {Rudak}, {Ruppel}, {Sahakian}, {Santangelo},
  {Schlickeiser}, {Sch{\"o}ck}, {Schr{\"o}der}, {Schwanke}, {Schwarzburg},
  {Schwemmer}, {Shalchi}, {Sol}, {Spangler}, {Stawarz}, {Steenkamp},
  {Stegmann}, {Superina}, {Tam}, {Tavernet}, {Terrier}, {van Eldik},
  {Vasileiadis}, {Venter}, {Vialle}, {Vincent}, {Vivier}, {V{\"o}lk}, {Volpe},
  {Wagner}, {Ward}, {Zdziarski}, \& {Zech}}]{REF::TEV_1ES_0229+20}
---. 2007{\natexlab{c}}, \aap, 475, L9

\bibitem[{{Aharonian} {et~al.}(2008{\natexlab{a}}){Aharonian}, {Akhperjanian},
  {Barres de Almeida}, {Bazer-Bachi}, {Behera}, {Beilicke}, {Benbow},
  {Bernl{\"o}hr}, {Boisson}, {Borrel}, {Braun}, {Brion}, {Brucker},
  {B{\"u}hler}, {Bulik}, {B{\"u}sching}, {Boutelier}, {Carrigan}, {Chadwick},
  {Chaves}, {Chounet}, {Clapson}, {Coignet}, {Cornils}, {Costamante}, {Dalton},
  {Degrange}, {Dickinson}, {Djannati-Ata{\"i}}, {Domainko}, {O.'c.~Drury},
  {Dubois}, {Dubus}, {Dyks}, {Egberts}, {Emmanoulopoulos}, {Espigat},
  {Farnier}, {Feinstein}, {Fiasson}, {F{\"o}rster}, {Fontaine},
  {F{\"u}{\ss}ling}, {Gabici}, {Gallant}, {Giebels}, {Glicenstein},
  {Gl{\"u}ck}, {Goret}, {Hadjichristidis}, {Hauser}, {Hauser}, {Heinzelmann},
  {Henri}, {Hermann}, {Hinton}, {Hoffmann}, {Hofmann}, {Holleran}, {Hoppe},
  {Horns}, {Jacholkowska}, {de Jager}, {Jung}, {Katarzy{\'n}ski}, {Kaufmann},
  {Kendziorra}, {Kerschhaggl}, {Khangulyan}, {Kh{\'e}lifi}, {Keogh}, {Komin},
  {Kosack}, {Lamanna}, {Latham}, {Lenain}, {Lohse}, {Martin},
  {Martineau-Huynh}, {Marcowith}, {Masterson}, {Maurin}, {McComb}, {Moderski},
  {Moulin}, {Naumann-Godo}, {de Naurois}, {Nedbal}, {Nekrassov}, {Nolan},
  {Ohm}, {Olive}, {de O{\~n}a Wilhelmi}, {Orford}, {Osborne}, {Ostrowski},
  {Panter}, {Pedaletti}, {Pelletier}, {Petrucci}, {Pita}, {P{\"u}hlhofer},
  {Punch}, {Quirrenbach}, {Raubenheimer}, {Raue}, {Rayner}, {Renaud}, {Rieger},
  {Ripken}, {Rob}, {Rosier-Lees}, {Rowell}, {Rudak}, {Ruppel}, {Sahakian},
  {Santangelo}, {Schlickeiser}, {Sch{\"o}ck}, {Schr{\"o}der}, {Schwanke},
  {Schwarzburg}, {Schwemmer}, {Shalchi}, {Sol}, {Spangler}, {Stawarz},
  {Steenkamp}, {Stegmann}, {Superina}, {Tam}, {Tavernet}, {Terrier}, {van
  Eldik}, {Vasileiadis}, {Venter}, {Vialle}, {Vincent}, {Vivier}, {V{\"o}lk},
  {Volpe}, {Wagner}, {Ward}, {Zdziarski}, \& {Zech}}]{REF::TEV_RGB_J0152+017}
---. 2008{\natexlab{a}}, \aap, 481, L103

\bibitem[{{Aharonian} {et~al.}(2008{\natexlab{b}}){Aharonian}, {Akhperjanian},
  {Barres de Almeida}, {Bazer-Bachi}, {Behera}, {Beilicke}, {Benbow},
  {Bernl{\"o}hr}, {Boisson}, {Bolz}, {Borrel}, {Braun}, {Brion}, {Brown},
  {B{\"u}hler}, {Bulik}, {B{\"u}sching}, {Boutelier}, {Carrigan}, {Chadwick},
  {Chounet}, {Clapson}, {Coignet}, {Cornils}, {Costamante}, {Dalton},
  {Degrange}, {Dickinson}, {Djannati-Ata{\"i}}, {Domainko}, {O'C.~Drury},
  {Dubois}, {Dubus}, {Dyks}, {Egberts}, {Emmanoulopoulos}, {Espigat},
  {Farnier}, {Feinstein}, {Fiasson}, {F{\"o}rster}, {Fontaine}, {Funk},
  {F{\"u}{\ss}ling}, {Gallant}, {Giebels}, {Glicenstein}, {Gl{\"u}ck}, {Goret},
  {Hadjichristidis}, {Hauser}, {Hauser}, {Heinzelmann}, {Henri}, {Hermann},
  {Hinton}, {Hoffmann}, {Hofmann}, {Holleran}, {Hoppe}, {Horns},
  {Jacholkowska}, {de Jager}, {Jung}, {Katarzy{\'n}ski}, {Kendziorra},
  {Kerschhaggl}, {Kh{\'e}lifi}, {Keogh}, {Komin}, {Kosack}, {Lamanna},
  {Latham}, {Lemi{\`e}re}, {Lemoine-Goumard}, {Lenain}, {Lohse}, {Martin},
  {Martineau-Huynh}, {Marcowith}, {Masterson}, {Maurin}, {Maurin}, {McComb},
  {Moderski}, {Moulin}, {de Naurois}, {Nedbal}, {Nolan}, {Ohm}, {Olive}, {de
  O{\~n}a Wilhelmi}, {Orford}, {Osborne}, {Ostrowski}, {Panter}, {Pedaletti},
  {Pelletier}, {Petrucci}, {Pita}, {P{\"u}hlhofer}, {Punch}, {Ranchon},
  {Raubenheimer}, {Raue}, {Rayner}, {Renaud}, {Ripken}, {Rob}, {Rolland},
  {Rosier-Lees}, {Rowell}, {Rudak}, {Ruppel}, {Sahakian}, {Santangelo},
  {Schlickeiser}, {Sch{\"o}ck}, {Schr{\"o}der}, {Schwanke}, {Schwarzburg},
  {Schwemmer}, {Shalchi}, {Sol}, {Spangler}, {Stawarz}, {Steenkamp},
  {Stegmann}, {Superina}, {Tam}, {Tavernet}, {Terrier}, {van Eldik},
  {Vasileiadis}, {Venter}, {Vialle}, {Vincent}, {Vivier}, {V{\"o}lk}, {Volpe},
  {Wagner}, {Ward}, {Zdziarski}, \& {Zech}}]{REF::UL_HESS_BENBOW08}
---. 2008{\natexlab{b}}, \aap, 478, 387

\bibitem[{{Aharonian} {et~al.}(2009{\natexlab{a}}){Aharonian}, {Akhperjanian},
  {Anton}, {de Almeida}, {Bazer-Bachi}, {Becherini}, {Behera}, {Benbow},
  {Bernl{\"o}hr}, {Boisson}, {Bochow}, {Borrel}, {Brion}, {Brucker}, {Brun},
  {B{\"u}hler}, {Bulik}, {B{\"u}sching}, {Boutelier}, {Chadwick},
  {Charbonnier}, {Chaves}, {Cheesebrough}, {Chounet}, {Clapson}, {Coignet},
  {Dalton}, {Daniel}, {Davids}, {Degrange}, {Deil}, {Dickinson},
  {Djannati-Ata{\"i}}, {Domainko}, {Drury}, {Dubois}, {Dubus}, {Dyks}, {Dyrda},
  {Egberts}, {Emmanoulopoulos}, {Espigat}, {Farnier}, {Feinstein}, {Fiasson},
  {F{\"o}rster}, {Fontaine}, {F{\"u}{\ss}ling}, {Gabici}, {Gallant},
  {G{\'e}rard}, {Giebels}, {Glicenstein}, {Gl{\"u}ck}, {Goret}, {G{\"o}hring},
  {Hauser}, {Hauser}, {Heinz}, {Heinzelmann}, {Henri}, {Hermann}, {Hinton},
  {Hoffmann}, {Hofmann}, {Holleran}, {Hoppe}, {Horns}, {Jacholkowska}, {de
  Jager}, {Jahn}, {Jung}, {Katarzy{\'n}ski}, {Katz}, {Kaufmann}, {Kendziorra},
  {Kerschhaggl}, {Khangulyan}, {Kh{\'e}lifi}, {Keogh}, {Klu{\'z}niak},
  {Kneiske}, {Komin}, {Kosack}, {Lamanna}, {Latham}, {Lenain}, {Lohse},
  {Marandon}, {Martin}, {Martineau-Huynh}, {Marcowith}, {Maurin}, {McComb},
  {Medina}, {Moderski}, {Moulin}, {Naumann-Godo}, {de Naurois}, {Nedbal},
  {Nekrassov}, {Niemiec}, {Nolan}, {Ohm}, {Olive}, {de O{\~n}a Wilhelmi},
  {Orford}, {Ostrowski}, {Panter}, {Arribas}, {Pedaletti}, {Pelletier},
  {Petrucci}, {Pita}, {P{\"u}hlhofer}, {Punch}, {Quirrenbach}, {Raubenheimer},
  {Raue}, {Rayner}, {Renaud}, {Rieger}, {Ripken}, {Rob}, {Rosier-Lees},
  {Rowell}, {Rudak}, {Rulten}, {Ruppel}, {Sahakian}, {Santangelo},
  {Schlickeiser}, {Sch{\"o}ck}, {Schr{\"o}der}, {Schwanke}, {Schwarzburg},
  {Schwemmer}, {Shalchi}, {Sikora}, {Skilton}, {Sol}, {Spangler}, {Stawarz},
  {Steenkamp}, {Stegmann}, {Superina}, {Szostek}, {Tam}, {Tavernet}, {Terrier},
  {Tibolla}, {Tluczykont}, {van Eldik}, {Vasileiadis}, {Venter}, {Venter},
  {Vialle}, {Vincent}, {Vink}, {Vivier}, {V{\"o}lk}, {Volpe}, {Wagner}, {Ward},
  {Zdziarski}, \& {Zech}}]{REF::TEV_CEN_A}
---. 2009{\natexlab{a}}, \apjl, 695, L40

\bibitem[{{Aharonian} {et~al.}(2009{\natexlab{b}}){Aharonian}, {Akhperjanian},
  {Anton}, {Barres de Almeida}, {Bazer-Bachi}, {Becherini}, {Behera}, {Benbow},
  {Bernl{\"o}hr}, {Boisson}, {Bochow}, {Borrel}, {Brion}, {Brucker}, {Brun},
  {B{\"u}hler}, {Bulik}, {B{\"u}sching}, {Boutelier}, {Chadwick},
  {Charbonnier}, {Chaves}, {Cheesebrough}, {Chounet}, {Clapson}, {Coignet},
  {Costamante}, {Dalton}, {Daniel}, {Davids}, {Degrange}, {Deil}, {Dickinson},
  {Djannati-Ata{\"i}}, {Domainko}, {O'C.~Drury}, {Dubois}, {Dubus}, {Dyks},
  {Dyrda}, {Egberts}, {Emmanoulopoulos}, {Espigat}, {Farnier}, {Feinstein},
  {Fiasson}, {F{\"o}rster}, {Fontaine}, {F{\"u}{\ss}ling}, {Gabici}, {Gallant},
  {G{\'e}rard}, {Giebels}, {Glicenstein}, {Gl{\"u}ck}, {Goret}, {G{\"o}hring},
  {Hauser}, {Hauser}, {Heinz}, {Heinzelmann}, {Henri}, {Hermann}, {Hinton},
  {Hoffmann}, {Hofmann}, {Holleran}, {Hoppe}, {Horns}, {Jacholkowska}, {de
  Jager}, {Jahn}, {Jung}, {Katarzy{\'n}ski}, {Katz}, {Kaufmann}, {Kendziorra},
  {Kerschhaggl}, {Khangulyan}, {Kh{\'e}lifi}, {Keogh}, {Klu{\'z}niak},
  {Kneiske}, {Komin}, {Kosack}, {Lamanna}, {Lenain}, {Lohse}, {Marandon},
  {Martin}, {Martineau-Huynh}, {Marcowith}, {Maurin}, {McComb}, {Medina},
  {Moderski}, {Monard}, {Moulin}, {Naumann-Godo}, {de Naurois}, {Nedbal},
  {Nekrassov}, {Niemiec}, {Nolan}, {Ohm}, {Olive}, {de O{\~n}a Wilhelmi},
  {Orford}, {Ostrowski}, {Panter}, {Paz Arribas}, {Pedaletti}, {Pelletier},
  {Petrucci}, {Pita}, {P{\"u}hlhofer}, {Punch}, {Quirrenbach}, {Raubenheimer},
  {Raue}, {Rayner}, {Renaud}, {Rieger}, {Ripken}, {Rob}, {Rosier-Lees},
  {Rowell}, {Rudak}, {Rulten}, {Ruppel}, {Sahakian}, {Santangelo},
  {Schlickeiser}, {Sch{\"o}ck}, {Schr{\"o}der}, {Schwanke}, {Schwarzburg},
  {Schwemmer}, {Shalchi}, {Sikora}, {Skilton}, {Sol}, {Spangler}, {Stawarz},
  {Steenkamp}, {Stegmann}, {Superina}, {Szostek}, {Tam}, {Tavernet}, {Terrier},
  {Tibolla}, {Tluczykont}, {van Eldik}, {Vasileiadis}, {Venter}, {Venter},
  {Vialle}, {Vincent}, {Vivier}, {V{\"o}lk}, {Volpe}, {Wagner}, {Ward},
  {Zdziarski}, \& {Zech}}]{REF::PKS2155_HESS_CHANDRA}
---. 2009{\natexlab{b}}, \aap, 502, 749

\bibitem[{{Aharonian} {et~al.}(2009{\natexlab{c}}){Aharonian}, {Akhperjanian},
  {Anton}, {Barres de Almeida}, {Bazer-Bachi}, {Becherini}, {Behera},
  {Bernl{\"o}hr}, {Boisson}, {Bochow}, {Borrel}, {Brion}, {Brucker}, {Brun},
  {B{\"u}hler}, {Bulik}, {B{\"u}sching}, {Boutelier}, {Chadwick},
  {Charbonnier}, {Chaves}, {Cheesebrough}, {Chounet}, {Clapson}, {Coignet},
  {Dalton}, {Daniel}, {Davids}, {Degrange}, {Deil}, {Dickinson},
  {Djannati-Ata{\"i}}, {Domainko}, {O'C.~Drury}, {Dubois}, {Dubus}, {Dyks},
  {Dyrda}, {Egberts}, {Emmanoulopoulos}, {Espigat}, {Farnier}, {Feinstein},
  {Fiasson}, {F{\"o}rster}, {Fontaine}, {F{\"u}{\ss}ling}, {Gabici}, {Gallant},
  {G{\'e}rard}, {Giebels}, {Glicenstein}, {Gl{\"u}ck}, {Goret}, {G{\"o}hring},
  {Hauser}, {Hauser}, {Heinz}, {Heinzelmann}, {Henri}, {Hermann}, {Hinton},
  {Hoffmann}, {Hofmann}, {Holleran}, {Hoppe}, {Horns}, {Jacholkowska}, {de
  Jager}, {Jahn}, {Jung}, {Katarzy{\'n}ski}, {Katz}, {Kaufmann}, {Kendziorra},
  {Kerschhaggl}, {Khangulyan}, {Kh{\'e}lifi}, {Keogh}, {Klu{\'z}niak}, {Komin},
  {Kosack}, {Lamanna}, {Lenain}, {Lohse}, {Marandon}, {Martin},
  {Martineau-Huynh}, {Marcowith}, {Maurin}, {McComb}, {Medina}, {Moderski},
  {Moulin}, {Naumann-Godo}, {de Naurois}, {Nedbal}, {Nekrassov}, {Niemiec},
  {Nolan}, {Ohm}, {Olive}, {de O{\~n}a Wilhelmi}, {Orford}, {Ostrowski},
  {Panter}, {Arribas}, {Pedaletti}, {Pelletier}, {Petrucci}, {Pita},
  {P{\"u}hlhofer}, {Punch}, {Quirrenbach}, {Raubenheimer}, {Raue}, {Rayner},
  {Renaud}, {Rieger}, {Ripken}, {Rob}, {Rosier-Lees}, {Rowell}, {Rudak},
  {Rulten}, {Ruppel}, {Sahakian}, {Santangelo}, {Schlickeiser}, {Sch{\"o}ck},
  {Schr{\"o}der}, {Schwanke}, {Schwarzburg}, {Schwemmer}, {Shalchi}, {Sikora},
  {Skilton}, {Sol}, {Spangler}, {Stawarz}, {Steenkamp}, {Stegmann}, {Superina},
  {Szostek}, {Tam}, {Tavernet}, {Terrier}, {Tibolla}, {van Eldik},
  {Vasileiadis}, {Venter}, {Venter}, {Vialle}, {Vincent}, {Vivier}, {V{\"o}lk},
  {Volpe}, {Wagner}, {Ward}, {Zdziarski}, {Zech}, {The HESS Collaboration},
  {Abdo}, {Ackermann}, {Ajello}, {Atwood}, {Axelsson}, {Baldini}, {Ballet},
  {Barbiellini}, {Baring}, {Bastieri}, {Battelino}, {Baughman}, {Bechtol},
  {Bellazzini}, {Berenji}, {Bloom}, {Bonamente}, {Borgland}, {Bregeon}, {Brez},
  {Brigida}, {Bruel}, {Caliandro}, {Cameron}, {Caraveo}, {Casandjian},
  {Cavazzuti}, {Cecchi}, {Charles}, {Chekhtman}, {Chen}, {Cheung}, {Chiang},
  {Ciprini}, {Claus}, {Cohen-Tanugi}, {Colafrancesco}, {Conrad}, {Costamante},
  {Cutini}, {Dermer}, {de Angelis}, {de Palma}, {Digel}, {do Couto e Silva},
  {Drell}, {Dubois}, {Dubus}, {Dumora}, {Farnier}, {Favuzzi}, {Fegan},
  {Ferrara}, {Fleury}, {Focke}, {Frailis}, {Fukazawa}, {Funk}, {Fusco},
  {Gargano}, {Gasparrini}, {Gehrels}, {Germani}, {Giebels}, {Giglietto},
  {Giordano}, {Grondin}, {Grove}, {Guillemot}, {Guiriec}, {Hanabata},
  {Harding}, {Hayashida}, {Hays}, {Horan}, {J{\'o}hannesson}, {Johnson},
  {Johnson}, {Johnson}, {Kadler}, {Kamae}, {Katagiri}, {Kataoka}, {Kerr},
  {Kn{\"o}dlseder}, {Kuehn}, {Kuss}, {Lande}, {Latronico}, {Lee},
  {Lemoine-Goumard}, {Longo}, {Loparco}, {Lott}, {Lovellette}, {Madejski},
  {Makeev}, {Mazziotta}, {McEnery}, {Meurer}, {Michelson}, {Mitthumsiri},
  {Mizuno}, {Moiseev}, {Monte}, {Monzani}, {Morselli}, {Moskalenko}, {Murgia},
  {Nolan}, {Nuss}, {Ohsugi}, {Omodei}, {Orlando}, {Ormes}, {Paneque},
  {Panetta}, {Parent}, {Pelassa}, {Pepe}, {Pesce-Rollins}, {Piron}, {Porter},
  {Rain{\`o}}, {Razzano}, {Reimer}, {Reimer}, {Reposeur}, {Ritz}, {Rodriguez},
  {Ryde}, {Sadrozinski}, {Sanchez}, {Sander}, {Scargle}, {Schalk},
  {Sellerholm}, {Sgr{\`o}}, {Shaw}, {Smith}, {Spandre}, {Spinelli}, {Starck},
  {Strickman}, {Tajima}, {Takahashi}, {Takahashi}, {Tanaka}, {Thayer},
  {Thompson}, {Tibaldo}, {Torres}, {Tosti}, {Tramacere}, {Uchiyama}, {Usher},
  {Vilchez}, {Villata}, {Vitale}, {Waite}, {Wood}, {Ylinen}, {Ziegler}, \& {The
  Fermi-LAT Collaboration}}]{REF::PKS2155_FERMI_MWCAMPAIGN}
---. 2009{\natexlab{c}}, \apjl, 696, L150

\bibitem[{{Aharonian} {et~al.}(2008{\natexlab{c}}){Aharonian}, {Khangulyan}, \&
  {Costamante}}]{REF::INTRINSIC_ABSORPTION_AHARONIAN}
{Aharonian}, F.~A., {Khangulyan}, D., \& {Costamante}, L. 2008{\natexlab{c}},
  \mnras, 387, 1206

\bibitem[{{Albert} {et~al.}(2006{\natexlab{a}}){Albert}, {Aliu}, {Anderhub},
  {Antoranz}, {Armada}, {Asensio}, {Baixeras}, {Barrio}, {Bartelt}, {Bartko},
  {Bastieri}, {Bavikadi}, {Bednarek}, {Berger}, {Bigongiari}, {Biland},
  {Bisesi}, {Bock}, {Bretz}, {Britvitch}, {Camara}, {Chilingarian}, {Ciprini},
  {Coarasa}, {Commichau}, {Contreras}, {Cortina}, {Curtef}, {Danielyan},
  {Dazzi}, {De Angelis}, {de los Reyes}, {De Lotto},
  {Domingo-Santamar{\'{\i}}a}, {Dorner}, {Doro}, {Errando}, {Fagiolini},
  {Ferenc}, {Fern{\'a}ndez}, {Firpo}, {Flix}, {Fonseca}, {Font}, {Galante},
  {Garczarczyk}, {Gaug}, {Giller}, {Goebel}, {Hakobyan}, {Hayashida},
  {Hengstebeck}, {H{\"o}hne}, {Hose}, {Jacon}, {Kalekin}, {Kranich}, {Laille},
  {Lenisa}, {Liebing}, {Lindfors}, {Longo}, {L{\'o}pez}, {L{\'o}pez}, {Lorenz},
  {Lucarelli}, {Majumdar}, {Maneva}, {Mannheim}, {Mariotti}, {Mart{\'{\i}}nez},
  {Mase}, {Mazin}, {Meucci}, {Meyer}, {Miranda}, {Mirzoyan}, {Mizobuchi},
  {Moralejo}, {Nilsson}, {O{\~n}a-Wilhelmi}, {Ordu{\~n}a}, {Otte}, {Oya},
  {Paneque}, {Paoletti}, {Pasanen}, {Pascoli}, {Pauss}, {Pavel}, {Pegna},
  {Persic}, {Peruzzo}, {Piccioli}, {Poller}, {Prandini}, {Rhode}, {Rico},
  {Riegel}, {Rissi}, {Robert}, {R{\"u}gamer}, {Saggion}, {S{\'a}nchez},
  {Sartori}, {Scalzotto}, {Schmitt}, {Schweizer}, {Shayduk}, {Shinozaki},
  {Shore}, {Sidro}, {Sillanp{\"a}{\"a}}, {Sobczy{\'n}ska}, {Stamerra}, {Stark},
  {Takalo}, {Temnikov}, {Tescaro}, {Teshima}, {Tonello}, {Torres}, {Torres},
  {Turini}, {Vankov}, {Vardanyan}, {Vitale}, {Wagner}, {Wibig}, {Wittek}, \&
  {Zapatero}}]{REF::TEV_1ES_1218+304}
{Albert}, J., {et~al.} 2006{\natexlab{a}}, \apjl, 642, L119

\bibitem[{{Albert} {et~al.}(2006{\natexlab{b}}){Albert}, {Aliu}, {Anderhub},
  {Antoranz}, {Armada}, {Asensio}, {Baixeras}, {Barrio}, {Bartko}, {Bastieri},
  {Becker}, {Bednarek}, {Berger}, {Bigongiari}, {Biland}, {Bisesi}, {Bock},
  {Bordas}, {Bosch-Ramon}, {Bretz}, {Britvitch}, {Camara}, {Carmona},
  {Chilingarian}, {Ciprini}, {Coarasa}, {Commichau}, {Contreras}, {Cortina},
  {Curtef}, {Danielyan}, {Dazzi}, {De Angelis}, {de los Reyes}, {De Lotto},
  {Domingo-Santamar{\'{\i}}a}, {Dorner}, {Doro}, {Errando}, {Fagiolini},
  {Ferenc}, {Fern{\'a}ndez}, {Firpo}, {Flix}, {Fonseca}, {Font}, {Fuchs},
  {Galante}, {Garczarczyk}, {Gaug}, {Giller}, {Goebel}, {Hakobyan},
  {Hayashida}, {Hengstebeck}, {H{\"o}hne}, {Hose}, {Hsu}, {Jacon}, {Kalekin},
  {Kosyra}, {Kranich}, {Laatiaoui}, {Laille}, {Lenisa}, {Liebing}, {Lindfors},
  {Lombardi}, {Longo}, {L{\'o}pez}, {L{\'o}pez}, {Lorenz}, {Majumdar},
  {Maneva}, {Mannheim}, {Mansutti}, {Mariotti}, {Mart{\'{\i}}nez}, {Mazin},
  {Merck}, {Meucci}, {Meyer}, {Miranda}, {Mirzoyan}, {Mizobuchi}, {Moralejo},
  {Nilsson}, {Ninkovic}, {O{\~n}a-Wilhelmi}, {Ordu{\~n}a}, {Otte}, {Oya},
  {Paneque}, {Paoletti}, {Paredes}, {Pasanen}, {Pascoli}, {Pauss}, {Pegna},
  {Persic}, {Peruzzo}, {Piccioli}, {Poller}, {Prandini}, {Raymers}, {Rhode},
  {Rib{\'o}}, {Rico}, {Riegel}, {Rissi}, {Robert}, {R{\"u}gamer}, {Saggion},
  {S{\'a}nchez}, {Sartori}, {Scalzotto}, {Scapin}, {Schmitt}, {Schweizer},
  {Shayduk}, {Shinozaki}, {Shore}, {Sidro}, {Sillanp{\"a}{\"a}}, {Sobczynska},
  {Stamerra}, {Stark}, {Takalo}, {Temnikov}, {Tescaro}, {Teshima}, {Tonello},
  {Torres}, {Torres}, {Turini}, {Vankov}, {Vitale}, {Wagner}, {Wibig},
  {Wittek}, {Zanin}, \& {Zapatero}}]{REF::TEV_MARKARIAN_180}
---. 2006{\natexlab{b}}, \apjl, 648, L105

\bibitem[{{Albert} {et~al.}(2007{\natexlab{a}}){Albert}, {Aliu}, {Anderhub},
  {Antoranz}, {Armada}, {Baixeras}, {Barrio}, {Bartko}, {Bastieri}, {Becker},
  {Bednarek}, {Berger}, {Bigongiari}, {Biland}, {Bock}, {Bordas},
  {Bosch-Ramon}, {Bretz}, {Britvitch}, {Camara}, {Carmona}, {Chilingarian},
  {Coarasa}, {Commichau}, {Contreras}, {Cortina}, {Costado}, {Curtef},
  {Danielyan}, {Dazzi}, {De Angelis}, {Delgado}, {de los Reyes}, {De Lotto},
  {Domingo-Santamar{\'{\i}}a}, {Dorner}, {Doro}, {Errando}, {Fagiolini},
  {Ferenc}, {Fern{\'a}ndez}, {Firpo}, {Flix}, {Fonseca}, {Font}, {Fuchs},
  {Galante}, {Garc{\'{\i}}a-L{\'o}pez}, {Garczarczyk}, {Gaug}, {Giller},
  {Goebel}, {Hakobyan}, {Hayashida}, {Hengstebeck}, {Herrero}, {H{\"o}hne},
  {Hose}, {Hsu}, {Jacon}, {Jogler}, {Kosyra}, {Kranich}, {Kritzer}, {Laille},
  {Lindfors}, {Lombardi}, {Longo}, {L{\'o}pez}, {L{\'o}pez}, {Lorenz},
  {Majumdar}, {Maneva}, {Mannheim}, {Mansutti}, {Mariotti}, {Mart{\'{\i}}nez},
  {Mazin}, {Merck}, {Meucci}, {Meyer}, {Miranda}, {Mirzoyan}, {Mizobuchi},
  {Moralejo}, {Nilsson}, {Ninkovic}, {O{\~n}a-Wilhelmi}, {Otte}, {Oya},
  {Paneque}, {Panniello}, {Paoletti}, {Paredes}, {Pasanen}, {Pascoli}, {Pauss},
  {Pegna}, {Persic}, {Peruzzo}, {Piccioli}, {Poller}, {Prandini}, {Puchades},
  {Raymers}, {Rhode}, {Rib{\'o}}, {Rico}, {Rissi}, {Robert}, {R{\"u}gamer},
  {Saggion}, {S{\'a}nchez}, {Sartori}, {Scalzotto}, {Scapin}, {Schmitt},
  {Schweizer}, {Shayduk}, {Shinozaki}, {Shore}, {Sidro}, {Sillanp{\"a}{\"a}},
  {Sobczynska}, {Stamerra}, {Stark}, {Takalo}, {Temnikov}, {Tescaro},
  {Teshima}, {Tonello}, {Torres}, {Turini}, {Vankov}, {Vitale}, {Wagner},
  {Wibig}, {Wittek}, {Zandanel}, {Zanin}, \&
  {Zapatero}}]{REF::TEV_BL_LACERTAE_MAGIC}
---. 2007{\natexlab{a}}, \apjl, 666, L17

\bibitem[{{Albert} {et~al.}(2007{\natexlab{b}}){Albert}, {Aliu}, {Anderhub},
  {Antoranz}, {Armada}, {Baixeras}, {Barrio}, {Bartko}, {Bastieri}, {Becker},
  {Bednarek}, {Berger}, {Bigongiari}, {Biland}, {Bock}, {Bordas},
  {Bosch-Ramon}, {Bretz}, {Britvitch}, {Camara}, {Carmona}, {Chilingarian},
  {Coarasa}, {Commichau}, {Contreras}, {Cortina}, {Costado}, {Curtef},
  {Danielyan}, {Dazzi}, {De Angelis}, {Delgado}, {de los Reyes}, {De Lotto},
  {Domingo-Santamar{\'{\i}}a}, {Dorner}, {Doro}, {Errando}, {Fagiolini},
  {Ferenc}, {Fern{\'a}ndez}, {Firpo}, {Flix}, {Fonseca}, {Font}, {Fuchs},
  {Galante}, {Garc{\'{\i}}a-L{\'o}pez}, {Garczarczyk}, {Gaug}, {Giller},
  {Goebel}, {Hakobyan}, {Hayashida}, {Hengstebeck}, {Herrero}, {H{\"o}hne},
  {Hose}, {Hsu}, {Jacon}, {Jogler}, {Kosyra}, {Kranich}, {Kritzer}, {Laille},
  {Lindfors}, {Lombardi}, {Longo}, {L{\'o}pez}, {L{\'o}pez}, {Lorenz},
  {Majumdar}, {Maneva}, {Mannheim}, {Mansutti}, {Mariotti}, {Mart{\'{\i}}nez},
  {Mazin}, {Merck}, {Meucci}, {Meyer}, {Miranda}, {Mirzoyan}, {Mizobuchi},
  {Moralejo}, {Nieto}, {Nilsson}, {Ninkovic}, {O{\~n}a-Wilhelmi}, {Otte},
  {Oya}, {Paneque}, {Panniello}, {Paoletti}, {Paredes}, {Pasanen}, {Pascoli},
  {Pauss}, {Pegna}, {Perlman}, {Persic}, {Peruzzo}, {Piccioli}, {Prandini},
  {Puchades}, {Raymers}, {Rhode}, {Rib{\'o}}, {Rico}, {Rissi}, {Robert},
  {R{\"u}gamer}, {Saggion}, {Saito}, {S{\'a}nchez}, {Sartori}, {Scalzotto},
  {Scapin}, {Schmitt}, {Schweizer}, {Shayduk}, {Shinozaki}, {Shore}, {Sidro},
  {Sillanp{\"a}{\"a}}, {Sobczynska}, {Stamerra}, {Stark}, {Takalo},
  {Tavecchio}, {Temnikov}, {Tescaro}, {Teshima}, {Torres}, {Turini}, {Vankov},
  {Vitale}, {Wagner}, {Wibig}, {Wittek}, {Zandanel}, {Zanin}, \&
  {Zapatero}}]{REF::TEV_1ES_1011+496}
---. 2007{\natexlab{b}}, \apjl, 667, L21

\bibitem[{{Albert} {et~al.}(2007{\natexlab{c}}){Albert}, {Aliu}, {Anderhub},
  {Antoranz}, {Armada}, {Baixeras}, {Barrio}, {Bartko}, {Bastieri}, {Becker},
  {Bednarek}, {Berger}, {Bigongiari}, {Biland}, {Bock}, {Bordas},
  {Bosch-Ramon}, {Bretz}, {Britvitch}, {Camara}, {Carmona}, {Chilingarian},
  {Ciprini}, {Coarasa}, {Commichau}, {Contreras}, {Cortina}, {Costado},
  {Curtef}, {Danielyan}, {Dazzi}, {De Angelis}, {Delgado}, {de los Reyes}, {De
  Lotto}, {Domingo-Santamar{\'{\i}}a}, {Dorner}, {Doro}, {Errando},
  {Fagiolini}, {De Angelis}, {Ferenc}, {Fern{\'a}ndez}, {Firpo}, {Flix},
  {Fonseca}, {Font}, {Fuchs}, {Galante}, {Garc{\'{\i}}a-L{\'o}pez},
  {Garczarczyk}, {Gaug}, {Giller}, {Goebel}, {Hakobyan}, {Hayashida},
  {Hengstebeck}, {Herrero}, {H{\"o}hne}, {Hose}, {Hsu}, {Jacon}, {Jogler},
  {Kalekin}, {Kosyra}, {Kranich}, {Kritzer}, {Laille}, {Liebing}, {Lindfors},
  {Lombardi}, {Longo}, {L{\'o}pez}, {L{\'o}pez}, {Lorenz}, {Majumdar},
  {Maneva}, {Mannheim}, {Mansutti}, {Mariotti}, {Mart{\'{\i}}nez}, {Mazin},
  {Merck}, {Meucci}, {Meyer}, {Miranda}, {Mirzoyan}, {Mizobuchi}, {Moralejo},
  {Nilsson}, {Ninkovic}, {On{\~n}a-Wilhelmi}, {Otte}, {Oya}, {Paneque},
  {Panniello}, {Paoletti}, {Paredes}, {Pasanen}, {Pascoli}, {Pauss}, {Pegna},
  {Persic}, {Peruzzo}, {Piccioli}, {Poller}, {Prandini}, {Puchades}, {Raymers},
  {Rhode}, {Rib{\'o}}, {Rico}, {Rissi}, {Robert}, {R{\"u}gamer}, {Saggion},
  {S{\'a}nchez}, {Sartori}, {Scalzotto}, {Scapin}, {Schmitt}, {Schweizer},
  {Shayduk}, {Shinozaki}, {Shore}, {Sidro}, {Sillanp{\"a}{\"a}}, {Sobczynska},
  {Stamerra}, {Stark}, {Takalo}, {Temnikov}, {Tescaro}, {Teshima}, {Tonello},
  {Torres}, {Turini}, {Vankov}, {Vitale}, {Wagner}, {Wibig}, {Wittek},
  {Zandanel}, {Zanin}, \& {Zapatero}}]{REF::TEV_1ES_2344+514_MAGIC_2007}
---. 2007{\natexlab{c}}, \apj, 662, 892

\bibitem[{{Albert} {et~al.}(2007{\natexlab{d}}){Albert}, {Aliu}, {Anderhub},
  {Antoranz}, {Armada}, {Asensio}, {Baixeras}, {Barrio}, {Bartko}, {Bastieri},
  {Becker}, {Bednarek}, {Berger}, {Bigongiari}, {Biland}, {Bock}, {Bordas},
  {Bosch-Ramon}, {Bretz}, {Britvitch}, {Camara}, {Carmona}, {Chilingarian},
  {Ciprini}, {Coarasa}, {Commichau}, {Contreras}, {Cortina}, {Curtef},
  {Danielyan}, {Dazzi}, {De Angelis}, {de los Reyes}, {De Lotto},
  {Domingo-Santamar{\'{\i}}a}, {Dorner}, {Doro}, {Errando}, {Fagiolini},
  {Ferenc}, {Fern{\'a}ndez}, {Firpo}, {Flix}, {Fonseca}, {Font}, {Fuchs},
  {Galante}, {Garczarczyk}, {Gaug}, {Giller}, {Goebel}, {Hakobyan},
  {Hayashida}, {Hengstebeck}, {H{\"o}hne}, {Hose}, {Hsu}, {Jacon}, {Jogler},
  {Kalekin}, {Kosyra}, {Kranich}, {Kritzer}, {Laatiaoui}, {Laille}, {Liebing},
  {Lindfors}, {Lombardi}, {Longo}, {L{\'o}pez}, {L{\'o}pez}, {Lorenz},
  {Majumdar}, {Maneva}, {Mannheim}, {Mansutti}, {Mariotti}, {Mart{\'{\i}}nez},
  {Mazin}, {Merck}, {Meucci}, {Meyer}, {Miranda}, {Mirzoyan}, {Mizobuchi},
  {Moralejo}, {Nilsson}, {Ninkovic}, {O{\~n}a-Wilhelmi}, {Ordu{\~n}a}, {Otte},
  {Oya}, {Paneque}, {Paoletti}, {Paredes}, {Pasanen}, {Pascoli}, {Pauss},
  {Pegna}, {Persic}, {Peruzzo}, {Piccioli}, {Poller}, {Prandini}, {Raymers},
  {Rhode}, {Rib{\'o}}, {Rico}, {Rissi}, {Robert}, {R{\"u}gamer}, {Saggion},
  {S{\'a}nchez}, {Sartori}, {Scalzotto}, {Scapin}, {Schmitt}, {Schweizer},
  {Shayduk}, {Shinozaki}, {Shore}, {Sidro}, {Sillanp{\"a}{\"a}}, {Sobczynska},
  {Stamerra}, {Stark}, {Takalo}, {Temnikov}, {Tescaro}, {Teshima}, {Tonello},
  {Torres}, {Torres}, {Turini}, {Vankov}, {Vitale}, {Wagner}, {Wibig},
  {Wittek}, {Zanin}, \& {Zapatero}}]{REF::TEV_MARKARIAN_421_BIS}
---. 2007{\natexlab{d}}, \apj, 663, 125

\bibitem[{{Albert} {et~al.}(2007{\natexlab{e}}){Albert}, {Aliu}, {Anderhub},
  {Antoranz}, {Armada}, {Baixeras}, {Barrio}, {Bartko}, {Bastieri}, {Becker},
  {Bednarek}, {Berger}, {Bigongiari}, {Biland}, {Bock}, {Bordas},
  {Bosch-Ramon}, {Bretz}, {Britvitch}, {Camara}, {Carmona}, {Chilingarian},
  {Coarasa}, {Commichau}, {Contreras}, {Cortina}, {Costado}, {Curtef},
  {Danielyan}, {Dazzi}, {De Angelis}, {Delgado}, {de los Reyes}, {De Lotto},
  {Domingo-Santamar{\'{\i}}a}, {Dorner}, {Doro}, {Errando}, {Fagiolini},
  {Ferenc}, {Fern{\'a}ndez}, {Firpo}, {Flix}, {Fonseca}, {Font}, {Fuchs},
  {Galante}, {Garc{\'{\i}}a-L{\'o}pez}, {Garczarczyk}, {Gaug}, {Giller},
  {Goebel}, {Hakobyan}, {Hayashida}, {Hengstebeck}, {Herrero}, {H{\"o}hne},
  {Hose}, {Hrupec}, {Hsu}, {Jacon}, {Jogler}, {Kosyra}, {Kranich}, {Kritzer},
  {Laille}, {Lindfors}, {Lombardi}, {Longo}, {L{\'o}pez}, {L{\'o}pez},
  {Lorenz}, {Majumdar}, {Maneva}, {Mannheim}, {Mansutti}, {Mariotti},
  {Mart{\'{\i}}nez}, {Mazin}, {Merck}, {Meucci}, {Meyer}, {Miranda},
  {Mirzoyan}, {Mizobuchi}, {Moralejo}, {Nieto}, {Nilsson}, {Ninkovic},
  {O{\~n}a-Wilhelmi}, {Otte}, {Oya}, {Paneque}, {Panniello}, {Paoletti},
  {Paredes}, {Pasanen}, {Pascoli}, {Pauss}, {Pegna}, {Persic}, {Peruzzo},
  {Piccioli}, {Prandini}, {Puchades}, {Raymers}, {Rhode}, {Rib{\'o}}, {Rico},
  {Rissi}, {Robert}, {R{\"u}gamer}, {Saggion}, {Saito}, {S{\'a}nchez},
  {Sartori}, {Scalzotto}, {Scapin}, {Schmitt}, {Schweizer}, {Shayduk},
  {Shinozaki}, {Shore}, {Sidro}, {Sillanp{\"a}{\"a}}, {Sobczynska}, {Stamerra},
  {Stark}, {Takalo}, {Tavecchio}, {Temnikov}, {Tescaro}, {Teshima}, {Torres},
  {Turini}, {Vankov}, {Vitale}, {Wagner}, {Wibig}, {Wittek}, {Zandanel},
  {Zanin}, \& {Zapatero}}]{REF::TEV_MARKARIAN_501_MAGIC}
---. 2007{\natexlab{e}}, \apj, 669, 862

\bibitem[{{Albert} {et~al.}(2008{\natexlab{a}}){Albert}, {Aliu}, {Anderhub},
  {Antoranz}, {Baixeras}, {Barrio}, {Bartko}, {Bastieri}, {Becker}, {Bednarek},
  {Berger}, {Bigongiari}, {Biland}, {Bock}, {Bordas}, {Bosch-Ramon}, {Bretz},
  {Britvitch}, {Camara}, {Carmona}, {Chilingarian}, {Coarasa}, {Commichau},
  {Contreras}, {Cortina}, {Costado}, {Curtef}, {Danielyan}, {Dazzi}, {De
  Angelis}, {Delgado}, {de los Reyes}, {De Lotto}, {Dorner}, {Doro}, {Errando},
  {Fagiolini}, {Ferenc}, {Fern{\'a}ndez}, {Firpo}, {Fonseca}, {Font}, {Fuchs},
  {Galante}, {Garc{\'{\i}}a-L{\'o}pez}, {Garczarczyk}, {Gaug}, {Giller},
  {Goebel}, {Hakobyan}, {Hayashida}, {Hengstebeck}, {Herrero}, {H{\"o}hne},
  {Hose}, {Huber}, {Hsu}, {Jacon}, {Jogler}, {Kosyra}, {Kranich}, {Kritzer},
  {Laille}, {Lindfors}, {Lombardi}, {Longo}, {L{\'o}pez}, {Lorenz}, {Majumdar},
  {Maneva}, {Mannheim}, {Mariotti}, {Mart{\'{\i}}nez}, {Mazin}, {Merck},
  {Meucci}, {Meyer}, {Miranda}, {Mirzoyan}, {Mizobuchi}, {Moralejo}, {Nieto},
  {Nilsson}, {Ninkovic}, {O{\~n}a-Wilhelmi}, {Otte}, {Oya}, {Panniello},
  {Paoletti}, {Paredes}, {Pasanen}, {Pascoli}, {Pauss}, {Pegna}, {Persic},
  {Peruzzo}, {Piccioli}, {Prandini}, {Puchades}, {Raymers}, {Rhode},
  {Rib{\'o}}, {Rico}, {Rissi}, {Robert}, {R{\"u}gamer}, {Saggion}, {Saito},
  {S{\'a}nchez}, {Sartori}, {Scalzotto}, {Scapin}, {Schmitt}, {Schweizer},
  {Shayduk}, {Shinozaki}, {Shore}, {Sidro}, {Sillanp{\"a}{\"a}}, {Sobczynska},
  {Spanier}, {Stamerra}, {Stark}, {Takalo}, {Temnikov}, {Tescaro}, {Teshima},
  {Torres}, {Turini}, {Vankov}, {Venturini}, {Vitale}, {Wagner}, {Wibig},
  {Wittek}, {Zandanel}, {Zanin}, \& {Zapatero}}]{REF::UL_MAGIC_ALBERT}
---. 2008{\natexlab{a}}, \apj, 681, 944

\bibitem[{{Albert} {et~al.}(2008{\natexlab{b}}){Albert}, {Aliu}, {Anderhub},
  {Antonelli}, {Antoranz}, {Backes}, {Baixeras}, {Barrio}, {Bartko},
  {Bastieri}, {Becker}, {Bednarek}, {Berger}, {Bernardini}, {Bigongiari},
  {Biland}, {Bock}, {Bonnoli}, {Bordas}, {Bosch-Ramon}, {Bretz}, {Britvitch},
  {Camara}, {Carmona}, {Chilingarian}, {Commichau}, {Contreras}, {Cortina},
  {Costado}, {Covino}, {Curtef}, {Dazzi}, {De Angelis}, {Cea del Pozo}, {de los
  Reyes}, {De Lotto}, {De Maria}, {De Sabata}, {Mendez}, {Dominguez}, {Dorner},
  {Doro}, {Errando}, {Fagiolini}, {Ferenc}, {Fern{\'a}ndez}, {Firpo},
  {Fonseca}, {Font}, {Galante}, {L{\'o}pez}, {Garczarczyk}, {Gaug}, {Goebel},
  {Hayashida}, {Herrero}, {H{\"o}hne}, {Hose}, {Hsu}, {Huber}, {Jogler},
  {Kneiske}, {Kranich}, {La Barbera}, {Laille}, {Leonardo}, {Lindfors},
  {Lombardi}, {Longo}, {L{\'o}pez}, {Lorenz}, {Majumdar}, {Maneva},
  {Mankuzhiyil}, {Mannheim}, {Maraschi}, {Mariotti}, {Mart{\'{\i}}nez},
  {Mazin}, {Meucci}, {Meyer}, {Miranda}, {Mirzoyan}, {Mizobuchi}, {Moles},
  {Moralejo}, {Nieto}, {Nilsson}, {Ninkovic}, {Otte}, {Oya}, {Panniello},
  {Paoletti}, {Paredes}, {Pasanen}, {Pascoli}, {Pauss}, {Pegna},
  {Perez-Torres}, {Persic}, {Peruzzo}, {Piccioli}, {Prada}, {Prandini},
  {Puchades}, {Raymers}, {Rhode}, {Rib{\'o}}, {Rico}, {Rissi}, {Robert},
  {R{\"u}gamer}, {Saggion}, {Saito}, {Salvati}, {Sanchez-Conde}, {Sartori},
  {Satalecka}, {Scalzotto}, {Scapin}, {Schmitt}, {Schweizer}, {Shayduk},
  {Shinozaki}, {Shore}, {Sidro}, {Sierpowska-Bartosik}, {Sillanp{\"a}{\"a}},
  {Sobczynska}, {Spanier}, {Stamerra}, {Stark}, {Takalo}, {Tavecchio},
  {Temnikov}, {Tescaro}, {Teshima}, {Tluczykont}, {Torres}, {Turini}, {Vankov},
  {Venturini}, {Vitale}, {Wagner}, {Wittek}, {Zabalza}, {Zandanel}, {Zanin}, \&
  {Zapatero}}]{REF::TEV_3C279}
---. 2008{\natexlab{b}}, Science, 320, 1752

\bibitem[{{Albert} {et~al.}(2009){Albert}, {Aliu}, {Anderhub}, {Antoranz},
  {Baixeras}, {Barrio}, {Bartko}, {Bastieri}, {Becker}, {Bednarek},
  {Berdyugin}, {Berger}, {Bigongiari}, {Biland}, {Bock}, {Bordas},
  {Bosch-Ramon}, {Bretz}, {Britvitch}, {Camara}, {Carmona}, {Chilingarian},
  {Commichau}, {Contreras}, {Cortina}, {Costado}, {Curtef}, {Danielyan},
  {Dazzi}, {de Angelis}, {Delgado}, {de Los Reyes}, {de Lotto}, {Dorner},
  {Doro}, {Errando}, {Fagiolini}, {Ferenc}, {Fern{\'a}ndez}, {Firpo},
  {Fonseca}, {Font}, {Fuchs}, {Galante}, {Garc{\'{\i}}a-L{\'o}pez},
  {Garczarczyk}, {Gaug}, {Goebel}, {Hakobyan}, {Hayashida}, {Hengstebeck},
  {Herrero}, {H{\"o}hne}, {Hose}, {Hsu}, {Huber}, {Jacon}, {Jogler}, {Kosyra},
  {Kranich}, {Kritzer}, {Laille}, {Lindfors}, {Lombardi}, {Longo}, {L{\'o}pez},
  {Lorenz}, {Majumdar}, {Maneva}, {Mannheim}, {Mariotti}, {Mart{\'{\i}}nez},
  {Mazin}, {Merck}, {Meucci}, {Meyer}, {Miranda}, {Mirzoyan}, {Mizobuchi},
  {Moralejo}, {Nieto}, {Nilsson}, {Ninkovic}, {O{\~n}a-Wilhelmi}, {Otte},
  {Oya}, {Panniello}, {Paoletti}, {Pasanen}, {Pascoli}, {Pauss}, {Pegna},
  {Persic}, {Peruzzo}, {Piccioli}, {Prandini}, {Puchades}, {Raymers}, {Rico},
  {Rhode}, {Rico}, {Rissi}, {Robert}, {R{\"u}gamer}, {Saggion}, {Saito},
  {S{\'a}nchez}, {Sartori}, {Scalzotto}, {Scapin}, {Schmitt}, {Schweizer},
  {Shayduk}, {Shinozaki}, {Shore}, {Sidro}, {Sillanp{\"a}{\"a}}, {Sobczynska},
  {Spanier}, {Stamerra}, {Stark}, {Takalo}, {Temnikov}, {Tescaro}, {Teshima},
  {Torres}, {Turini}, {Vankov}, {Venturini}, {Vitale}, {Wagner}, {Wibig},
  {Wittek}, {Zandanel}, {Zanin}, \& {Zapatero}}]{REF::TEV_PG_1553+113_MAGIC}
---. 2009, \aap, 493, 467

\bibitem[{{Aliu} {et~al.}(2009){Aliu}, {Anderhub}, {Antonelli}, {Antoranz},
  {Backes}, {Baixeras}, {Balestra}, {Barrio}, {Bartko}, {Bastieri}, {Becerra
  Gonz{\'a}lez}, {Becker}, {Bednarek}, {Berger}, {Bernardini}, {Biland},
  {Bock}, {Bonnoli}, {Bordas}, {Borla Tridon}, {Bosch-Ramon}, {Bretz},
  {Britvitch}, {Camara}, {Carmona}, {Chilingarian}, {Commichau}, {Contreras},
  {Cortina}, {Costado}, {Covino}, {Curtef}, {Dazzi}, {DeAngelis}, {DeCea del
  Pozo}, {de los Reyes}, {DeLotto}, {DeMaria}, {DeSabata}, {Delgado Mendez},
  {Dominguez}, {Dorner}, {Doro}, {Elsaesser}, {Errando}, {Ferenc},
  {Fern{\'a}ndez}, {Firpo}, {Fonseca}, {Font}, {Galante}, {Garc{\'{\i}}a
  L{\'o}pez}, {Garczarczyk}, {Gaug}, {Goebel}, {Hadasch}, {Hayashida},
  {Herrero}, {H{\"o}hne-M{\"o}nch}, {Hose}, {Hsu}, {Huber}, {Jogler},
  {Kranich}, {La Barbera}, {Laille}, {Leonardo}, {Lindfors}, {Lombardi},
  {Longo}, {L{\'o}pez}, {Lorenz}, {Majumdar}, {Maneva}, {Mankuzhiyil},
  {Mannheim}, {Maraschi}, {Mariotti}, {Mart{\'{\i}}nez}, {Mazin}, {Meucci},
  {Meyer}, {Miranda}, {Mirzoyan}, {Mold{\'o}n}, {Moles}, {Moralejo}, {Nieto},
  {Nilsson}, {Ninkovic}, {Otte}, {Oya}, {Paoletti}, {Paredes}, {Pasanen},
  {Pascoli}, {Pauss}, {Pegna}, {Perez-Torres}, {Persic}, {Peruzzo}, {Prada},
  {Prandini}, {Puchades}, {Raymers}, {Rhode}, {Rib{\'o}}, {Rico}, {Rissi},
  {Robert}, {R{\"u}gamer}, {Saggion}, {Saito}, {Salvati}, {Sanchez-Conde},
  {Sartori}, {Satalecka}, {Scalzotto}, {Scapin}, {Schweizer}, {Shayduk},
  {Shinozaki}, {Shore}, {Sidro}, {Sierpowska-Bartosik}, {Sillanp{\"a}{\"a}},
  {Sitarek}, {Sobczynska}, {Spanier}, {Stamerra}, {Stark}, {Takalo},
  {Tavecchio}, {Temnikov}, {Tescaro}, {Teshima}, {Tluczykont}, {Torres},
  {Turini}, {Vankov}, {Venturini}, {Vitale}, {Wagner}, {Wittek}, {Zabalza},
  {Zandanel}, {Zanin}, \& {Zapatero}}]{REF::TEV_3C66B_MAGIC}
{Aliu}, E., {et~al.} 2009, \apjl, 692, L29

\bibitem[{{Atwood} {et~al.}(2009)}]{REF::LAT_INSTRUMENT}
{Atwood}, B., {et~al.} 2009, \apj, {in press}, {arXiv:0902.1089}

\bibitem[{{B{\"o}ttcher}(2007)}]{REF::BOTTCHER_BLAZAR_EMISSION}
{B{\"o}ttcher}, M. 2007, \apss, 309, 95

\bibitem[{{Buckley} {et~al.}(1996){Buckley}, {Akerlof}, {Biller},
  {Carter-Lewis}, {Catanese}, {Cawley}, {Connaughton}, {Fegan}, {Finley},
  {Gaidos}, {Hillas}, {Kartje}, {Koenigl}, {Krennrich}, {Lamb}, {Lessard},
  {Macomb}, {Mattox}, {McEnery}, {Mohanty}, {Quinn}, {Rodgers}, {Rose},
  {Schubnel}, {Sembroski}, {Smith}, {Weekes}, {Wilson}, \&
  {Zweerink}}]{REF::MRK_MWCAMPAIGN}
{Buckley}, J.~H., {et~al.} 1996, \apjl, 472, L9+

\bibitem[{Cash(1979)}]{REF::CASH_LIKE}
Cash, W. 1979, \apj, 228, 939

\bibitem[{{Catanese} {et~al.}(1998){Catanese}, {Akerlof}, {Badran}, {Biller},
  {Bond}, {Boyle}, {Bradbury}, {Buckley}, {Burdett}, {Bussons Gordo},
  {Carter-Lewis}, {Cawley}, {Connaughton}, {Fegan}, {Finley}, {Gaidos}, {Hall},
  {Hillas}, {Krennrich}, {Lamb}, {Lessard}, {Masterson}, {McEnery}, {Mohanty},
  {Quinn}, {Rodgers}, {Rose}, {Samuelson}, {Schubnell}, {Sembroski},
  {Srinivasan}, {Weekes}, {Wilson}, \& {Zweerink}}]{REF::TEV_1ES_2344+514}
{Catanese}, M., {et~al.} 1998, \apj, 501, 616

\bibitem[{{Chadwick} {et~al.}(1999){Chadwick}, {Lyons}, {McComb}, {Orford},
  {Osborne}, {Rayner}, {Shaw}, {Turver}, \&
  {Wieczorek}}]{REF::TEV_PKS_2155-304}
{Chadwick}, P.~M., {et~al.} 1999, Astroparticle Physics, 11, 145

\bibitem[{{Costamante} {et~al.}(2001){Costamante}, {Ghisellini}, {Giommi},
  {Tagliaferri}, {Celotti}, {Chiaberge}, {Fossati}, {Maraschi}, {Tavecchio},
  {Treves}, \& {Wolter}}]{REF::COSTAMENTE_1426}
{Costamante}, L., {et~al.} 2001, \aap, 371, 512

\bibitem[{{de la Calle P{\'e}rez} {et~al.}(2003){de la Calle P{\'e}rez},
  {Bond}, {Boyle}, {Bradbury}, {Buckley}, {Carter-Lewis}, {Celik}, {Cui},
  {Dowdall}, {Duke}, {Falcone}, {Fegan}, {Fegan}, {Finley}, {Fortson},
  {Gaidos}, {Gibbs}, {Gammell}, {Hall}, {Hall}, {Hillas}, {Holder}, {Horan},
  {Jordan}, {Kertzman}, {Kieda}, {Kildea}, {Knapp}, {Kosack}, {Krawczynski},
  {Krennrich}, {LeBohec}, {Linton}, {Lloyd-Evans}, {Moriarty}, {M{\"u}ller},
  {Nagai}, {Ong}, {Page}, {Pallassini}, {Petry}, {Power-Mooney}, {Quinn},
  {Rebillot}, {Reynolds}, {Rose}, {Schroedter}, {Sembroski}, {Swordy},
  {Vassiliev}, {Wakely}, {Walker}, \& {Weekes}}]{REF::UL_WHIPPLE_DLCP}
{de la Calle P{\'e}rez}, I., {et~al.} 2003, \apj, 599, 909

\bibitem[{{Fabian} {et~al.}(1999){Fabian}, {Celotti}, {Pooley}, {Iwasawa},
  {Brandt}, {McMahon}, \& {Hoenig}}]{REF::FABIAN_1428}
{Fabian}, A.~C., {Celotti}, A., {Pooley}, G., {Iwasawa}, K., {Brandt}, W.~N.,
  {McMahon}, R.~G., \& {Hoenig}, M.~D. 1999, \mnras, 308, L6

\bibitem[{{Falcone} {et~al.}(2004){Falcone}, {Bond}, {Boyle}, {Bradbury},
  {Buckley}, {Carter-Lewis}, {Celik}, {Cui}, {Daniel}, {D'Vali}, {de la Calle
  Perez}, {Duke}, {Fegan}, {Fegan}, {Finley}, {Fortson}, {Gaidos}, {Gammell},
  {Gibbs}, {Gillanders}, {Grube}, {Hall}, {Hall}, {Hanna}, {Hillas}, {Holder},
  {Horan}, {Jarvis}, {Kenny}, {Kertzman}, {Kieda}, {Kildea}, {Knapp}, {Kosack},
  {Krawczynski}, {Krennrich}, {Lang}, {LeBohec}, {Linton}, {Lloyd-Evans},
  {Milovanovic}, {Moriarty}, {Muller}, {Nagai}, {Nolan}, {Ong}, {Pallassini},
  {Petry}, {Pizlo}, {Power-Mooney}, {Quinn}, {Quinn}, {Ragan}, {Rebillot},
  {Reynolds}, {Rose}, {Schroedter}, {Sembroski}, {Swordy}, {Syson}, {Tyler},
  {Vassiliev}, {Wakely}, {Walker}, {Weekes}, \&
  {Zweerink}}]{REF::UL_WHIPPLE_FALCONE}
{Falcone}, A.~D., {et~al.} 2004, \apj, 613, 710

\bibitem[{{Falomo} \& {Kotilainen}(1999)}]{REF::FALOMO_BLLACS}
{Falomo}, R., \& {Kotilainen}, J.~K. 1999, \aap, 352, 85

\bibitem[{{Fegan} {et~al.}(2005){Fegan}, {Badran}, {Bond}, {Boyle}, {Bradbury},
  {Buckley}, {Carter-Lewis}, {Catanese}, {Celik}, {Cui}, {Daniel}, {D'Vali},
  {de la Calle Perez}, {Duke}, {Falcone}, {Fegan}, {Finley}, {Fortson},
  {Gaidos}, {Gammell}, {Gibbs}, {Gillanders}, {Grube}, {Hall}, {Hall}, {Hanna},
  {Hillas}, {Holder}, {Horan}, {Jarvis}, {Jordan}, {Kenny}, {Kertzman},
  {Kieda}, {Kildea}, {Knapp}, {Kosack}, {Krawczynski}, {Krennrich}, {Lang}, {Le
  Bohec}, {Lessard}, {Linton}, {Lloyd-Evans}, {Milovanovic}, {McEnery},
  {Moriarty}, {Mukherjee}, {Muller}, {Nagai}, {Nolan}, {Ong}, {Pallassini},
  {Petry}, {Power-Mooney}, {Quinn}, {Quinn}, {Ragan}, {Rebillot}, {Reynolds},
  {Rose}, {Schroedter}, {Sembroski}, {Swordy}, {Syson}, {Vassiliev}, {Wakely},
  {Walker}, {Weekes}, \& {Zweerink}}]{REF::TEV_SURVEY_OF_EGRET}
{Fegan}, S.~J., {et~al.} 2005, \apj, 624, 638

\bibitem[{{Fossati} {et~al.}(1999){Fossati}, {Celotti}, {Ghisellini}, \&
  {Maraschi}}]{REF::BLASAR_SEQ}
{Fossati}, G., {Celotti}, A., {Ghisellini}, G., \& {Maraschi}, L. 1999, in
  Astronomical Society of the Pacific Conference Series, Vol. 159, BL Lac
  Phenomenon, ed. L.~O. {Takalo} \& A.~{Sillanp{\"a}{\"a}}, 351--+

\bibitem[{{Franceschini} {et~al.}(2008){Franceschini}, {Rodighiero}, \&
  {Vaccari}}]{REF::EBL_MODEL_FRANCESCHINI}
{Franceschini}, A., {Rodighiero}, G., \& {Vaccari}, M. 2008, \aap, 487, 837

\bibitem[{{Gaidos} {et~al.}(1996){Gaidos}, {Akerlof}, {Biller}, {Boyle},
  {Breslin}, {Buckley}, {Carter-Lewis}, {Catanese}, {Cawley}, {Fegan},
  {Finley}, {Hillas}, {Krennrich}, {Lamb}, {Lessard}, {McEnery}, {Mohanty},
  {Moriarty}, {Quinn}, {Rodgers}, {Rose}, {Samuelson}, {Schubnell},
  {Sembroski}, {Srinivasan}, {Weekes}, {Wilson}, \&
  {Zweerink}}]{REF::MRK_FLARE_NATURE}
{Gaidos}, J.~A., {et~al.} 1996, \nat, 383, 319

\bibitem[{{Giommi} {et~al.}(2008){Giommi}, {Perri}, {Verrecchia}, {Pittori},
  {Tavani}, {Gehrels}, \& {Chester}}]{REF::SWIFT_S5_0716+714}
{Giommi}, P., {Perri}, M., {Verrecchia}, F., {Pittori}, C., {Tavani}, M.,
  {Gehrels}, N., \& {Chester}, M. 2008, The Astronomer's Telegram, 1495, 1

\bibitem[{{Gleissner} {et~al.}(2004){Gleissner}, {Wilms}, {Pottschmidt},
  {Uttley}, {Nowak}, \& {Staubert}}]{REF::KENDAL}
{Gleissner}, T., {Wilms}, J., {Pottschmidt}, K., {Uttley}, P., {Nowak}, M.~A.,
  \& {Staubert}, R. 2004, \aap, 414, 1091

\bibitem[{{Gould} \& {Schr{\'e}der}(1967)}]{REF::EBL_TEV_THEORY}
{Gould}, R.~J., \& {Schr{\'e}der}, G.~P. 1967, Physical Review, 155, 1408

\bibitem[{{Hartman} {et~al.}(1999){Hartman}, {Bertsch}, {Bloom}, {Chen},
  {Deines-Jones}, {Esposito}, {Fichtel}, {Friedlander}, {Hunter}, {McDonald},
  {Sreekumar}, {Thompson}, {Jones}, {Lin}, {Michelson}, {Nolan}, {Tompkins},
  {Kanbach}, {Mayer-Hasselwander}, {M{\"u}cke}, {Pohl}, {Reimer}, {Kniffen},
  {Schneid}, {von Montigny}, {Mukherjee}, \& {Dingus}}]{REF::3EG}
{Hartman}, R.~C., {et~al.} 1999, \apjs, 123, 79

\bibitem[{{Hillas} {et~al.}(1998){Hillas}, {Akerlof}, {Biller}, {Buckley},
  {Carter-Lewis}, {Catanese}, {Cawley}, {Fegan}, {Finley}, {Gaidos},
  {Krennrich}, {Lamb}, {Lang}, {Mohanty}, {Punch}, {Reynolds}, {Rodgers},
  {Rose}, {Rovero}, {Schubnell}, {Sembroski}, {Vacanti}, {Weekes}, {West}, \&
  {Zweerink}}]{REF::CRAB_SPECTRUM_HILLAS}
{Hillas}, A.~M., {et~al.} 1998, \apj, 503, 744

\bibitem[{{Holder} {et~al.}(2008){Holder}, {Acciari}, {Aliu}, {Arlen},
  {Beilicke}, {Benbow}, {Bradbury}, {Buckley}, {Bugaev}, {Butt}, {Byrum},
  {Cannon}, {Celik}, {Cesarini}, {Ciupik}, {Chow}, {Cogan}, {Colin}, {Cui},
  {Daniel}, {Ergin}, {Falcone}, {Fegan}, {Finley}, {Finnegan}, {Fortin},
  {Fortson}, {Furniss}, {Gillanders}, {Grube}, {Guenette}, {Gyuk}, {Hanna},
  {Hays}, {Horan}, {Hui}, {Humensky}, {Imran}, {Kaaret}, {Karlsson},
  {Kertzman}, {Kieda}, {Kildea}, {Konopelko}, {Krawczynski}, {Krennrich},
  {Lang}, {Lebohec}, {Maier}, {McCann}, {McCutcheon}, {Moriarty}, {Mukherjee},
  {Nagai}, {Niemiec}, {Ong}, {Pandel}, {Perkins}, {Pohl}, {Quinn}, {Ragan},
  {Reyes}, {Reynolds}, {Rose}, {Schroedter}, {Sembroski}, {Smith}, {Steele},
  {Swordy}, {Toner}, {Valcarcel}, {Vassiliev}, {Wagner}, {Wakely}, {Ward},
  {Weekes}, {Weinstein}, {White}, {Williams}, {Wissel}, {Wood}, \&
  {Zitzer}}]{REF::VERITAS_STATUS}
{Holder}, J., {et~al.} 2008, in American Institute of Physics Conference
  Series, Vol. 1085, American Institute of Physics Conference Series, ed. F.~A.
  {Aharonian}, W.~{Hofmann}, \& F.~{Rieger}, 657--660

\bibitem[{{Horan} {et~al.}(2002){Horan}, {Badran}, {Bond}, {Bradbury},
  {Buckley}, {Carson}, {Carter-Lewis}, {Catanese}, {Cui}, {Dunlea}, {Das}, {de
  la Calle Perez}, {D'Vali}, {Fegan}, {Fegan}, {Finley}, {Gaidos}, {Gibbs},
  {Gillanders}, {Hall}, {Hillas}, {Holder}, {Jordan}, {Kertzman}, {Kieda},
  {Kildea}, {Knapp}, {Kosack}, {Krennrich}, {Lang}, {LeBohec}, {Lessard},
  {Lloyd-Evans}, {McKernan}, {Moriarty}, {Muller}, {Ong}, {Pallassini},
  {Petry}, {Quinn}, {Reay}, {Reynolds}, {Rose}, {Sembroski}, {Sidwell},
  {Stanton}, {Swordy}, {Vassiliev}, {Wakely}, \&
  {Weekes}}]{REF::TEV_H_1426+428}
{Horan}, D., {et~al.} 2002, \apj, 571, 753

\bibitem[{{Horan} {et~al.}(2004){Horan}, {Badran}, {Bond}, {Boyle}, {Bradbury},
  {Buckley}, {Carter-Lewis}, {Catanese}, {Celik}, {Cui}, {Daniel}, {D'Vali},
  {de la Calle Perez}, {Duke}, {Falcone}, {Fegan}, {Fegan}, {Finley},
  {Fortson}, {Gaidos}, {Gammell}, {Gibbs}, {Gillanders}, {Grube}, {Hall},
  {Hall}, {Hanna}, {Hillas}, {Holder}, {Jarvis}, {Jordan}, {Kenny}, {Kertzman},
  {Kieda}, {Kildea}, {Knapp}, {Kosack}, {Krawczynski}, {Krennrich}, {Lang}, {Le
  Bohec}, {Linton}, {Lloyd-Evans}, {Milovanovic}, {Moriarty}, {Muller},
  {Nagai}, {Nolan}, {Ong}, {Pallassini}, {Petry}, {Power-Mooney}, {Quinn},
  {Quinn}, {Ragan}, {Rebillot}, {Reynolds}, {Rose}, {Schroedter}, {Sembroski},
  {Swordy}, {Syson}, {Vassiliev}, {Wakely}, {Walker}, {Weekes}, \&
  {Zweerink}}]{REF::UL_WHIPPLE_HORAN}
---. 2004, \apj, 603, 51

\bibitem[{{Imran} {et~al.}(2009)}]{REF::1ES_1218+304_VERITAS_ICRC2009}
{Imran}, A., {et~al.} 2009, in Proc. 31st ICRC ({in press}), {arXiv:0908.0142}

\bibitem[{{Kanbach} {et~al.}(1988)}]{REF::EGRET_INSTRUMENT}
{Kanbach}, G., {et~al.} 1988, \ssr, 49, 61

\bibitem[{{Katarzy{\'n}ski} {et~al.}(2005){Katarzy{\'n}ski}, {Ghisellini},
  {Tavecchio}, {Maraschi}, {Fossati}, \&
  {Mastichiadis}}]{REF::XRAY_TEV_CORRELATION}
{Katarzy{\'n}ski}, K., {Ghisellini}, G., {Tavecchio}, F., {Maraschi}, L.,
  {Fossati}, G., \& {Mastichiadis}, A. 2005, \aap, 433, 479

\bibitem[{{Krawczynski} {et~al.}(2004){Krawczynski}, {Hughes}, {Horan},
  {Aharonian}, {Aller}, {Aller}, {Boltwood}, {Buckley}, {Coppi}, {Fossati},
  {G{\"o}tting}, {Holder}, {Horns}, {Kurtanidze}, {Marscher}, {Nikolashvili},
  {Remillard}, {Sadun}, \& {Schr{\"o}der}}]{REF::ORPHAN_FLARE}
{Krawczynski}, H., {et~al.} 2004, \apj, 601, 151

\bibitem[{{Krennrich} {et~al.}(2008){Krennrich}, {Dwek}, \&
  {Imran}}]{REF::KRENNRICH_DWEK_IMRAN_CONSTRAINTS}
{Krennrich}, F., {Dwek}, E., \& {Imran}, A. 2008, \apjl, 689, L93

\bibitem[{{Krennrich} {et~al.}(2002{\natexlab{a}}){Krennrich}, {Bond},
  {Bradbury}, {Buckley}, {Carter-Lewis}, {Cui}, {de la Calle Perez}, {Fegan},
  {Fegan}, {Finley}, {Gaidos}, {Gibbs}, {Gillanders}, {Hall}, {Hillas},
  {Holder}, {Horan}, {Jordan}, {Kertzman}, {Kieda}, {Kildea}, {Knapp},
  {Kosack}, {Lang}, {LeBohec}, {Moriarty}, {M{\"u}ller}, {Ong}, {Pallassini},
  {Petry}, {Quinn}, {Reay}, {Reynolds}, {Rose}, {Sembroski}, {Sidwell},
  {Stanton}, {Swordy}, {Vassiliev}, {Wakely}, \&
  {Weekes}}]{REF::MRK_SPECTRAL_VAR}
{Krennrich}, F., {et~al.} 2002{\natexlab{a}}, \apjl, 575, L9

\bibitem[{{Krennrich} {et~al.}(2002{\natexlab{b}}){Krennrich}, {Bond},
  {Bradbury}, {Buckley}, {Carter-Lewis}, {Cui}, {de la Calle Perez}, {Fegan},
  {Fegan}, {Finley}, {Gaidos}, {Gibbs}, {Gillanders}, {Hall}, {Hillas},
  {Holder}, {Horan}, {Jordan}, {Kertzman}, {Kieda}, {Kildea}, {Knapp},
  {Kosack}, {Lang}, {LeBohec}, {Moriarty}, {M{\"u}ller}, {Ong}, {Pallassini},
  {Petry}, {Quinn}, {Reay}, {Reynolds}, {Rose}, {Sembroski}, {Sidwell},
  {Stanton}, {Swordy}, {Vassiliev}, {Wakely}, \&
  {Weekes}}]{REF::TEV_MRK421_WHIPPLE_HS}
---. 2002{\natexlab{b}}, \apjl, 575, L9

\bibitem[{{Kuiper} {et~al.}(2000){Kuiper}, {Hermsen}, {Verbunt}, {Thompson},
  {Stairs}, {Lyne}, {Strickman}, \& {Cusumano}}]{REF::PSR0218_EGRET}
{Kuiper}, L., {Hermsen}, W., {Verbunt}, F., {Thompson}, D.~J., {Stairs}, I.~H.,
  {Lyne}, A.~G., {Strickman}, M.~S., \& {Cusumano}, G. 2000, \aap, 359, 615

\bibitem[{{Mattox} {et~al.}(1996){Mattox}, {Bertsch}, \&
  {Chiang}}]{REF::MATOX_LIKE}
{Mattox}, J.~R., {Bertsch}, D.~L., \& {Chiang}, J. 1996, \apj, 461, 396

\bibitem[{{Miller} {et~al.}(1978){Miller}, {French}, \&
  {Hawley}}]{REF::3C66A_REDSHIFT}
{Miller}, J.~S., {French}, H.~B., \& {Hawley}, S.~A. 1978, in BL Lac Objects,
  ed. A.~M. {Wolfe}, Vol. 176

\bibitem[{{Neshpor} {et~al.}(2001){Neshpor}, {Chalenko}, {Stepanian},
  {Kalekin}, {Jogolev}, {Fomin}, \& {Shitov}}]{REF::TEV_BL_LACERTAE_CRIMEA}
{Neshpor}, Y.~I., {Chalenko}, N.~N., {Stepanian}, A.~A., {Kalekin}, O.~R.,
  {Jogolev}, N.~A., {Fomin}, V.~P., \& {Shitov}, V.~G. 2001, Astronomy Reports,
  45, 249

\bibitem[{{Neshpor} {et~al.}(1998){Neshpor}, {Stepanyan}, {Kalekin}, {Fomin},
  {Chalenko}, \& {Shitov}}]{REF::TEV_3C66A_CRIMEA}
{Neshpor}, Y.~I., {Stepanyan}, A.~A., {Kalekin}, O.~P., {Fomin}, V.~P.,
  {Chalenko}, N.~N., \& {Shitov}, V.~G. 1998, Astronomy Letters, 24, 134

\bibitem[{{Nishiyama}(1999)}]{REF::TEV_1ES_1959+650}
{Nishiyama}, T. 1999, in Proc. 26th ICRC, Vol.~3, 370--+

\bibitem[{{Ong} {et~al.}(2009{\natexlab{a}})}]{REF::TEV_PKS_1424}
{Ong}, R., {et~al.} 2009{\natexlab{a}}, The Astronomer's Telegram, 2098, 1

\bibitem[{{Ong} {et~al.}(2009{\natexlab{b}})}]{REF::TEV_RGB_J0710+591}
---. 2009{\natexlab{b}}, The Astronomer's Telegram, 1941, 1

\bibitem[{{Punch} {et~al.}(1992){Punch}, {Akerlof}, {Cawley}, {Chantell},
  {Fegan}, {Fennell}, {Gaidos}, {Hagan}, {Hillas}, {Jiang}, {Kerrick}, {Lamb},
  {Lawrence}, {Lewis}, {Meyer}, {Mohanty}, {O'Flaherty}, {Reynolds}, {Rovero},
  {Schubnell}, {Sembroski}, {Weekes}, \& {Wilson}}]{REF::TEV_MARKARIAN_421}
{Punch}, M., {et~al.} 1992, \nat, 358, 477

\bibitem[{{Quinn} {et~al.}(1996){Quinn}, {Akerlof}, {Biller}, {Buckley},
  {Carter-Lewis}, {Cawley}, {Catanese}, {Connaughton}, {Fegan}, {Finley},
  {Gaidos}, {Hillas}, {Lamb}, {Krennrich}, {Lessard}, {McEnery}, {Meyer},
  {Mohanty}, {Rodgers}, {Rose}, {Sembroski}, {Schubnell}, {Weekes}, {Wilson},
  \& {Zweerink}}]{REF::TEV_MARKARIAN_501}
{Quinn}, J., {et~al.} 1996, \apjl, 456, L83+

\bibitem[{{Reyes} {et~al.}(2009)}]{REF::3C66A_FERMI_REYES_ICRC}
{Reyes}, L.~C., {et~al.} 2009, in Proc. 31st ICRC ({in press}),
  {arXiv:0907.5175}

\bibitem[{{Samuelson} {et~al.}(1998){Samuelson}, {Biller}, {Bond}, {Boyle},
  {Bradbury}, {Breslin}, {Buckley}, {Burdett}, {Buss'ons Gordo},
  {Carter-Lewis}, {Cantanese}, {Cawley}, {Fegan}, {Finley}, {Gaidos}, {Hall},
  {Hillas}, {Krennrich}, {Lamb}, {Lessard}, {McEnery}, {Masterson}, {Quinn},
  {Rodgers}, {Rose}, {Sembroski}, {Srinivasan}, {Vassiliev}, {Weekes}, \&
  {Zweerink}}]{REF::TEV_MARKARIAN_501_WHIPPLE_SPECT}
{Samuelson}, F.~W., {et~al.} 1998, \apjl, 501, L17+

\bibitem[{{Sbarufatti} {et~al.}(2005){Sbarufatti}, {Treves}, \&
  {Falomo}}]{REF::SBARUFATTI_REDSHIFTS}
{Sbarufatti}, B., {Treves}, A., \& {Falomo}, R. 2005, \apj, 635, 173

\bibitem[{{Stecker} \& {de Jager}(1993)}]{REF::EBL_TEV_MRK}
{Stecker}, F.~W., \& {de Jager}, O.~C. 1993, \apjl, 415, L71+

\bibitem[{{Stecker} \& {Scully}(2006)}]{REF::EBL_STECKER_SCULLY_LINEAR}
{Stecker}, F.~W., \& {Scully}, S.~T. 2006, \apjl, 652, L9

\bibitem[{{Strong} {et~al.}(2004{\natexlab{a}}){Strong}, {Moskalenko}, \&
  {Reimer}}]{REF::GALPROP}
{Strong}, {Moskalenko}, \& {Reimer}. 2004{\natexlab{a}}, \apj, 613, 962

\bibitem[{{Strong} {et~al.}(2004{\natexlab{b}}){Strong}, {Moskalenko},
  {Reimer}, {Digel}, \& {Diehl}}]{REF::GALPROP2}
{Strong}, A.~W., {Moskalenko}, I.~V., {Reimer}, O., {Digel}, S., \& {Diehl}, R.
  2004{\natexlab{b}}, A\&A, L47, 422

\bibitem[{{Superina} {et~al.}(2008){Superina}, {Benbow}, {Boutelier},
  {et~al.}}]{REF::TEV_PKS_0548-322}
{Superina}, G., {Benbow}, W., {Boutelier}, T., {et~al.} 2008, in Proc. 30th
  ICRC, Vol.~3, 913--916

\bibitem[{{Tagliaferri} {et~al.}(2008){Tagliaferri}, {Foschini}, {Ghisellini},
  {Maraschi}, {Tosti}, {Albert}, {Aliu}, {Anderhub}, {Antoranz}, {Baixeras},
  {Barrio}, {Bartko}, {Bastieri}, {Becker}, {Bednarek}, {Bedyugin}, {Berger},
  {Bigongiari}, {Biland}, {Bock}, {Bordas}, {Bosch-Ramon}, {Bretz},
  {Britvitch}, {Camara}, {Carmona}, {Chilingarian}, {Coarasa}, {Commichau},
  {Contreras}, {Cortina}, {Costado}, {Curtef}, {Danielyan}, {Dazzi}, {De
  Angelis}, {Delgado}, {de los Reyes}, {De Lotto}, {Dorner}, {Doro}, {Errando},
  {Fagiolini}, {Ferenc}, {Fern{\'a}ndez}, {Firpo}, {Fonseca}, {Font}, {Fuchs},
  {Galante}, {Garc{\'{\i}}a-L{\'o}pez}, {Garczarczyk}, {Gaug}, {Giller},
  {Goebel}, {Hakobyan}, {Hayashida}, {Hengstebeck}, {Herrero}, {H{\"o}hne},
  {Hose}, {Huber}, {Hsu}, {Jacon}, {Jogler}, {Kosyra}, {Kranich}, {Kritzer},
  {Laille}, {Lindfors}, {Lombardi}, {Longo}, {L{\'o}pez}, {Lorenz}, {Majumdar},
  {Maneva}, {Mannheim}, {Mariotti}, {Mart{\'{\i}}nez}, {Mazin}, {Merck},
  {Meucci}, {Meyer}, {Miranda}, {Mirzoyan}, {Mizobuchi}, {Moralejo}, {Nieto},
  {Nilsson}, {Ninkovic}, {O{\~n}a-Wilhelmi}, {Otte}, {Oya}, {Panniello},
  {Paoletti}, {Paredes}, {Pasanen}, {Pascoli}, {Pauss}, {Pegna}, {Persic},
  {Peruzzo}, {Piccioli}, {Prandini}, {Puchades}, {Raymers}, {Rhode},
  {Rib{\'o}}, {Rico}, {Rissi}, {Robert}, {R{\"u}gamer}, {Saggion}, {Saito},
  {S{\'a}nchez}, {Sartori}, {Scalzotto}, {Scapin}, {Schmitt}, {Schweizer},
  {Shayduk}, {Shinozaki}, {Shore}, {Sidro}, {Sillanp{\"a}{\"a}}, {Sobczynska},
  {Spanier}, {Stamerra}, {Stark}, {Takalo}, {Tavecchio}, {Temnikov}, {Tescaro},
  {Teshima}, {Torres}, {Turini}, {Vankov}, {Venturini}, {Vitale}, {Wagner},
  {Wibig}, {Wittek}, {Zandanel}, {Zanin}, \&
  {Zapatero}}]{REF::TEV_1ES_1959+650_MAGIC_2008}
{Tagliaferri}, G., {et~al.} 2008, \apj, 679, 1029

\bibitem[{{Teshima} {et~al.}(2008)}]{REF::TEV_S5_0716+714}
{Teshima}, M., {et~al.} 2008, The Astronomer's Telegram, 1500, 1

\bibitem[{{Thompson} {et~al.}(1993){Thompson}, {Bertsch}, {Fichtel}, {Hartman},
  {Hofstadter}, {Hughes}, {Hunter}, {Hughlock}, {Kanbach}, {Kniffen}, {Lin},
  {Mattox}, {Mayer-Hasselwander}, {von Montigny}, {Nolan}, {Nel}, {Pinkau},
  {Rothermel}, {Schneid}, {Sommer}, {Sreekumar}, {Tieger}, \&
  {Walker}}]{REF::EGRET_CALIBRATION}
{Thompson}, D.~J., {et~al.} 1993, \apjs, 86, 629

\bibitem[{{Vassiliev}(2000)}]{REF::EBL_VASSILIEV}
{Vassiliev}, V.~V. 2000, Astroparticle Physics, 12, 217

\bibitem[{{Wakely} \& {Horan}(2008)}]{REF::TEVCAT}
{Wakely}, S.~P., \& {Horan}, D. 2008, in International Cosmic Ray Conference,
  Vol.~3, International Cosmic Ray Conference, 1341--1344

\bibitem[{{Weekes}(2008)}]{REF::TEV_REVIEW}
{Weekes}, T.~C. 2008, in AIP conf. series, ed. F.~A. {Aharonian}, W.~{Hofmann},
  \& F.~{Rieger}, Vol. 1085, 3--17

\end{thebibliography}


\clearpage

%
%
\newcommand{\HMS}[3]{$#1\hmsh\,#2\hmsm\,#3\hmss$}
\newcommand{\DMS}[3]{$#1\arcdeg\,#2\arcmin\,#3\arcsec$}

\newcounter{refctr}
\setcounter{refctr}{1}
\begin{deluxetable}{llllllll}
\tabletypesize{\scriptsize}
\tablecaption{AGN detected at TeV energies\label{TAB::OBJ_DET}.}
\tablewidth{0pt}
\tablehead{
\colhead{Name} &
\colhead{$\alpha_{\mathrm{J2000}}$} & 
\colhead{$\delta_{\mathrm{J2000}}$} &
\colhead{Type\tablenotemark{a}} &
\colhead{$z$} &
\colhead{Ref}}
\startdata
\multicolumn{6}{l}{\textbf{Blazars:}} \\
\objectname{RGB J0152+017} & \HMS{01}{52}{39.6} & \DMS{+01}{47}{17} & HBL  & $0.080$    & 1  \\
\objectname{3C 66A}        & \HMS{02}{22}{39.6} & \DMS{+43}{02}{08} & IBL  & $0.444$\tablenotemark{b} & 2,3\tablenotemark{c} \\
\objectname{1ES 0229+200}  & \HMS{02}{32}{48.6} & \DMS{+20}{17}{17} & HBL  & $0.140$    & 4  \\
\objectname{1ES 0347-121}  & \HMS{03}{49}{23.2} & \DMS{-11}{59}{27} & HBL  & $0.188$    & 5  \\
\objectname{PKS 0548-322}  & \HMS{05}{50}{40.8} & \DMS{-32}{16}{18} & HBL  & $0.069$    & 6  \\
\objectname{RGB J0710+591} & \HMS{07}{10}{30.1} & \DMS{+59}{08}{20} & HBL  & $0.125$    & 7  \\
\objectname{S5 0716+714}   & \HMS{07}{21}{53.4} & \DMS{+71}{20}{36} & LBL  & $0.300$    & 8  \\
\objectname{1ES 0806+524}  & \HMS{08}{09}{49.2} & \DMS{+52}{18}{58} & HBL  & $0.138$    & 9  \\
\objectname{1ES 1011+496}  & \HMS{10}{15}{04.1} & \DMS{+49}{26}{01} & HBL  & $0.212$    & 10 \\
\objectname{1ES 1101-232}  & \HMS{11}{03}{37.6} & \DMS{-23}{29}{30} & HBL  & $0.186$    & 11 \\
\objectname{Markarian 421} & \HMS{11}{04}{27.3} & \DMS{+38}{12}{32} & HBL  & $0.031$    & 12 \\
\objectname{Markarian 180} & \HMS{11}{36}{26.4} & \DMS{+70}{09}{27} & HBL  & $0.046$    & 13 \\
\objectname{1ES 1218+304}  & \HMS{12}{21}{21.9} & \DMS{+30}{10}{37} & HBL  & $0.182$    & 14 \\
\objectname{W Comae}       & \HMS{12}{21}{31.7} & \DMS{+28}{13}{59} & IBL  & $0.102$    & 15 \\
\objectname{3C 279}        & \HMS{12}{56}{11.2} & \DMS{-05}{47}{22} & FSRQ & $0.536$    & 16 \\
\objectname{PKS 1424+240}  & \HMS{14}{27}{00.4} & \DMS{+23}{48}{00} & IBL  & \nodata    & 17 \\
\objectname{H 1426+428}    & \HMS{14}{28}{32.7} & \DMS{+42}{40}{21} & HBL  & $0.129$    & 18 \\
\objectname{PG 1553+113}   & \HMS{15}{55}{43.0} & \DMS{+11}{11}{24} & HBL  & $0.09-0.78$& 19 \\
\objectname{Markarian 501} & \HMS{16}{53}{52.2} & \DMS{+39}{45}{37} & HBL  & $0.034$    & 20 \\
\objectname{1ES 1959+650}  & \HMS{19}{59}{59.9} & \DMS{+65}{08}{55} & HBL  & $0.048$    & 21 \\
\objectname{PKS 2005-489}  & \HMS{20}{09}{25.4} & \DMS{-48}{49}{54} & HBL  & $0.071$    & 22 \\
\objectname{PKS 2155-304}  & \HMS{21}{58}{52.1} & \DMS{-30}{13}{32} & HBL  & $0.117$    & 23 \\
\objectname{BL Lacertae}   & \HMS{22}{02}{43.3} & \DMS{+42}{16}{40} & LBL  & $0.069$    & 24,25\tablenotemark{c} \\
\objectname{1ES 2344+514}  & \HMS{23}{47}{04.8} & \DMS{+51}{42}{18} & HBL  & $0.044$    & 26 \\
\objectname{H 2356-309}    & \HMS{23}{59}{07.9} & \DMS{-30}{37}{41} & HBL  & $0.167$    & 27 \\
\\
\multicolumn{6}{l}{\textbf{Others}} \\
\objectname{3C 66B}        & \HMS{02}{23}{11.4} & \DMS{+42}{59}{31} & FR1  & $0.02106$  & 28 \\
\objectname{M 87}          & \HMS{12}{30}{49.4} & \DMS{+12}{23}{28} & FR1  & $0.004233$ & 29 \\
\objectname{Centaurus A}   & \HMS{13}{25}{27.6} & \DMS{-43}{01}{09} & FR1  & $0.00183$  & 30 \\
\enddata

\tablenotetext{a}{See notes for Table~\ref{TAB::OBJ_UL} for explanation of object types.}
\tablenotetext{b}{The redshift of 3C~66A is considered to be uncertain.}
\tablenotetext{c}{Detection of E$>$1\,TeV emission from 3C~66A and
BL~Lacertae was first claimed by
\citet{REF::TEV_3C66A_CRIMEA,REF::TEV_BL_LACERTAE_CRIMEA}. 
The measured fluxes are not consistent with the later measurements
made with more sensitive instruments.}
\tablerefs{(1) \citet{REF::TEV_RGB_J0152+017};
(2) \citet{REF::TEV_3C66A_CRIMEA};
(3) \citet{REF::TEV_3C66A_VERITAS};
(4) \citet{REF::TEV_1ES_0229+20};
(5) \citet{REF::TEV_1ES_0347-121};
(6) \citet{REF::TEV_PKS_0548-322};
(7) \citet{REF::TEV_RGB_J0710+591};
(8) \citet{REF::TEV_S5_0716+714};
(9) \citet{REF::TEV_1ES_0806+524};
(10) \citet{REF::TEV_1ES_1011+496};
(11) \citet{REF::TEV_1ES_1101-232};
(12) \citet{REF::TEV_MARKARIAN_421};
(13) \citet{REF::TEV_MARKARIAN_180};
(14) \citet{REF::TEV_1ES_1218+304};
(15) \citet{REF::TEV_W_COMAE};
(16) \citet{REF::TEV_3C279};
(17) \citet{REF::TEV_PKS_1424}
(18) \citet{REF::TEV_H_1426+428};
(19) \citet{REF::TEV_PG_1553+113};
(20) \citet{REF::TEV_MARKARIAN_501};
(21) \citet{REF::TEV_1ES_1959+650};
(22) \citet{REF::TEV_PKS_2005-489};
(23) \citet{REF::TEV_PKS_2155-304};
(24) \citet{REF::TEV_BL_LACERTAE_CRIMEA};
(25) \citet{REF::TEV_BL_LACERTAE_MAGIC};
(26) \citet{REF::TEV_1ES_2344+514};
(27) \citet{REF::TEV_H_2356-309};
(28) \citet{REF::TEV_3C66B_MAGIC};
(29) \citet{REF::TEV_M_87};
(30) \citet{REF::TEV_CEN_A}.
}
\end{deluxetable}

%
%

\begin{deluxetable}{llcccll}
\tabletypesize{\scriptsize}
\tablecaption{\label{TAB::OBJ_DET_SPEC} Flux ($\phi$), photon index 
($\Gamma$) from measurements of AGN with TeV instruments, along with
threshold energy ($E_\mathrm{thres}$) of the observation. The
differential flux at 200\,GeV ($F_{200}$) is also calculated from the
TeV spectrum for comparison between objects.  See text for further
details.}
\tablewidth{0pt}
\tablehead{
\colhead{Name} &
\colhead{$E_\mathrm{thres}$} &
\colhead{$\phi(>E_\mathrm{thres})$} &
\colhead{$\Gamma$} &
\colhead{$F_{200}$} &
\colhead{Note} &
\colhead{Ref}\\
&
\colhead{[GeV]}  &
\colhead{[$10^{-11}$\cmsc]} &
\colhead{[1]}& 
\colhead{[$10^{-9}$\cmsc TeV$^{-1}$]} & 
&
}
\startdata
\multicolumn{2}{l}{\textbf{Blazars:}} \\
\objectname{RGB J0152+017} & 300 &  $ 0.27   \pm 0.05  $  & $2.95 \pm 0.36$ & 0.058   & No variability        &1  \\
\objectname{3C 66A}        & 200 &  $ 1.3    \pm 0.1   $  & $4.1  \pm 0.4 $ & 0.201   & Flaring state         &2  \\
\objectname{1ES 0229+200}  & 580 &  $ 0.094  \pm 0.015 $  & $2.50 \pm 0.19$ & 0.003   & No variability        &3  \\
\objectname{1ES 0347-121}  & 250 &  $ 0.39   \pm 0.01  $  & $3.10 \pm 0.23$ & 0.065   & No variability        &4  \\
\objectname{PKS 0548-322}  & 200 &  $ 0.33   \pm 0.07  $  & $2.8  \pm 0.3 $ & 0.030   &                       &5  \\
\objectname{RGB J0710+591} & 300 &$\approx 0.016$\,\Icrabc&\nodata          & \nodata & ATEL 1941             &6  \\
\objectname{S5 0716+714}   & 400 &  $\approx 1\,$\Icrabc  &\nodata          & \nodata & ATEL 1502             &7  \\
\objectname{1ES 0806+524}  & 300 &  $ 0.22   \pm 0.05  $  & $3.6  \pm 1.0 $ & 2.231   & Low flux state?       &8  \\
\objectname{1ES 1011+496}  & 200 &  $ 1.58   \pm 0.32  $  & $4.0  \pm 0.5 $ & 0.237   & Flaring state         &9  \\
\objectname{1ES 1101-232}  & 225 &  $ 0.52   \pm 0.14  $  & $2.94 \pm 0.20$ & 0.063   & No variability        &10 \\
\objectname{Markarian 421} & 383 &  $ 573.3  \pm 57.9  $  & $2.31 \pm 0.04$ & 87.96   & Flaring state         &11 \\
                           & 200 &  $ 26.2   \pm 2.1   $  & $2.20 \pm 0.08$ & 1.572   & Lowest flux state     &12 \\
\objectname{Markarian 180} & 200 &  $ 2.3    \pm 0.7   $  & $3.3  \pm 0.7 $ & 0.264   & Flaring state         &13 \\
\objectname{1ES 1218+304}  & 200 &  $ 1.22   \pm 0.26  $  & $3.08 \pm 0.34$ & 0.125   &                       &14 \\
\objectname{W Comae}       & 200 &  $ 1.99   \pm 0.07  $  & $3.81 \pm 0.35$ & 0.280   & Flaring state         &15 \\
\objectname{3C 279}        & 100 &  $51.5    \pm 8.2   $  & $4.11  \pm 0.68 $ & 0.931 & Flaring state         &16 \\
\objectname{PKS 1424+240}  & 200 &$\approx 0.02$\,\Icrabc & \nodata         & \nodata & ATEL 2084             &17 \\
\objectname{H 1426+428}    & 280 &  $ 2.4    \pm 4.1   $  & $3.50 \pm 0.35$ & 0.696   &                       &18,19\tablenotemark{a}\\
\objectname{PG 1553+113}   & 200 &  $ 4.8    \pm 1.3   $  & $4.0  \pm 0.6 $ & 0.720   & Low flux state?       &20 \\
\objectname{Markarian 501} & 300 &  $30.3   \pm 1.9  $  & $2.22 \pm 0.04$ & 3.03   & Flaring state         &21 \\
                           & 150 &  $12.4    \pm 0.8   $  & $2.45 \pm 0.07$ & 0.592   & Lowest flux state     &22 \\
\objectname{1ES 1959+650}  & 1300 &  $ 2.50  \pm 0.46  $  & $2.83 \pm 0.14$ & 7.030   & Orphan flare          &23 \\
                           & 150 &  $ 3.42   \pm 0.92  $  & $2.58 \pm 0.18$ & 0.171   & Low flux state        &24 \\
\objectname{PKS 2005-489}  & 200 &  $ 0.62   \pm 0.1   $  & $4.0  \pm 0.4 $ & 0.093   &                       &25 \\
\objectname{PKS 2155-304}  & 200 &  $ 172    \pm 0.05  $  & $3.19 \pm 0.02$\tablenotemark{b} & 20.6  & Flaring state  &26 \\
                           & 300 &  $ 4.2    \pm 0.75  $  & $3.32 \pm 0.06$ & 1.248   & Low flux state        &27 \\
\objectname{BL Lacertae}   & 200 &  $ 0.6    \pm 0.2   $  & $3.6  \pm 0.5 $ & 0.078   &                       &28 \\
\objectname{1ES 2344+514}  & 200 &  $ 2.39   \pm 0.3   $  & $2.95 \pm 0.12$ & 0.233   & Low flux state        &29 \\
\objectname{H 2356-309}    & 200 &  $ 0.41   \pm 0.05  $  & $3.09 \pm 0.24$ & 0.043   & No variability        &30 \\
\\
\multicolumn{6}{l}{\textbf{Others}} \\
\objectname{3C 66B}        & 150 &  $ 0.73   \pm 0.15  $  & $3.1  \pm 0.31$ & 0.042   & Flaring state         &31 \\
\objectname{M 87}          & 730 &  $ 0.025  \pm 0.03  $  & $2.62 \pm 0.35$ & 0.016   &                       &32 \\
\objectname{Centaurus A}   & 250 &  $ 0.156  \pm 0.067 $  & $2.73 \pm 0.45$ & 0.019   & Low flux state?       &33 \\
\enddata
\tablenotetext{a}{The flux reported by the HEGRA collaboration is 
0.08\,\Icrab above 1 TeV.}
\tablenotetext{b}{A broken power law with $\Gamma_1=2.71 \pm 0.06$, $
\Gamma_2=3.53\pm0.05$ and $E_\mathrm{break}=430 \pm 22$\,GeV was 
preferred.}

\tablerefs{(1) \citet{REF::TEV_RGB_J0152+017};
(2) \citet{REF::TEV_3C66A_VERITAS};
(3) \citet{REF::TEV_1ES_0229+20};
(4) \citet{REF::TEV_1ES_0347-121};
(5) \citet{REF::TEV_PKS_0548-322};
(6) \citet{REF::TEV_RGB_J0710+591};
(7) \citet{REF::TEV_S5_0716+714};
(8) \citet{REF::TEV_1ES_0806+524};
(9) \citet{REF::TEV_1ES_1011+496};
(10) \citet{REF::TEV_1ES_1101-232};
(11) \citet{REF::TEV_MRK421_WHIPPLE_HS};
(12) \citet{REF::TEV_MARKARIAN_421_BIS};
(13) \citet{REF::TEV_MARKARIAN_180};
(14) \citet{REF::TEV_1ES_1218+304_VERITAS};
(15) \citet{REF::TEV_W_COMAE};
(16) \citet{REF::TEV_3C279};
(17) \citet{REF::TEV_PKS_1424}
(18) \citet{REF::TEV_H_1426+428};
(19) \citet{REF::TEV_H_1426+428_HEGRA};
(20) \citet{REF::TEV_PG_1553+113};
(21) \citet{REF::TEV_MARKARIAN_501_WHIPPLE_SPECT}
(22) \citet{REF::TEV_MARKARIAN_501_MAGIC};
(23) \citet{REF::ORPHAN_FLARE};
(24) \citet{REF::TEV_1ES_1959+650_MAGIC_2008};
(25) \citet{REF::TEV_PKS_2005-489};
(26) \citet{REF::TEV_PKS_FLARE};
(27) \citet{REF::TEV_PKS_2155-304_HESS_2003};
(28) \citet{REF::TEV_BL_LACERTAE_MAGIC};
(29) \citet{REF::TEV_1ES_2344+514_MAGIC_2007};
(30) \citet{REF::TEV_H_2356-309};
(31) \citet{REF::TEV_3C66B_MAGIC};
(32) \citet{REF::TEV_M_87_HESS_2006};
(33) \citet{REF::TEV_CEN_A}.
}
\end{deluxetable}

%
%

\newcommand{\ULRBENBEIGHT}{1}
\newcommand{\ULRDLCP}{2}
\newcommand{\ULRTLUCZ}{3}
\newcommand{\ULRALBERT}{4}
\newcommand{\ULRBENBFIVE}{5}
\newcommand{\ULRHORAN}{6}
\newcommand{\ULRFALCONE}{7}

\begin{deluxetable}{llllllllll}
\tabletypesize{\scriptsize}
\tablecaption{AGN with published upper limits at TeV energies\label{TAB::OBJ_UL}.}
\tablewidth{0pt}
\tablehead{
&&&&& \multicolumn{2}{c}{TeV limit} & \\
\colhead{Name} &
\colhead{$\alpha_{\mathrm{J2000}}$} & 
\colhead{$\delta_{\mathrm{J2000}}$} &
\colhead{Type\tablenotemark{a}} &
\colhead{$z$} &
\colhead{Flux} & \colhead{Energy} &
\colhead{Ref} \\
&&&&& \colhead{[\Icrabc]} & \colhead{[GeV]} & }
\startdata

\objectname{III Zw 2}         & \HMS{00}{10}{31.0} & \DMS{+10}{58}{30} & Sy1    & $0.089$    & $<$0.027 & $>$430  & \ULRBENBEIGHT \\
\objectname{1ES 0033+595}     & \HMS{00}{35}{52.6} & \DMS{+59}{50}{05} & HBL    & $0.086$\tablenotemark{b}    & $<$0.11  & $>$390  & \ULRDLCP \\
\objectname{NGC 315}          & \HMS{00}{57}{48.9} & \DMS{+30}{21}{09} & FR1    & $0.016$    & $<$0.05  & $>$860  & \ULRTLUCZ \\
\objectname{4C+31.04}         & \HMS{01}{19}{35.0} & \DMS{+32}{10}{50} & RG     & $0.060$    & $<$0.14  & $>$760  & \ULRTLUCZ \\
\objectname{1ES 0120+340}     & \HMS{01}{23}{08.6} & \DMS{+34}{20}{49} & HBL    & $0.272$    & $<$0.032 & $>$190  & \ULRALBERT \\
\objectname{1ES 0145+138}     & \HMS{01}{48}{29.8} & \DMS{+14}{02}{19} & HBL    & $0.125$    & $<$0.015 & $>$310  & \ULRBENBFIVE \\
\objectname{UGC 01651}        & \HMS{02}{09}{38.6} & \DMS{+35}{47}{50} & RG     & $0.038$    & $<$0.07  & $>$790  & \ULRTLUCZ \\
\objectname{BWE 0210+1159}    & \HMS{02}{13}{05.2} & \DMS{+12}{13}{11} & LBL    & $0.250$    & $<$0.012 & $>$530  & \ULRBENBEIGHT \\
\objectname{RGB J0214+517}    & \HMS{02}{14}{17.9} & \DMS{+51}{44}{52} & HBL    & $0.049$    & $<$0.17  & $>$430  & \ULRHORAN \\
\objectname{PKS 0219-164}     & \HMS{02}{22}{01.0} & \DMS{-16}{15}{17} & LBL & $0.698$       & $<$0.27  & $>$1780 & \ULRTLUCZ \\
\objectname{NGC 1054}         & \HMS{02}{42}{15.7} & \DMS{+18}{13}{02} & Sy     & $0.033$    & $<$0.02  & $>$860  & \ULRTLUCZ \\
\objectname{NGC 1068}         & \HMS{02}{42}{40.7} & \DMS{-00}{00}{48} & Sy2    & $0.004$    & $<$0.013 & $>$210  & \ULRBENBFIVE \\
\objectname{V Zw 331}         & \HMS{03}{13}{57.6} & \DMS{+41}{15}{24} & LBL    & $0.029$    & $<$0.09  & $>$870  & \ULRTLUCZ \\ 
\objectname{NGC 1275}         & \HMS{03}{19}{48.2} & \DMS{+41}{30}{42} & FR1    & $0.018$    & $<$0.03  & $>$850  & \ULRTLUCZ \\
\objectname{RX J0319.8+1845}  & \HMS{03}{19}{51.8} & \DMS{+18}{45}{34} & HBL    & $0.190$    & $<$0.033 & $>$190  & \ULRALBERT \\
\objectname{B2 0321+33B}      & \HMS{03}{24}{41.2} & \DMS{+34}{10}{46} & NLSy1  & $0.063$    & $<$0.10  & $>$400  & \ULRFALCONE \\
\objectname{1ES 0323+022}     & \HMS{03}{26}{13.9} & \DMS{+02}{25}{15} & HBL    & $0.147$    & $<$0.015 & $>$210  & \ULRBENBFIVE \\
\objectname{4C +37.11}        & \HMS{04}{05}{49.3} & \DMS{+38}{03}{32} & RG     & $0.055$    & $<$0.05  & $>$800  & \ULRTLUCZ \\
\objectname{1ES 0414+009}     & \HMS{04}{16}{52.3} & \DMS{+01}{05}{54} & HBL    & $0.287$    & $<$0.057 & $>$230  & \ULRALBERT \\
\objectname{3C 120}           & \HMS{04}{33}{11.1} & \DMS{+05}{21}{16} & RG     & $0.033$    & $<$0.004 & $>$230  & \ULRBENBFIVE \\
\objectname{MG J0509+0541}    & \HMS{05}{09}{26.0} & \DMS{+05}{41}{35} & IBL    & \nodata    & $<$0.11  & $>$960  & \ULRTLUCZ \\
\objectname{4C+01.13}         & \HMS{05}{13}{52.5} & \DMS{+01}{57}{10} & BL~Lac & $0.084$    & $<$0.10  & $>$1010 & \ULRTLUCZ \\
\objectname{Pictor A}         & \HMS{05}{19}{49.7} & \DMS{-45}{46}{45} & FR2    & $0.034$    & $<$0.014 & $>$220  & \ULRBENBFIVE \\
\objectname{PKS B0521-365}    & \HMS{05}{22}{58.0} & \DMS{-36}{27}{31} & LBL    & $0.055$    & $<$0.042 & $>$310  & \ULRBENBEIGHT \\
\objectname{EXO 0556.4-3838}  & \HMS{05}{58}{06.2} & \DMS{-38}{38}{27} & HBL    & $0.302$    & $<$0.051 & $>$220  & \ULRBENBFIVE \\
\objectname{PKS 0558-504}     & \HMS{05}{59}{47.4} & \DMS{-50}{26}{52} & NLSy1  & $0.137$    & $<$0.018 & $>$310  & \ULRBENBEIGHT \\
\objectname{1ES 0647+250}     & \HMS{06}{50}{46.6} & \DMS{+25}{03}{00} & HBL    & $0.203$    & $<$0.13  & $>$780  & \ULRTLUCZ \\
\objectname{UGC 03927}        & \HMS{07}{37}{30.1} & \DMS{+59}{41}{03} & UnC    & $0.041$    & $<$0.09  & $>$1090 & \ULRTLUCZ \\
\objectname{3C 192.0}         & \HMS{08}{05}{35.0} & \DMS{+24}{09}{50} & RG     & $0.060$    & $<$0.20  & $>$930  & \ULRTLUCZ \\
\objectname{RGB J0812+026}    & \HMS{08}{12}{01.9} & \DMS{+02}{37}{33} & BL~Lac & \nodata    & $<$0.031 & $>$220  & \ULRBENBFIVE \\
\objectname{3C 197.1}         & \HMS{08}{21}{33.6} & \DMS{+47}{02}{37} & FR2    & $0.130$    & $<$0.05  & $>$960  & \ULRTLUCZ \\
\objectname{PKS 0829+046}     & \HMS{08}{31}{48.9} & \DMS{+04}{29}{39} & LBL    & $0.174$    & $<$0.06  & $>$1000 & \ULRTLUCZ \\
\objectname{NGC 2622}         & \HMS{08}{38}{11.0} & \DMS{+24}{53}{43} & Sy1    & $0.028$    & $<$0.05  & $>$400  & \ULRFALCONE \\
\objectname{1ES 0927+500}     & \HMS{09}{30}{37.6} & \DMS{+49}{50}{26} & HBL    & $0.188$    & $<$0.052 & $>$230  & \ULRALBERT \\
\objectname{S4 0954+65}       & \HMS{09}{58}{47.2} & \DMS{+65}{33}{55} & LBL    & $0.368$    & $<$0.096 & $>$300  & \ULRHORAN \\
\objectname{MS1019.0+5139}    & \HMS{10}{22}{12.6} & \DMS{+51}{24}{00} & BL~Lac & $0.141$    & $<$0.07  & $>$920  & \ULRTLUCZ \\
\objectname{1ES 1028+511}     & \HMS{10}{31}{18.5} & \DMS{+50}{53}{36} & HBL    & $0.361$    & $<$0.29  & $>$400  & \ULRHORAN \\
\objectname{RGB J1117+202}    & \HMS{11}{17}{06.3} & \DMS{+20}{14}{07} & HBL    & $0.139$    & $<$0.030 & $>$610  & \ULRBENBFIVE \\
\objectname{1ES 1118+424}     & \HMS{11}{20}{48.0} & \DMS{+42}{12}{12} & HBL    & $0.124$    & $<$0.12  & $>$430-500 & \ULRHORAN \\
\objectname{Markarian 40}     & \HMS{11}{25}{36.2} & \DMS{+54}{22}{57} & Sy1    & $0.021$    & $<$0.21  & $>$430  & \ULRHORAN \\
\objectname{NGC 3783}         & \HMS{11}{39}{01.7} & \DMS{-37}{44}{19} & Sy1    & $0.010$    & $<$0.025 & $>$220  & \ULRBENBFIVE \\
\objectname{NGC 4151}         & \HMS{12}{10}{32.6} & \DMS{+39}{24}{21} & Sy1.5  & $0.003$    & $<$0.07  & $>$790  & \ULRTLUCZ \\
\objectname{1ES 1212+078}     & \HMS{12}{15}{11.2} & \DMS{+07}{32}{05} & HBL    & $0.136$    & $<$0.17  & $>$920  & \ULRTLUCZ \\
\objectname{ON 325}           & \HMS{12}{17}{52.1} & \DMS{+30}{07}{01} & LBL    & $0.130$    & $<$0.22  & $>$400-430 & \ULRHORAN \\
\objectname{3C 273}           & \HMS{12}{29}{06.7} & \DMS{+02}{03}{09} & FSRQ   & $0.158$    & $<$0.014 & $>$300  & \ULRBENBEIGHT \\
\objectname{MS 1229.2+6430}   & \HMS{12}{31}{31.4} & \DMS{+64}{14}{18} & HBL    & $0.164$    & $<$0.17  & $>$300-430 & \ULRHORAN \\
\objectname{1ES 1239+069}     & \HMS{12}{41}{48.3} & \DMS{+06}{36}{01} & HBL    & $0.150$    & $<$0.20  & $>$400-430 & \ULRHORAN \\
\objectname{1ES 1255+244}     & \HMS{12}{57}{31.9} & \DMS{+24}{12}{40} & HBL    & $0.141$    & $<$0.11  & $>$350-500 & \ULRHORAN \\
\objectname{RGB J1413+436}    & \HMS{14}{13}{43.7} & \DMS{+43}{39}{45} & RG     & $0.089$    & $<$0.06  & $>$400  & \ULRFALCONE \\
\objectname{RX J1417.9+2543}  & \HMS{14}{17}{56.7} & \DMS{+25}{43}{26} & HBL    & $0.237$    & $<$0.023 & $>$190  & \ULRALBERT \\
\objectname{OQ530}            & \HMS{14}{19}{46.6} & \DMS{+54}{23}{15} & LBL    & $0.151$    & $<$0.058 & $>$300  & \ULRHORAN \\
\objectname{1ES 1440+122}     & \HMS{14}{42}{48.3} & \DMS{+12}{00}{40} & HBL    & $0.162$    & $<$0.033 & $>$290  & \ULRBENBFIVE \\
\objectname{RGB J1629+401}    & \HMS{16}{29}{01.3} & \DMS{+40}{08}{00} & NLSy1  & $0.271$    & $<$0.09  & $>$400  & \ULRFALCONE \\
\objectname{RX J1725.0+1152}  & \HMS{17}{25}{04.4} & \DMS{+11}{52}{15} & HBL    & $0.018$    & $<$0.046 & $>$190  & \ULRALBERT \\
\objectname{I Zw 187}         & \HMS{17}{28}{18.6} & \DMS{+50}{13}{10} & HBL    & $0.055$    & $<$0.086 & $>$300-350 & \ULRHORAN \\
\objectname{1ES 1741+196}     & \HMS{17}{43}{57.8} & \DMS{+19}{35}{09} & HBL    & $0.083$    & $<$0.053 & $>$350-500 & \ULRHORAN \\
\objectname{3C 371}           & \HMS{18}{06}{50.7} & \DMS{+69}{49}{28} & LBL    & $0.051$    & $<$0.19  & $>$300  & \ULRHORAN \\
\objectname{Cyg A}            & \HMS{19}{59}{28.4} & \DMS{+40}{44}{02} & FR2    & $0.056$    & $<$0.03  & $>$910  & \ULRTLUCZ \\
\objectname{PKS 2201+04}      & \HMS{22}{04}{17.7} & \DMS{+04}{40}{02} & BL~Lac & $0.028$    & $<$0.08  & $>$950  & \ULRTLUCZ \\
\objectname{PG 2209+184}      & \HMS{22}{11}{53.9} & \DMS{+18}{41}{50} & RG     & $0.070$    & $<$0.13  & $>$400  & \ULRFALCONE \\
\objectname{RBS 1888}         & \HMS{22}{43}{41.6} & \DMS{-12}{31}{38} & HBL    & $0.226$    & $<$0.009 & $>$170  & \ULRBENBFIVE \\
\objectname{HS 2250+1926}     & \HMS{22}{53}{07.4} & \DMS{+19}{42}{35} & FSRQ   & $0.284$    & $<$0.009 & $>$590  & \ULRBENBEIGHT \\
\objectname{2QZ J225453.2-272509}&\HMS{22}{54}{53.2}&\DMS{-27}{25}{09} & BL~Lac & $0.333$    & $<$0.016 & $>$170  & \ULRBENBFIVE \\
\objectname{PKS 2254+074}     & \HMS{22}{57}{17.3} & \DMS{+07}{43}{12} & LBL    & $0.193$    & $<$0.05  & $>$900  & \ULRTLUCZ \\
\objectname{NGC 7469}         & \HMS{23}{03}{15.6} & \DMS{+08}{52}{26} & Sy1    & $0.017$    & $<$0.006 & $>$250  & \ULRBENBFIVE \\
\objectname{PKS 2316-423}     & \HMS{23}{19}{05.8} & \DMS{-42}{06}{49} & HBL    & $0.055$    & $<$0.014 & $>$190  & \ULRBENBFIVE \\
\objectname{1ES 2321+419}     & \HMS{23}{23}{52.5} & \DMS{+42}{10}{55} & HBL    & $0.059$    & $<$0.03  & $>$890  & \ULRTLUCZ \\
\objectname{1ES 2343-151}     & \HMS{23}{45}{38.4} & \DMS{-14}{49}{29} & IBL    & $0.224$    & $<$0.012 & $>$230  & \ULRBENBEIGHT \\
\enddata
\tablenotetext{a}{BL~Lac, LBL, IBL, HBL -- BL~Lac object;
FSRQ -- Flat Spectrum Radio Quasar;
FR1 \& 2 -- Fanaroff-Riley 1 \& 2 galaxy;
Sy 1, 1.5 \& 2 -- Seyfert 1, 1.5 \& 2;
RG -- Radio Galaxy;
NLSy1 -- Narrow-Line Seyfert 1;
UnC -- Unclassified;
see the SIMBAD (\url{http://simbad.u-strasbg.fr/simbad}) and 
NED (\url{http://nedwww.ipac.caltech.edu}) databases.}
\tablenotetext{b}{Tentative measurement, see comment in text.}

\tablerefs{(\ULRBENBEIGHT) \citet{REF::UL_HESS_BENBOW08};
(\ULRDLCP) \citet{REF::UL_WHIPPLE_DLCP};
(\ULRTLUCZ) \citet{REF::UL_HEGRA_TLUCZYKONT};
(\ULRALBERT) \citet{REF::UL_MAGIC_ALBERT};
(\ULRBENBFIVE) \citet{REF::UL_HESS_BENBOW05};
(\ULRHORAN) \citet{REF::UL_WHIPPLE_HORAN};
(\ULRFALCONE) \citet{REF::UL_WHIPPLE_FALCONE}.}
\end{deluxetable}

%
%

\begin{deluxetable}{lll}
\tablecaption{Target objects with associations in the \ZFGL
source list and 3EG catalog.\label{TAB::OBJ_LBAS}.}
\tablewidth{0pt}
\tablehead{\colhead{TeV name} &\colhead{\ZFGL name} &\colhead{3EG name}}
\startdata
\multicolumn{2}{l}{\textbf{TeV detected:}} \\
\objectname{3C 66A}          & \objectname{0FGL J0222.6$+$4302} & \objectname{3EG J0222$+$4253} \\
\objectname{S5 0716+714}     & \objectname{0FGL J0722.0$+$7120} & \objectname{3EG J0721$+$7120} \\
\objectname{1ES 1011+496}    & \objectname{0FGL J1015.2$+$4927} & \nodata                       \\
\objectname{Markarian 421}   & \objectname{0FGL J1104.5$+$3811} & \objectname{3EG J1104$+$3809}\tablenotemark{a} \\
\objectname{W Comae}         & \objectname{0FGL J1221.7$+$2814} & \objectname{3EG J1222$+$2841} \\
\objectname{3C 279}          & \objectname{0FGL J1256.1$-$0547} & \objectname{3EG J1255$-$0549} \\
\objectname{PKS 1424+240}    & \objectname{0FGL J1427.1+2347}   & \nodata                       \\
\objectname{PG 1553+113}     & \objectname{0FGL J1555.8$+$1110} & \nodata                       \\
\objectname{Markarian 501}   & \objectname{0FGL J1653.9$+$3946} & \nodata                       \\
\objectname{1ES 1959+650}    & \objectname{0FGL J2000.2$+$6506} & \nodata                       \\
\objectname{PKS 2005-489}    & \objectname{0FGL J2009.4$-$4850} & \nodata                       \\
\objectname{PKS 2155-304}    & \objectname{0FGL J2158.8$-$3014} & \objectname{3EG J2158$-$3023}\tablenotemark{a} \\
\objectname{BL Lacertae}     & \objectname{0FGL J2202.4$+$4217} & \objectname{3EG J2202$+$4217} \\
\objectname{Centaurus A}     & \objectname{0FGL J1325.4$-$4303} & \objectname{3EG J1324$-$4314} \\
\\
\multicolumn{2}{l}{\textbf{TeV non-detected:}} \\
\objectname{1ES 0033+595}    & \objectname{0FGL J0036.7$+$5951} & \nodata                       \\
\objectname{ON 325}          & \objectname{0FGL J1218.0$+$3006} & \nodata                       \\
\objectname{3C 273}          & \objectname{0FGL J1229.1$+$0202} & \objectname{3EG J1229$+$0210} \\
\objectname{NGC 1275}        & \objectname{0FGL J0320.0$+$4131} & \nodata                       
\enddata
\tablenotetext{a}{A well established TeV source at the time of the 3EG catalog.}
\end{deluxetable}

%
%

\begin{deluxetable}{lrccccccc}
\tabletypesize{\scriptsize}
\tablecaption{\AFermiLAT detections (0.2\,GeV--300\,GeV)\label{TAB::OBJ_RES}.}
\tablewidth{0pt}
\tablehead{
& \multicolumn{4}{c}{Parameters of fitted power-law spectrum} &
\multicolumn{2}{c}{Highest energy} & \multicolumn{2}{c}{Probability
of} \\
\colhead{Name} & \colhead{TS} & \colhead{Flux ($>$200\,MeV)} 
& \colhead{Photon Index} & \colhead{Decorr.}
& \multicolumn{2}{c}{photons} & \multicolumn{2}{c}{constant flux} \\
& & \colhead{$F\pm\ \Delta F_\mathrm{stat}\pm\Delta F_\mathrm{sys}$} 
& \colhead{$\Gamma\pm\Delta\Gamma_\mathrm{stat}\pm\Delta\Gamma_\mathrm{sys}$}
& \colhead{energy}
& \colhead{$1^\mathrm{st}$} & \colhead{$5^\mathrm{th}$}
& \colhead{10 day} & \colhead{28 day} \\
& [1] & \colhead{[$10^{-9}$\cmsc]} & [1] 
& \colhead{[GeV]} & \colhead{[GeV]} & \colhead{[GeV]} & [1] & [1]
}
\startdata
\textbf{TeV detected:} \\
\objectname{3C 66A}           & 2221 &  96.7 $\pm$ 5.82 $\pm$ 3.39  &  1.93 $\pm$ 0.04 $\pm$ 0.04  & 1.54 & 111\tablenotemark{a} &  54 & $<0.01$  & $<0.01$\\
\objectname{RGB J0710+591}    &   42 & 0.087 $\pm$0.049 $\pm$ 0.076 &  1.21 $\pm$ 0.25 $\pm$ 0.02  &15.29 &  74 &   4 &  $0.98$  &  $0.94$\\
\objectname{S5 0716+714}      & 1668 &  79.9 $\pm$ 4.17 $\pm$ 2.84  &  2.16 $\pm$ 0.04 $\pm$ 0.05  & 0.82 &  63 &   9 & $<0.01$  & $<0.01$\\
\objectname{1ES 0806+524}     &  102 &  2.07 $\pm$ 0.38 $\pm$ 0.71  &  2.04 $\pm$ 0.14 $\pm$ 0.03  & 1.54 &  30 &   4 &  $0.05$  & $<0.01$\\
\objectname{1ES 1011+496}     &  889 &  32.0 $\pm$ 0.27 $\pm$ 0.29  &  1.82 $\pm$ 0.05 $\pm$ 0.03  & 1.50 & 168 &  32 &  $0.54$  &  $0.50$\\
\objectname{Markarian 421}    & 3980 &  94.3 $\pm$ 3.88 $\pm$ 2.60  &  1.78 $\pm$ 0.03 $\pm$ 0.04  & 1.35 & 801 & 155 &  $0.06$  &  $0.02$\\
\objectname{Markarian 180}    &   50 &  5.41 $\pm$ 1.69 $\pm$ 0.91  &  1.91 $\pm$ 0.18 $\pm$ 0.09  & 1.95 &  14 &   2 &  $0.98$  &  $0.54$\\
\objectname{1ES 1218+304}     &  147 &  7.56 $\pm$ 2.16 $\pm$ 0.67  &  1.63 $\pm$ 0.12 $\pm$ 0.04  & 5.17 & 356 &  31 &  $0.53$  &  $0.06$\\
\objectname{W Comae}          &  754 &  41.7 $\pm$ 3.40 $\pm$ 2.46  &  2.02 $\pm$ 0.06 $\pm$ 0.05  & 1.13 &  26 &  18 &  $0.01$  & $<0.01$\\
\objectname{3C 279}           & 6865 &  287  $\pm$ 7.13 $\pm$ 10.2  &  2.34 $\pm$ 0.03 $\pm$ 0.04  & 0.59 &  28 &  21 & $<0.01$  & $<0.01$\\
\objectname{PKS 1424+240}     &  800 & 34.35 $\pm$ 2.60 $\pm$ 1.37  &  1.85 $\pm$ 0.05 $\pm$ 0.04  & 1.50 & 137 &  30 & $<0.01$  &  $0.16$\\
\objectname{H 1426+428}       &   38 &  1.56 $\pm$ 1.05 $\pm$ 0.29  &  1.47 $\pm$ 0.30 $\pm$ 0.11  & 8.33 &  19 &   3 &  $0.83$  &  $0.39$\\
\objectname{PG 1553+113}      & 2009 &  54.8 $\pm$ 3.63 $\pm$ 0.85  &  1.69 $\pm$ 0.04 $\pm$ 0.04  & 2.32 & 157 &  76 &  $0.40$  &  $0.54$\\
\objectname{Markarian 501}    &  649 &  22.4 $\pm$ 2.52 $\pm$ 0.13  &  1.73 $\pm$ 0.06 $\pm$ 0.04  & 2.22 & 127 &  50 &  $0.57$  &  $0.18$\\
\objectname{1ES 1959+650}     &  306 &  25.1 $\pm$ 3.49 $\pm$ 2.83  &  1.99 $\pm$ 0.09 $\pm$ 0.07  & 1.60 &  75 &  21 &  $0.91$  &  $0.29$\\
\objectname{PKS 2005-489}     &  246 &  22.3 $\pm$ 3.09 $\pm$ 2.14  &  1.91 $\pm$ 0.09 $\pm$ 0.08  & 1.01 &  71 &   8 &  $0.86$  &  $0.97$\\
\objectname{PKS 2155-304}     & 3354 &  109  $\pm$ 4.45 $\pm$ 3.18  &  1.87 $\pm$ 0.03 $\pm$ 0.04  & 1.13 & 299 &  46 & $<0.01$  & $<0.01$\\
\objectname{BL Lacertae}      &  310 &  51.6 $\pm$ 5.81 $\pm$ 12.2  &  2.43 $\pm$ 0.10 $\pm$ 0.08  & 0.85 &  70 &   4 &  $0.61$  &  $0.23$\\
\objectname{1ES 2344+514}     &   37 &  3.67 $\pm$ 2.35 $\pm$ 1.62  &  1.76 $\pm$ 0.27 $\pm$ 0.23  & 5.28 &  53 &   3 &  $0.76$  &  $0.46$\\
\objectname{M 87}             &   31 &  7.56 $\pm$ 2.70 $\pm$ 2.24  &  2.30 $\pm$ 0.26 $\pm$ 0.14  & 1.11 &   8 &   1 &  $0.43$  &  $0.57$\\
\objectname{Centaurus A}      &  308 &  70.8 $\pm$ 5.97 $\pm$ 5.80  &  2.90 $\pm$ 0.11 $\pm$ 0.07  & 0.47 &   6 &   4 &  $0.38$  &  $0.97$\\
\\
\textbf{TeV non-detected:} \\
\objectname{1ES 0033+595}     &  137 &  20.3 $\pm$ 5.11 $\pm$ 1.74  &  2.00 $\pm$ 0.13 $\pm$ 0.07  & 2.58 & 150 &  16 &  $0.40$  &  $0.01$\\
\objectname{MG J0509+0541}    &  217 &  19.7 $\pm$ 3.78 $\pm$ 0.70  &  2.01 $\pm$ 0.11 $\pm$ 0.06  & 1.95 &  31 &  12 &  $0.73$  &  $0.23$\\
\objectname{PKS B0521-365}    &  148 &  26.6 $\pm$ 3.50 $\pm$ 3.34  &  2.52 $\pm$ 0.13 $\pm$ 0.10  & 0.64 &   7 &   2 &  $0.03$  &  $0.11$\\
\objectname{1ES 0647+250}     &   95 &  4.09 $\pm$ 1.39 $\pm$ 1.01  &  1.66 $\pm$ 0.15 $\pm$ 0.09  & 4.54 & 247 &  16 &  $0.30$  &  $0.72$\\
\objectname{PKS 0829+046}     &  187 &  27.3 $\pm$ 3.37 $\pm$ 1.08  &  2.43 $\pm$ 0.11 $\pm$ 0.04  & 0.70 &   4 &   2 &  $0.38$  &  $0.11$\\
\objectname{1ES 1028+511}     &   52 &  3.88 $\pm$ 1.43 $\pm$ 0.57  &  1.72 $\pm$ 0.19 $\pm$ 0.08  & 3.07 &  48 &   2 &  $0.85$  &  $0.31$\\
\objectname{RGB J1117+202}    &  116 &  7.12 $\pm$ 1.75 $\pm$ 0.36  &  1.79 $\pm$ 0.13 $\pm$ 0.06  & 2.37 &  46 &   6 &  $0.63$  &  $0.97$\\
\objectname{1ES 1118+424}     &   33 &  2.31 $\pm$ 1.27 $\pm$ 0.41  &  1.71 $\pm$ 0.26 $\pm$ 0.08  & 3.99 &  27 &   3 &  $0.64$  &  $0.69$\\
\objectname{ON 325}           &  761 &  42.3 $\pm$ 3.68 $\pm$ 2.98  &  1.99 $\pm$ 0.06 $\pm$ 0.06  & 1.26 &  45 &  12 & $<0.01$  & $<0.01$\\
\objectname{3C 273}           & 3569 &  224. $\pm$ 6.78 $\pm$ 8.49  &  2.79 $\pm$ 0.04 $\pm$ 0.04  & 0.45 &  11 &   5 & $<0.01$  & $<0.01$\\
\objectname{RX J1417+2543}    &   31 &  2.56 $\pm$ 2.14 $\pm$ 0.65  &  1.68 $\pm$ 0.39 $\pm$ 0.08  & 6.02 &  41 &   1 &  $0.95$  &  $0.32$\\
\objectname{1ES 1440+122}     &   33 &  1.05 $\pm$ 0.06 $\pm$ 0.10  &  1.18 $\pm$ 0.27 $\pm$ 0.03  &17.04 &  19 &   2 &  $0.68$  &  $0.86$\\
\objectname{RX J1725.0+1152}  &  152 &  18.1 $\pm$ 3.93 $\pm$ 0.86  &  2.01 $\pm$ 0.13 $\pm$ 0.05  & 1.87 &  39 &  11 & $<0.01$  &  $0.02$\\
\objectname{I Zw 187}         &   31 &  5.41 $\pm$ 2.23 $\pm$ 0.68  &  1.95 $\pm$ 0.23 $\pm$ 0.03  & 2.22 &  77\tablenotemark{a} &   3 &  $0.78$  &  $0.41$\\
\objectname{1ES 1741+196}     &   46 &  4.93 $\pm$ 2.17 $\pm$ 0.17  &  1.80 $\pm$ 0.22 $\pm$ 0.03  & 3.58 &  37 &   3 &  $0.81$  &  $0.99$\\
\objectname{1ES 2321+419}     &   88 &  6.76 $\pm$ 2.77 $\pm$ 0.79  &  1.78 $\pm$ 0.20 $\pm$ 0.09  & 3.90 &  42 &  14 &  $0.34$  &  $0.13$\\
\objectname{NGC 1275}         & 1351 &  99.1 $\pm$ 5.13 $\pm$ 3.87  &  2.20 $\pm$ 0.04 $\pm$ 0.06  & 0.80 &  18 &  13 & $<0.01$  & $<0.01$\\
\enddata
\tablenotetext{a}{Photon-like events were selected for study using
the so-called ``diffuse'' class cuts. Using a stricter set of cuts
(``extradiffuse''), developed to study the extragalactic diffuse
radiation \citep{REF::FERMI_EXTRAGALACTIC}, the highest energy photons
from 3C~66A and I~Zw~187 were eliminated, giving 
$E_\mathrm{max}=90$\,GeV and $4$\,GeV respectively for the two
sources.}
\end{deluxetable}

%
%

\begin{deluxetable}{lcccccc}
\tabletypesize{\scriptsize}
\tablecaption{\label{TAB::OBJ_RES_HE}
Parameters of fitted power-law spectra in low-energy
(0.2\,GeV--1\,GeV) and high-energy bands (1\,GeV--300\,GeV). Only
\Fermi sources detected with $TS>100$ in each band are listed.}
\tablewidth{0pt}
\tablehead{
& \multicolumn{3}{c}{Low-energy band (0.2\,GeV--1\,GeV)} 
& \multicolumn{3}{c}{High-energy band (1\,GeV--300\,GeV)} \\
\colhead{Name} & \colhead{Flux } & \colhead{Photon Index} & \colhead{Decorr.} & \colhead{Flux } & \colhead{Photon Index} & \colhead{Decorr.} \\
& \colhead{$F\pm\ \Delta F_\mathrm{stat}$} 
& \colhead{$\Gamma\pm\Delta\Gamma_\mathrm{stat}$}
& \colhead{energy}& \colhead{$F\pm\ \Delta F_\mathrm{stat}$} 
& \colhead{$\Gamma\pm\Delta\Gamma_\mathrm{stat}$}
& \colhead{energy} \\
& \colhead{[$10^{-9}$\cmsc]} & [1] & \colhead{[GeV]}
& \colhead{[$10^{-9}$\cmsc]} & [1] & \colhead{[GeV]}
}
\startdata
\textbf{TeV detected:} \\
\objectname{3C 66A}             & 80.5 $\pm$ 7.2 & 1.97 $\pm$ 0.16 & 0.52 &  17.3 $\pm$  1.0  &  1.98 $\pm$ 0.04  & 2.47 \\
\objectname{S5 0716+714}        & 65.8 $\pm$ 1.7 & 2.20 $\pm$ 0.05 & 0.37 &  12.8 $\pm$  2.2  &  2.37 $\pm$ 0.09  & 2.17 \\
\objectname{1ES 1011+496}       & 23.2 $\pm$ 3.1 & 2.11 $\pm$ 0.25 & 0.47 &   4.6 $\pm$  0.8  &  1.96 $\pm$ 0.09  & 3.08 \\
\objectname{Markarian 421}      & 74.4 $\pm$ 4.2 & 1.93 $\pm$ 0.11 & 0.47 &   9.6 $\pm$  0.9  &  1.78 $\pm$ 0.04  & 3.71 \\
\objectname{W Comae}            & 33.7 $\pm$ 3.9 & 1.92 $\pm$ 0.21 & 0.51 &   5.5 $\pm$  1.1  &  2.16 $\pm$ 0.10  & 2.52 \\
\objectname{3C 279}\tablenotemark{a}& 142 $\pm$ 7.61 & 2.49 $\pm$ 0.11 & 0.40 & 16.2 $\pm$ 1.5 & 2.55 $\pm$ 0.12 & 1.98 \\
\objectname{3C 279}\tablenotemark{b}& 512.6 $\pm$ 18.3 & 2.00 $\pm$ 0.08 & 0.43 & 67.9 $\pm$ 4.5 & 2.49 $\pm$ 0.09 & 1.98 \\
\objectname{PKS 1424+240}       &25.9 $\pm$ 3.0  &1.84 $\pm$ 0.22& 0.50 & 8.41 $\pm$ 0.63&1.82 $\pm$ 0.05& 1.77 \\
\objectname{PG 1553+113}        & 34.8 $\pm$ 4.6 & 1.52 $\pm$ 0.25 & 0.58 &   5.6 $\pm$  0.6  &  1.70 $\pm$ 0.05  & 4.14 \\
\objectname{PKS 2155-304}       & 78.6 $\pm$ 4.5 & 1.72 $\pm$ 0.11 & 0.49 &  13.2 $\pm$  1.2  &  1.96 $\pm$ 0.04  & 2.71 \\
\\
\textbf{TeV non-detected:} \\
\objectname{ON 325}             & 32.3 $\pm$ 4.4 & 1.98 $\pm$ 0.23 & 0.48 &   8.5 $\pm$  1.7  &  2.32 $\pm$ 0.11  & 2.25 \\
\objectname{3C 273}             & 206.5 $\pm$ 6.8 & 2.66 $\pm$ 0.07 & 0.38 &  62.5 $\pm$ 17.7  &  3.42 $\pm$ 0.17  & 1.47 \\
\objectname{NGC 1275}           & 81.6 $\pm$ 5.4 & 2.09 $\pm$ 0.13 & 0.46 &  14.5 $\pm$  2.9  &  2.40 $\pm$ 0.11  & 2.25 \\

\enddata

\tablenotetext{a}{Pre-flaring period (MJD$<54780$).}
\tablenotetext{b}{Peak flaring ($54790<$MJD$<54830$).}
\end{deluxetable}

%
%

\begin{deluxetable}{lcc}
\tablecaption{\AFermiLAT 95\% flux upper-limits (0.2\,GeV--300\,GeV) assuming
spectral indices of $\Gamma=1.5$ and $\Gamma=2.0$.\label{TAB::OBJ_RES_UL}.}
\tablewidth{0pt}
\tablehead{
& \multicolumn{2}{c}{Flux limit, assuming} \\
\colhead{Name} & $\Gamma=1.5$ & $\Gamma=2.0$ \\
& \multicolumn{2}{c}{[$10^{-9}$\cmsc]}
}
\startdata
\textbf{TeV detected:} \\
\objectname{RGB J0152+017}      &  2.02  &  5.01 \\
\objectname{1ES 0229+200}       &  1.94  &  5.12 \\
\objectname{1ES 0347-121}       &  0.80  &  1.81 \\
\objectname{PKS 0548-322}       &  0.59  &  3.14 \\
\objectname{1ES 1101-232}       &  0.83  &  4.40 \\
\objectname{H 2356-309}         &  0.28  &  7.25 \\
\enddata
\end{deluxetable}

%
%

\begin{deluxetable}{lccc}
\tabletypesize{\scriptsize}
\tablecaption{Extrapolation of measured GeV spectrum into TeV 
regime.\label{TAB::OBJ_PRED}}
\tablewidth{0pt}
\tablehead{
&
\colhead{Extrapolation at 100\,GeV} & 
\multicolumn{2}{c}{Extrapolation over 0.2\,TeV to 10\,TeV band}\\
\colhead{Name} &
\colhead{{$dF/dE(100\,\mathrm{GeV})$}} & 
\colhead{Photon index - $\Gamma_\mathrm{ext}$} & 
\colhead{Integral flux - $\phi_\mathrm{ext}$}\\
 & 
\colhead{[$10^{-9}$\cmsc TeV$^{-1}$]} & 
\colhead{[1]} & 
\colhead{[\Icrabc]}
}
\startdata

\multicolumn{1}{l}{\textbf{TeV detected:}} &&\\
\objectname{3C 66A}             &        1.98 $\pm$ 0.34        &       4.26 $\pm$ 0.02         &$0.0750 \pm   0.0142$\\
\objectname{RGB J0710+591}      &        0.61 $\pm$ 0.38        &       1.74 $\pm$ 0.11         &$0.2110 \pm   0.2015$\\
\objectname{S5 0716+714}        &        0.34 $\pm$ 0.12        &       3.20 $\pm$ 0.03         &$0.0135 \pm  0.0060$\\ 
\objectname{1ES 0806+524}       &        0.18 $\pm$ 0.11        &       2.65  $\pm$ 0.10        &$0.0184 \pm   0.0143$\\  
\objectname{1ES 1011+496}       &        1.05 $\pm$ 0.33        &       2.82 $\pm$ 0.04         &$0.0913 \pm  0.0368$\\ 
\objectname{Markarian 421}      &        5.86 $\pm$ 0.92        &       1.90 $\pm$ 0.02         &$1.4351 \pm   0.3265$\\  
\objectname{Markarian 180}      &        0.17 $\pm$ 0.12        &       2.09 $\pm$ 0.13         &$0.0315 \pm   0.0307$\\ 
\objectname{1ES 1218+304}       &        0.94 $\pm$ 0.38        &       2.46 $\pm$ 0.07         &$0.1354 \pm   0.0739$\\  
\objectname{W Comae}            &        0.46 $\pm$ 0.18        &       2.44 $\pm$ 0.04         &$0.0483 \pm  0.0251$\\ 
\objectname{3C 279}             &        0.35 $\pm$ 0.09        &       5.25 $\pm$ 0.02         &$0.0058 \pm  0.0017$\\ 
\objectname{PKS 1424+240}       &        1.47 $\pm$ 0.30        &       1.85 $\pm$ 0.03         &$0.4187 \pm 0.1236$\\
\objectname{H 1426+428}         &        0.42 $\pm$ 0.32        &       1.85 $\pm$ 0.18         &$0.0885 \pm  0.0108 $\\  
\objectname{PG 1553+113}        &        3.32 $\pm$ 0.56        &       5.27\tablenotemark{a} $\pm$ 0.02        &$0.0423 \pm 0.0074$\\ 
\nodata                         &        \nodata                &       2.06\tablenotemark{b} $\pm$ 0.03        &$0.9960 \pm  0.2327$\\ 
\objectname{Markarian 501}      &        1.71 $\pm$ 0.44        &       1.86 $\pm$ 0.05         &$0.4414 \pm   0.1572$\\   
\objectname{1ES 1959+650}       &        0.53 $\pm$ 0.20        &       2.17 $\pm$ 0.06         &$0.0867 \pm   0.0433$\\  
\objectname{PKS 2005-489}       &        0.68 $\pm$ 0.26        &       2.20 $\pm$ 0.06         &$0.1121 \pm   0.0555$\\  
\objectname{PKS 2155-304}       &        3.11 $\pm$ 0.54        &       2.37 $\pm$ 0.02         &$0.3942 \pm   0.0882$\\  
\objectname{BL Lacertae }       &        0.10 $\pm$ 0.05        &       2.72 $\pm$ 0.08         &$0.0081 \pm   0.0048$\\ 
\objectname{1ES 2344+514}       &        0.23 $\pm$ 0.20        &       1.94 $\pm$ 0.16         &$0.0542 \pm   0.0690$\\ 
\objectname{Centaurus A}        &        0.010 $\pm$ 0.006      &       2.90 $\pm$ 0.11         &$0.0006 \pm   0.0004$\\  
\objectname{M 87}               &        0.027 $\pm$ 0.033      &       2.33 $\pm$ 0.22         &$0.0033 \pm   0.0052$\\ 
&&\\

\multicolumn{1}{l}{\textbf{TeV non-detected:}} &&\\
\objectname{1ES 0033+595}       &        0.39 $\pm$ 0.19        &       2.35 $\pm$ 0.08         &$0.0532 \pm   0.0351$\\ 
\objectname{MG J0509+0541}      &        0.36 $\pm$ 0.17        &       2.01 $\pm$ 0.07         &$0.0708 \pm   0.0446$\\  
\objectname{PKS B0521-365}      &        0.03 $\pm$ 0.02        &       2.74 $\pm$ 0.14         &$0.0025 \pm   0.0021$\\ 
\objectname{1ES 0647+250}       &        0.43 $\pm$ 0.22        &       2.61 $\pm$ 0.08         &$0.0543 \pm   0.0363$\\  
\objectname{PKS 0829+046}       &        0.05 $\pm$ 0.03        &       3.24 $\pm$ 0.09         &$0.0029 \pm   0.0020$\\ 
\objectname{1ES 1028+511}       &        0.28 $\pm$ 0.20        &       3.56 $\pm$ 0.10         &$0.0180 \pm   0.0162$\\
\objectname{RGB J1117+202}      &        0.40 $\pm$ 0.22        &       2.40 $\pm$ 0.09         &$0.0572 \pm   0.0398$\\ 
\objectname{1ES 1118+424}       &        0.19 $\pm$ 0.17        &       2.25 $\pm$ 0.15         &$0.0319 \pm   0.0391$\\  
\objectname{ON 325}             &        0.28 $\pm$ 0.12        &       3.12 $\pm$ 0.04         &$0.0151 \pm   0.0079$\\  
\objectname{3C 273}             &        0.05 $\pm$ 0.01        &       3.52 $\pm$ 0.03         &$0.0001 \pm   0.0001$\\  
\objectname{RX J1417+2543}      &        0.24 $\pm$ 0.28        &       2.79 $\pm$ 0.18         &$0.0262 \pm   0.0413$\\  
\objectname{1ES 1440+122}       &        0.80 $\pm$ 0.48        &       1.68 $\pm$ 0.11         &$0.2310 \pm   0.2178$\\  
\objectname{RXJ 1725.0+1152}    &        0.33 $\pm$ 0.18        &       2.08 $\pm$ 0.10         &$0.0601 \pm   0.0452$\\  
\objectname{I Zw 187}           &        0.14 $\pm$ 0.13        &       2.17 $\pm$ 0.16         &$0.0235 \pm   0.0287$\\  
\objectname{1ES 1741+196}       &        0.27 $\pm$ 0.20        &       2.14 $\pm$ 0.13         &$0.0484 \pm   0.0511$\\  
\objectname{1ES 2321+419}       &        0.41 $\pm$ 0.29        &       2.01 $\pm$ 0.13         &$0.0868 \pm   0.0865$\\  
\objectname{NGC 1275}           &        0.34 $\pm$ 0.14        &       2.26 $\pm$ 0.04         &$0.0355 \pm   0.0200$\\  

\enddata

\tablenotetext{a}{Extrapolated assuming $z=0.09$.}
\tablenotetext{b}{Extrapolated assuming $z=0.78$.}
\end{deluxetable}


\clearpage

\newlength{\tswidth}
\setlength{\tswidth}{0.4\textwidth}
\newcommand{\includeTS}[1]{\includegraphics[bb=154   261   483   562,clip,width=\tswidth]{#1}}

\begin{figure}[p]
\centering
\includeTS{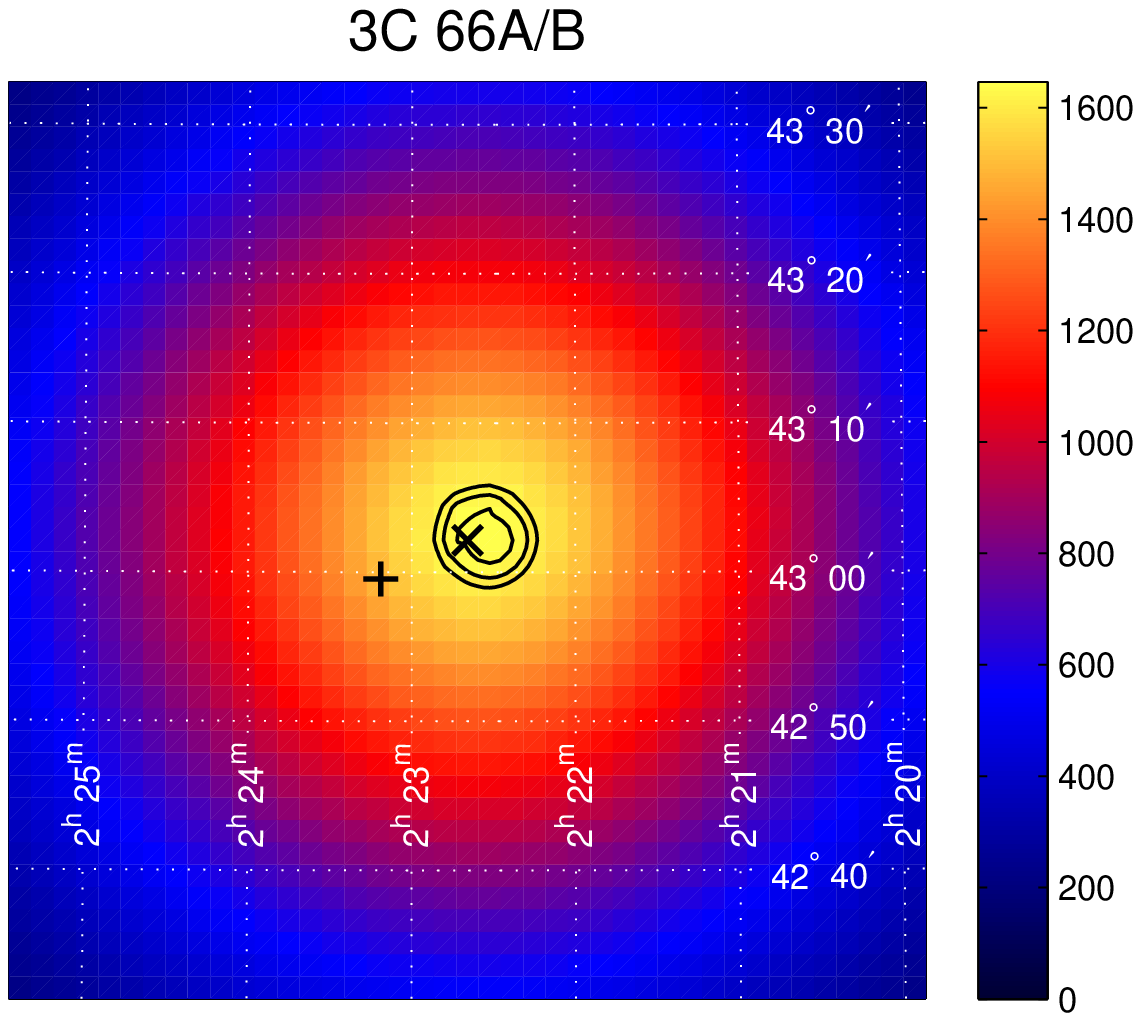}%
\includeTS{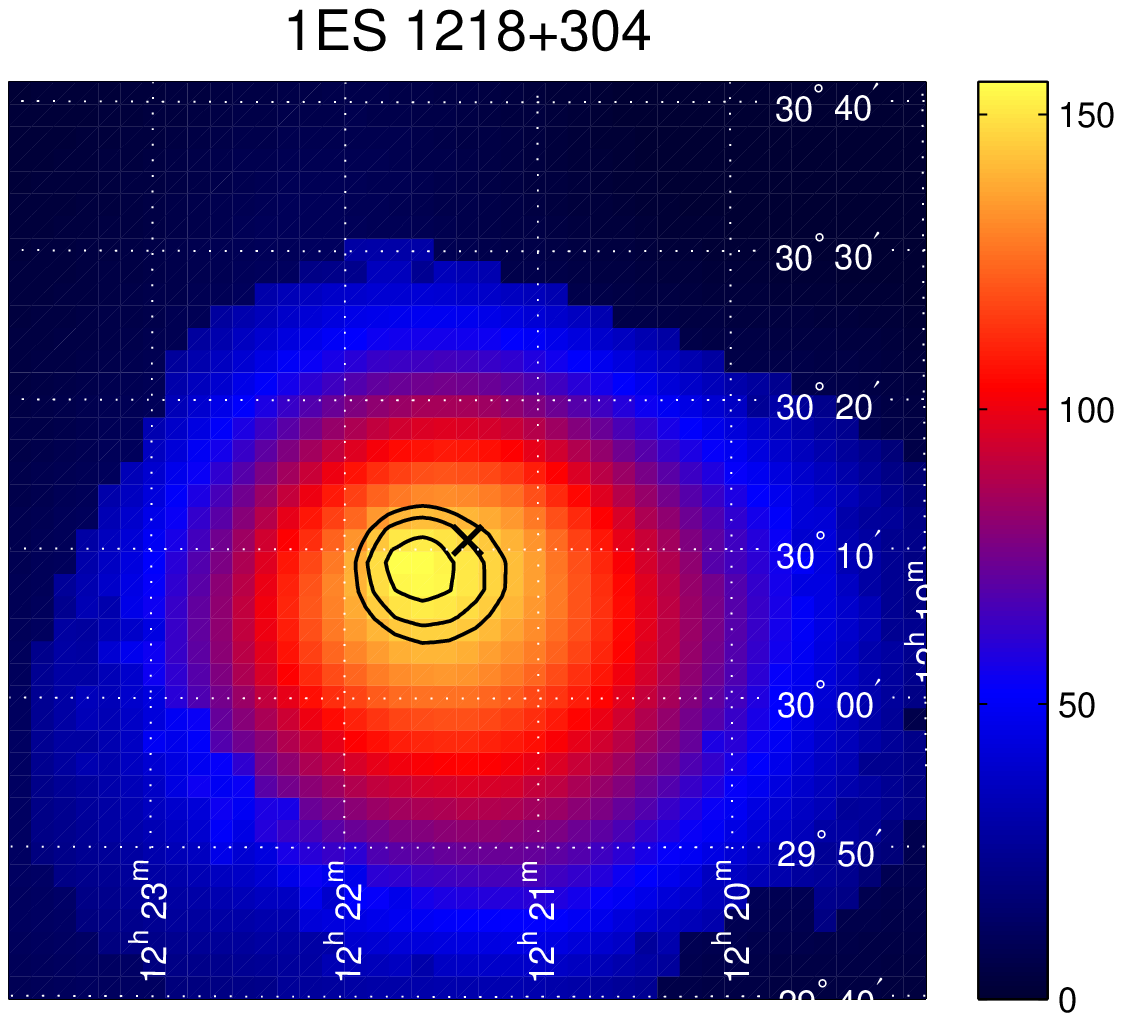}

\includeTS{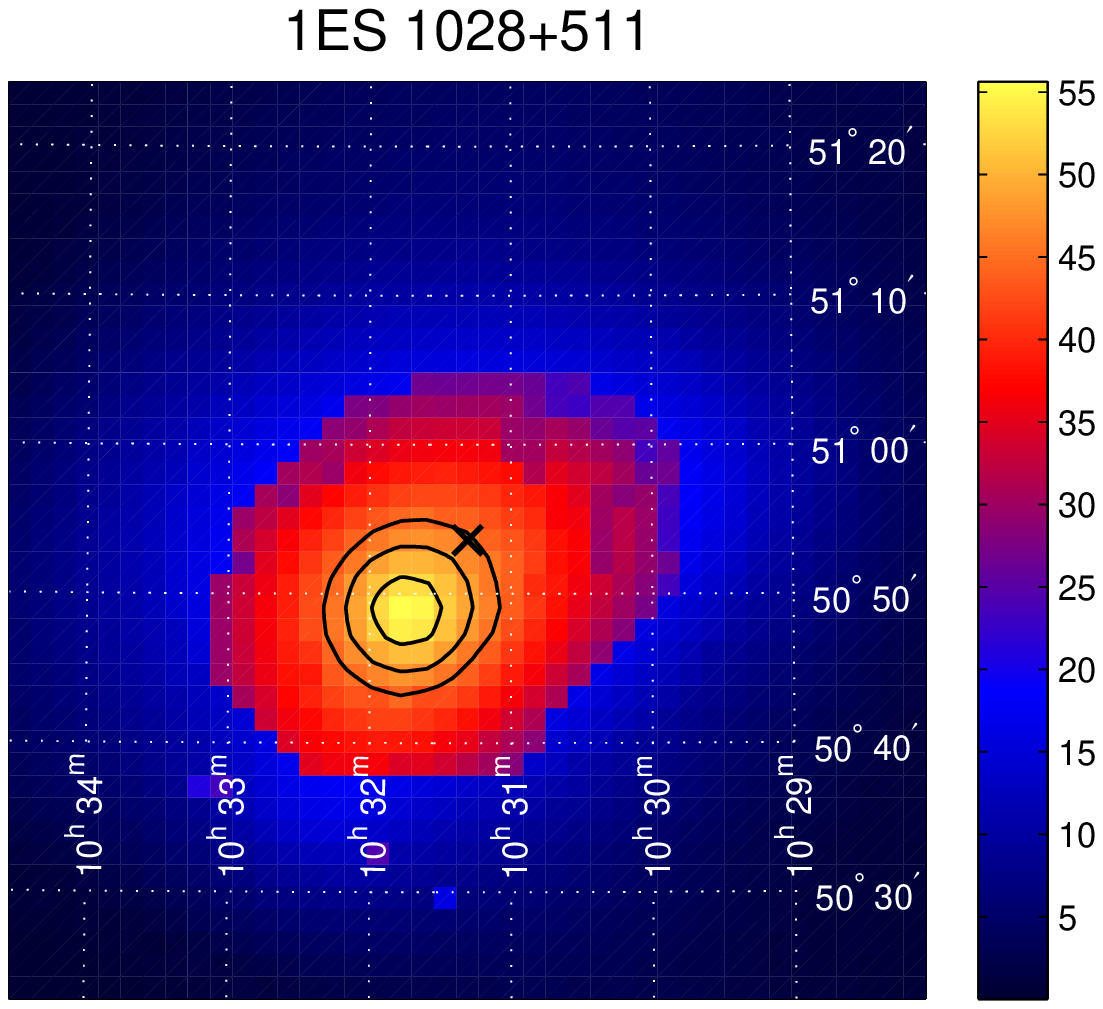}%
\includeTS{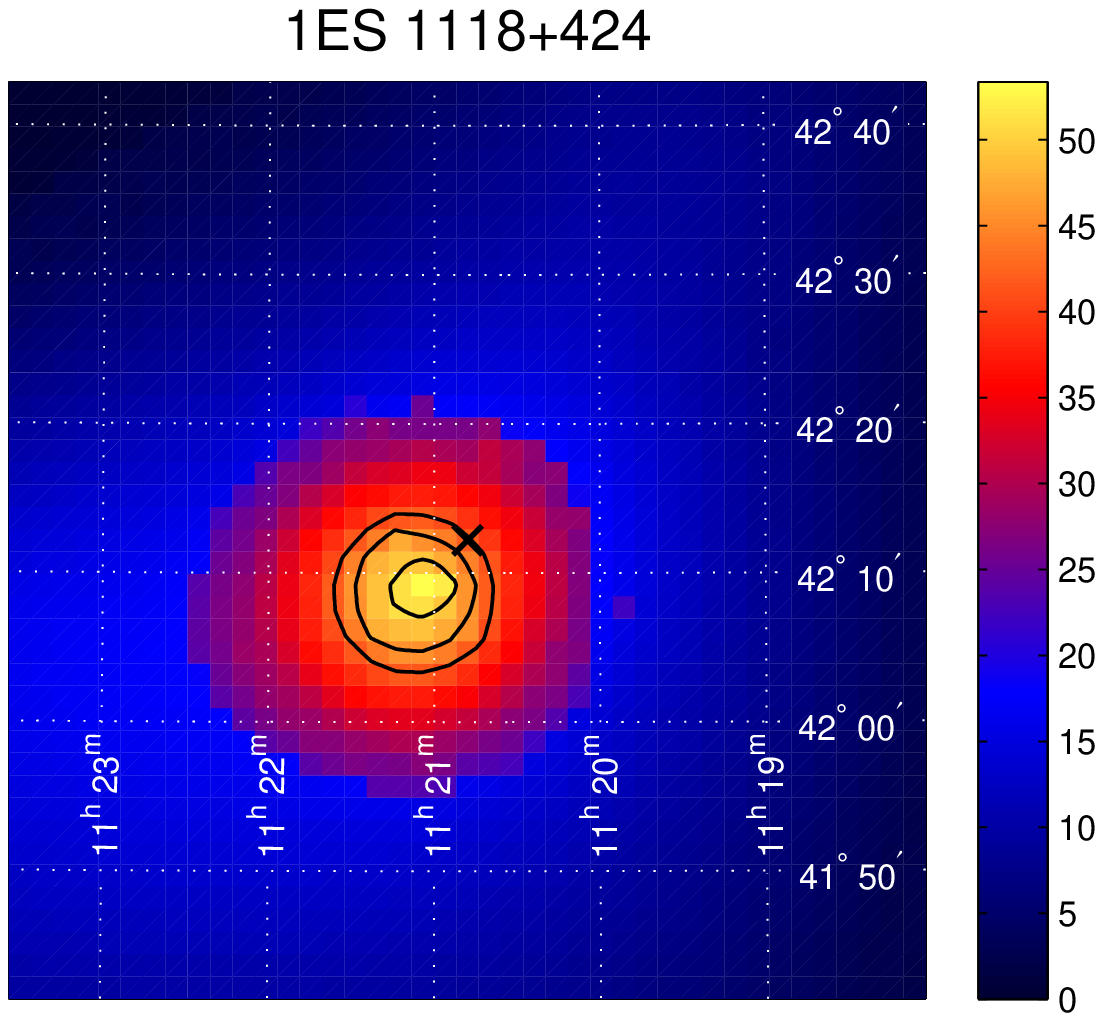}

\includeTS{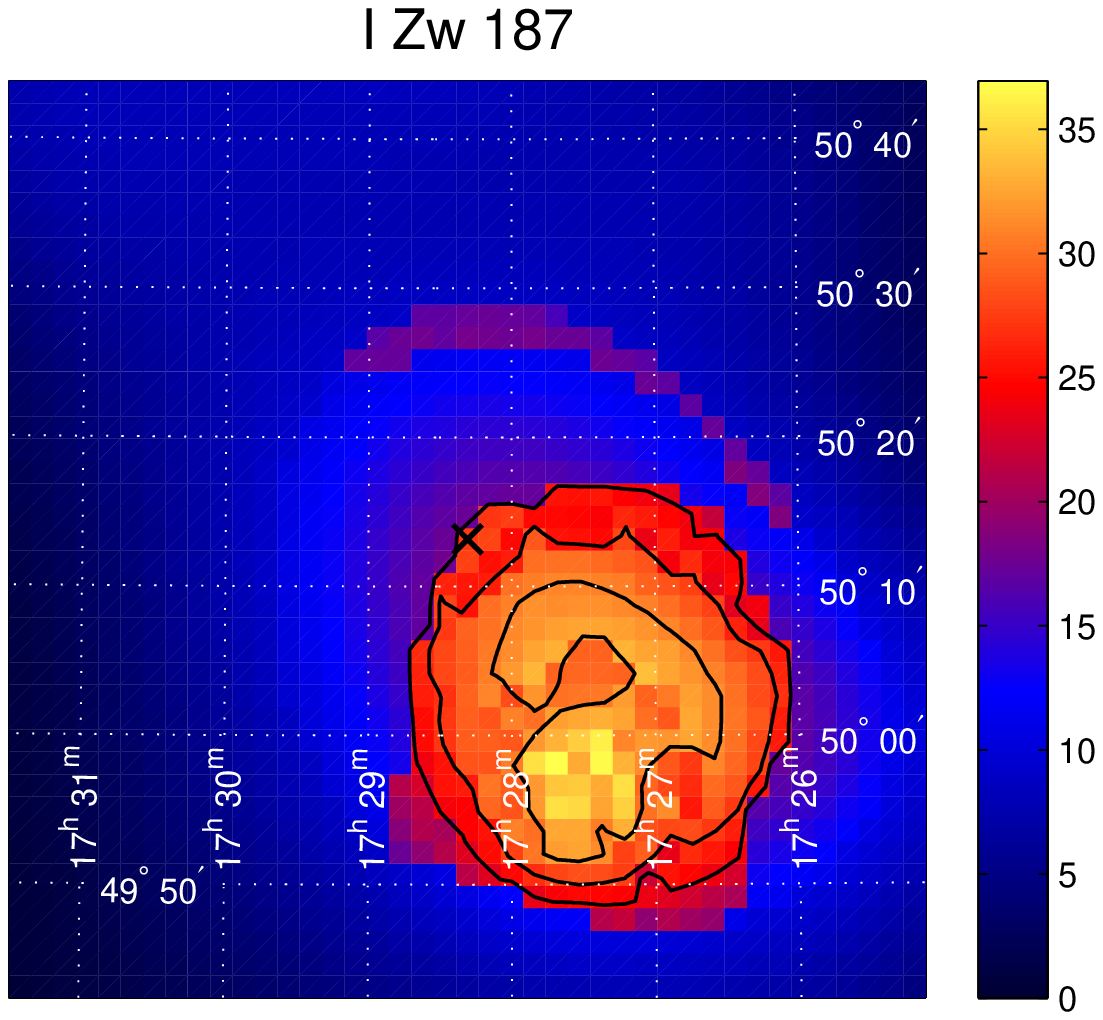}%
\hspace*{\tswidth}

\caption{\label{FIG::TS} Selected TS maps, covering a 
$1^\circ\times 1^\circ$ region around the source of interest. In each
case, the location of the AGN is indicated with a ``$\times$''. In the
case of the 3C~66A/B field (top left), the location of 3C~66A is
indicated with a ``$\times$'' and that of 3C~66B with a ``$+$''. TS
maps for all the targets in the study are available in the online
materials.}
\end{figure}

\clearpage

\newlength{\specwidth}
\setlength{\specwidth}{0.4\textwidth}
\newcommand{\includeSpec}[1]{\includegraphics[bb=25 0 527 374,clip,width=\specwidth]{#1}}


\begin{figure}[p]
\centering
\includeSpec{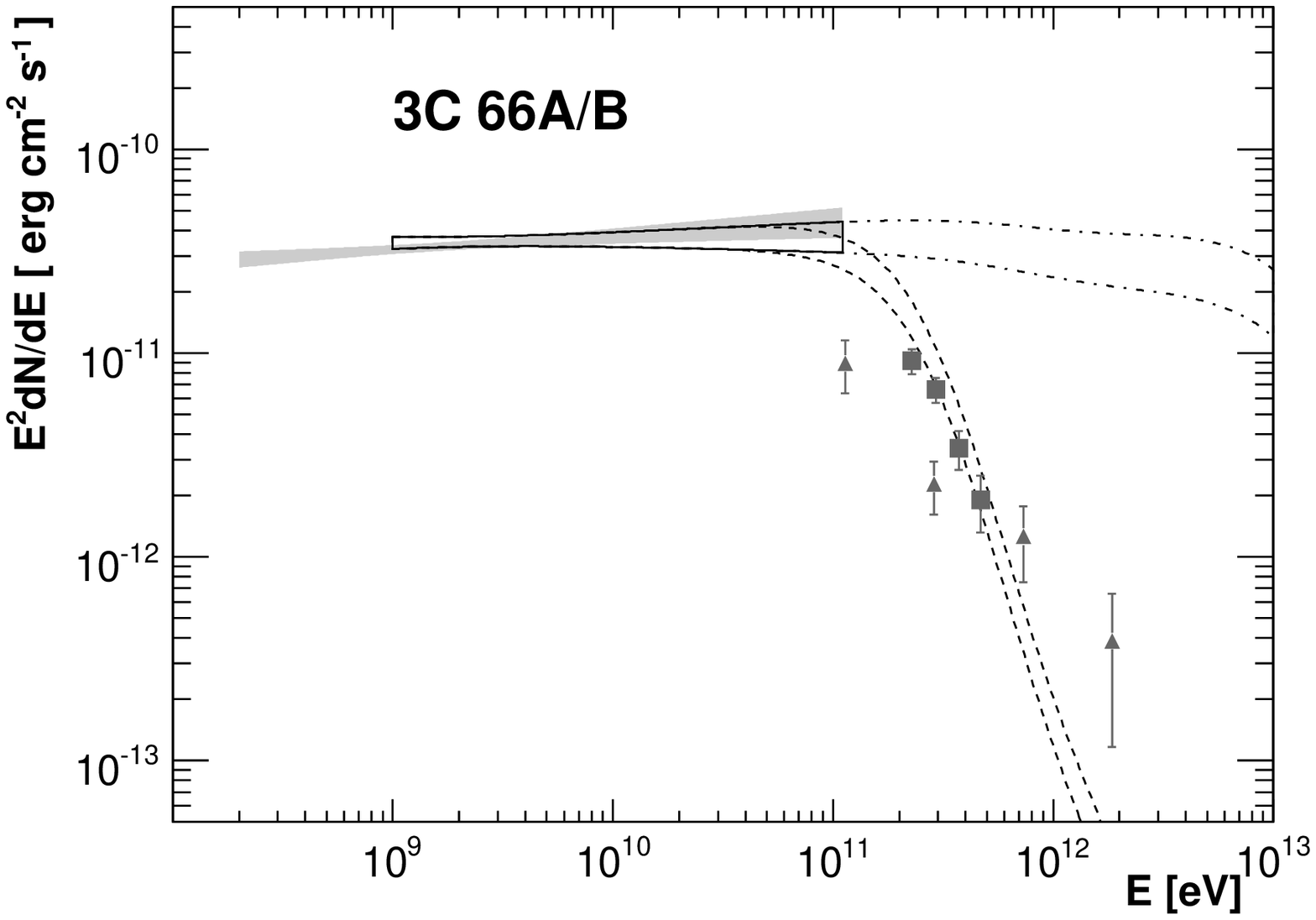}%
\includeSpec{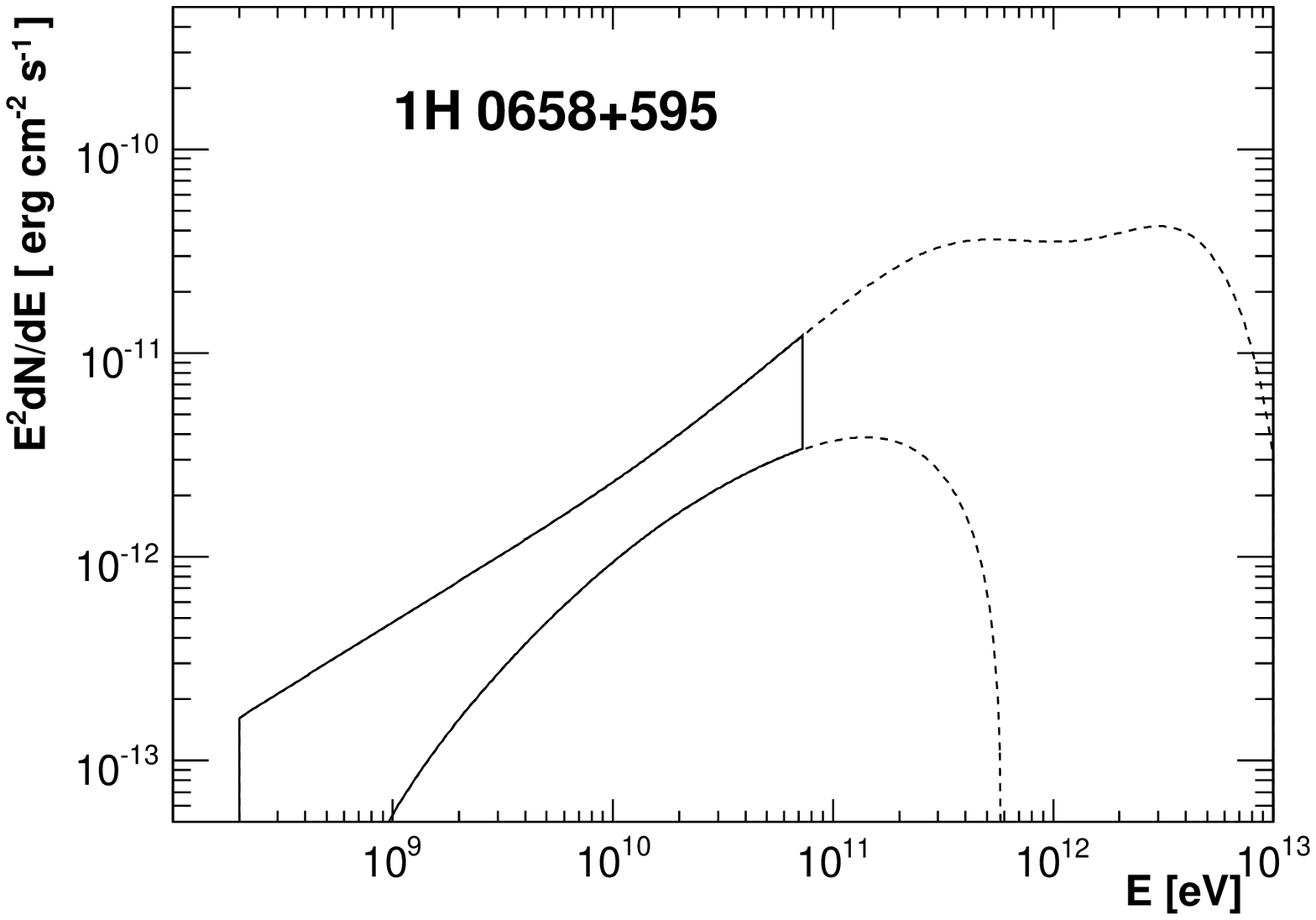}

\includeSpec{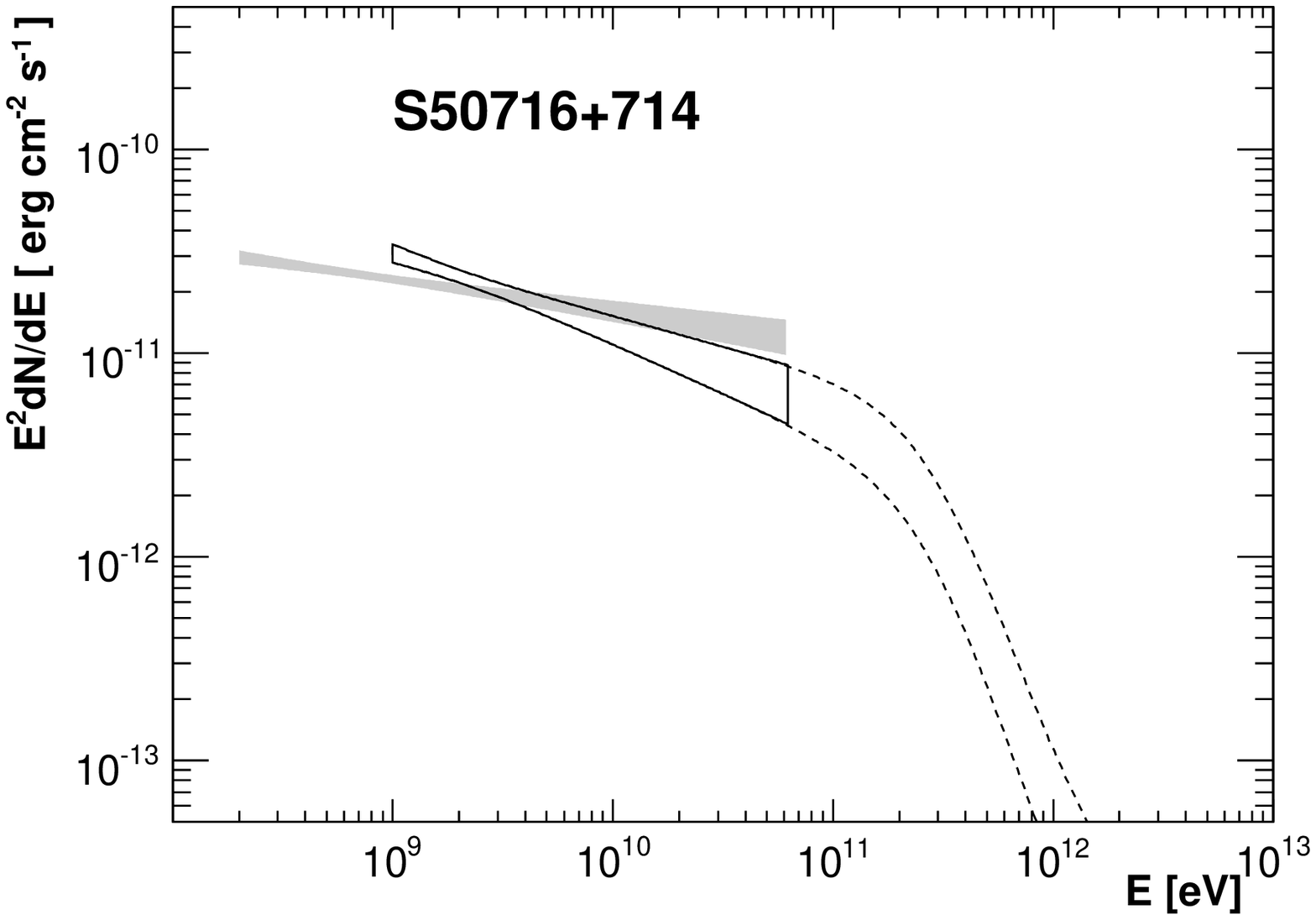}%
\includeSpec{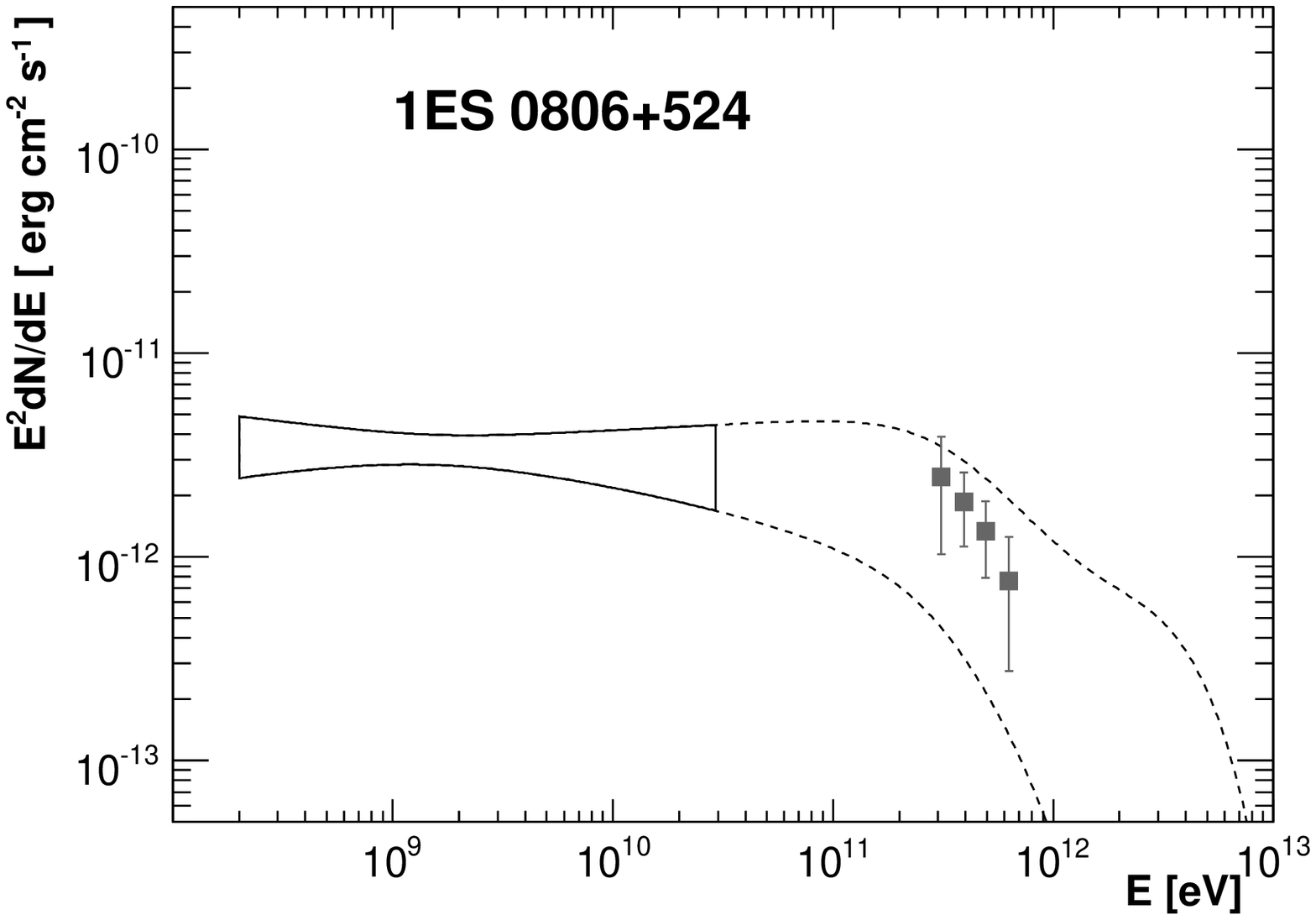}

\includeSpec{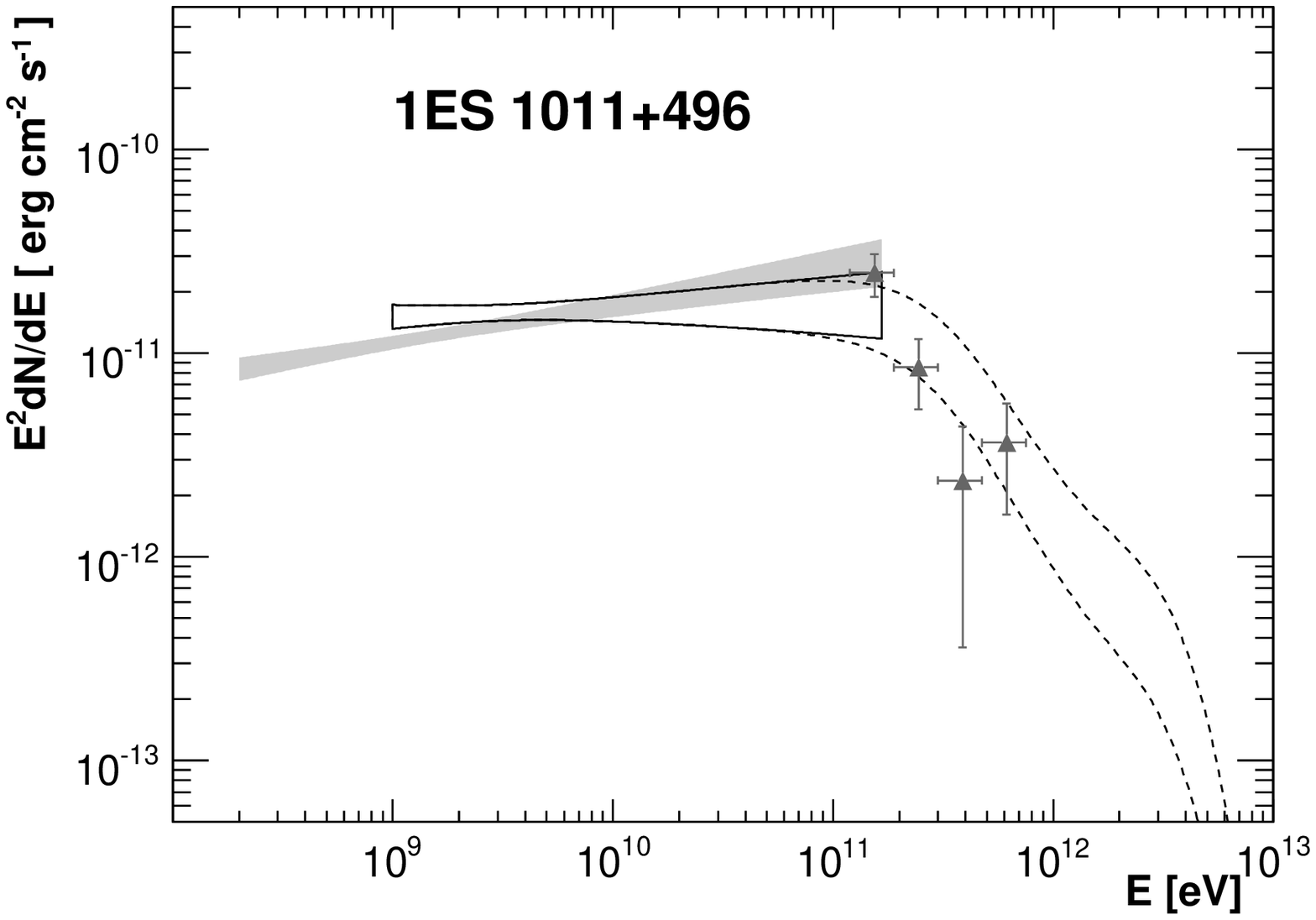}%
\includeSpec{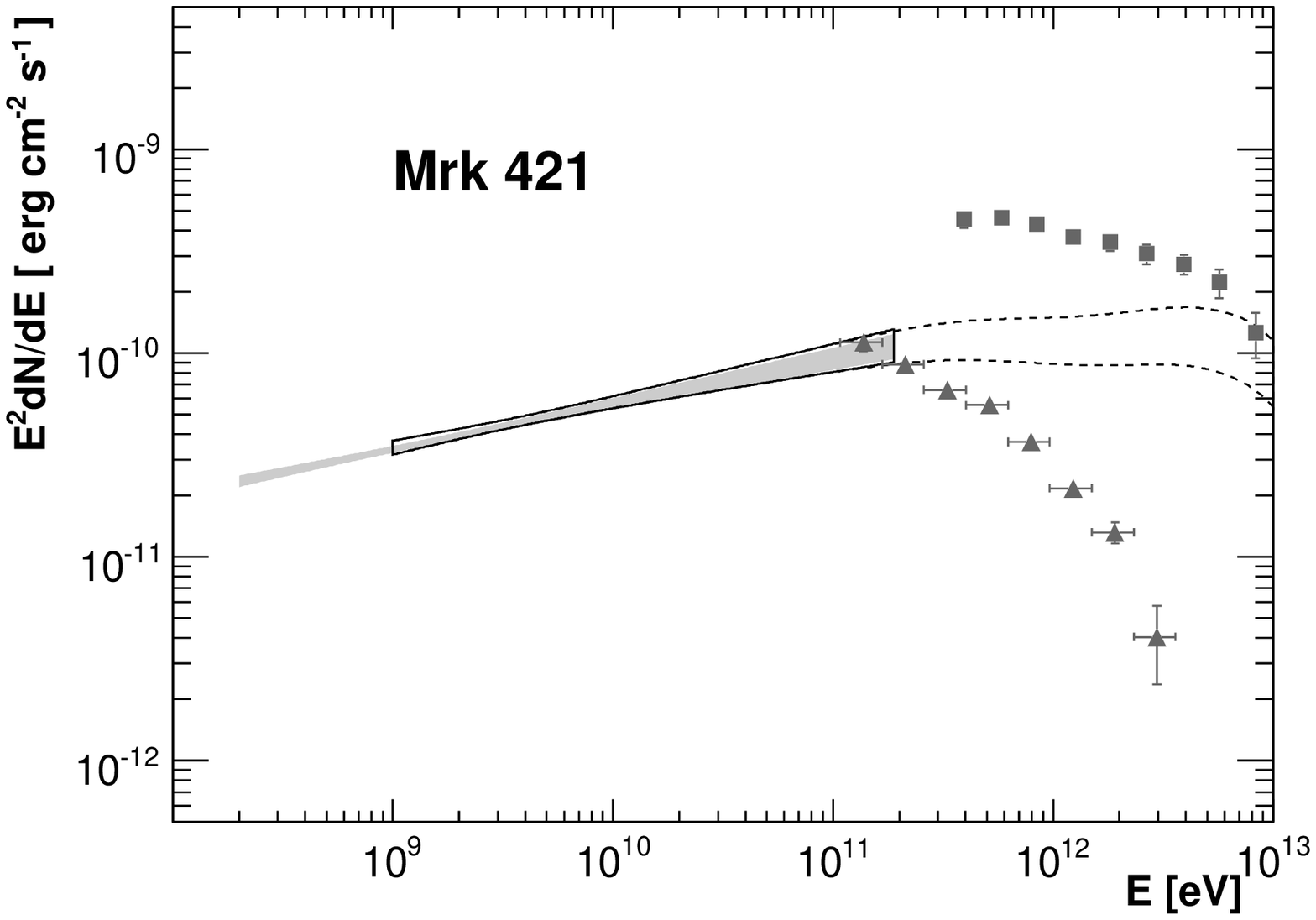}

\caption{\label{FIG::SPEC_TEVDET_1} Spectra for the \NDetTeVSrc\ 
GeV--TeV detected objects. The GeV spectrum derived from \AFermiLAT
observations is indicated as a ``butterfly'' contour (solid line). For
brighter sources (those in Table~\ref{TAB::OBJ_RES_HE}) the contours
correspond to the high-energy band ($E>1$\,GeV), with the fits over
the full energy range shown as gray bands. For the weaker sources,
only the fits over the full range (given in Table~\ref{TAB::OBJ_RES})
are shown. TeV spectral measurements published by H.E.S.S.\ (circles),
VERITAS/Whipple (squares) and MAGIC (triangles) are also shown.  An
extrapolation of the \Fermi spectrum to the TeV regime is shown
(dashed line), assuming absorption with the EBL as described in the
text. In the panel for the 3C~66A/B region, the extrapolation is shown
for $z=0.444$ (3C~66A -- dashed line) and $z=0.021$ (3C~66B --
dash-dotted line). In the case of PG~1553+113 extrapolations with
$z=0.78$ (dashed line) and $z=0.09$ (dash-dotted line) are shown.}
\end{figure}

\addtocounter{figure}{-1}
\begin{figure}[p]
\centering
\includeSpec{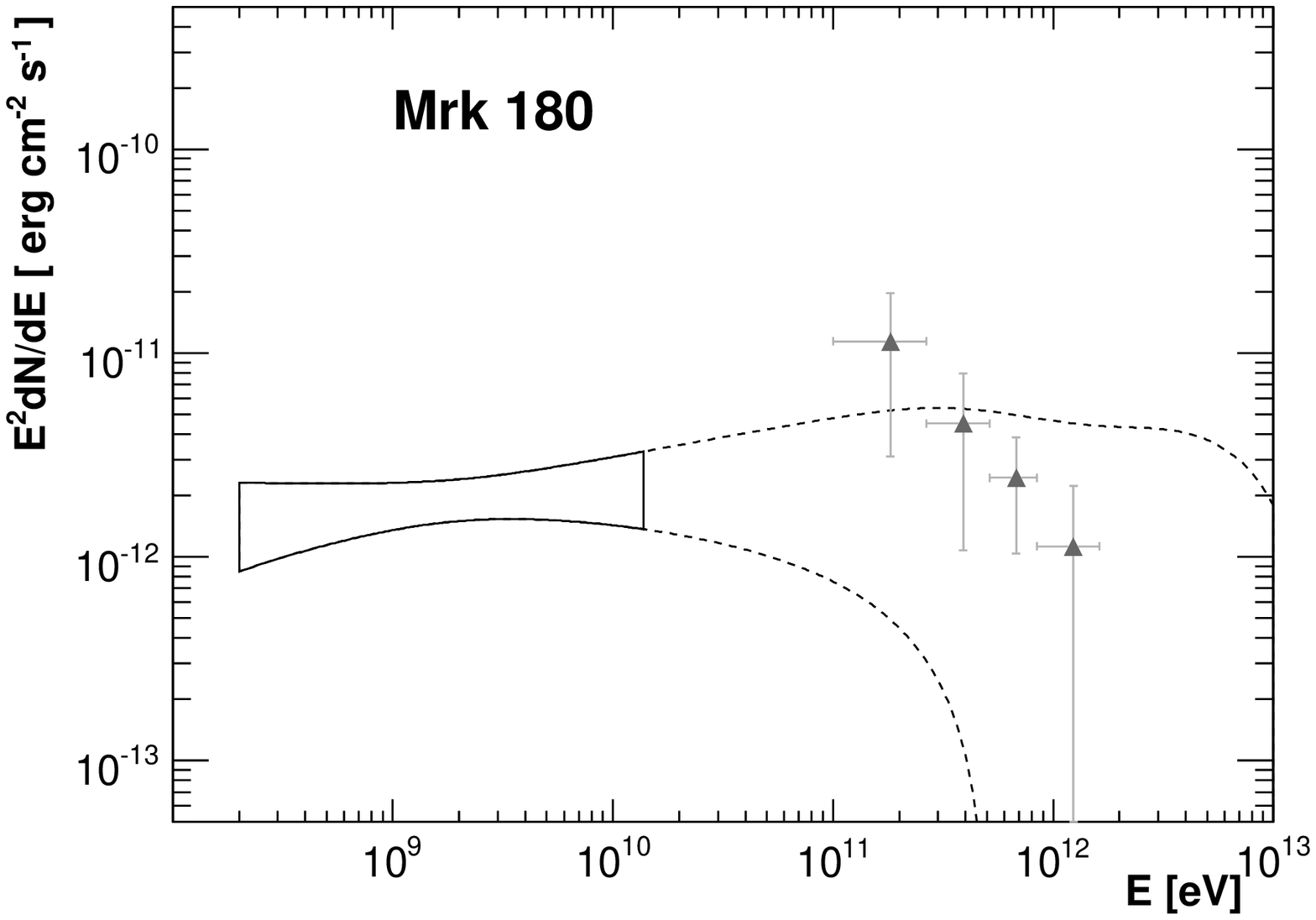}%
\includeSpec{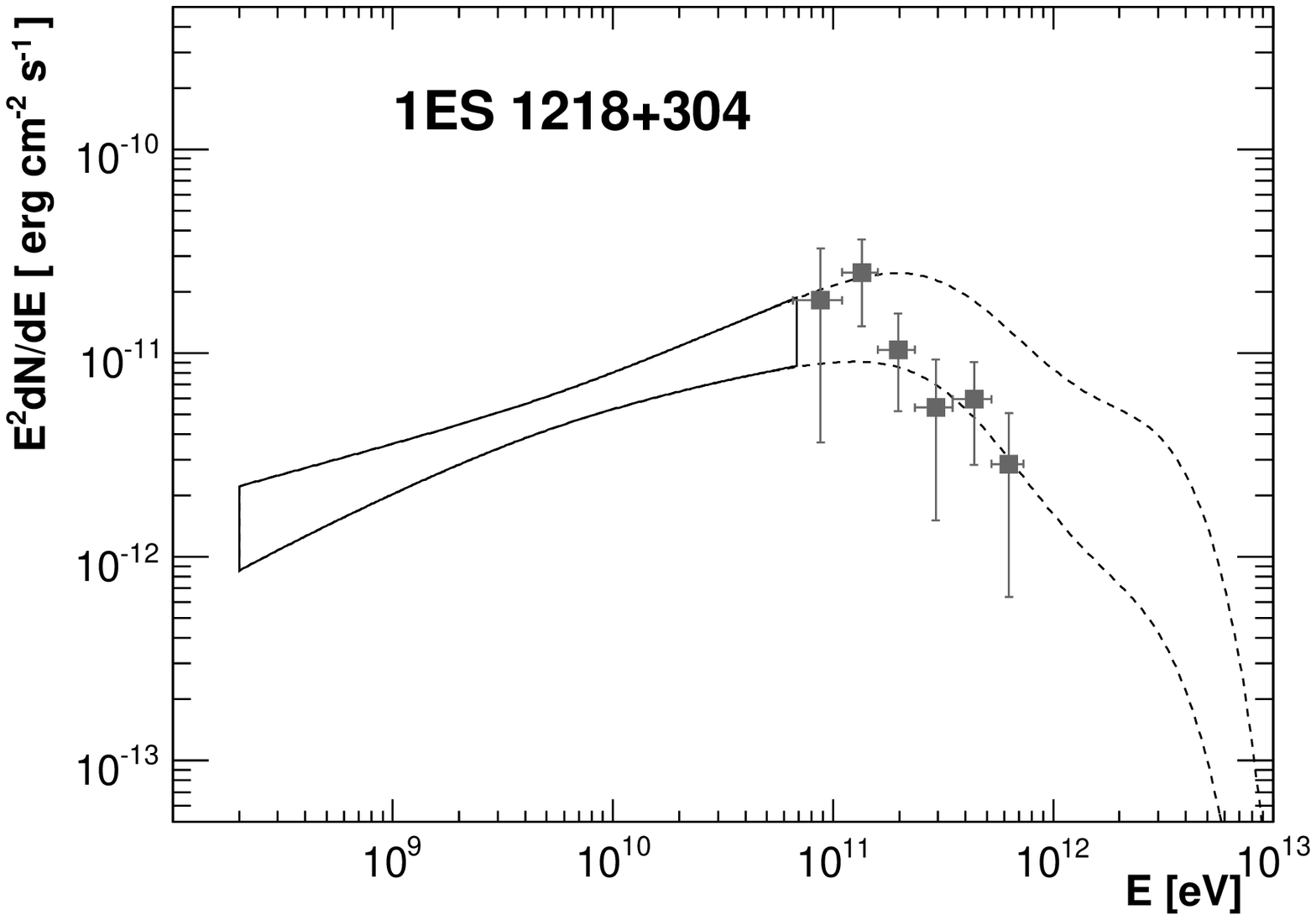}

\includeSpec{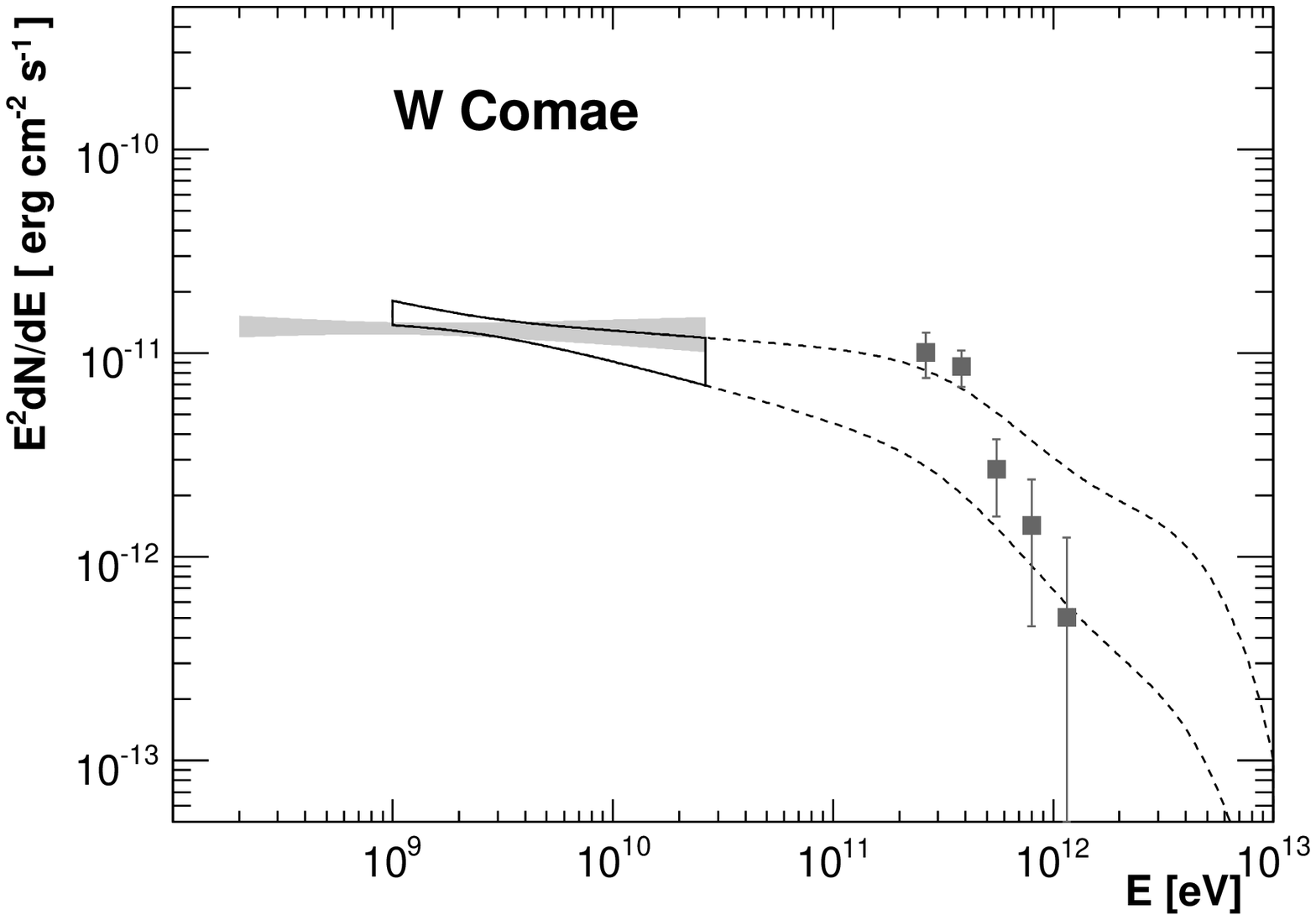}%
\includeSpec{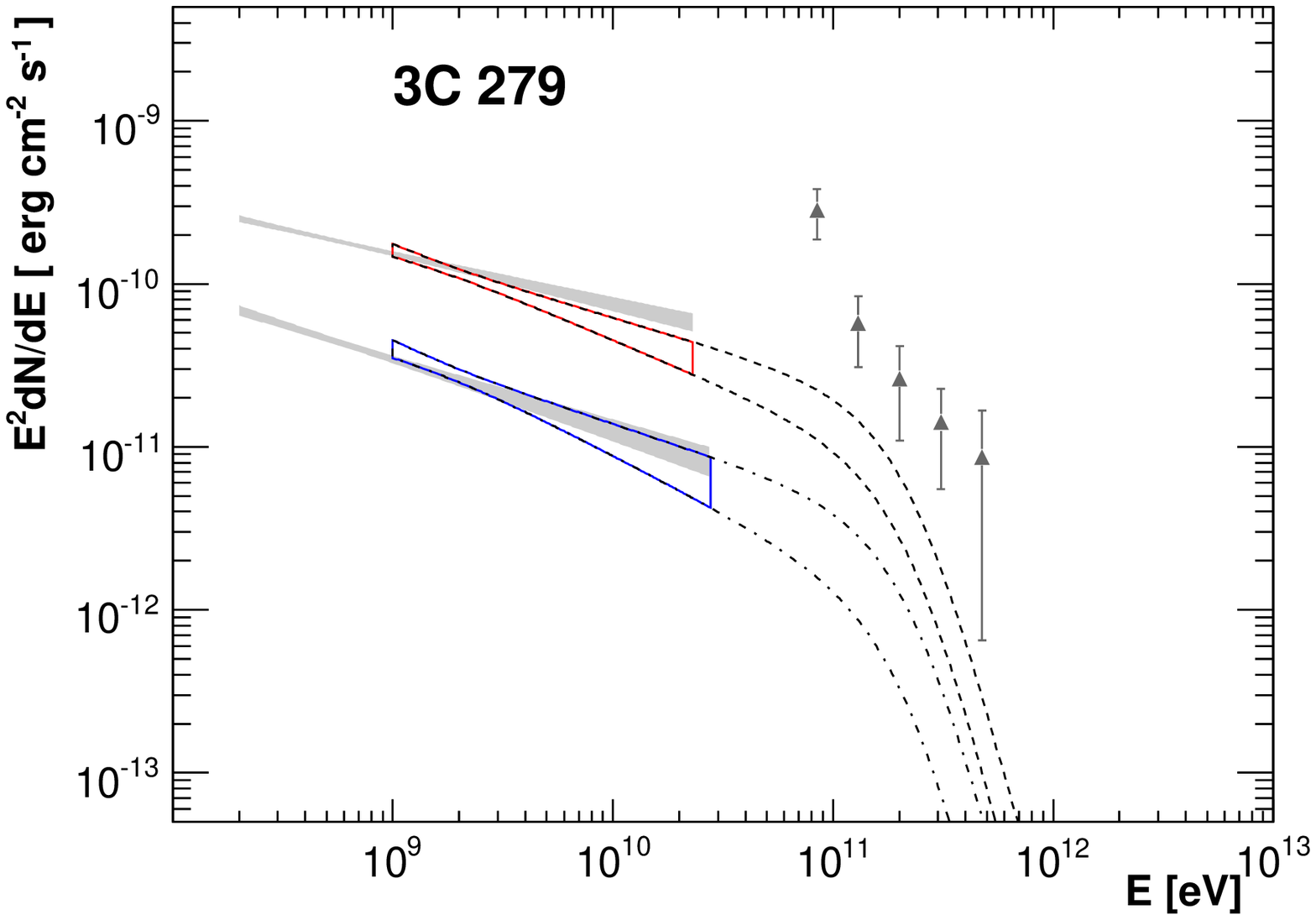}

\includeSpec{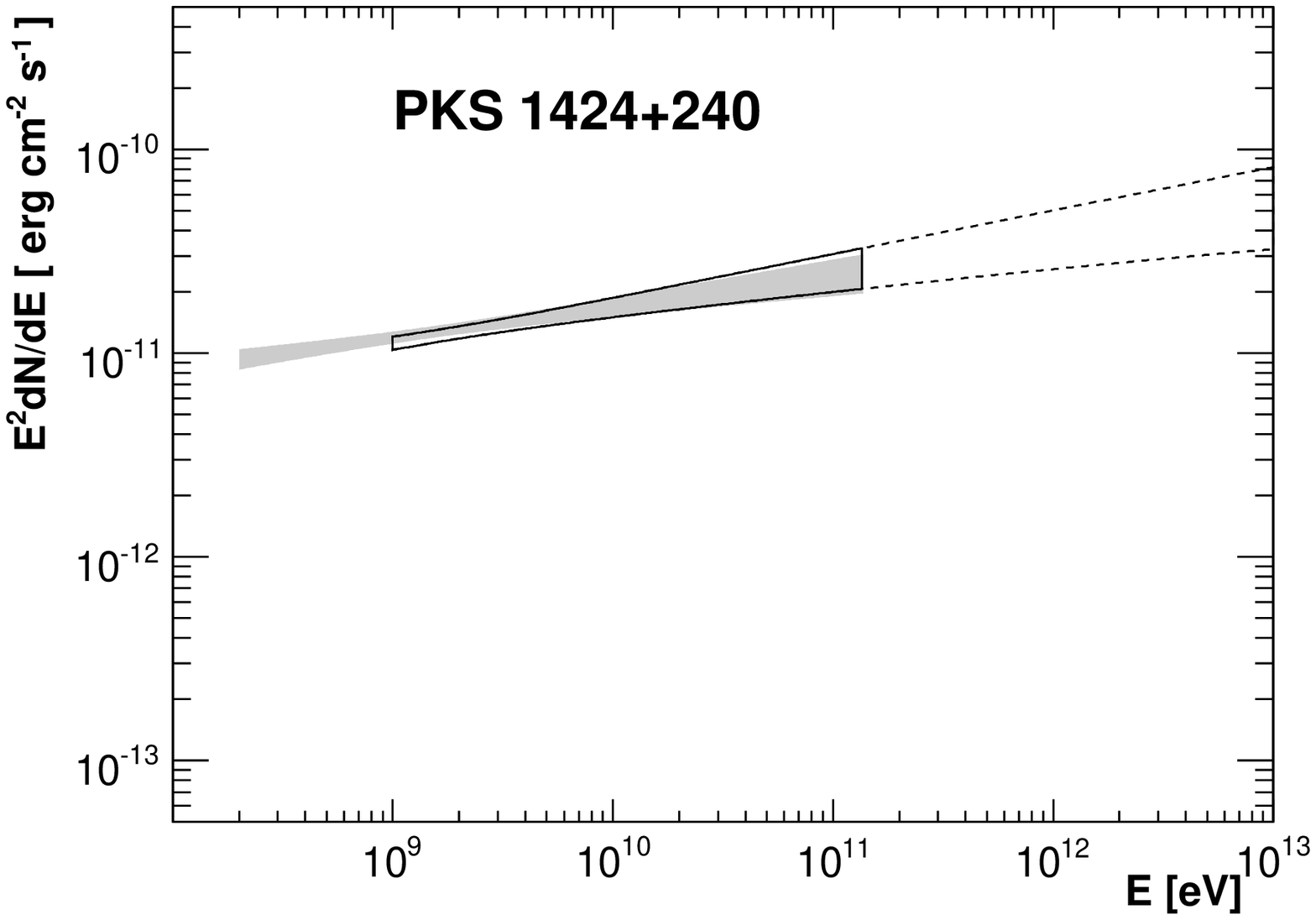}%
\includeSpec{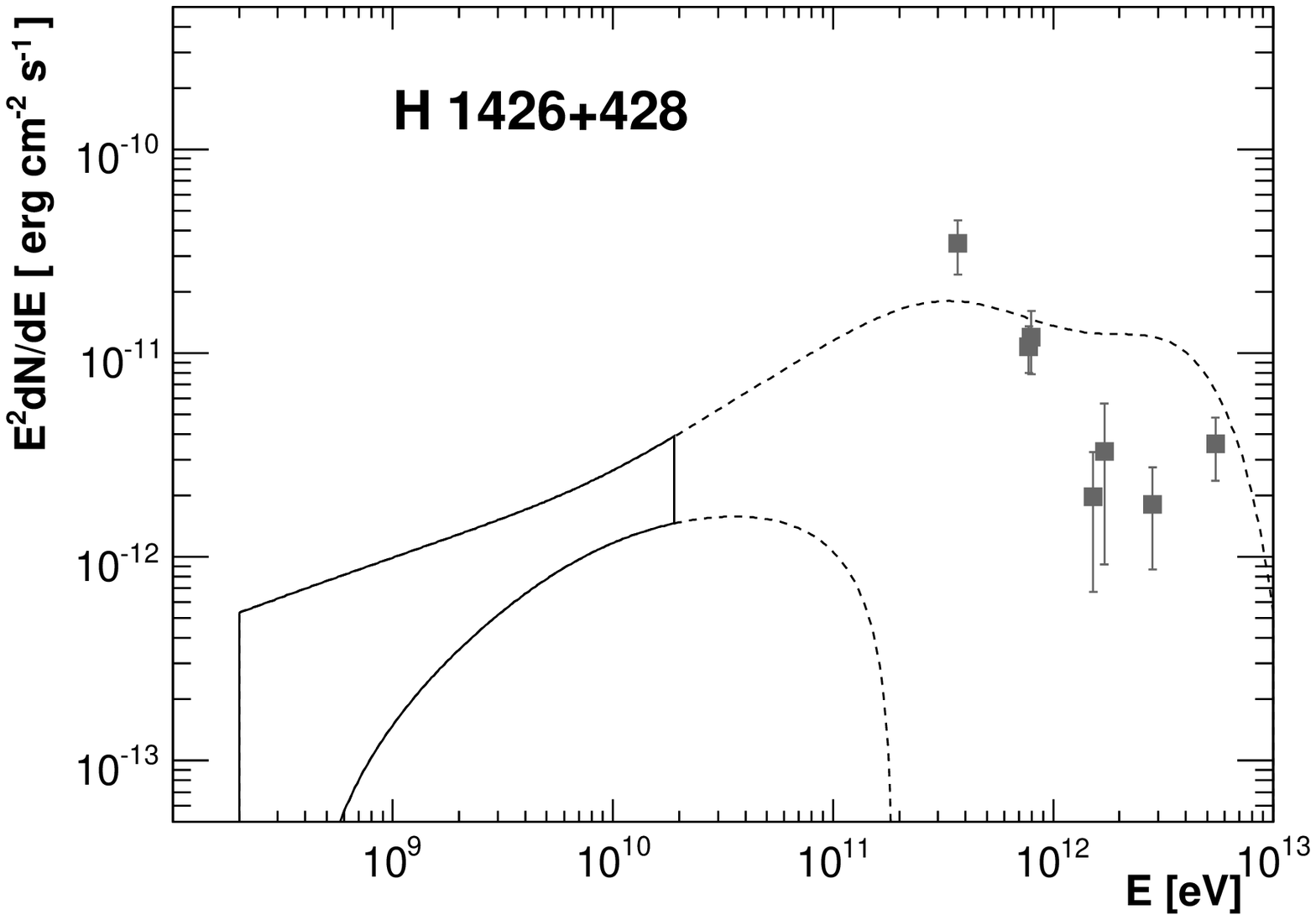}

\includeSpec{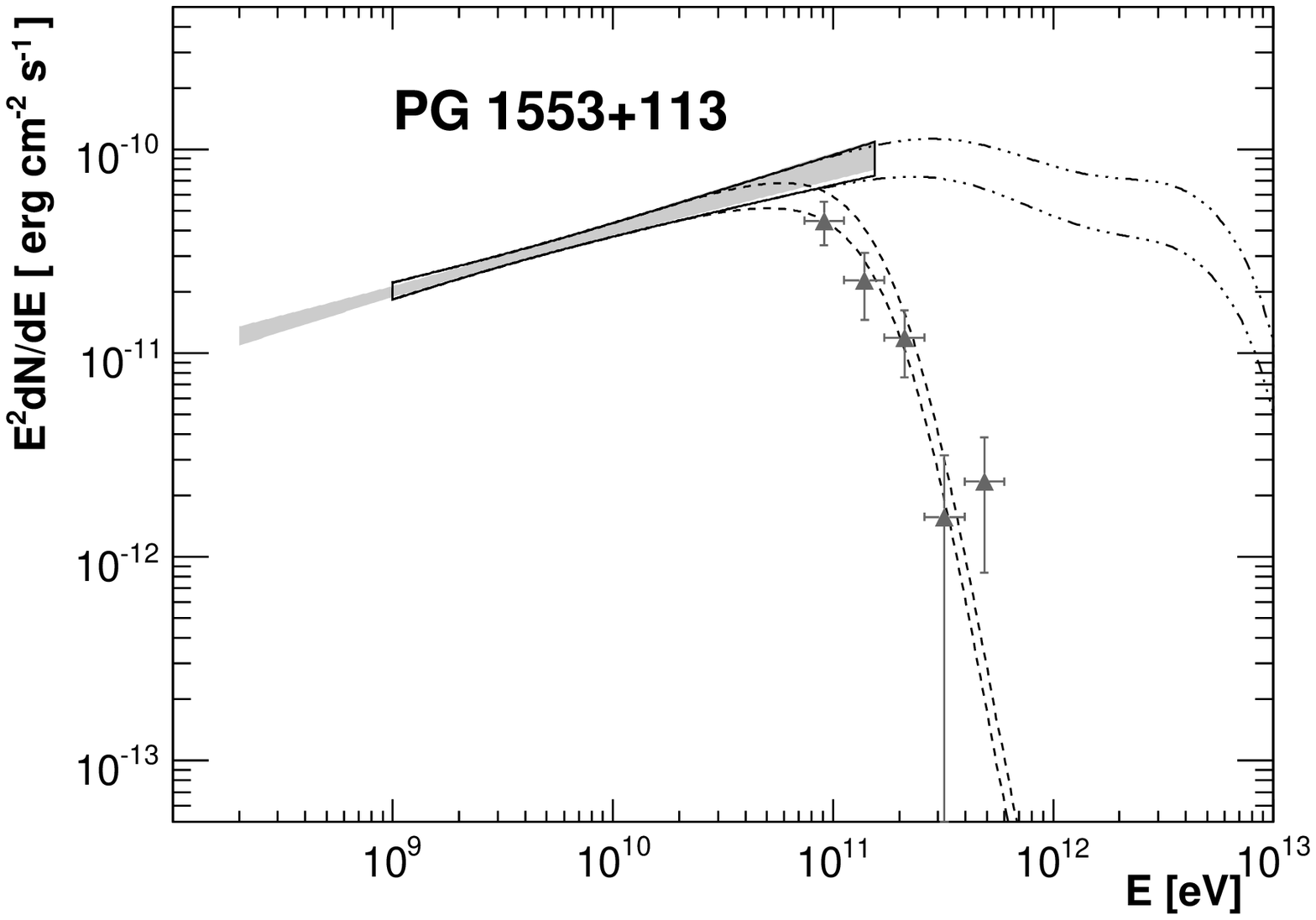}%
\includeSpec{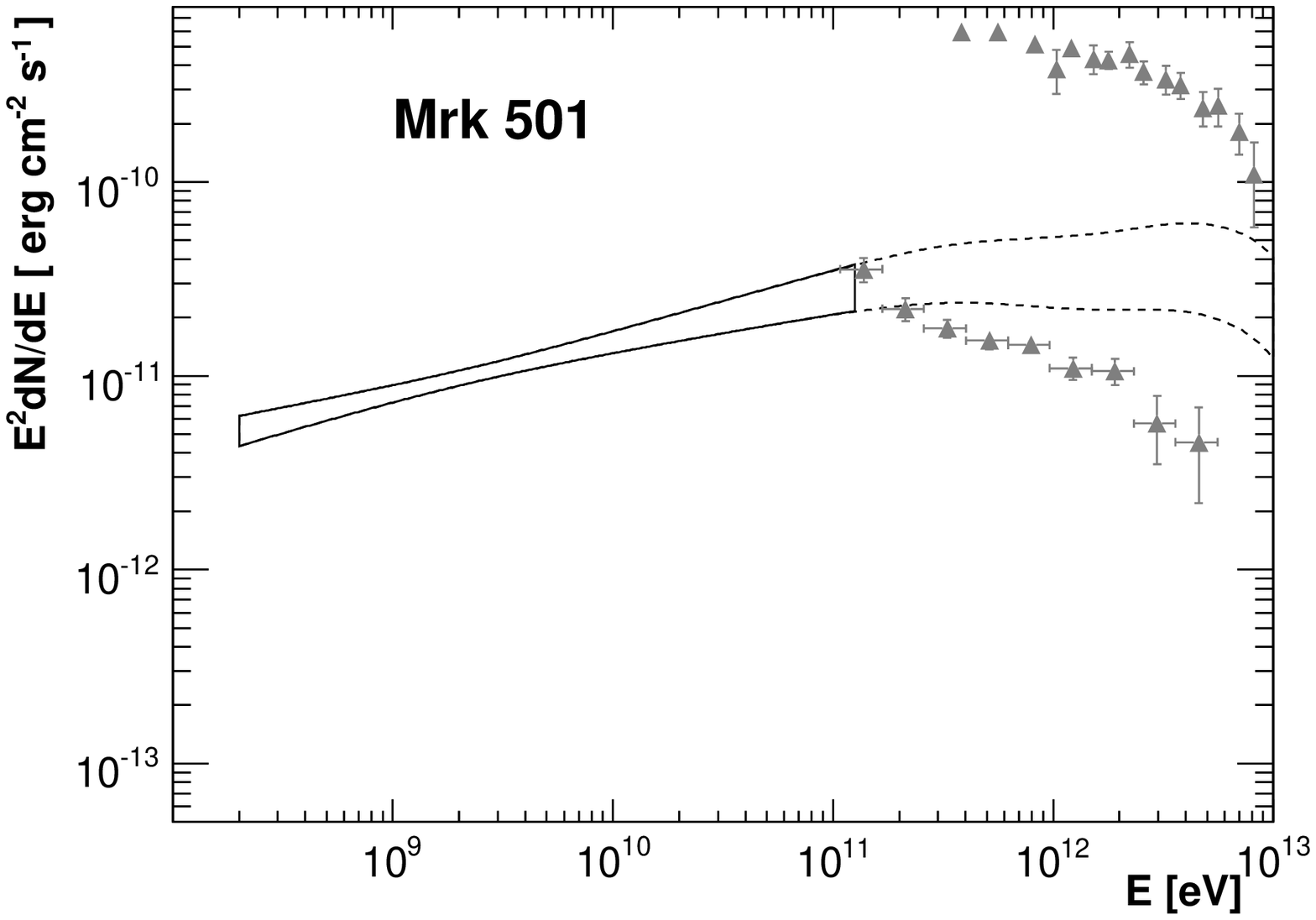}

\caption{\label{FIG::SPEC_TEVDET_2}Continued}
\end{figure}

\addtocounter{figure}{-1}
\begin{figure}[p]
\centering
\includeSpec{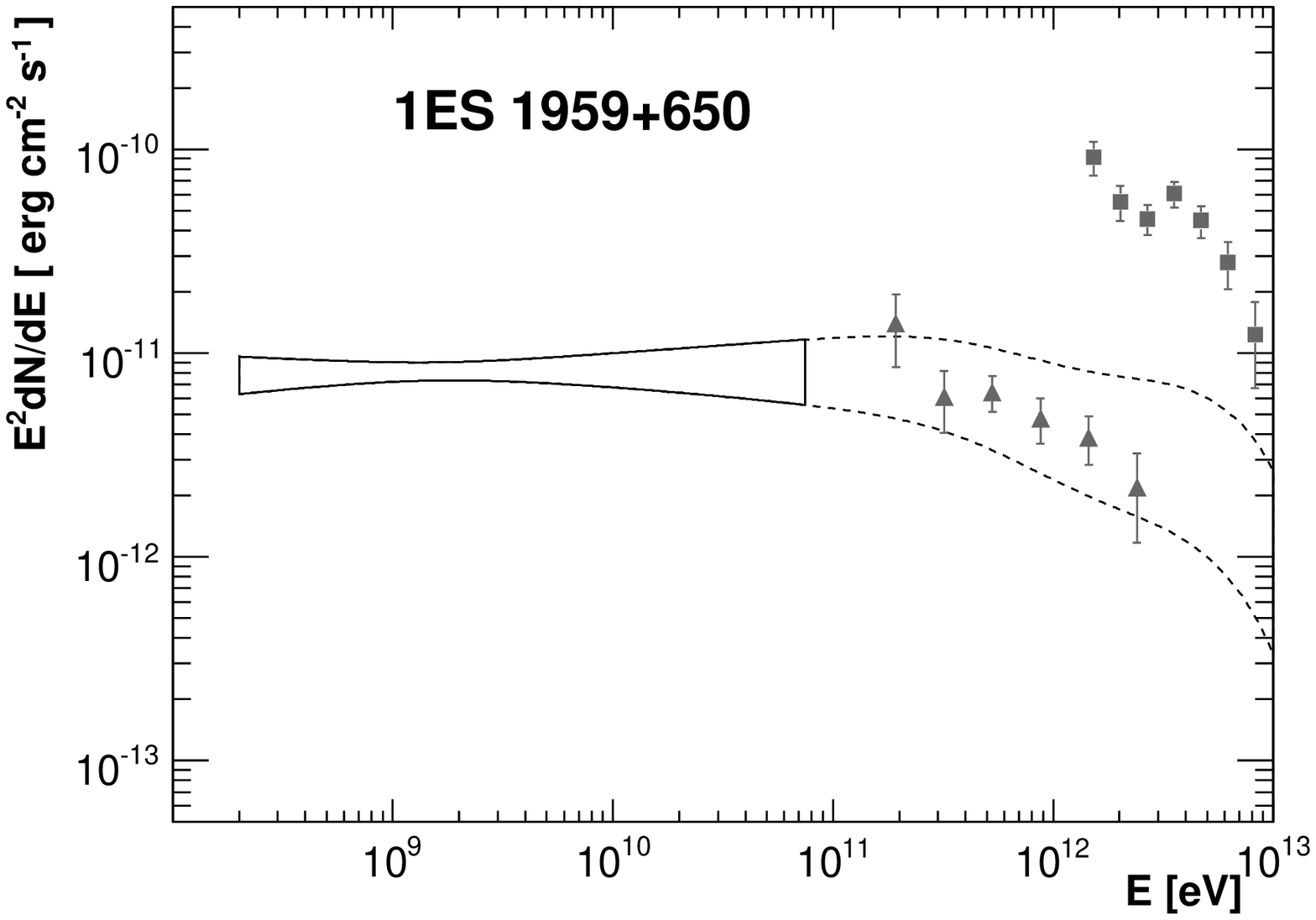}%
\includeSpec{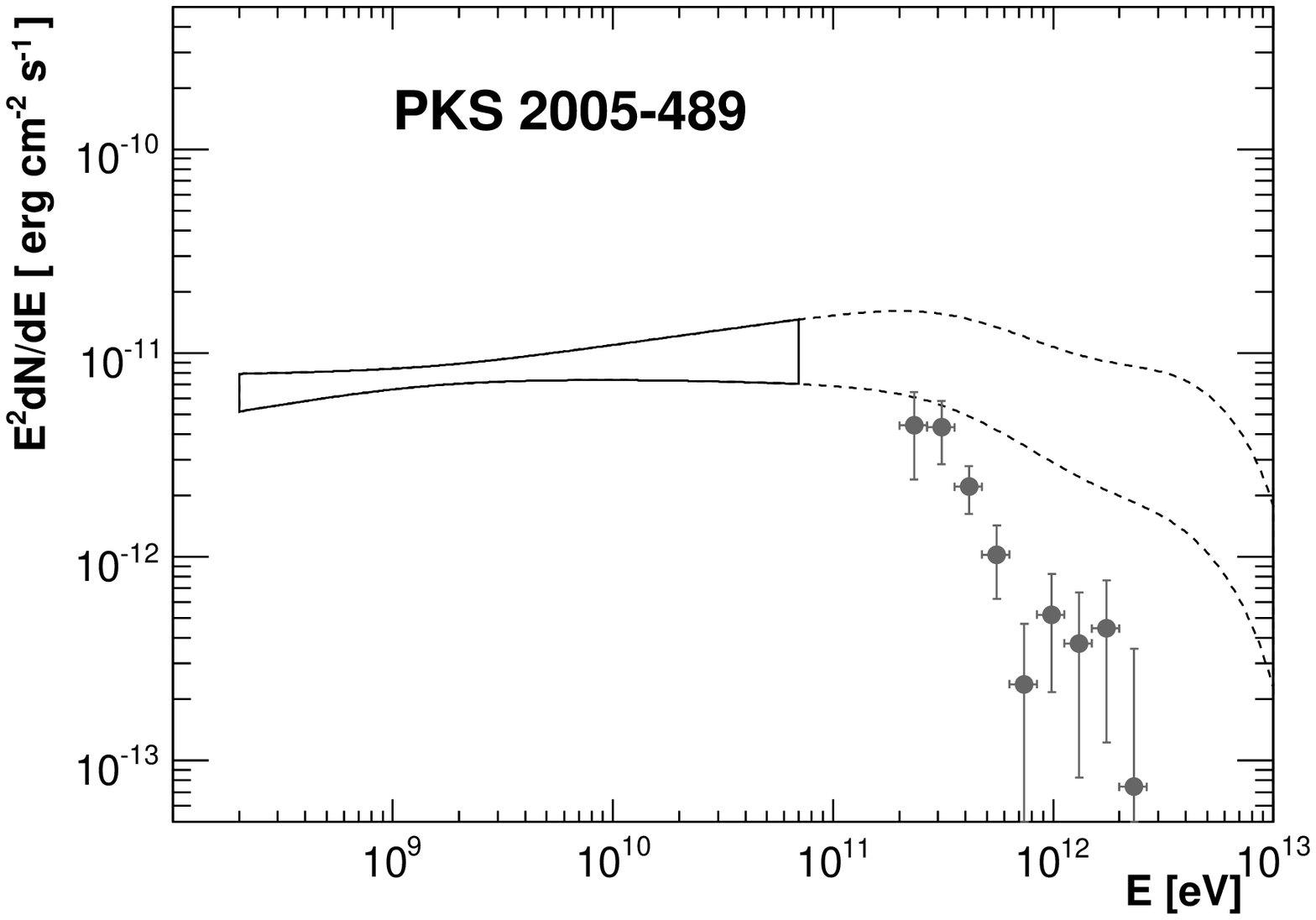}

\includeSpec{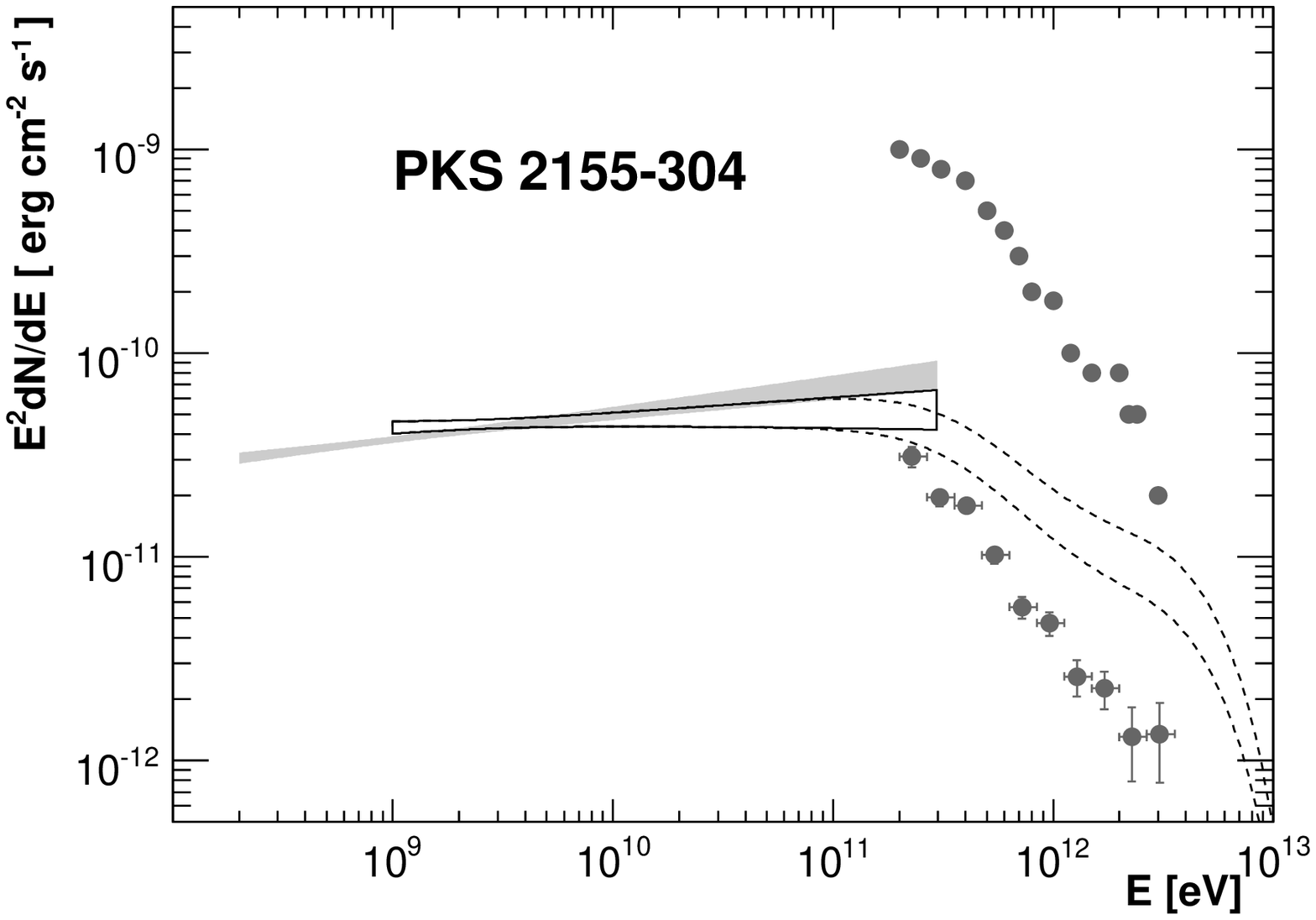}%
\includeSpec{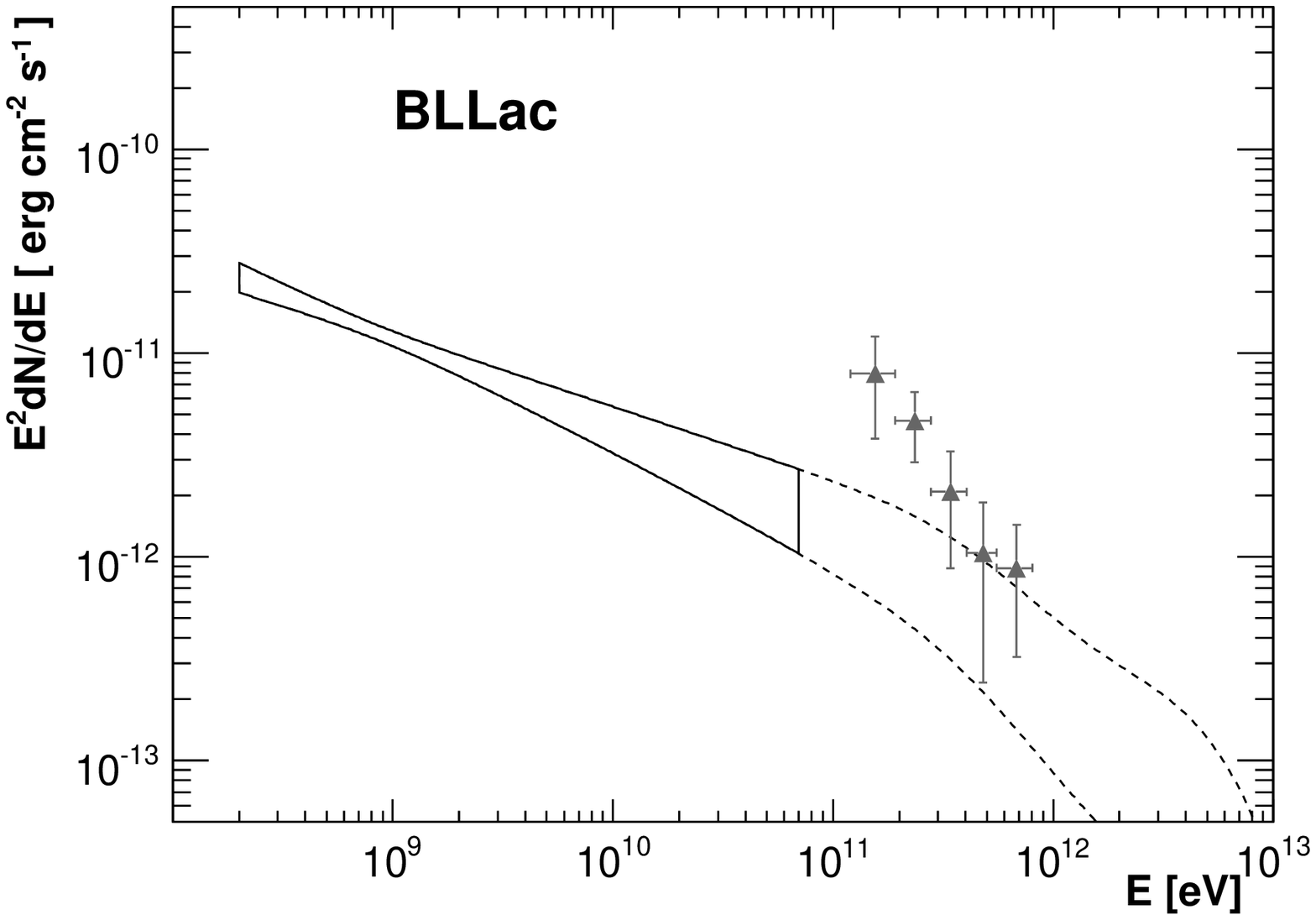}

\includeSpec{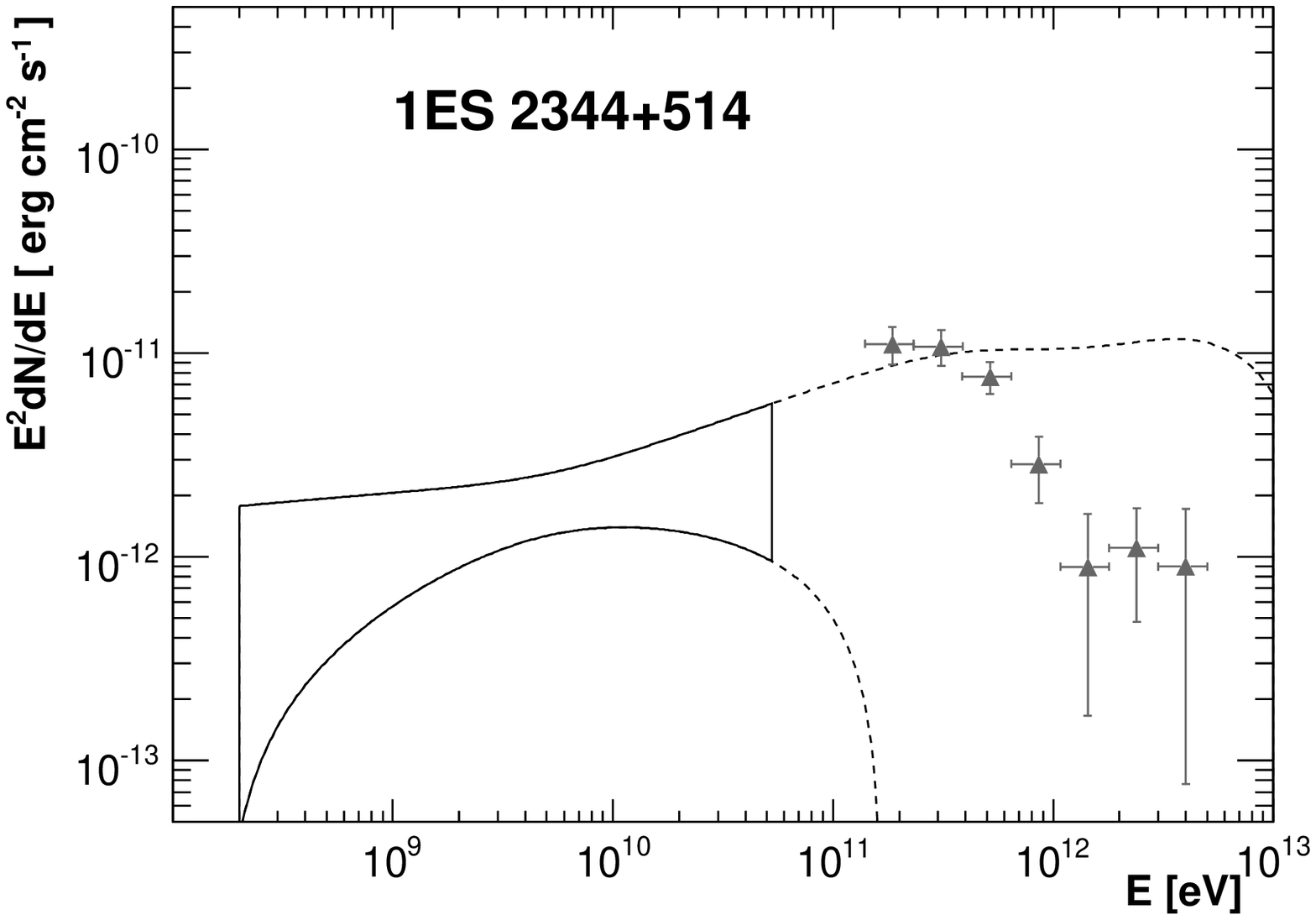}%
\includeSpec{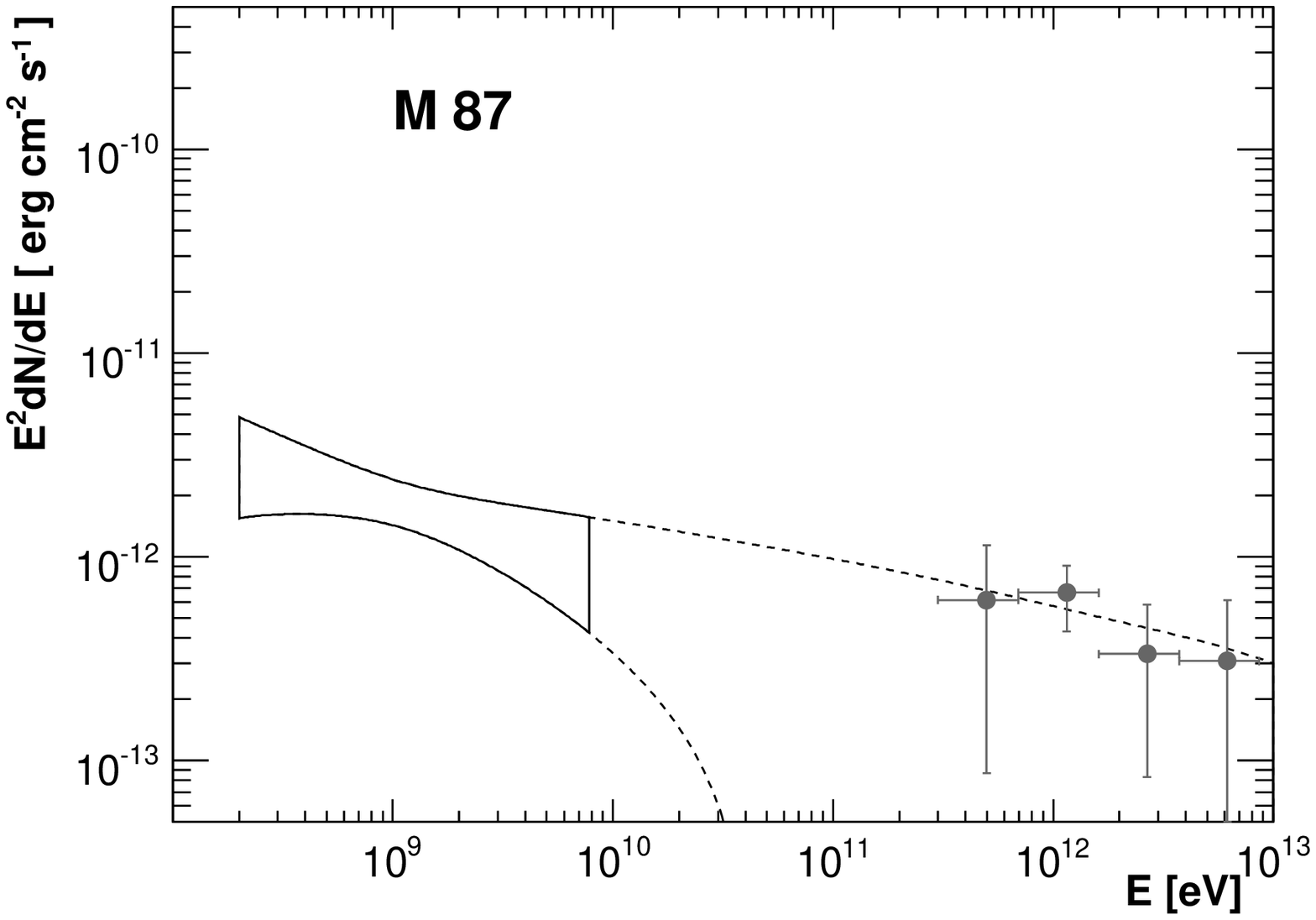}

\includeSpec{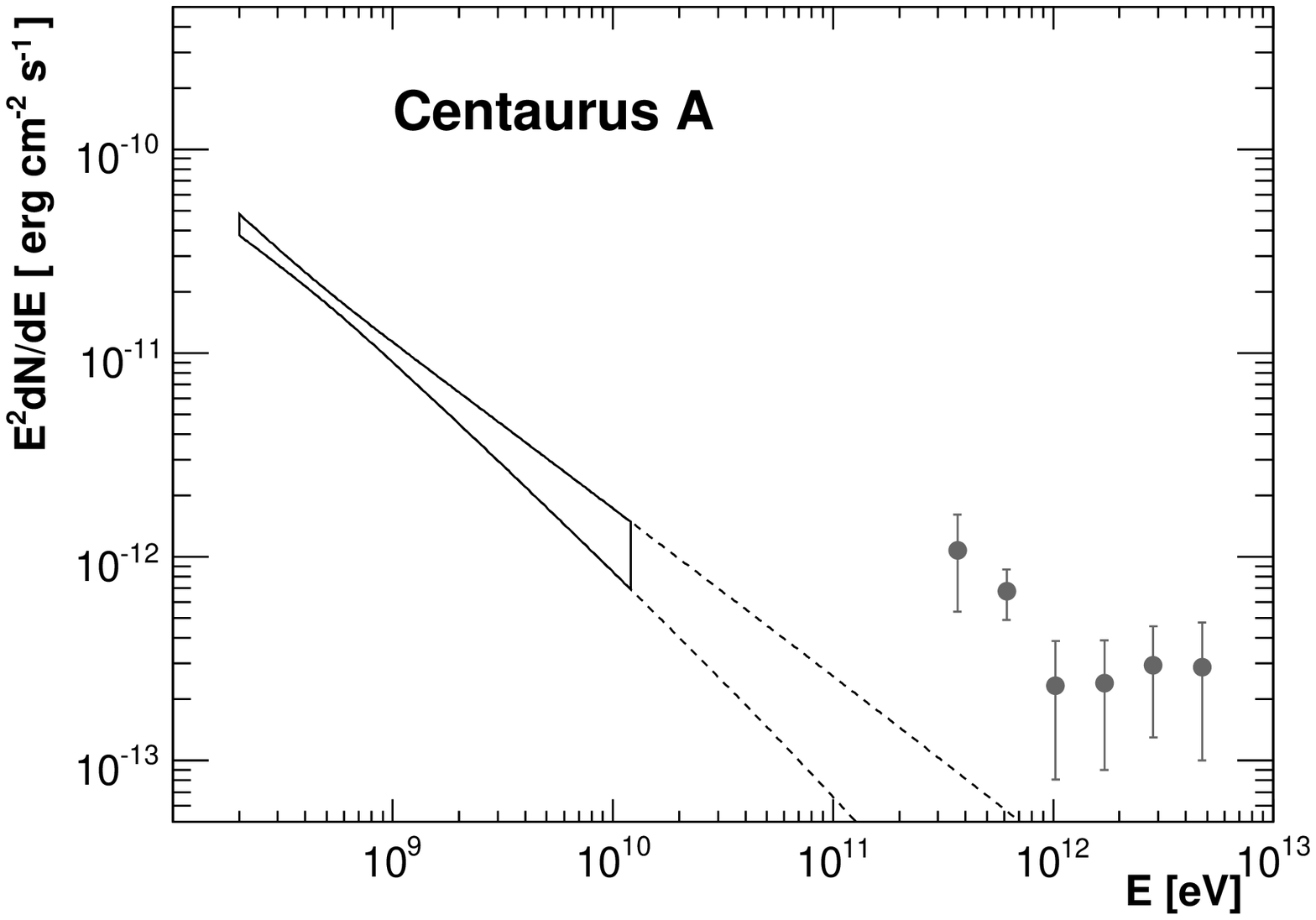}%
\hspace*{\specwidth}

\caption{\label{FIG::SPEC_TEVDET_3}Continued}
\end{figure}


\begin{figure}[p]
\centering
\includeSpec{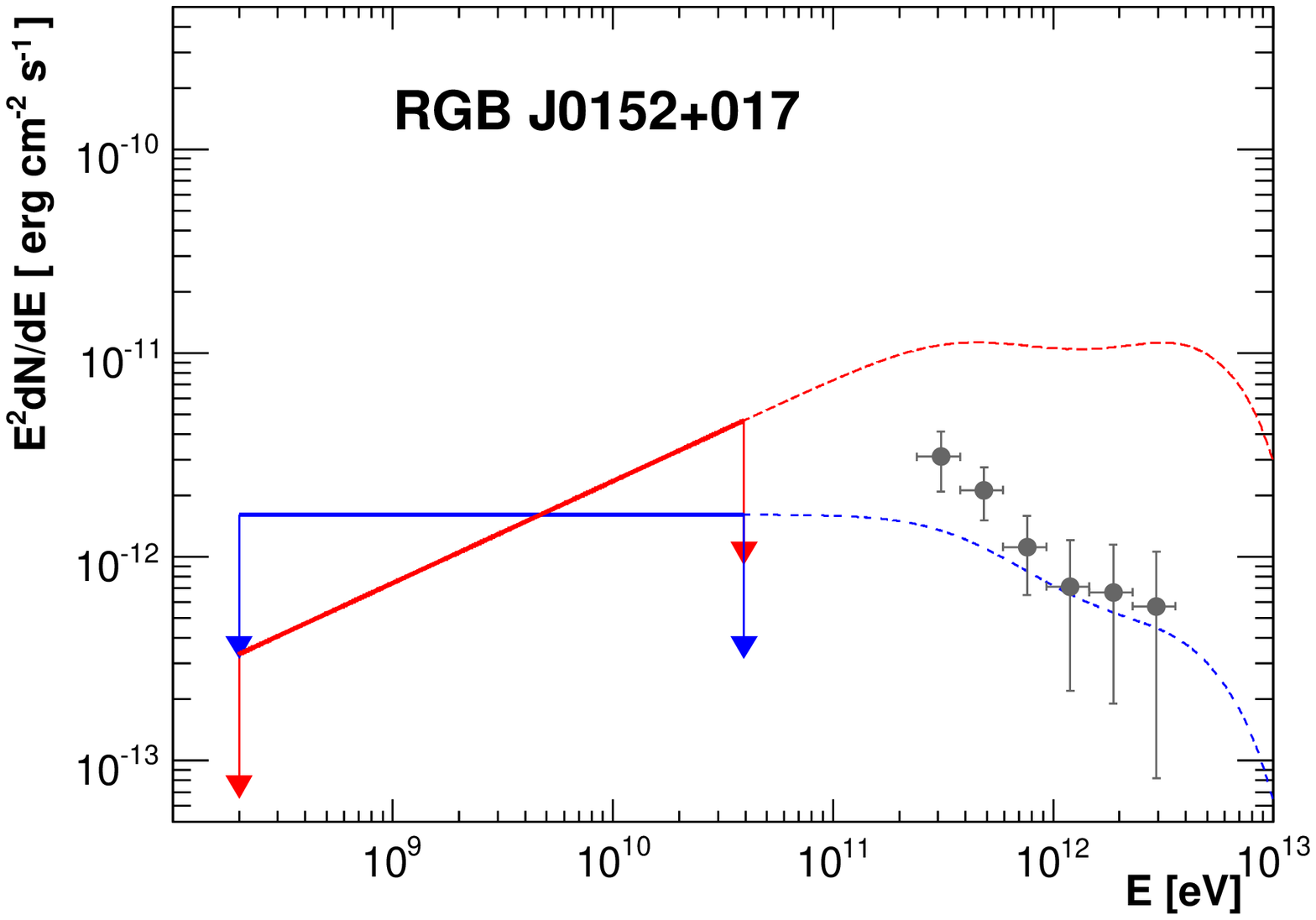}%
\includeSpec{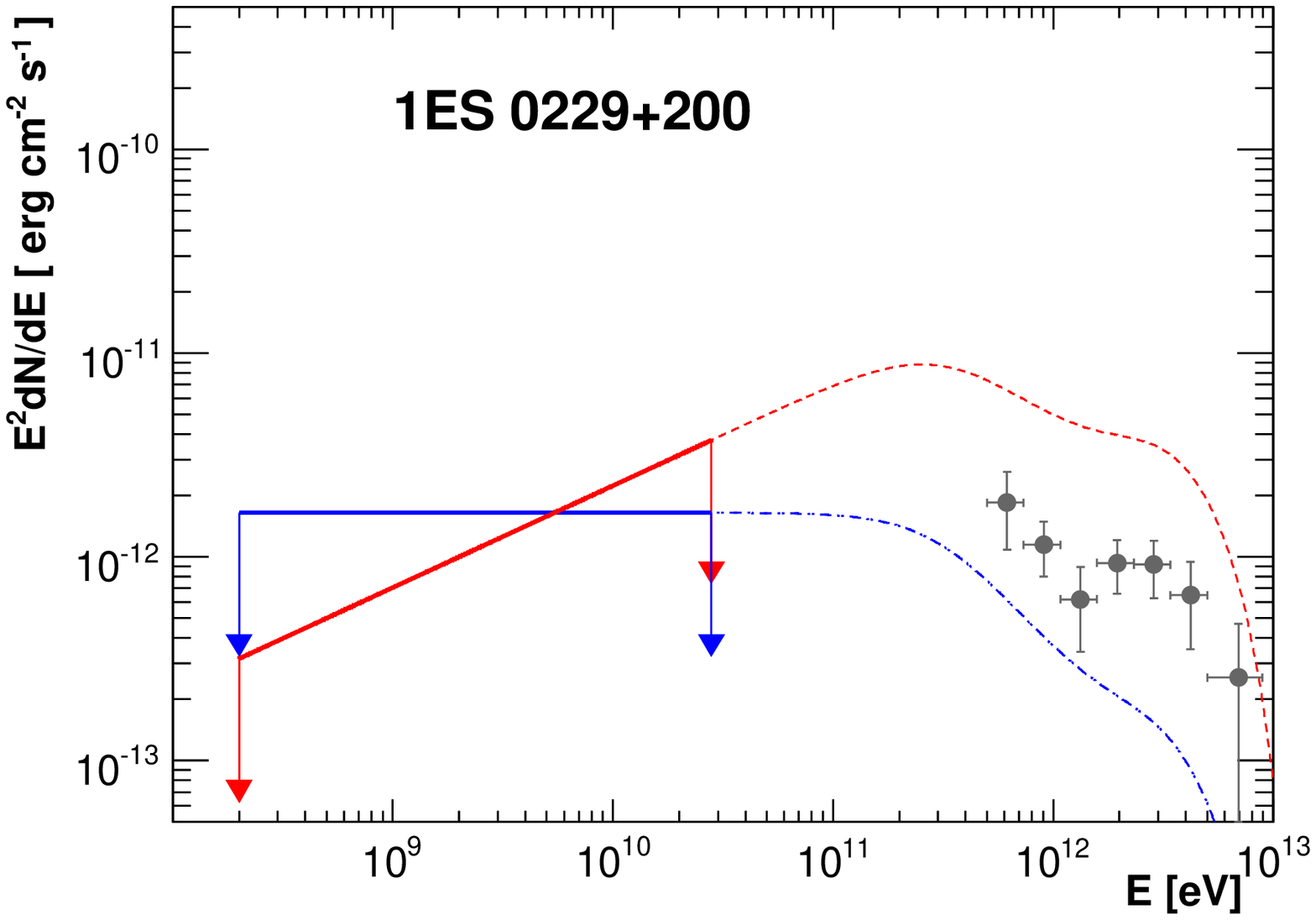}

\includeSpec{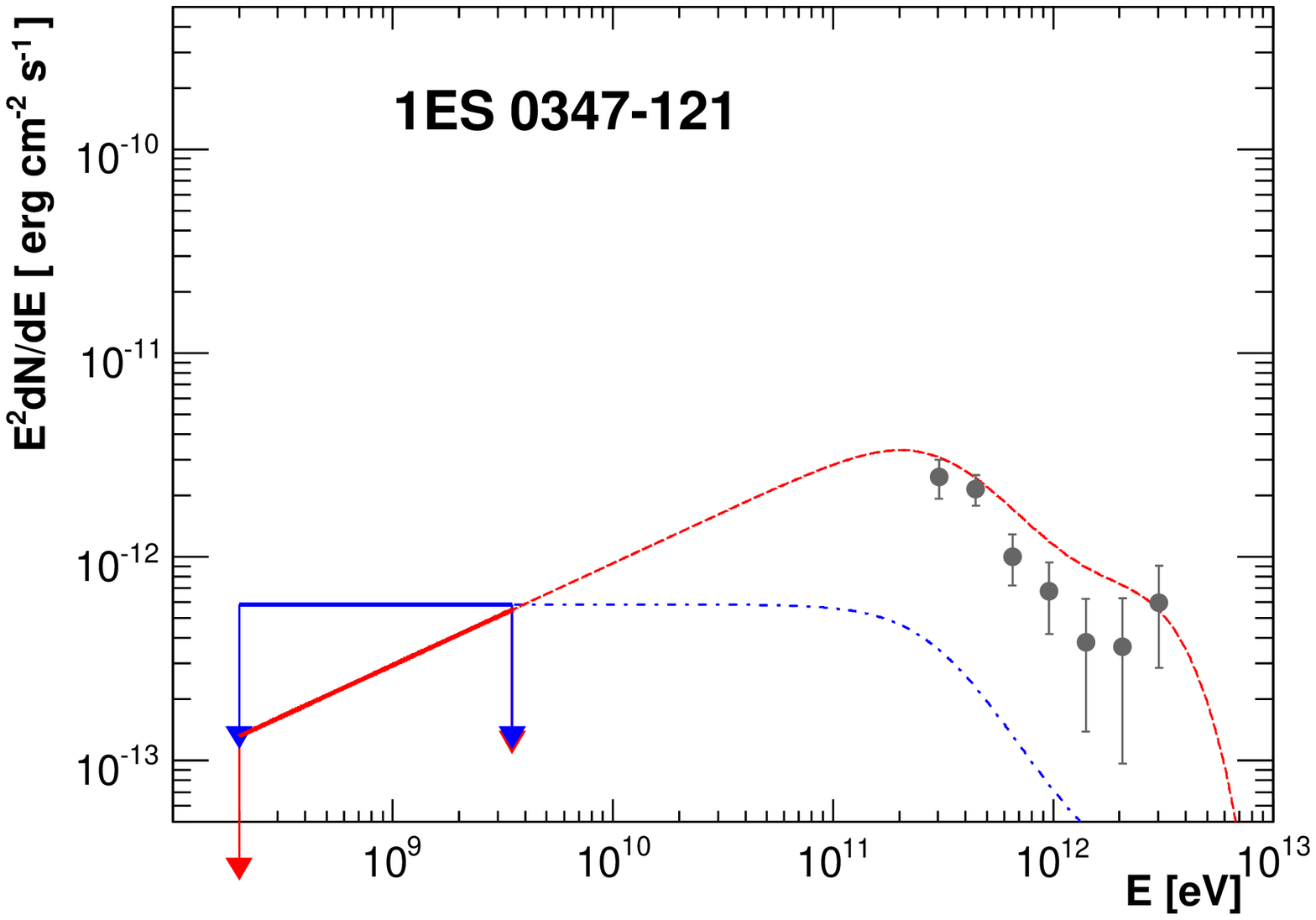}%
\includeSpec{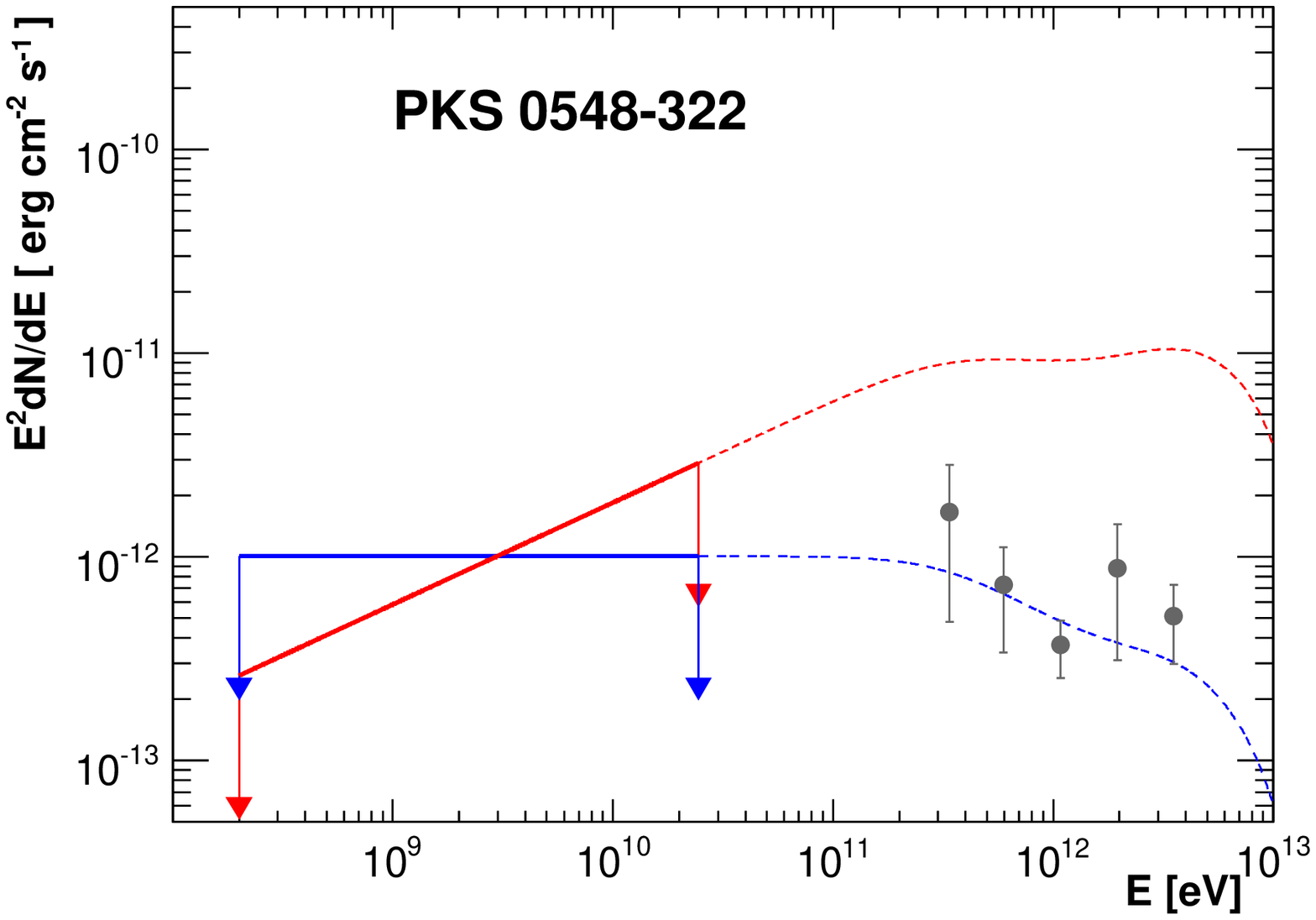}

\includeSpec{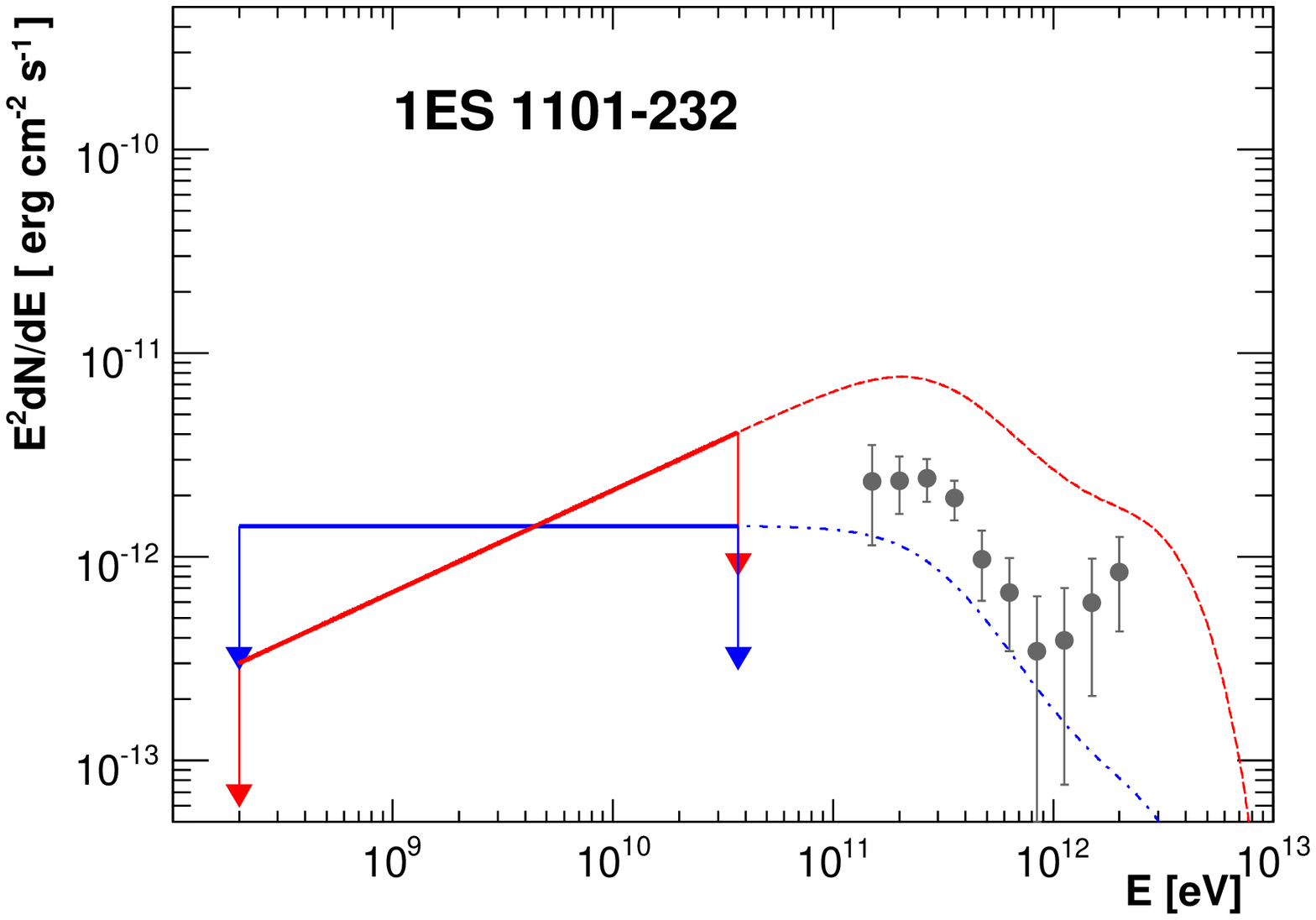}%
\includeSpec{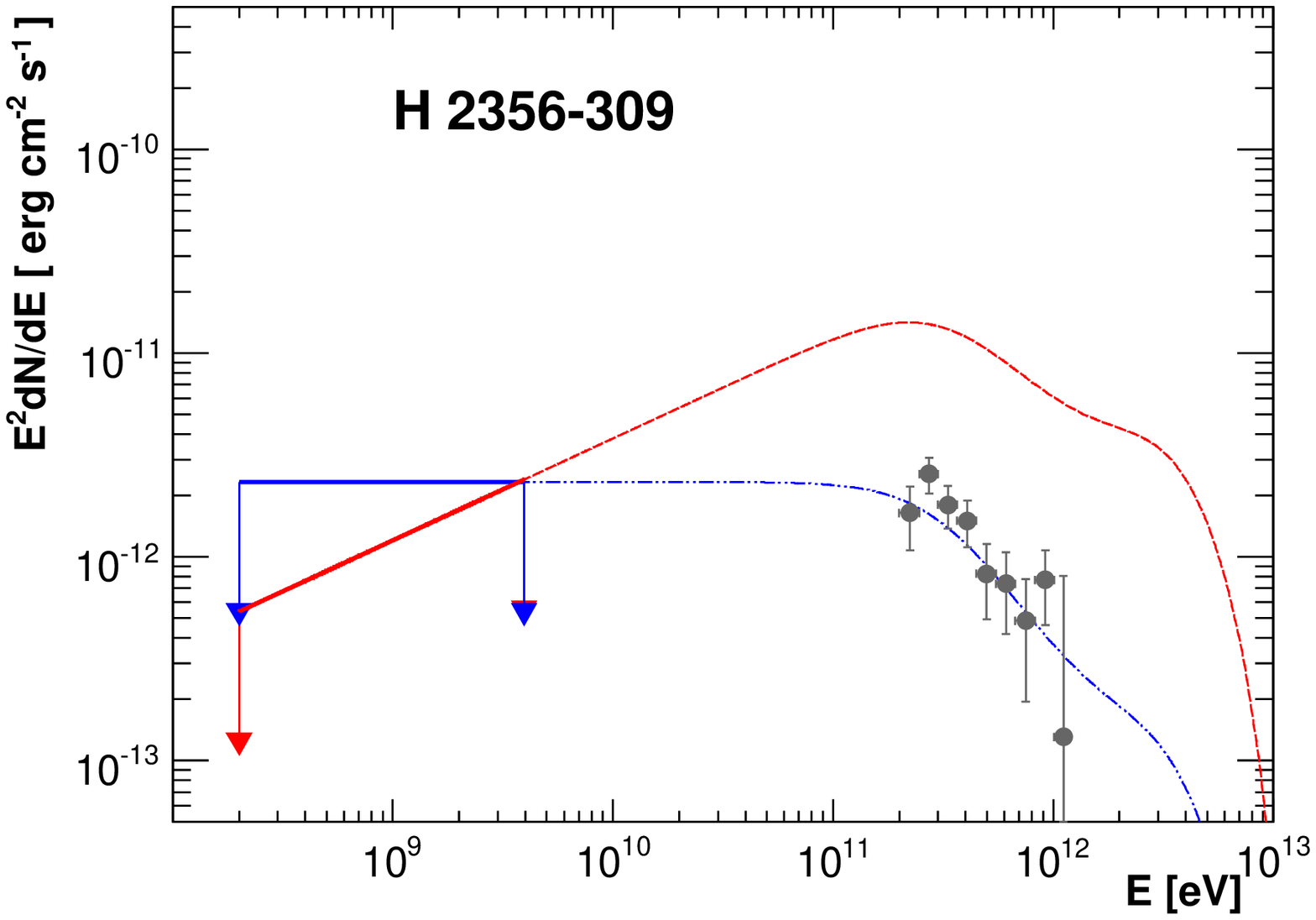}
\caption{\label{FIG::SPEC_LIMITS}\Fermi upper-limits with a 
spectral index frozen at 1.5 and 2 for the TeV detected sources. See
caption of Figure \ref{FIG::SPEC_TEVDET_1} for details.}
\end{figure}

\clearpage

\newlength{\lcLuneWidth}
\setlength{\lcLuneWidth}{0.3\textwidth}
\newcommand{\includeLCLune}[1]{\includegraphics[bb=25 0 510 345,clip,width=\lcLuneWidth]{#1}}


\begin{figure}[p]
\centering

\rotatebox{90}{\makebox[0mm][r]{Flux -- $F(>200\mathrm{MeV})$ [$10^{-9}$\cmsc]}}%
\begin{minipage}[t]{0.91\textwidth}
\ 

\includeLCLune{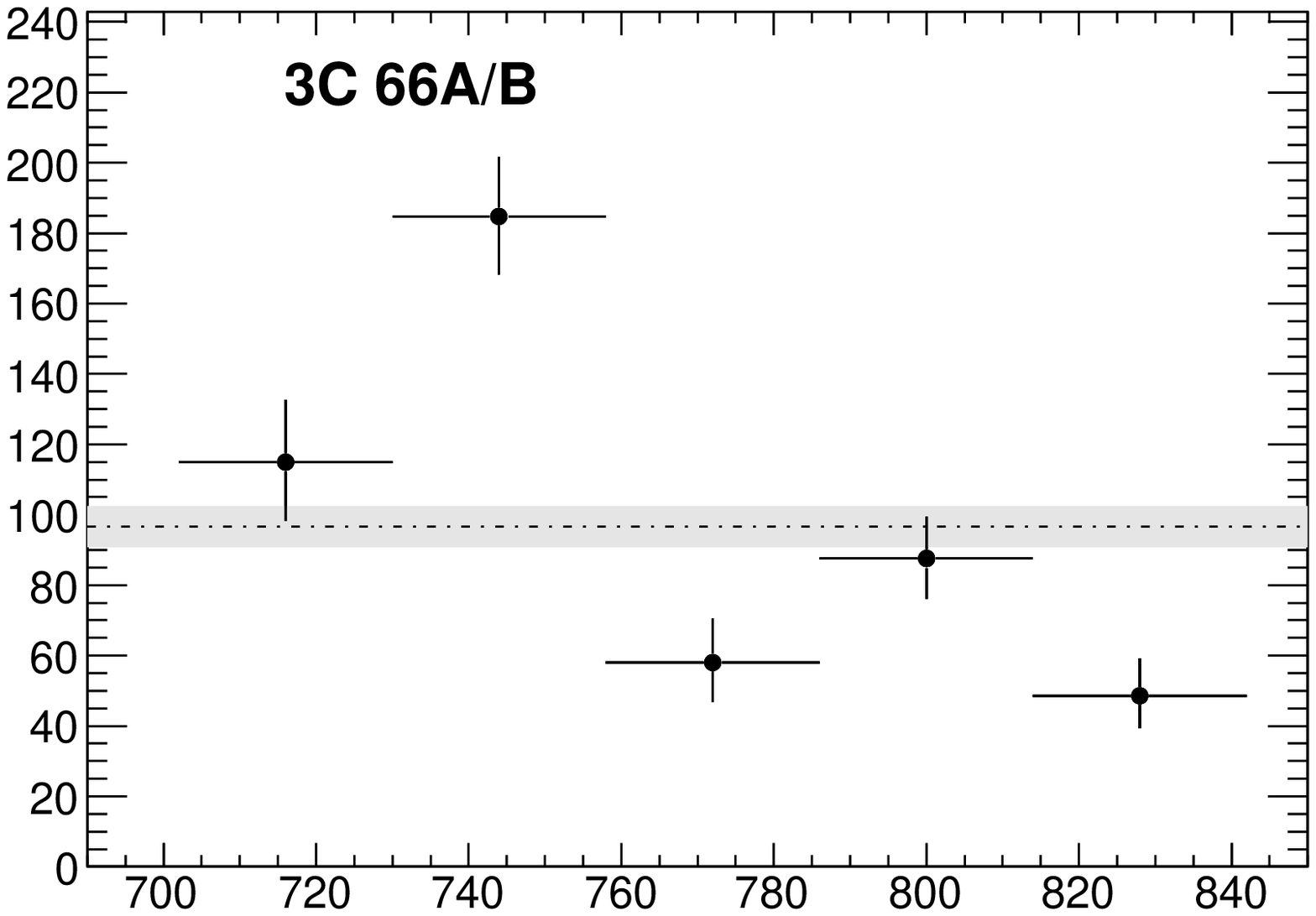}%
\includeLCLune{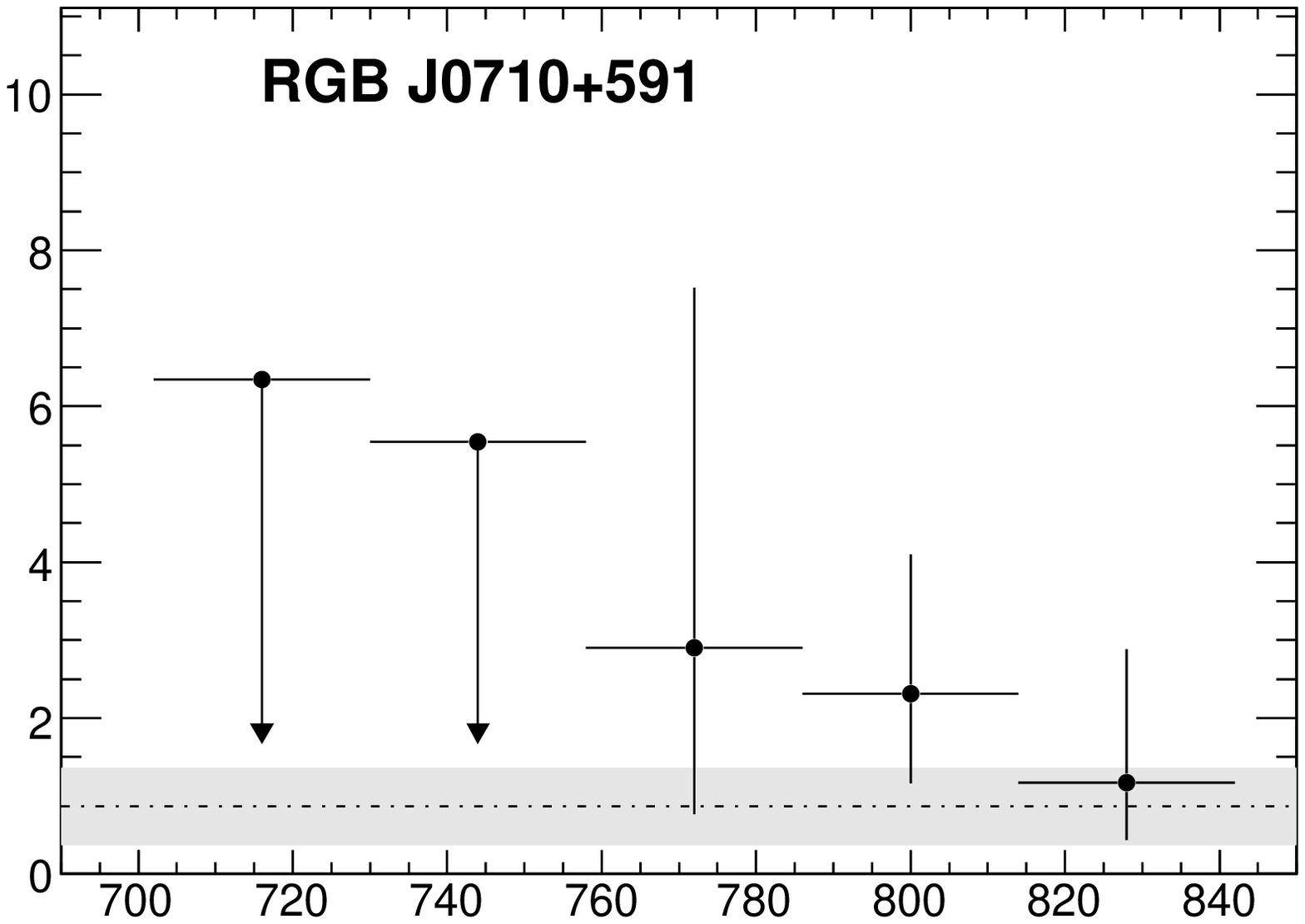}%
\includeLCLune{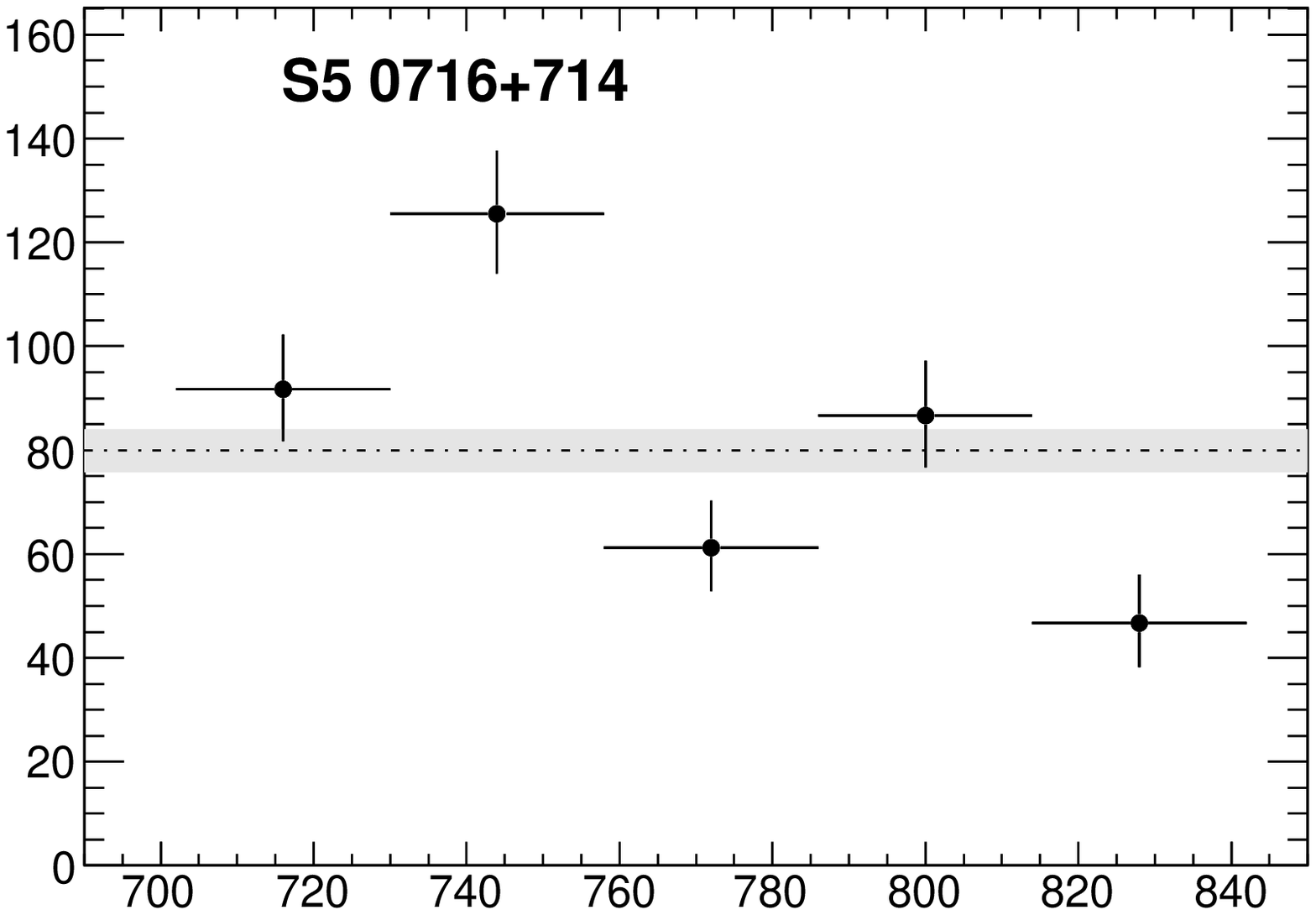}

\includeLCLune{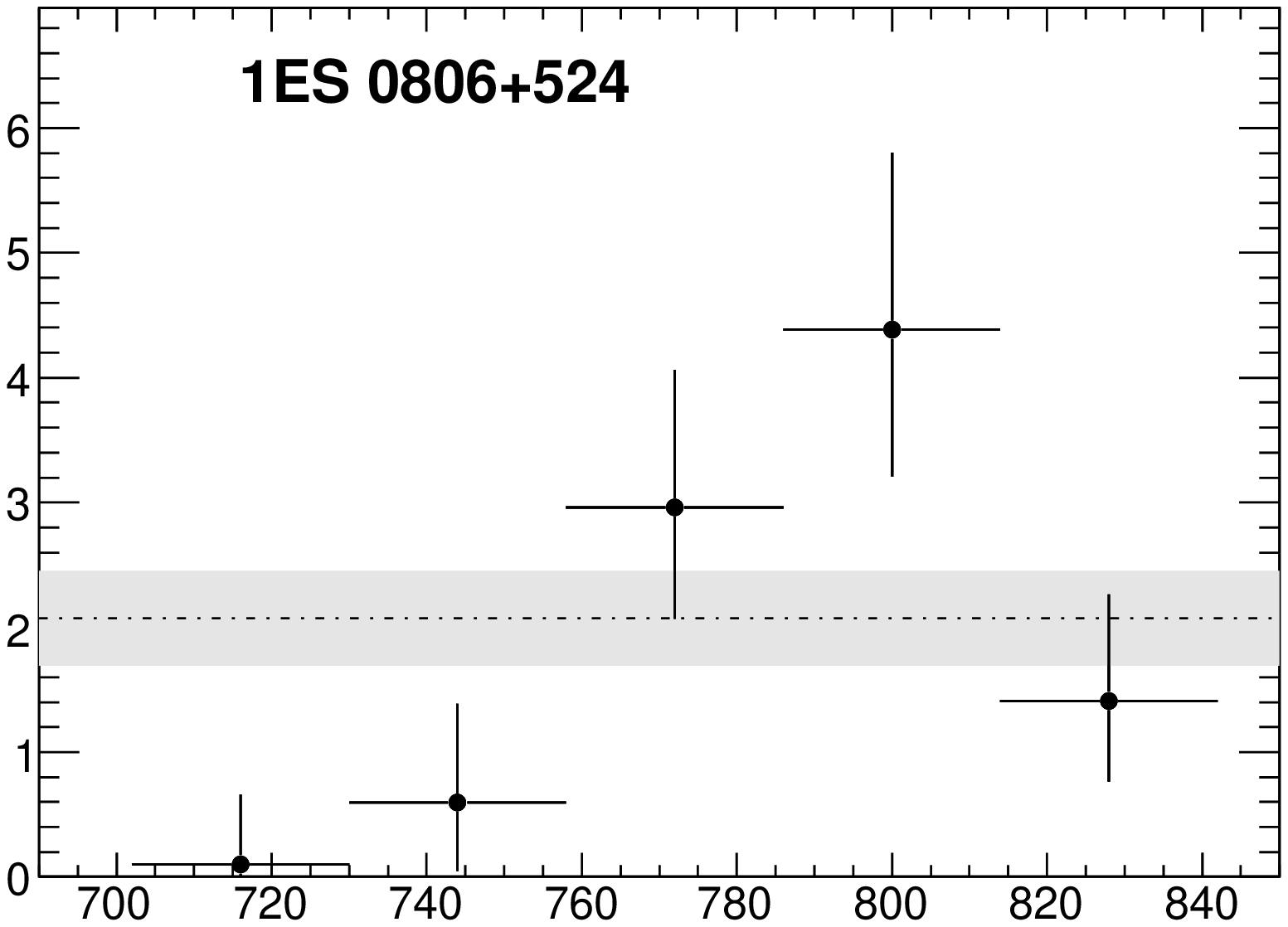}%
\includeLCLune{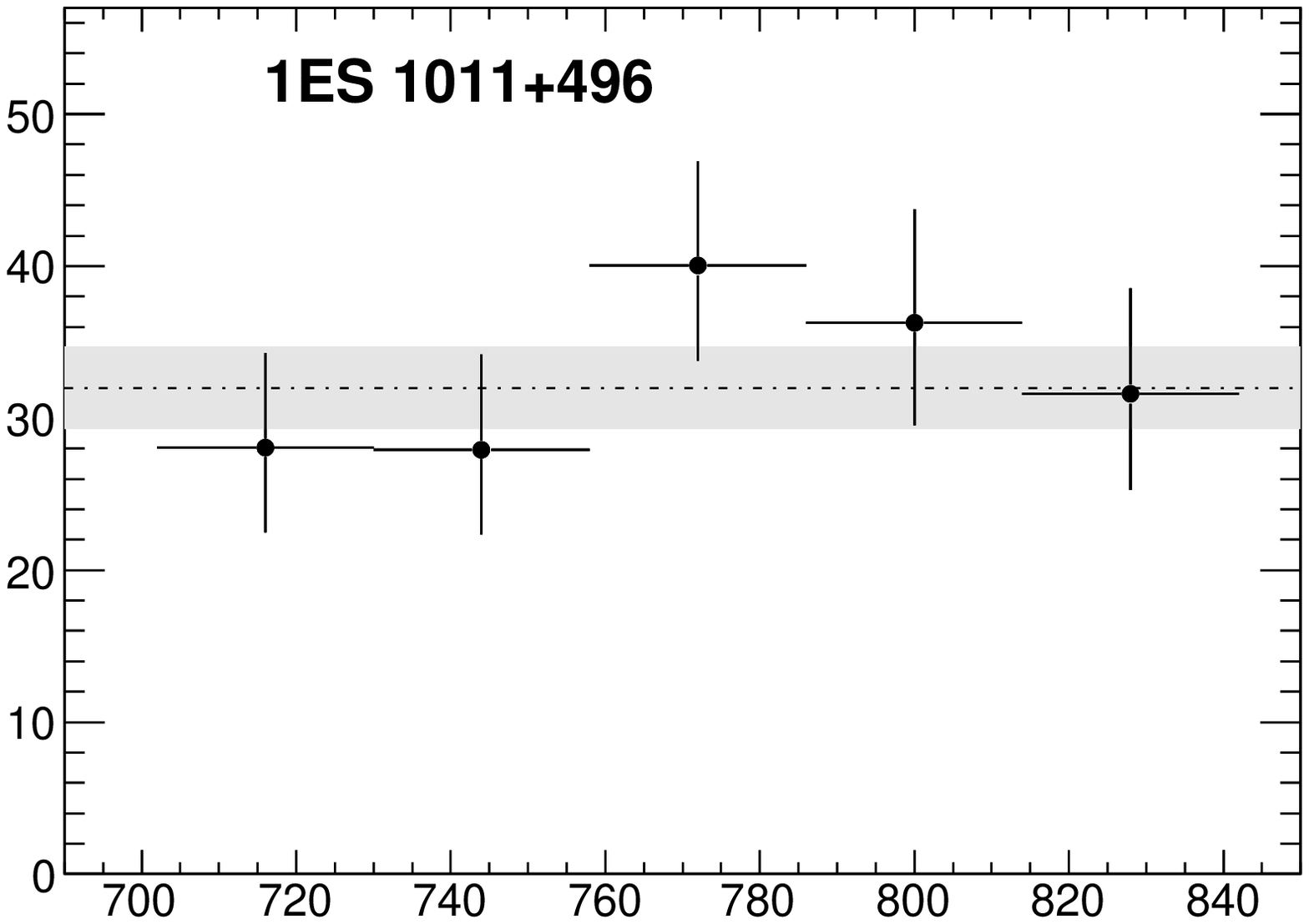}%
\includeLCLune{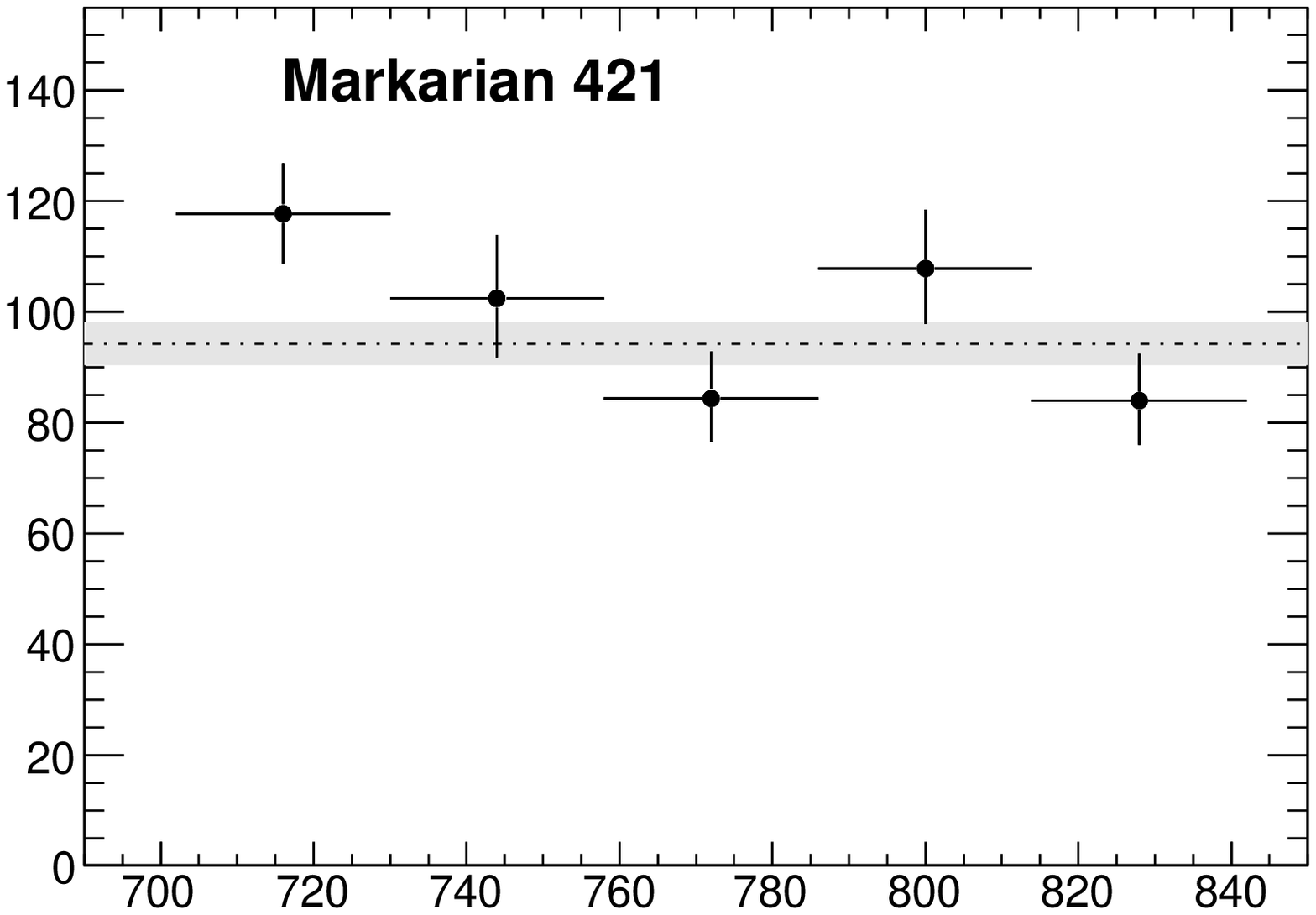}

\includeLCLune{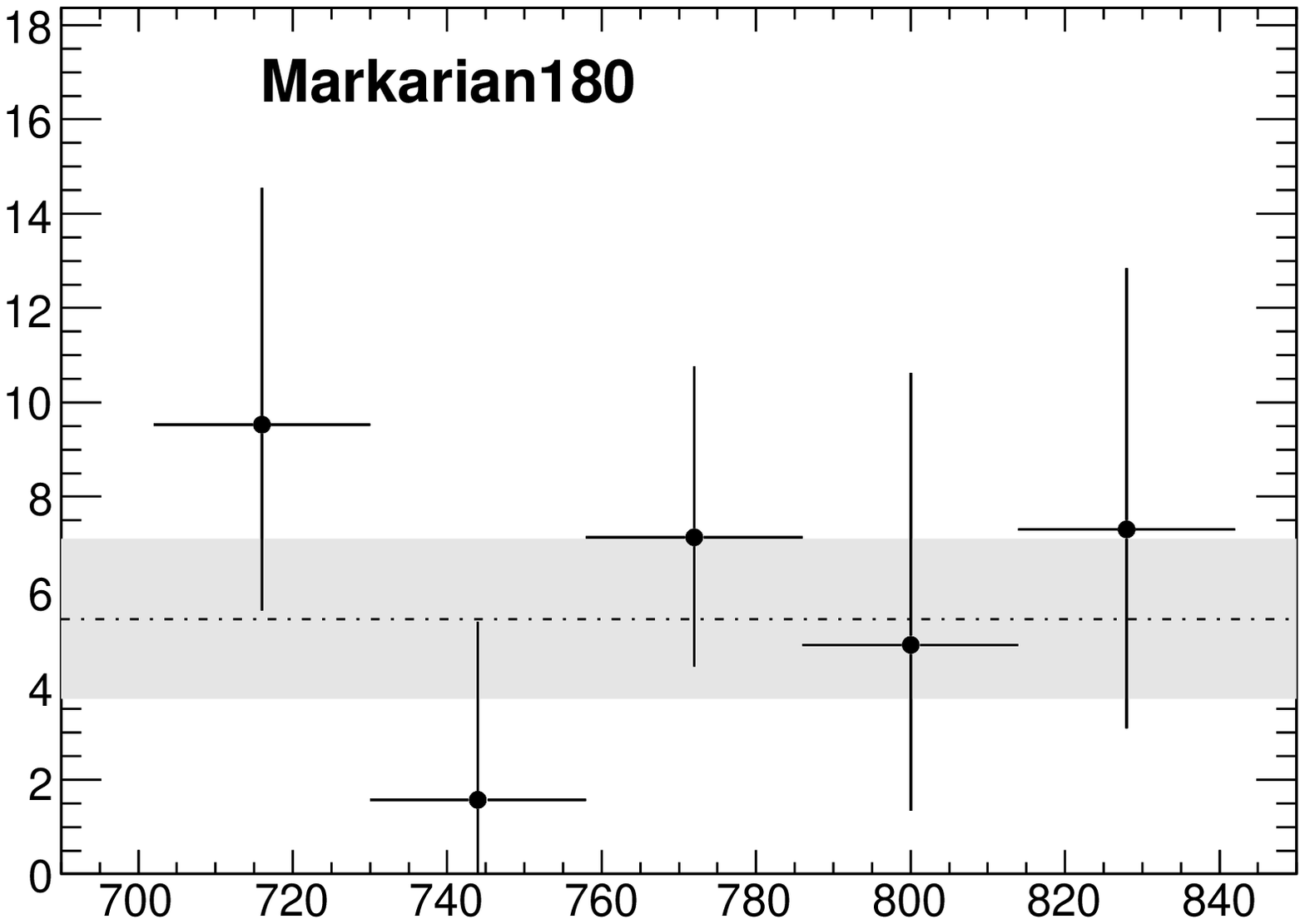}%
\includeLCLune{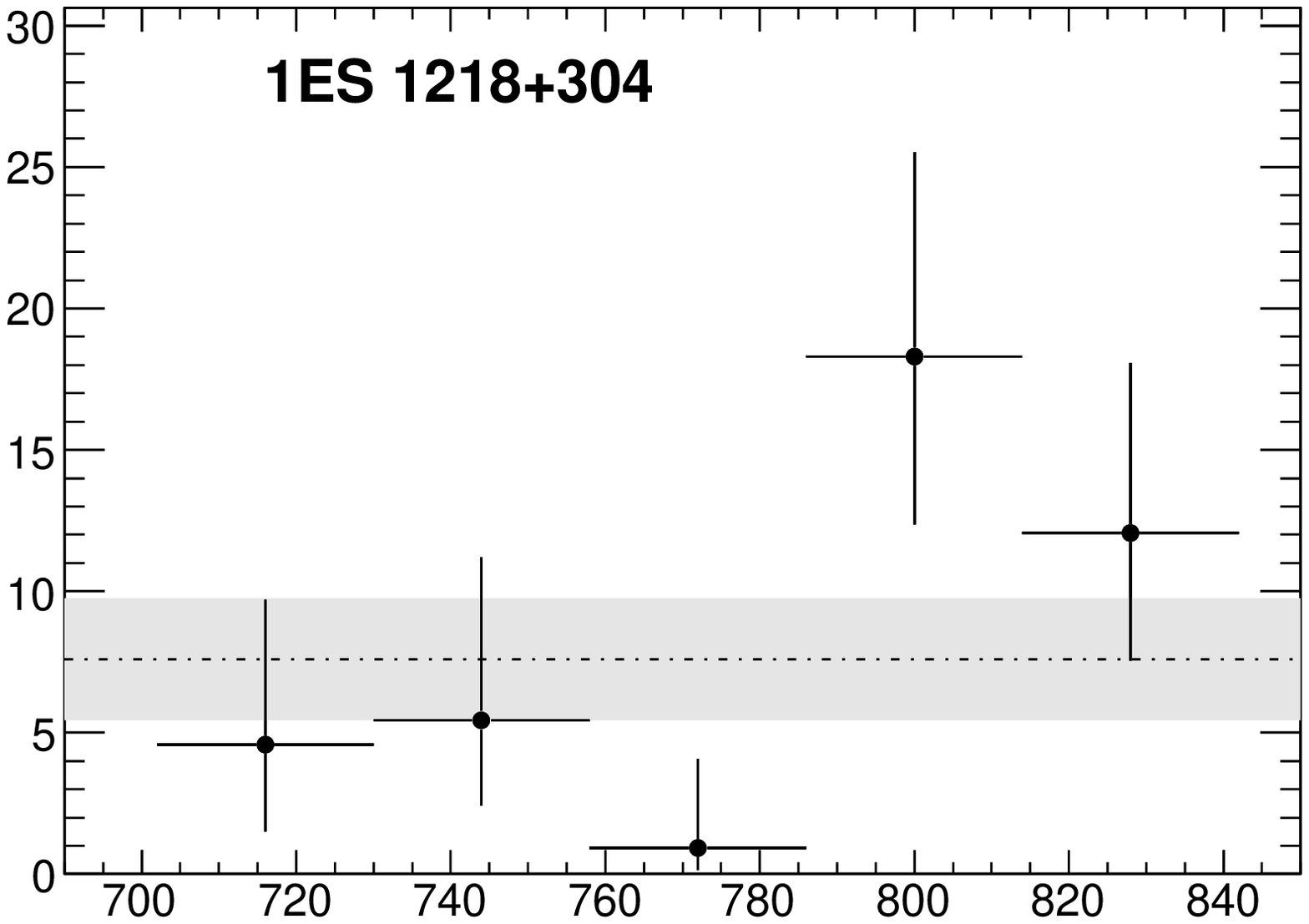}%
\includeLCLune{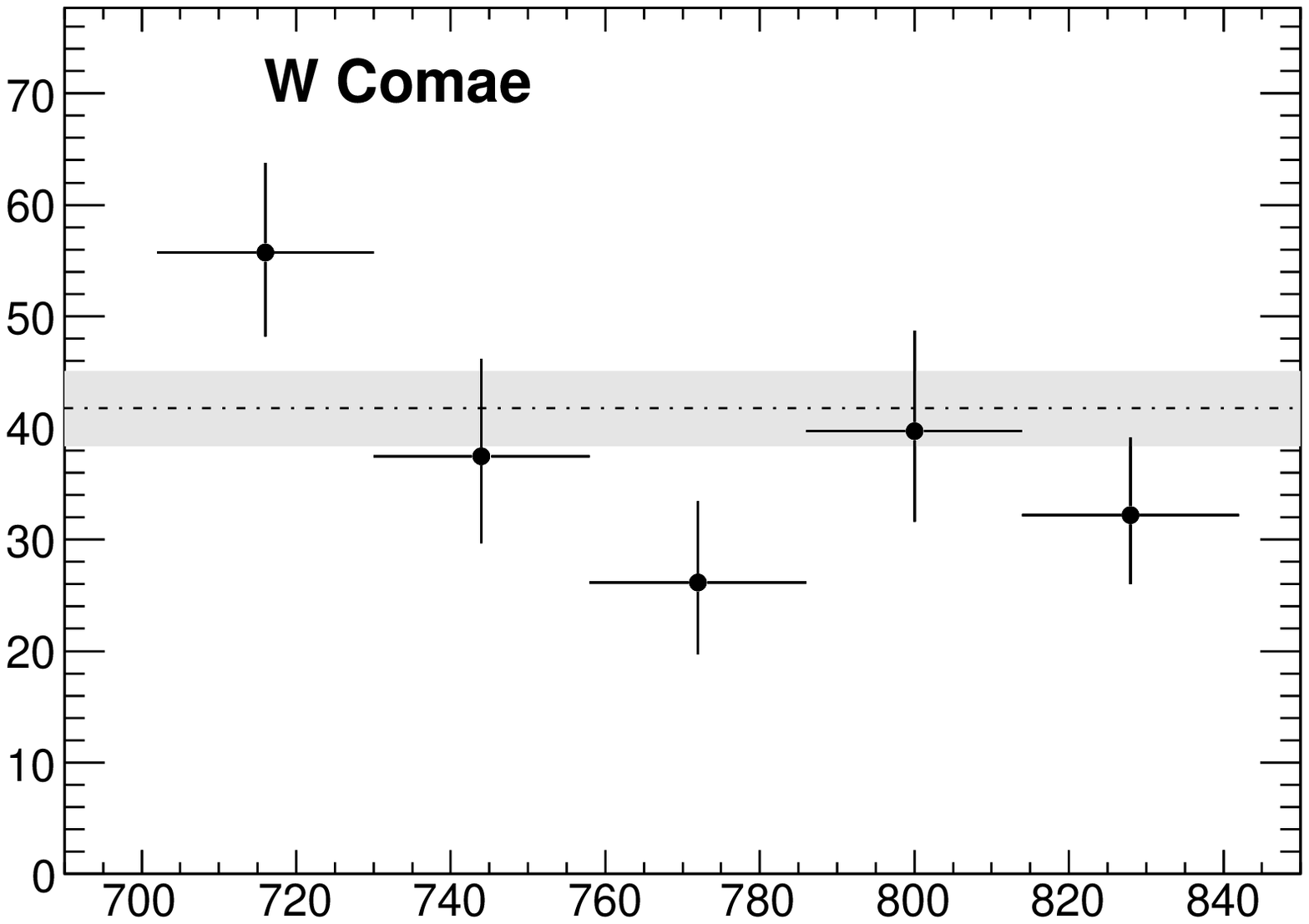}

\includeLCLune{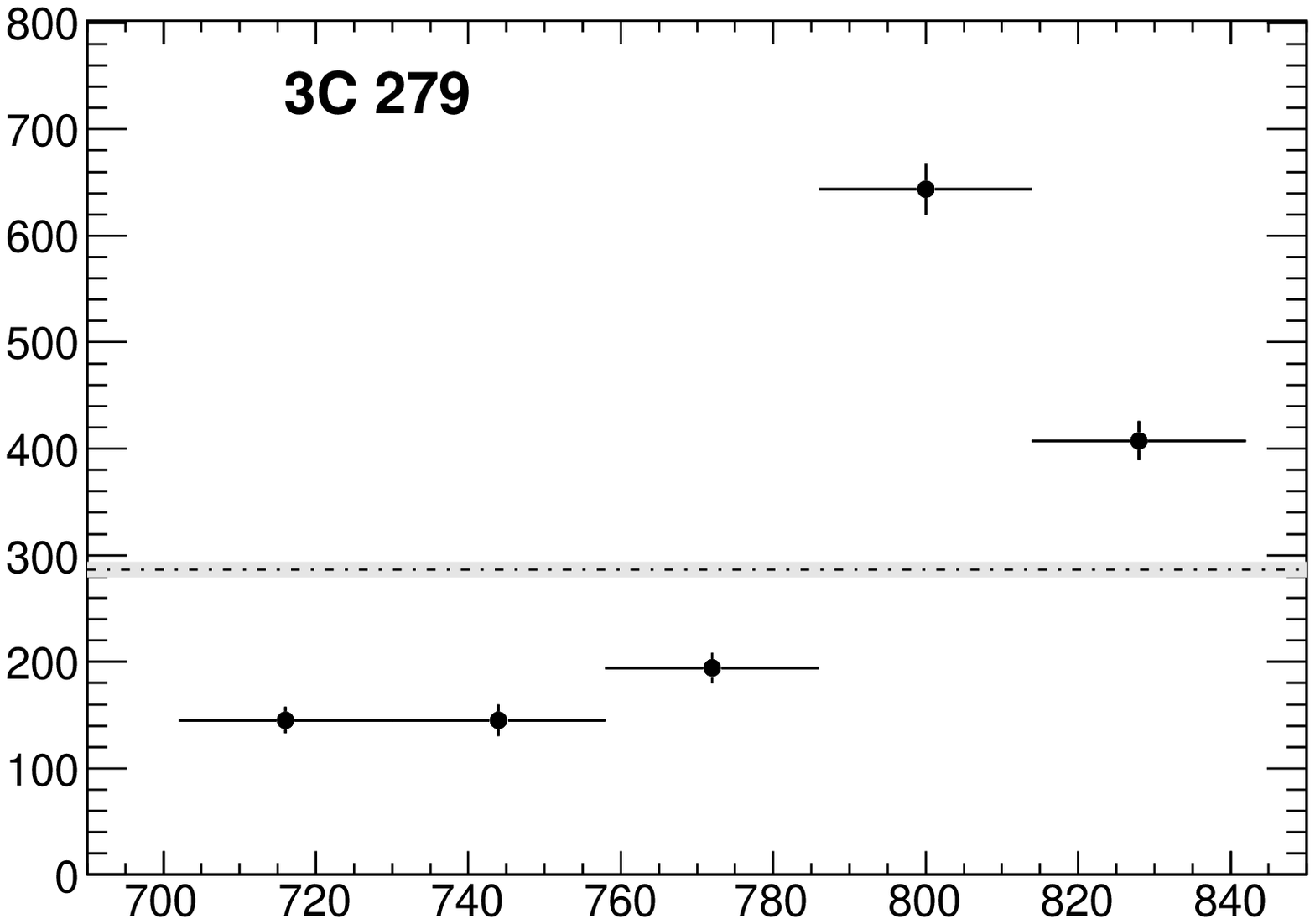}%
\includeLCLune{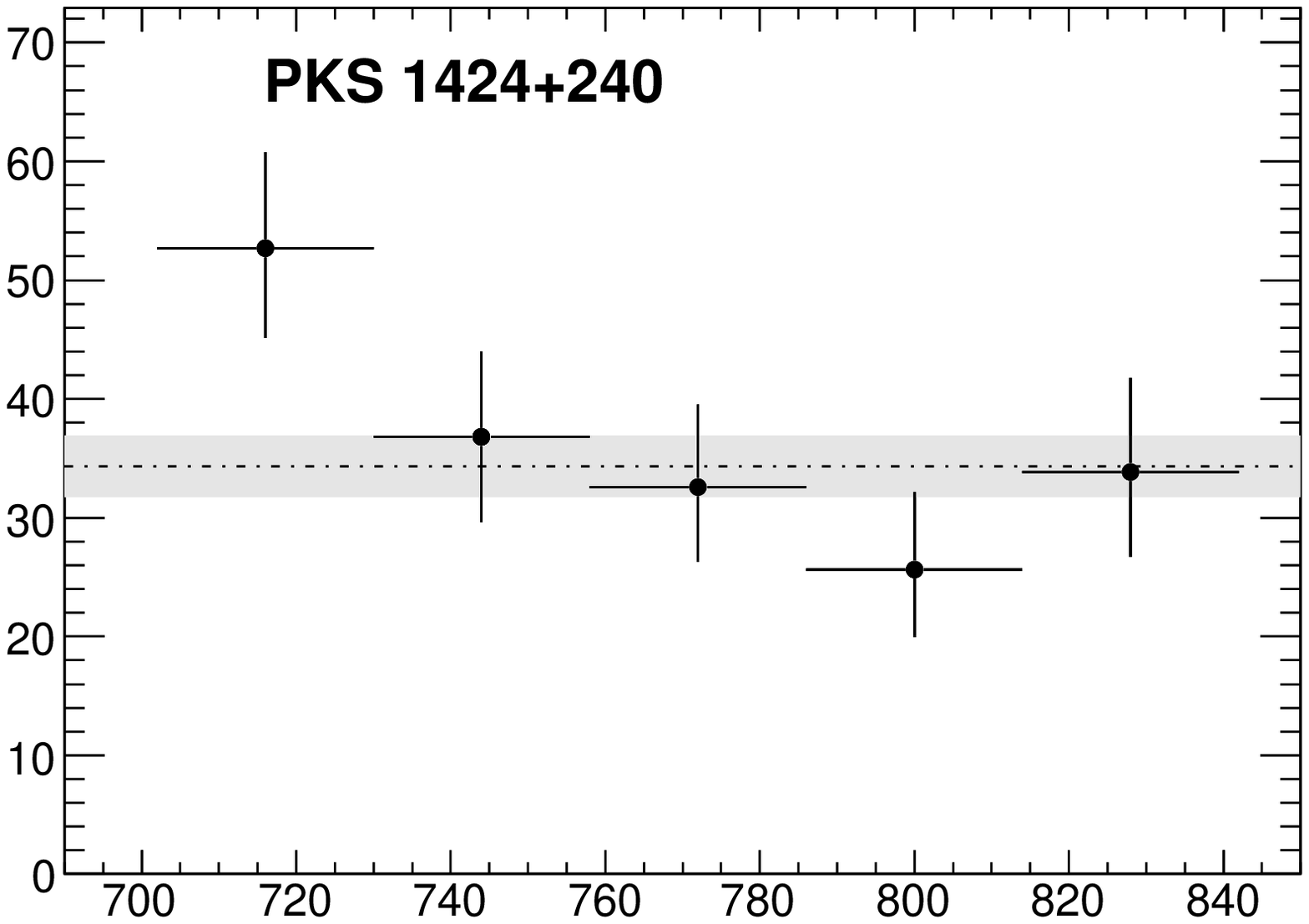}%
\includeLCLune{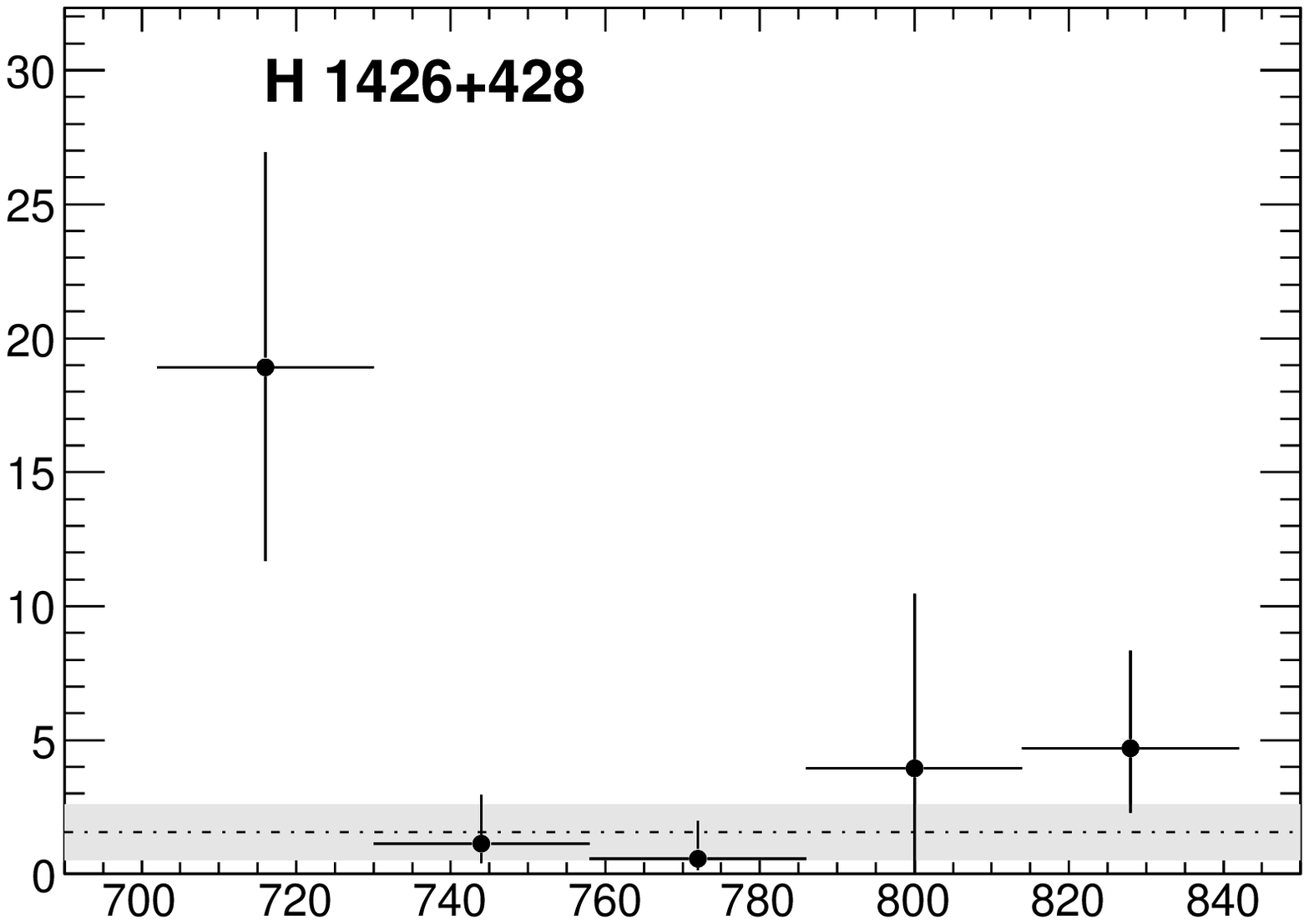}

\includeLCLune{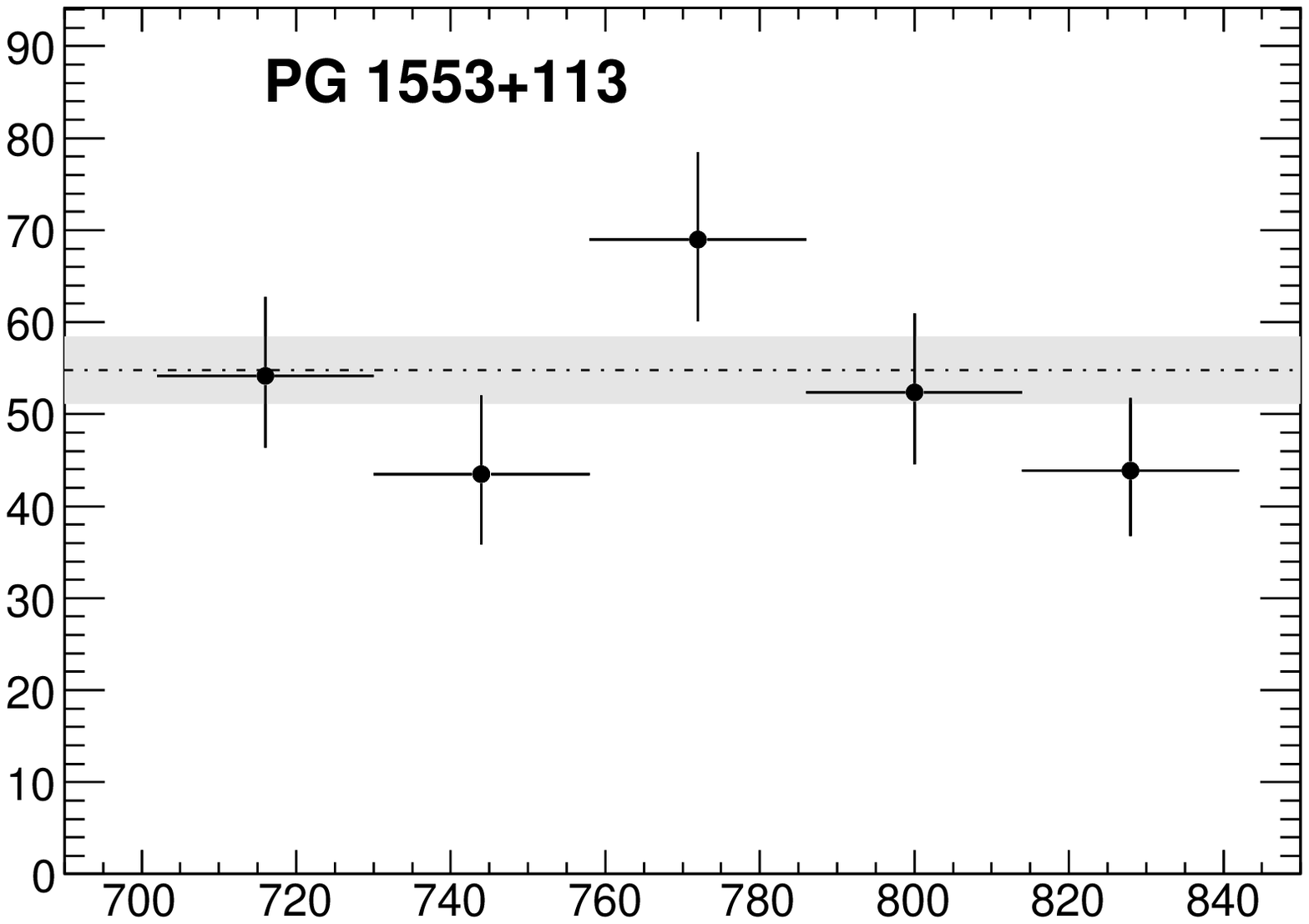}%
\includeLCLune{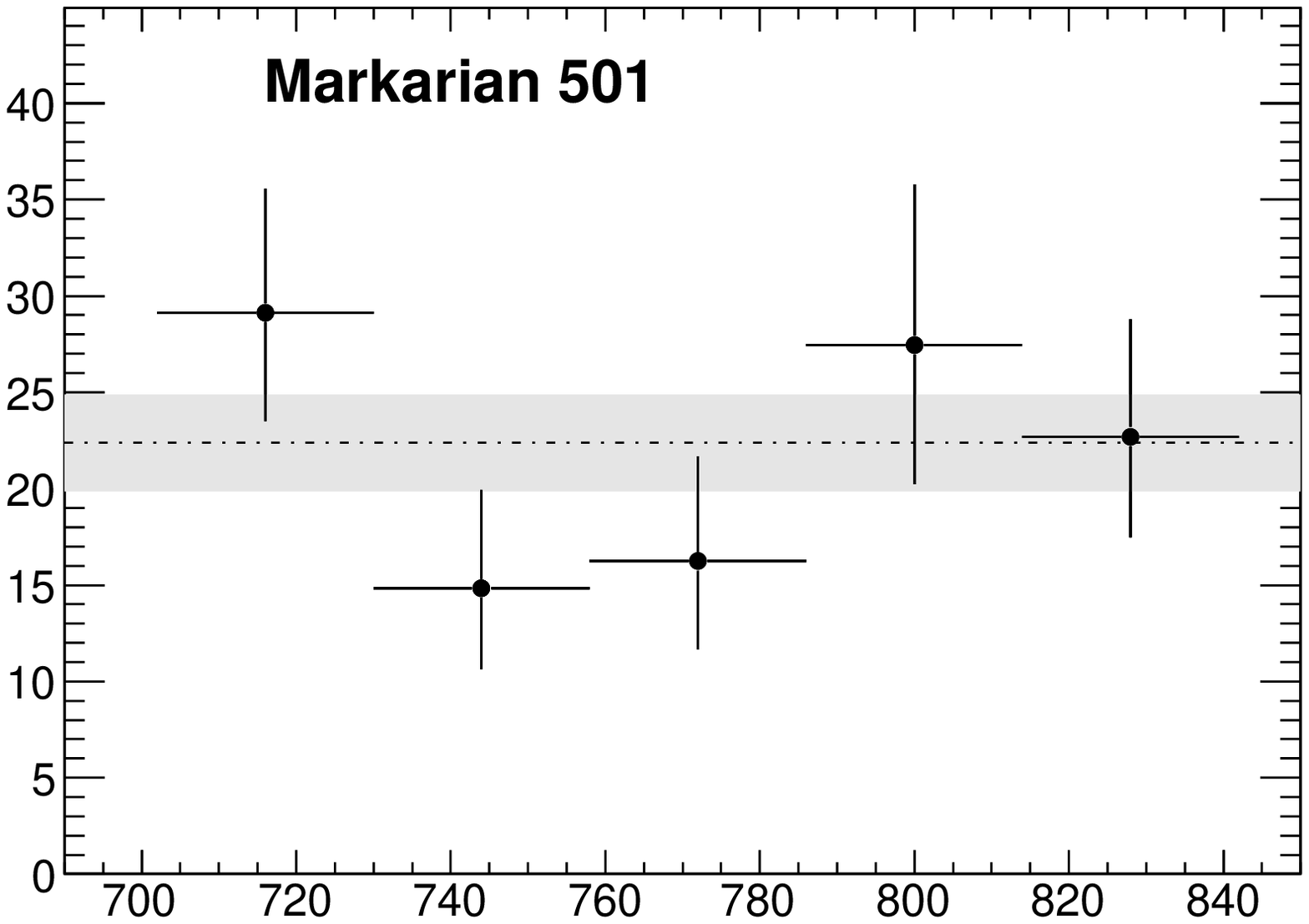}%
\includeLCLune{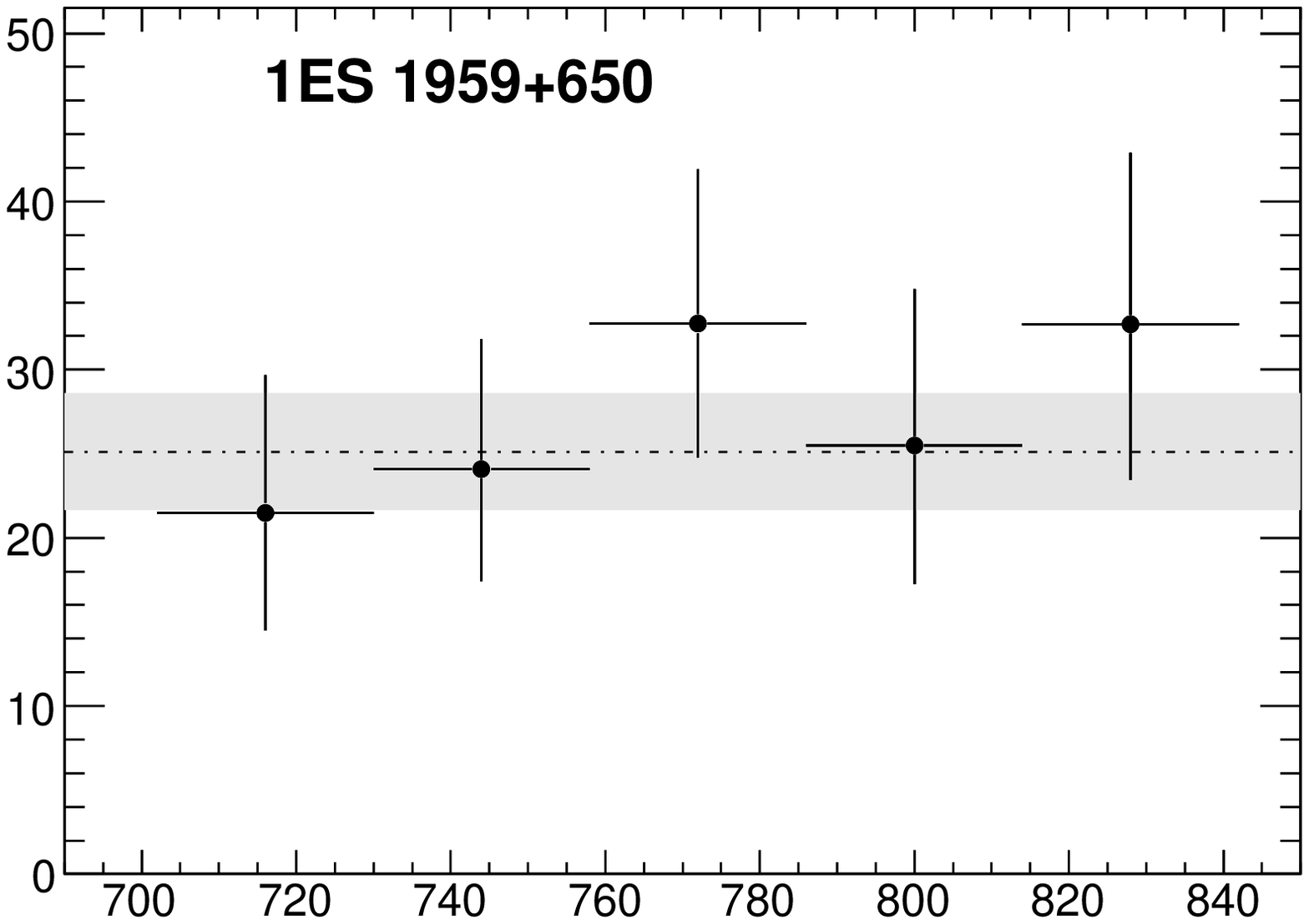}

\end{minipage}

Date -- MJD-54000 [days]

\caption{\label{FIG::LC28_1} 28-day light curves for \Fermic-detected
sources, centered on the new moon. For each flux point, the vertical
bar shows the statistical error (only) and the horizontal bar
indicates the duration of integration. For each source, the mean flux
over the full duration of the study is shown as a dashed line, and the
systematic uncertainty in the flux points, which is estimated to be
3\% of the flux, is shown as a gray band.}
\end{figure}


\addtocounter{figure}{-1}
\begin{figure}[p]
\centering
\rotatebox{90}{\makebox[0mm][r]{Flux -- $F(>200\mathrm{MeV})$ [$10^{-9}$\cmsc]}}%
\begin{minipage}[t]{0.91\textwidth}
\

\includeLCLune{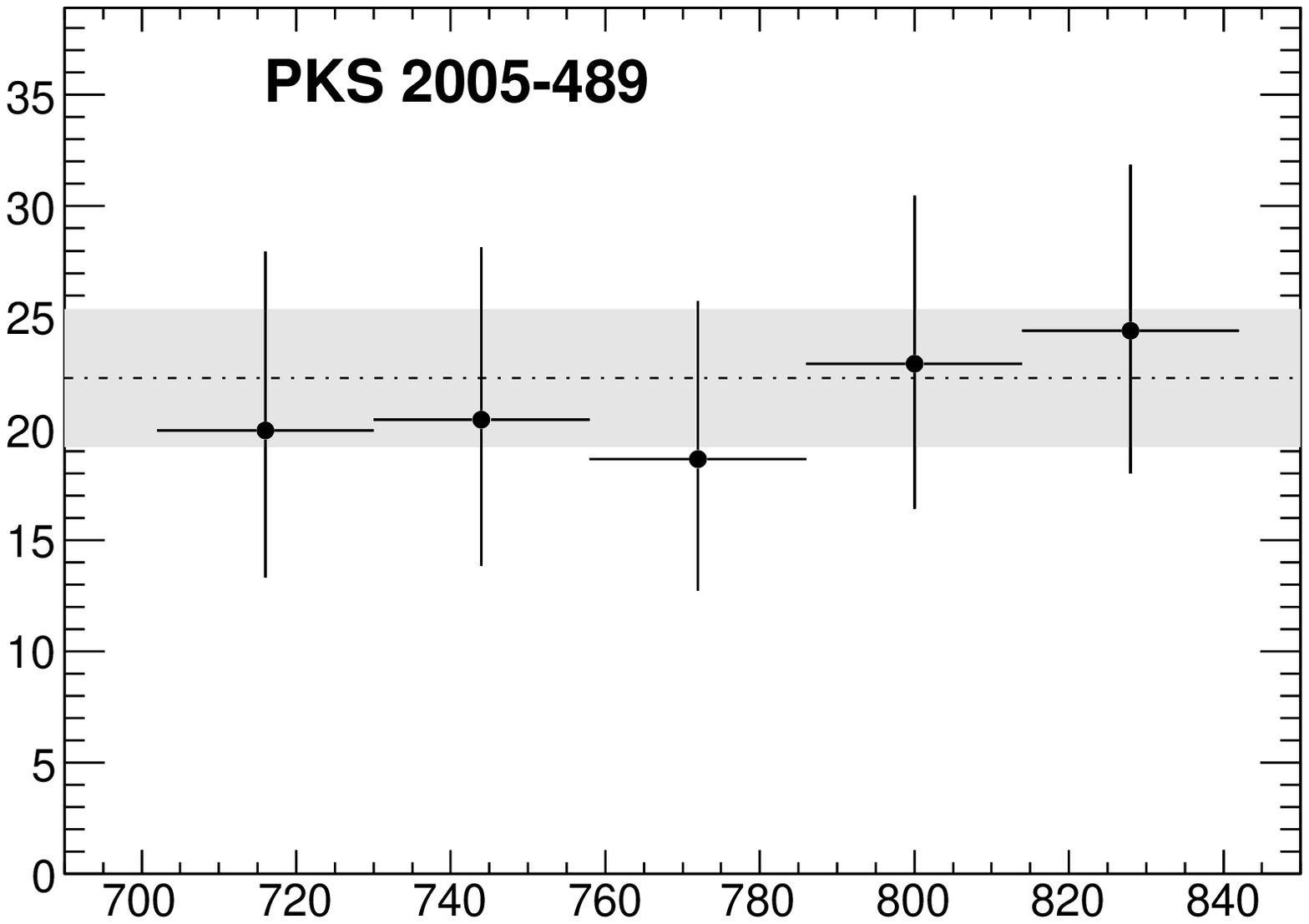}%
\includeLCLune{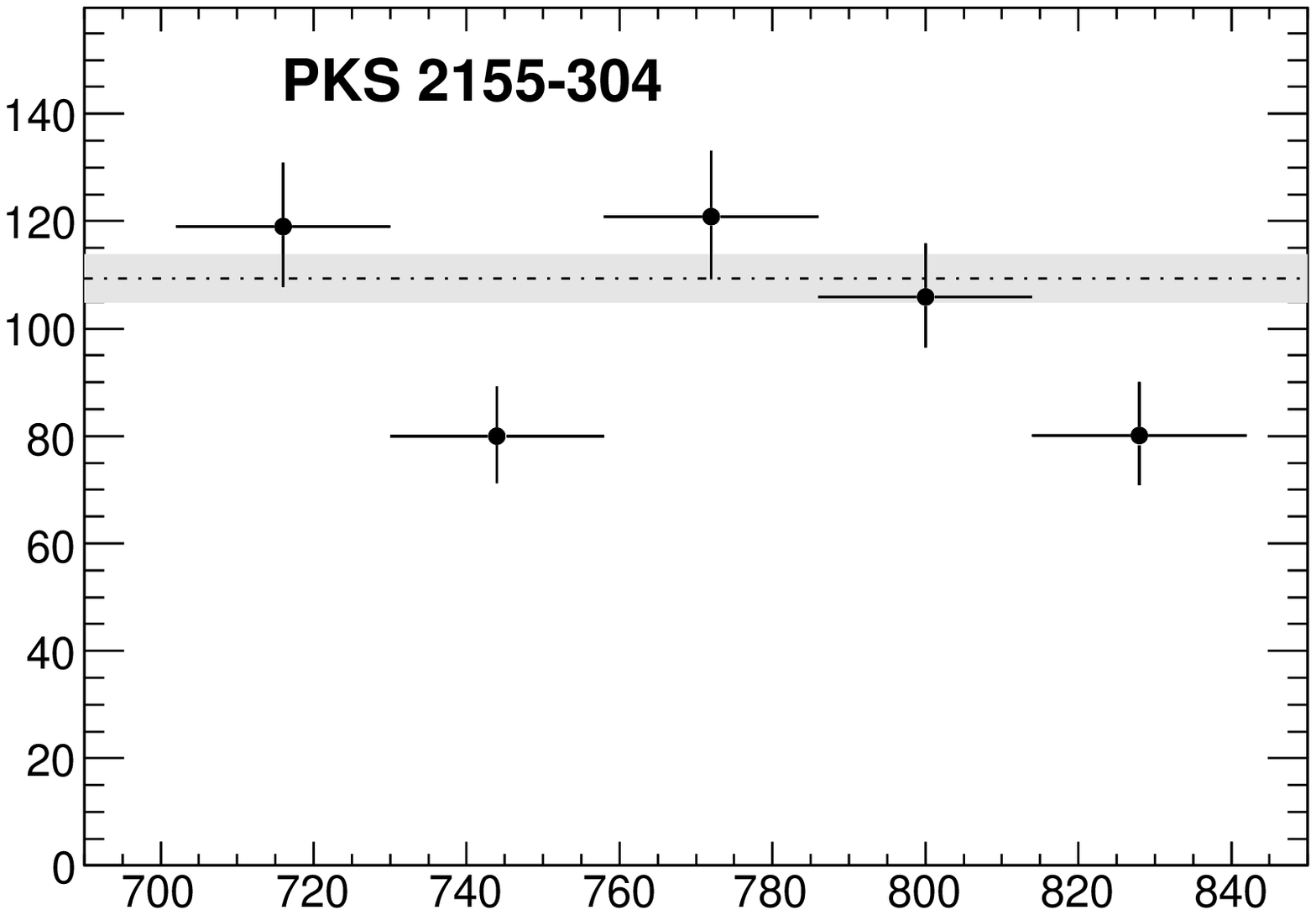}%
\includeLCLune{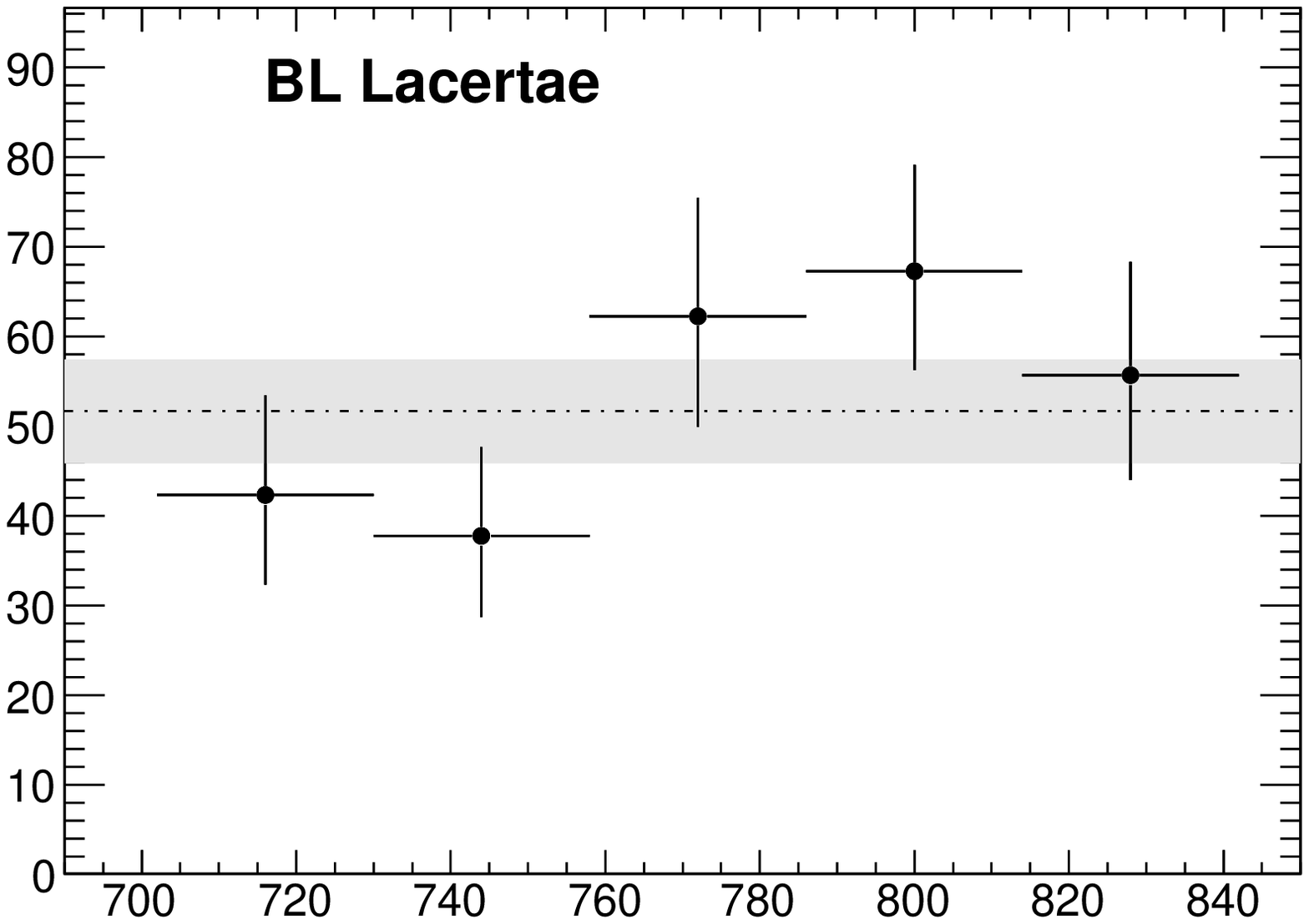}

\includeLCLune{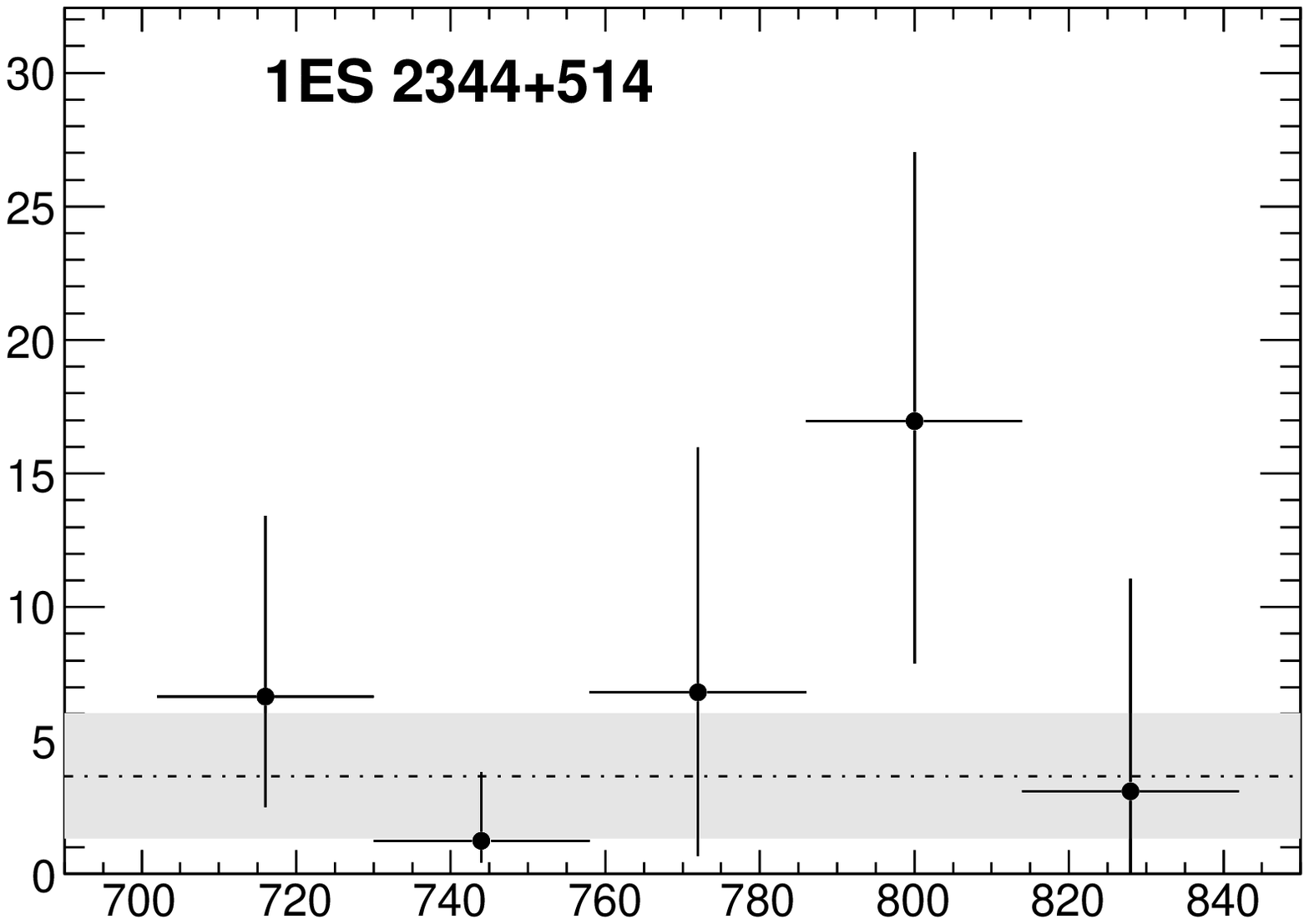}%
\includeLCLune{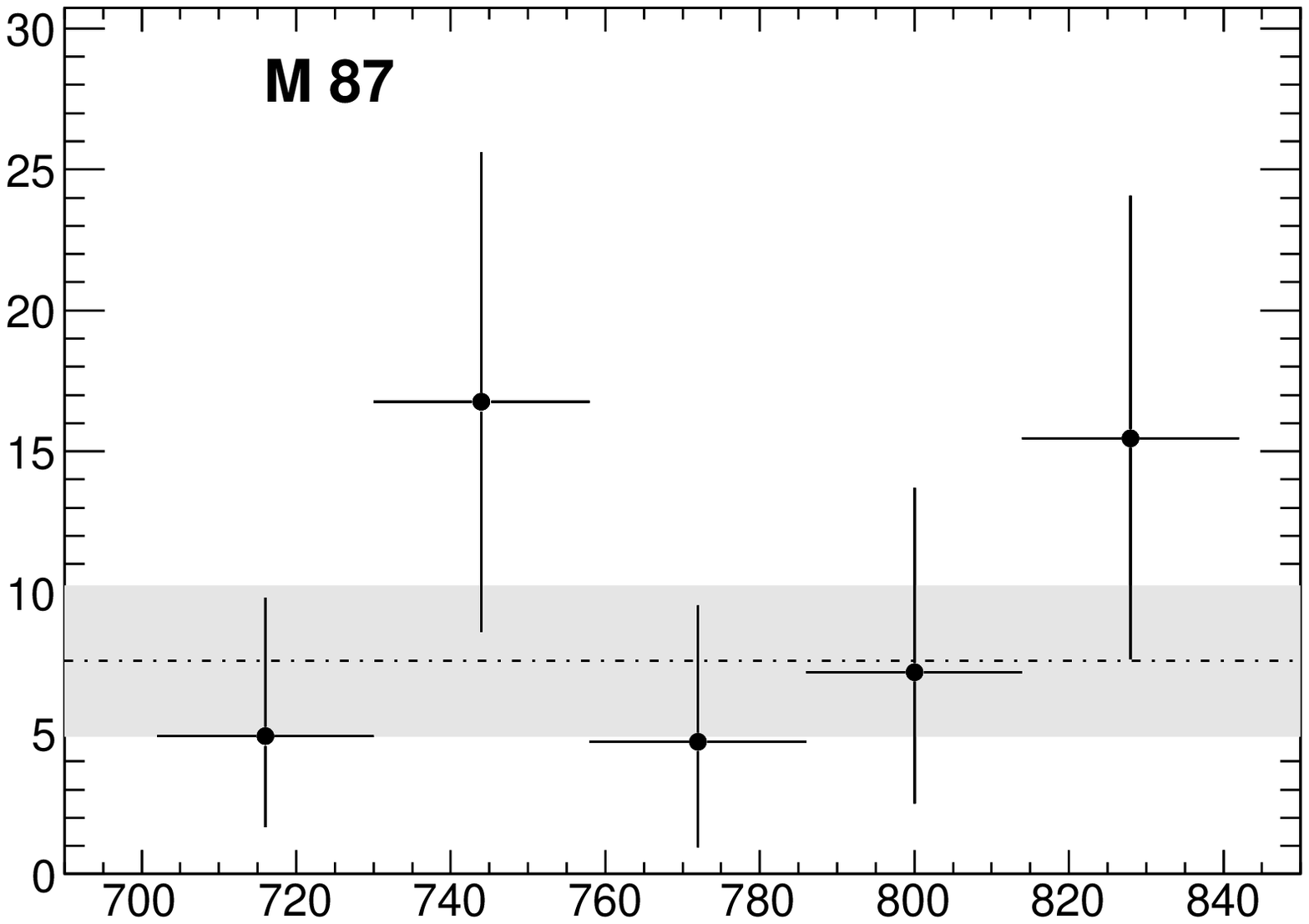}%
\includeLCLune{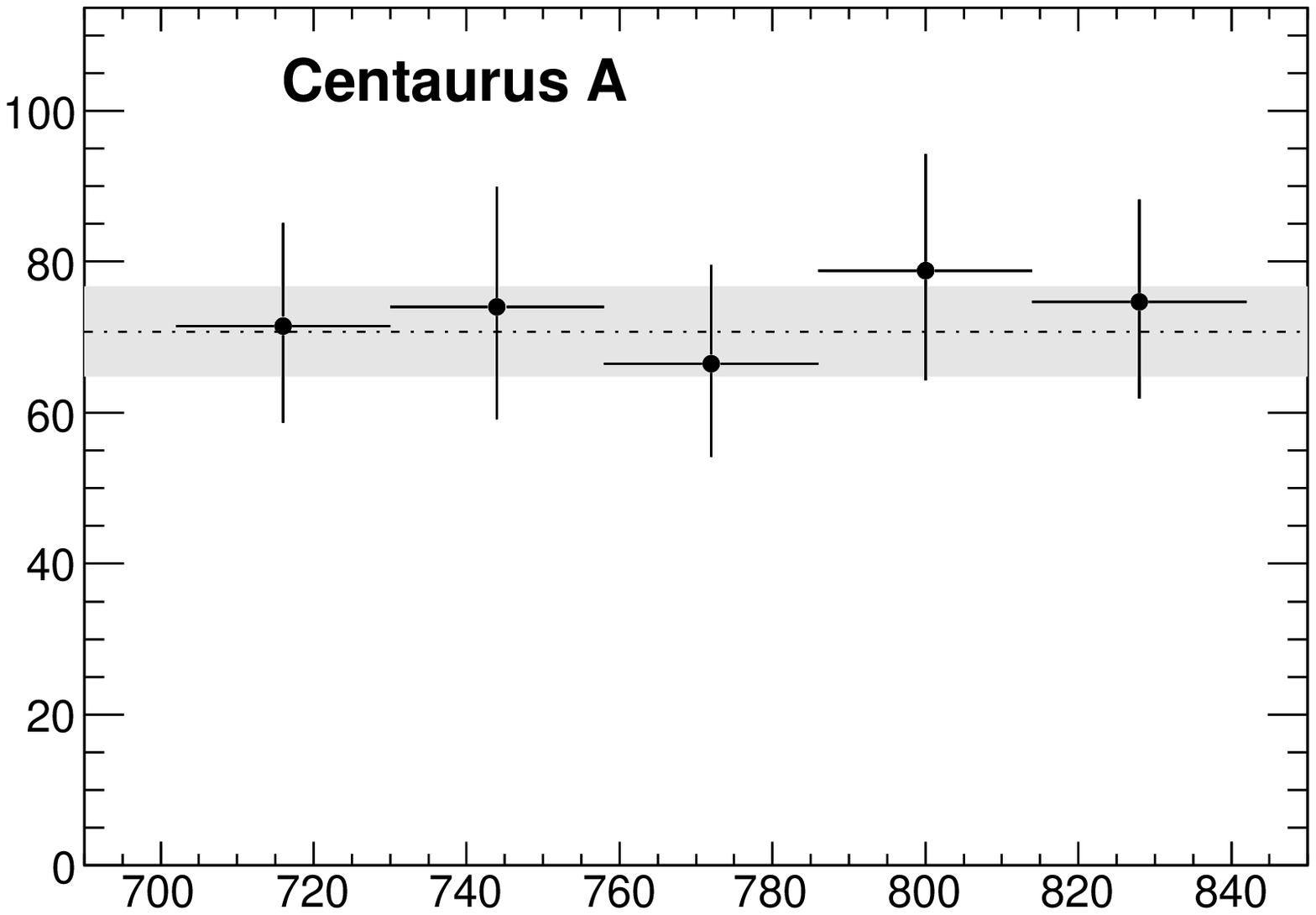}

\includeLCLune{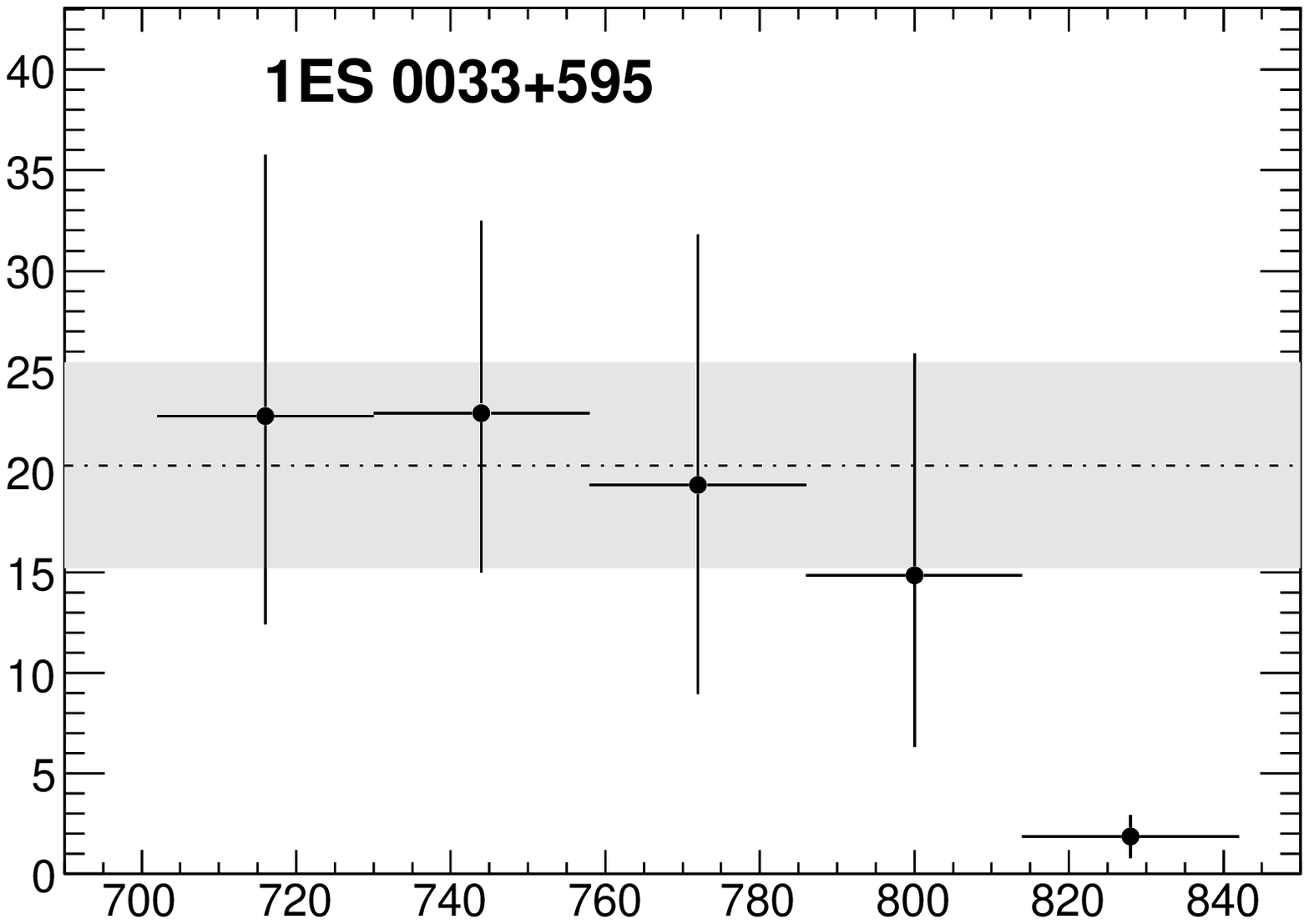}%
\includeLCLune{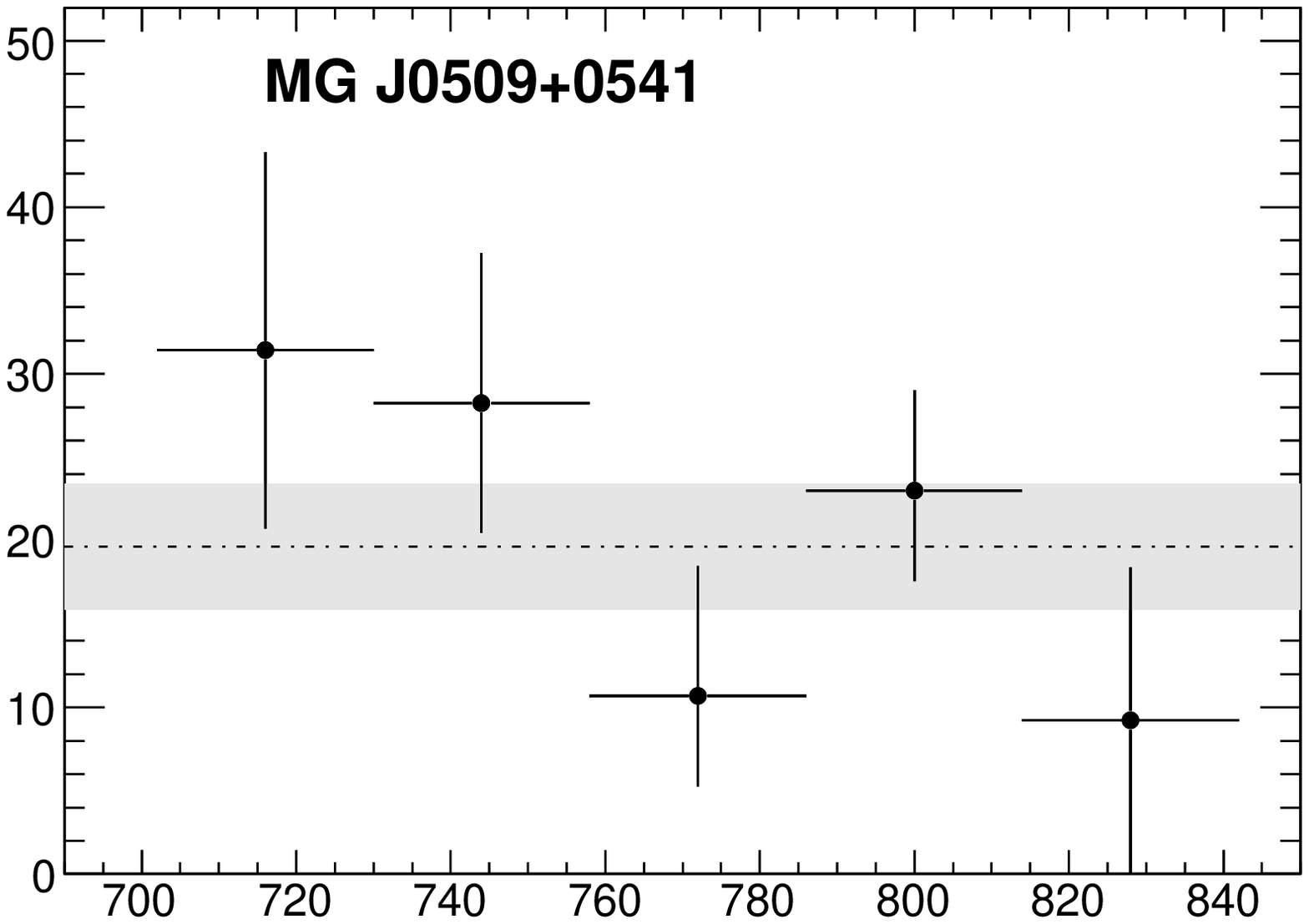}%
\includeLCLune{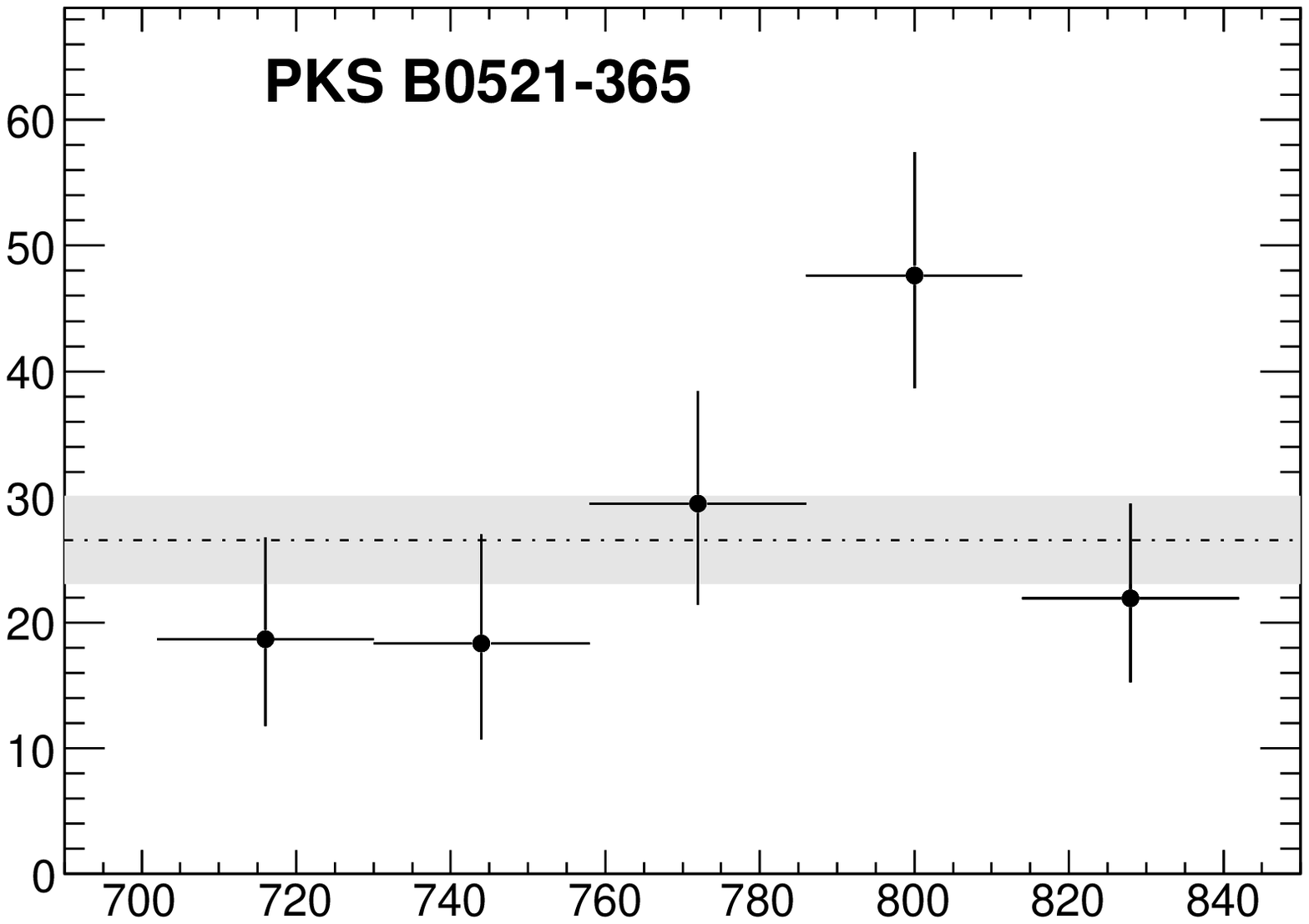}

\includeLCLune{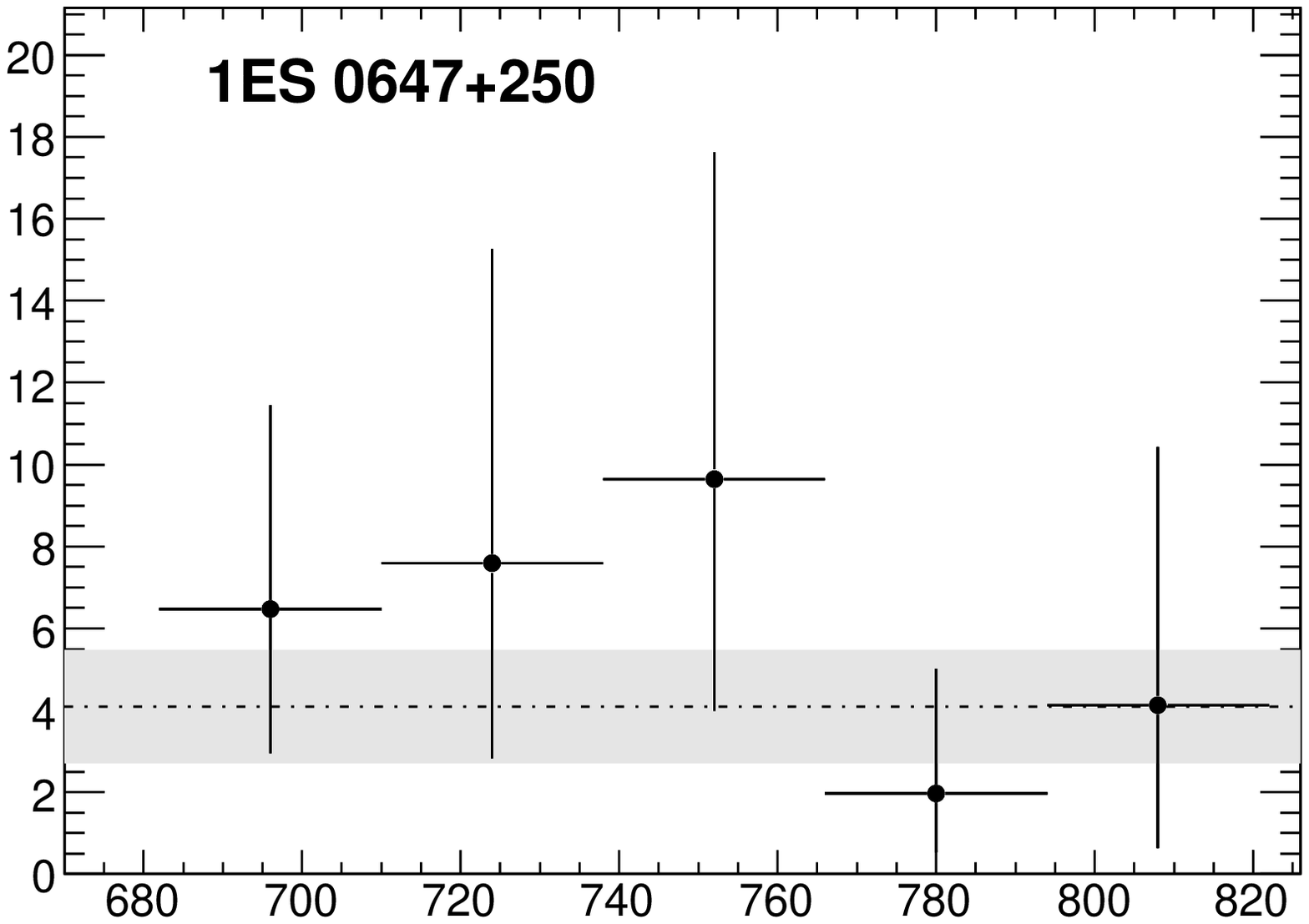}%
\includeLCLune{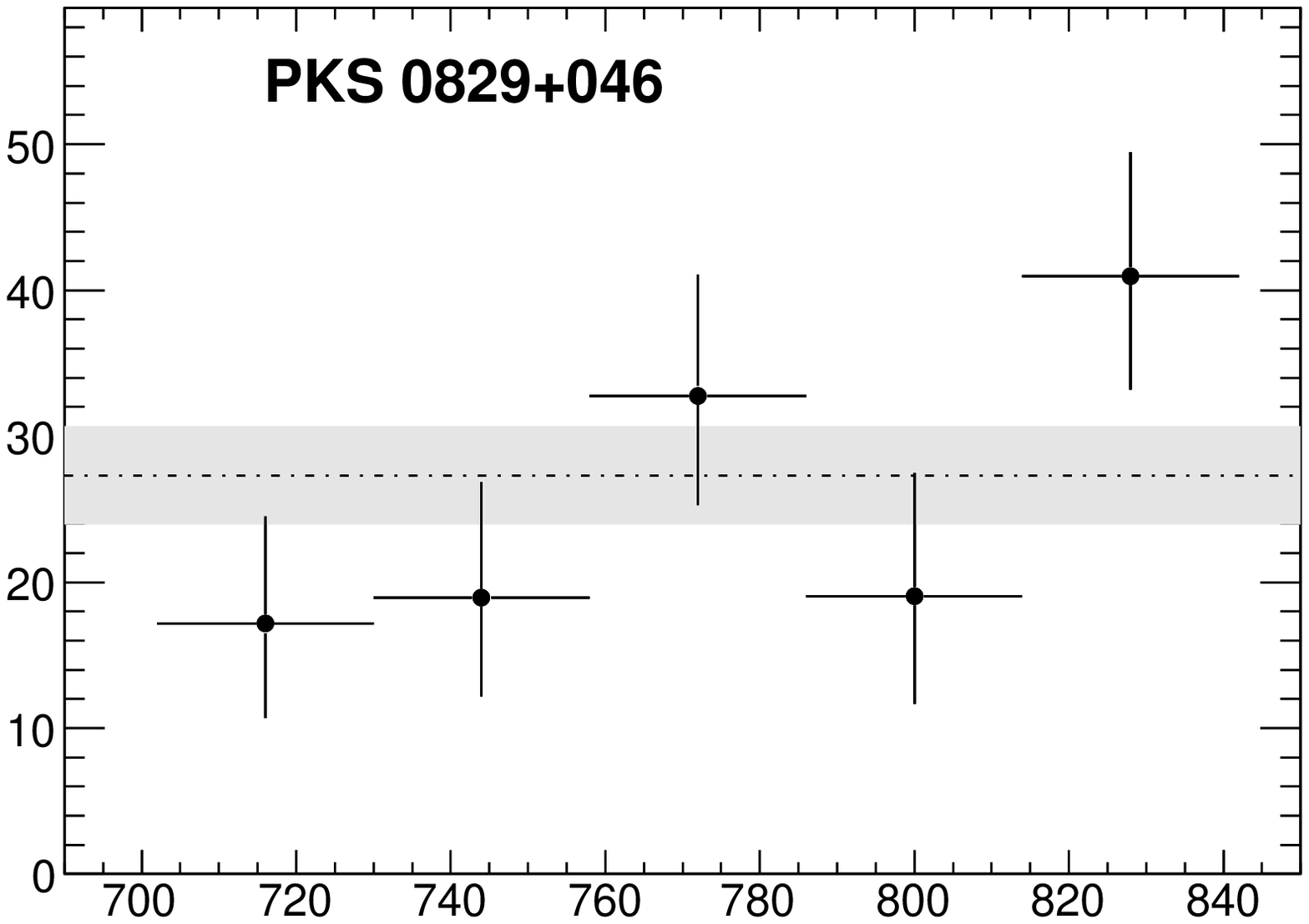}%
\includeLCLune{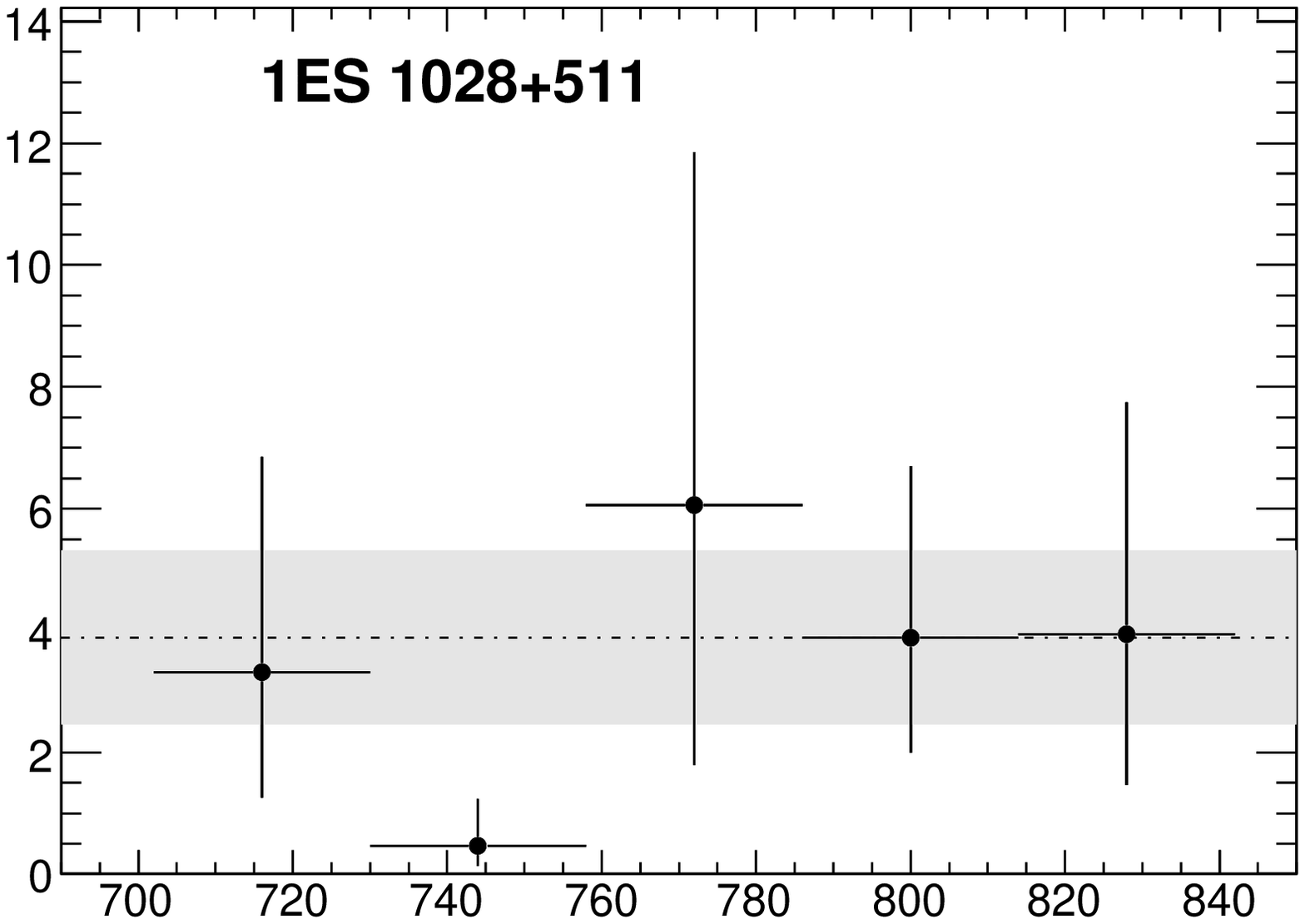}

\includeLCLune{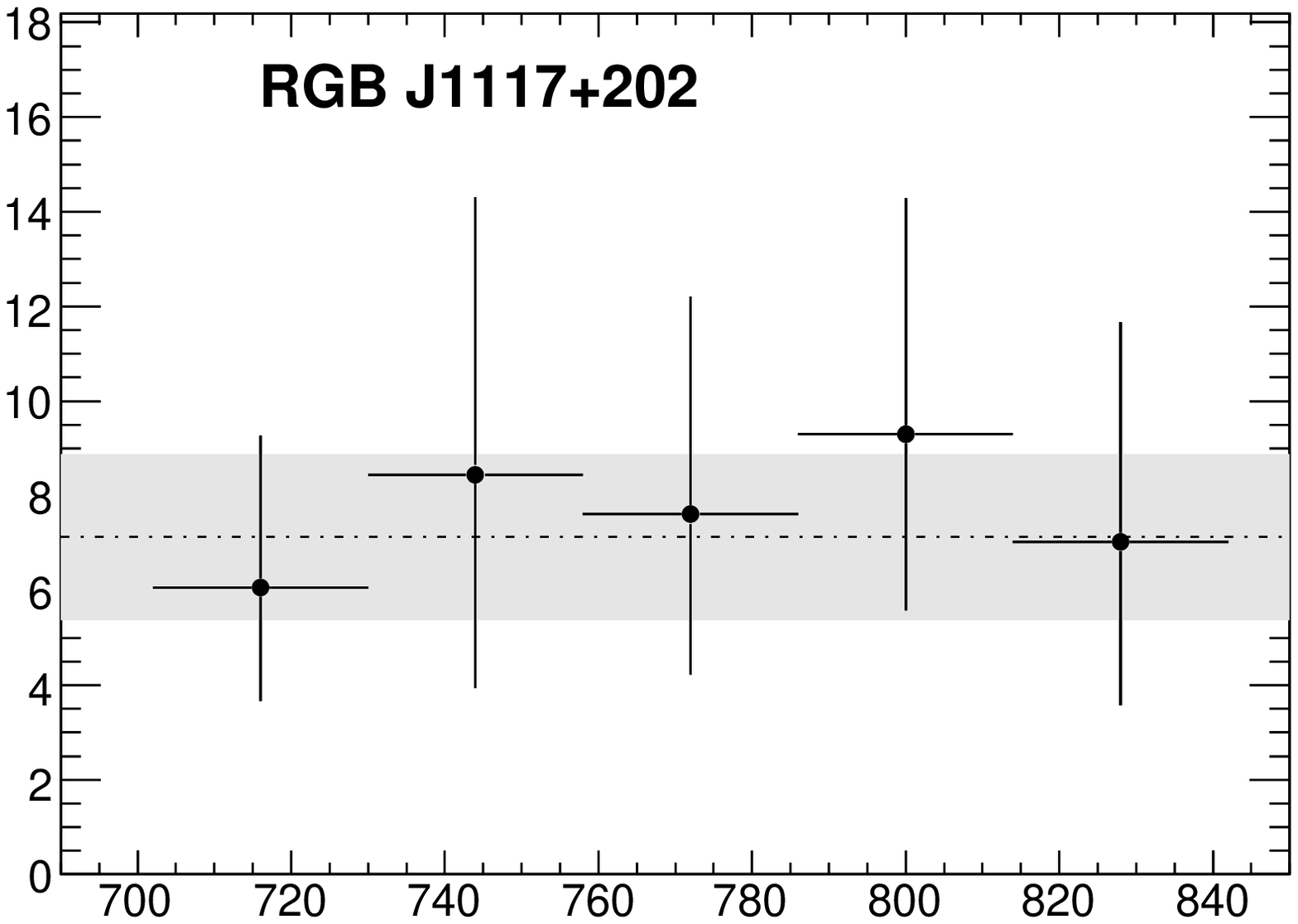}
\includeLCLune{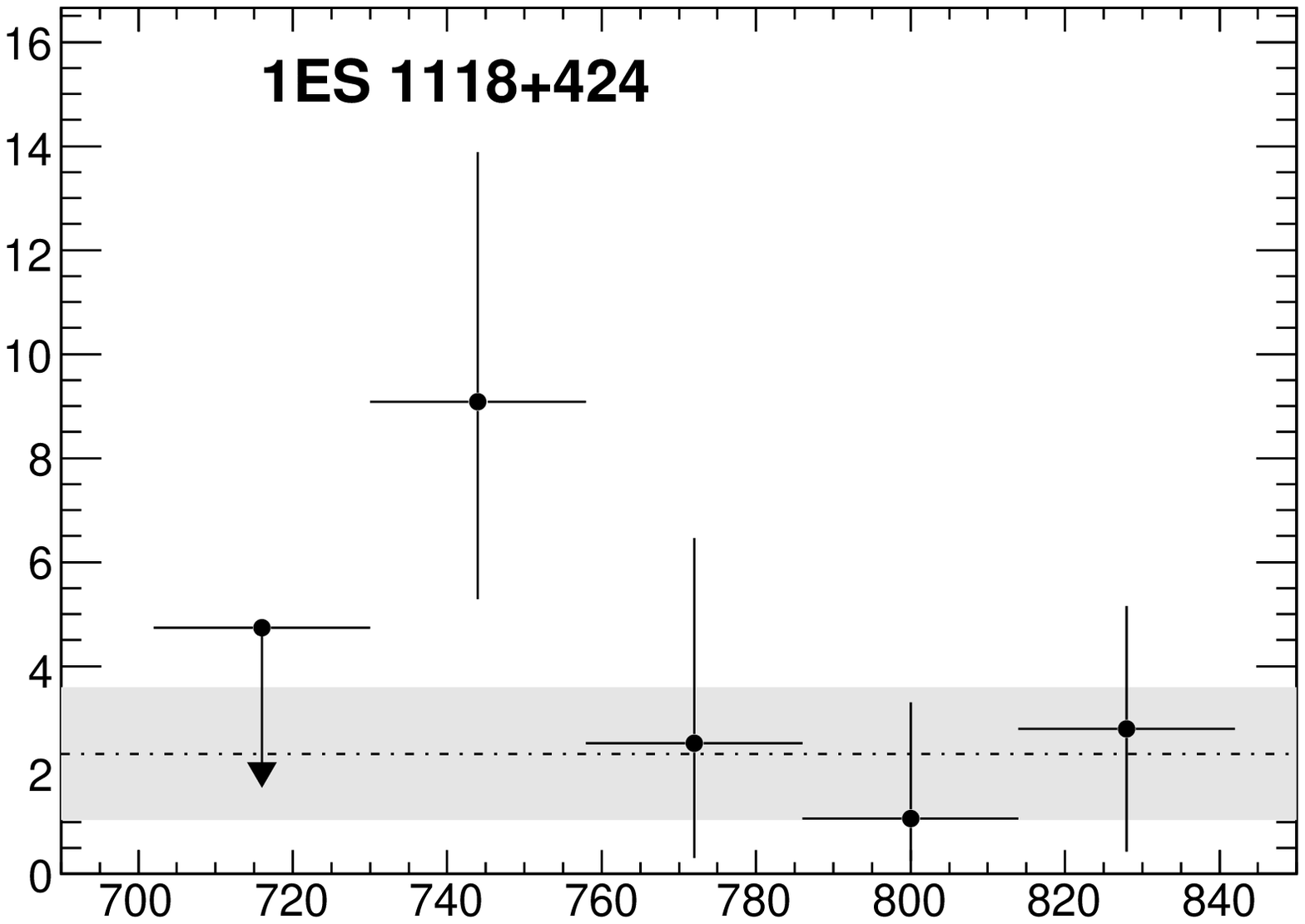}%
\includeLCLune{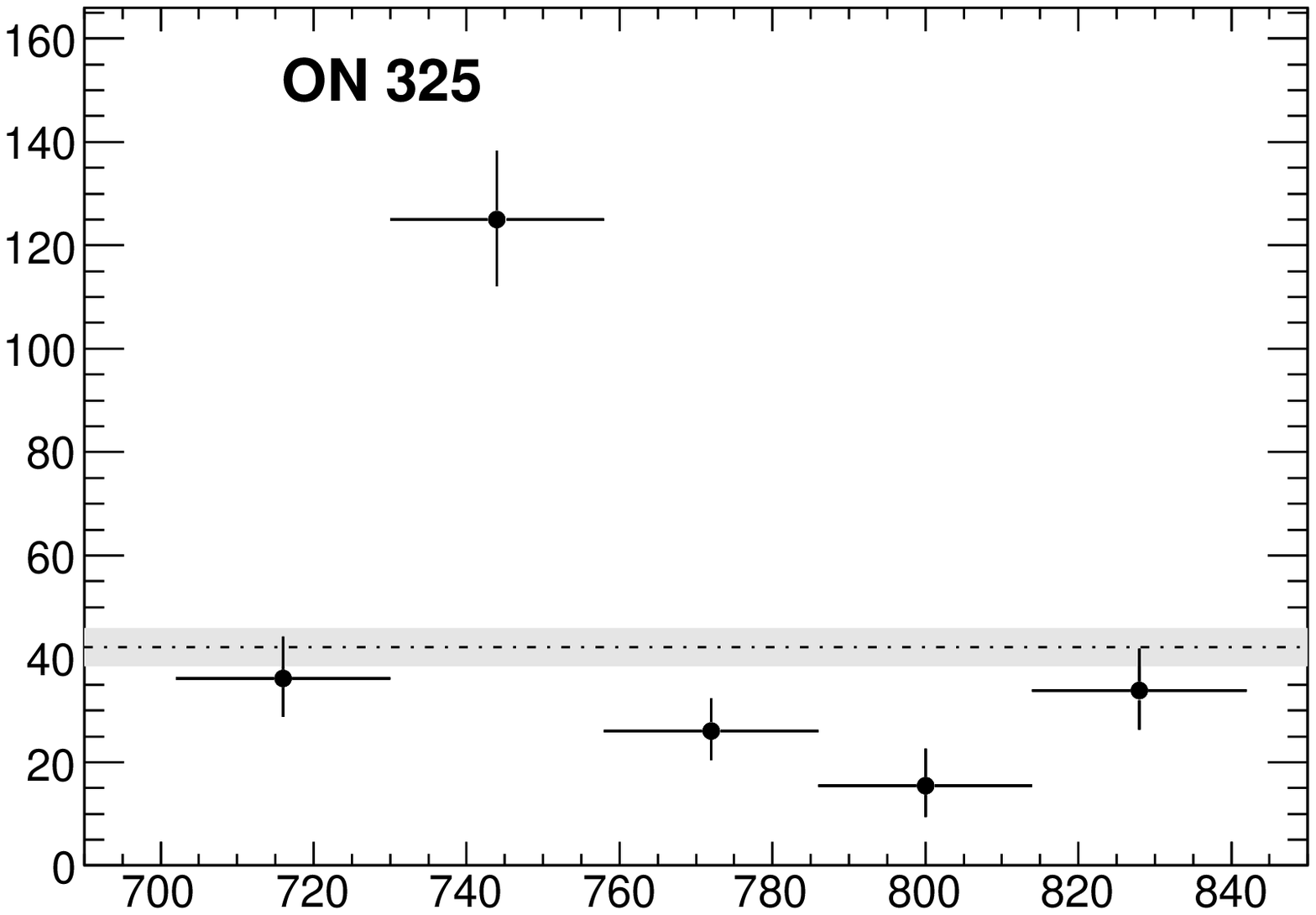}

\end{minipage}

Date -- MJD-54000 [days]

\caption{\label{FIG::LC28_2}Continued}
\end{figure}


\addtocounter{figure}{-1}
\begin{figure}[p]
\centering
\rotatebox{90}{\makebox[0mm][r]{Flux -- $F(>200\mathrm{MeV})$ [$10^{-9}$\cmsc]}}%
\begin{minipage}[t]{0.91\textwidth}
\

\includeLCLune{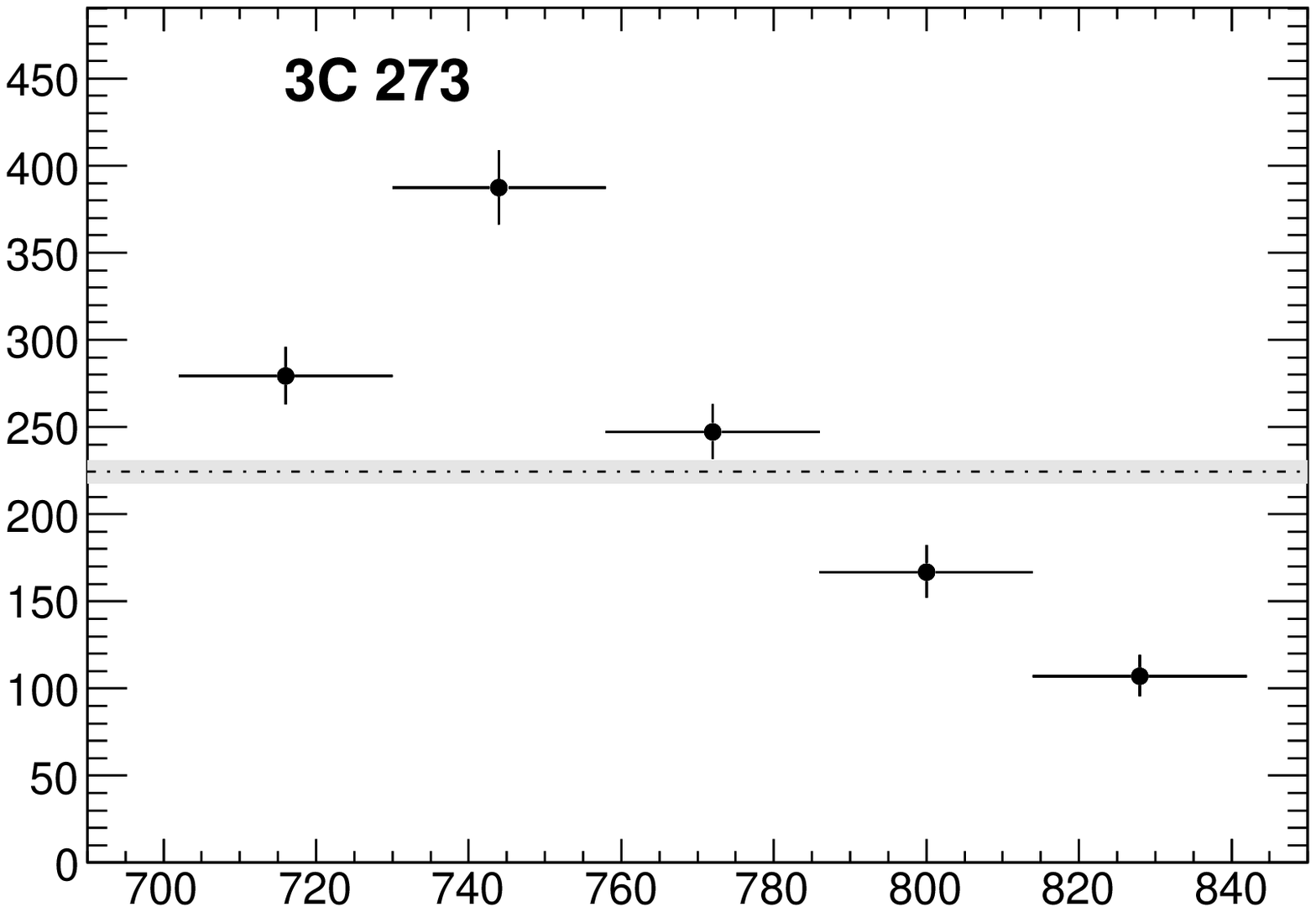}%
\includeLCLune{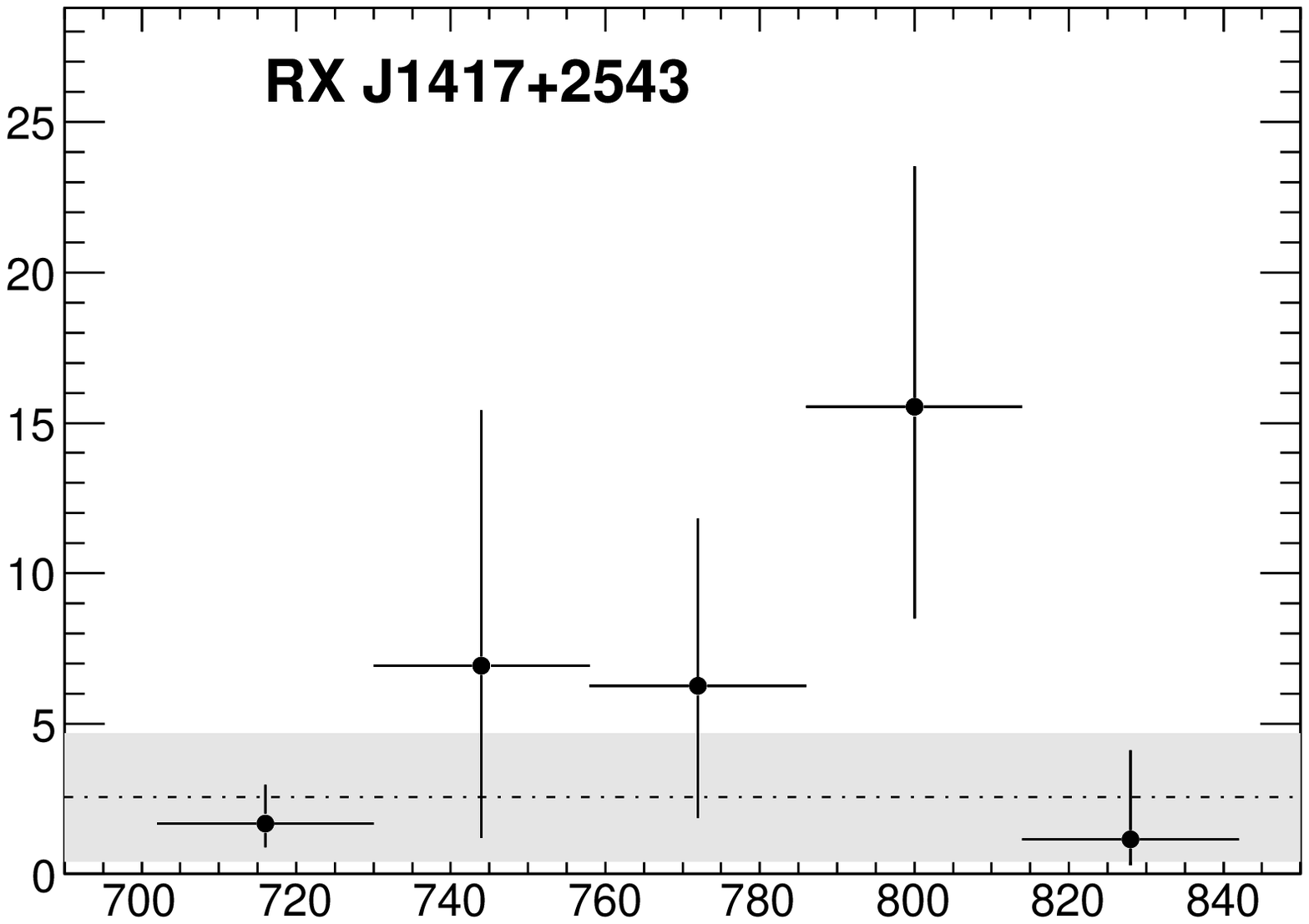}%
\includeLCLune{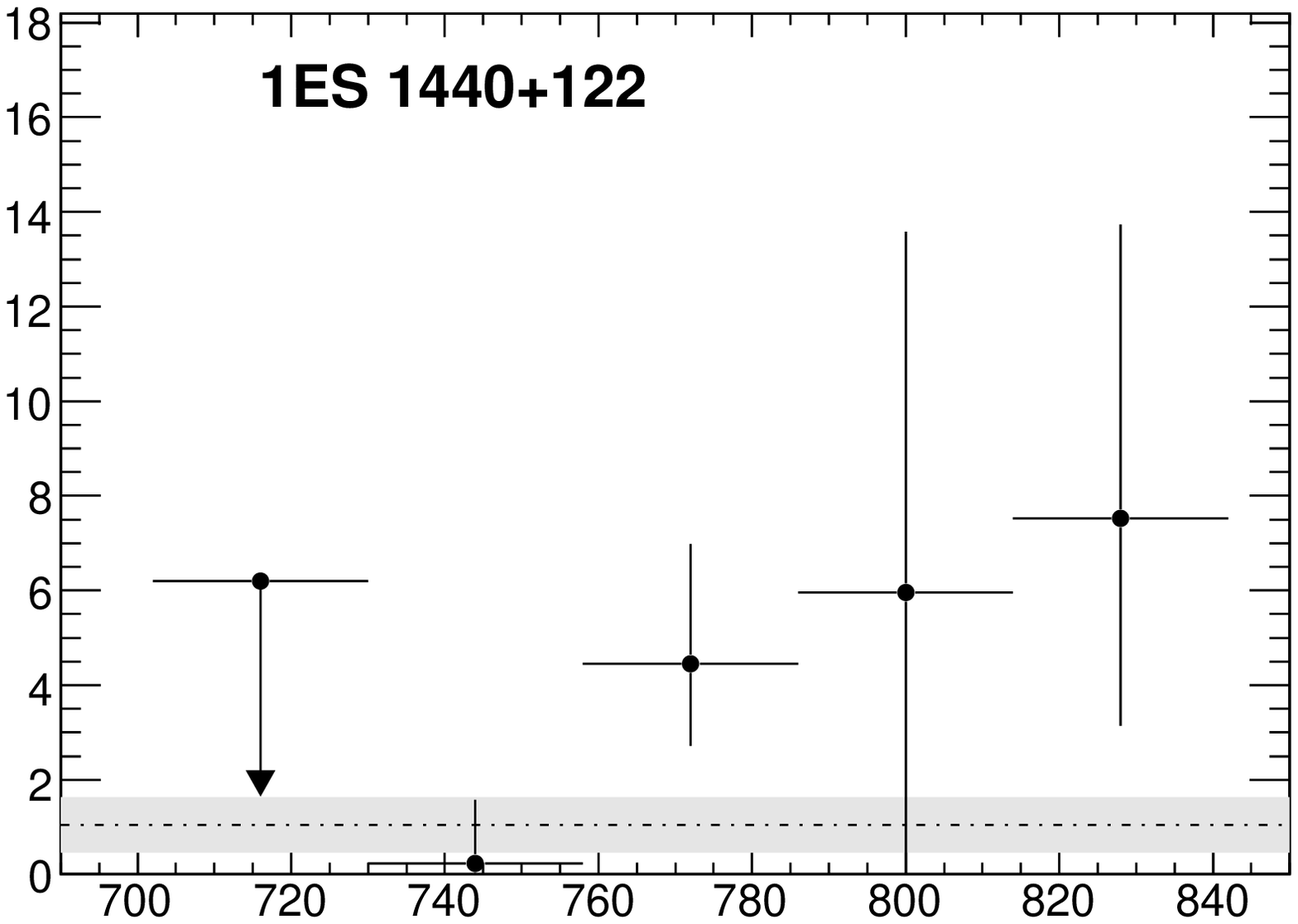}

\includeLCLune{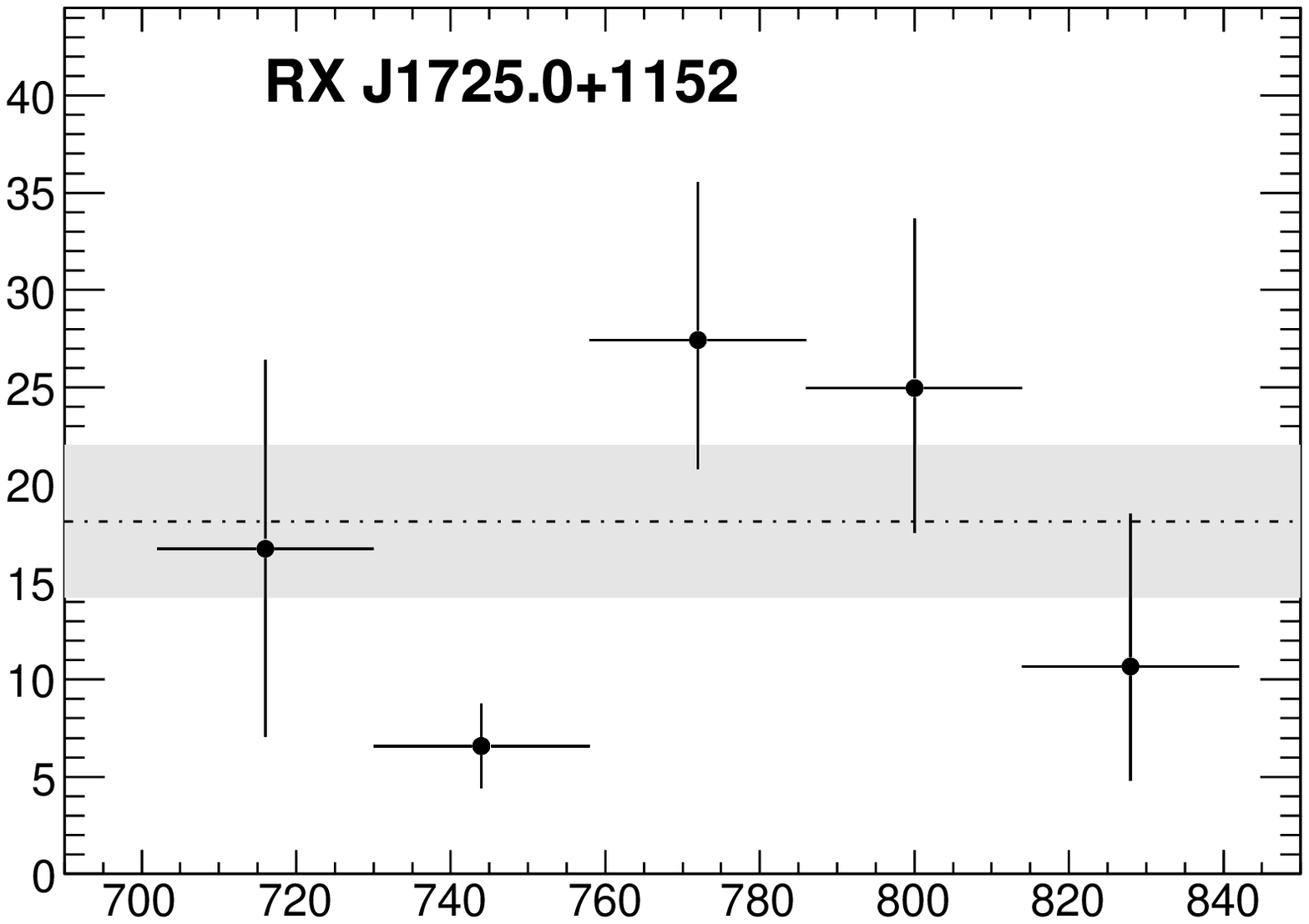}%
\includeLCLune{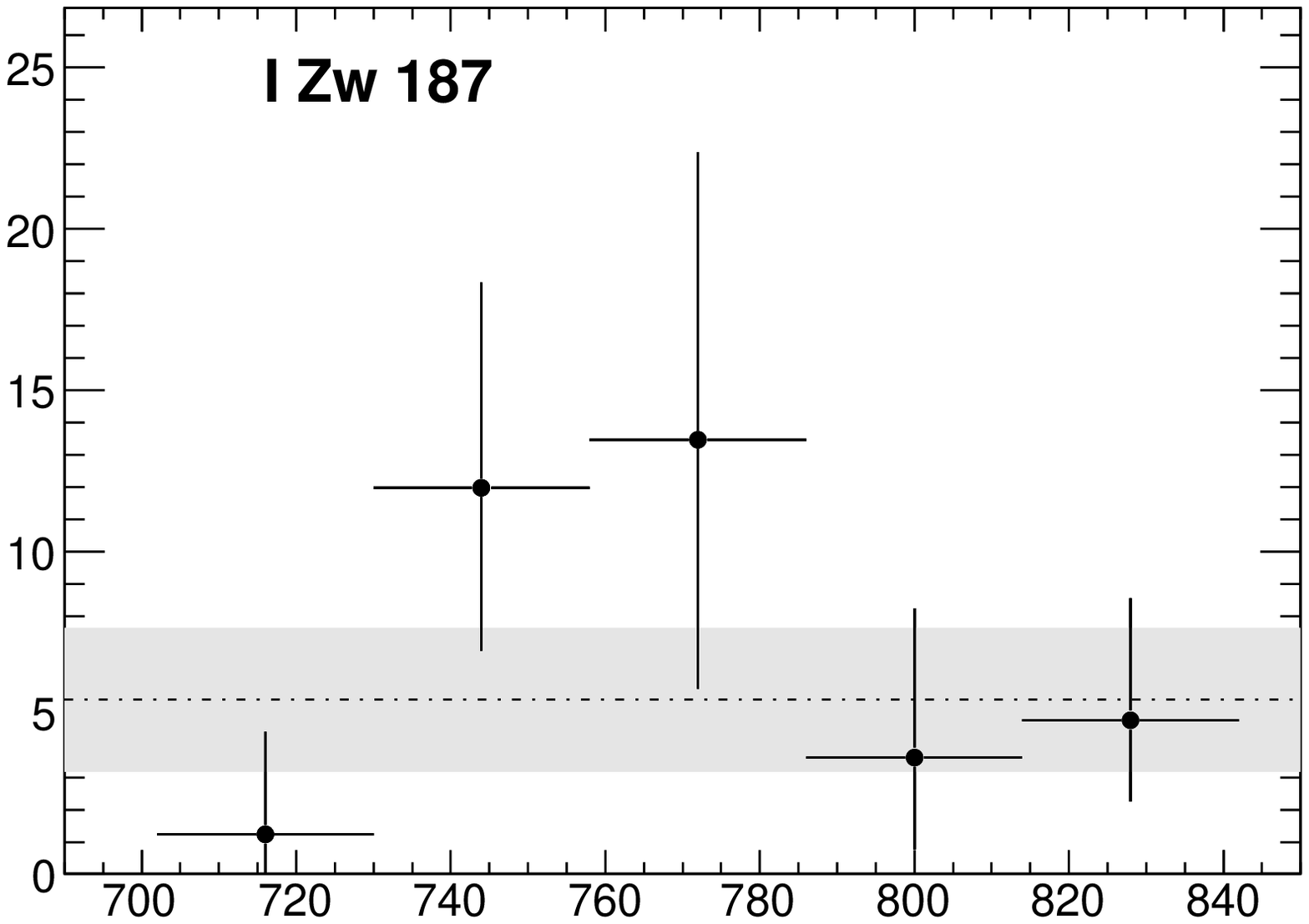}%
\includeLCLune{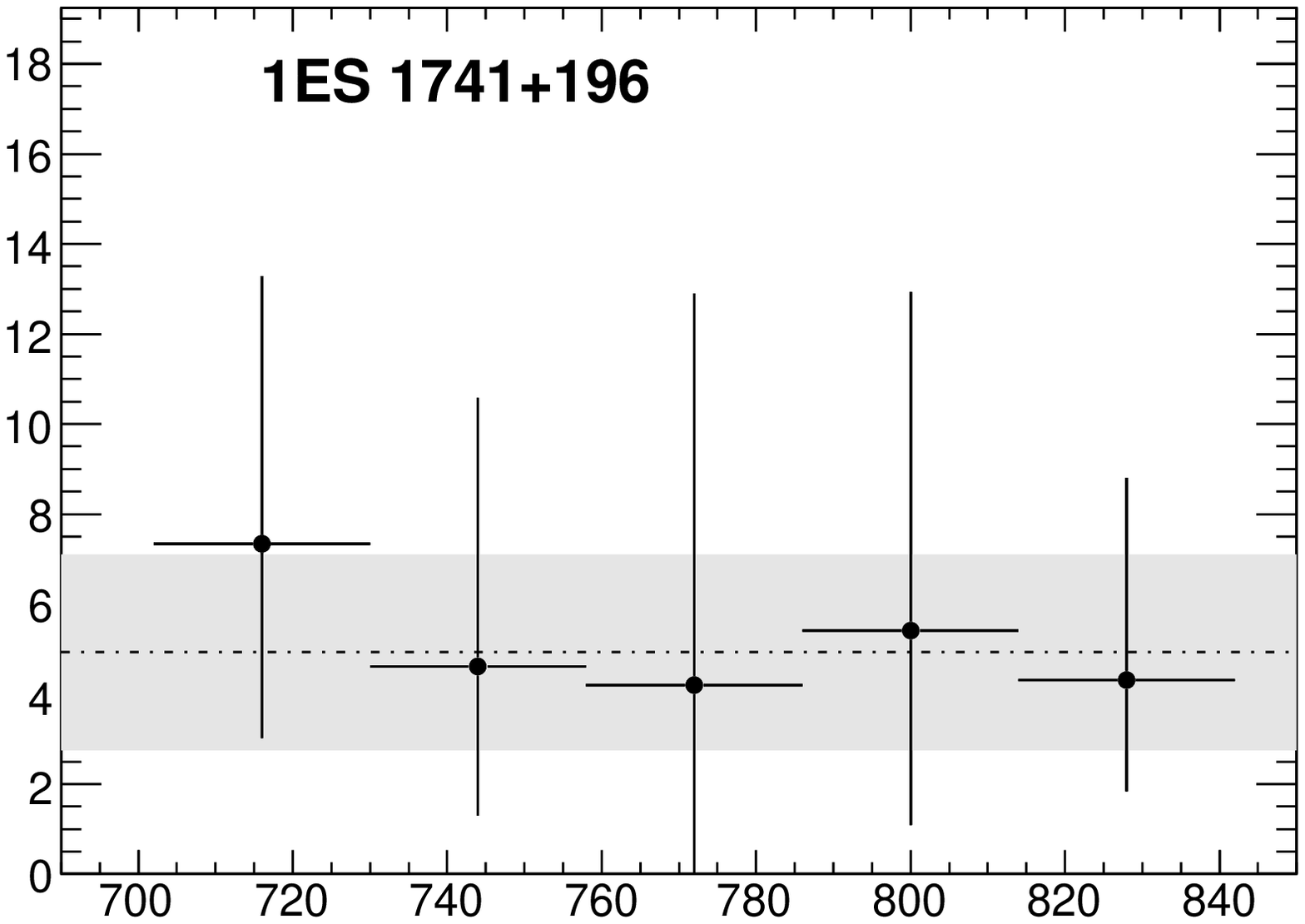}

\includeLCLune{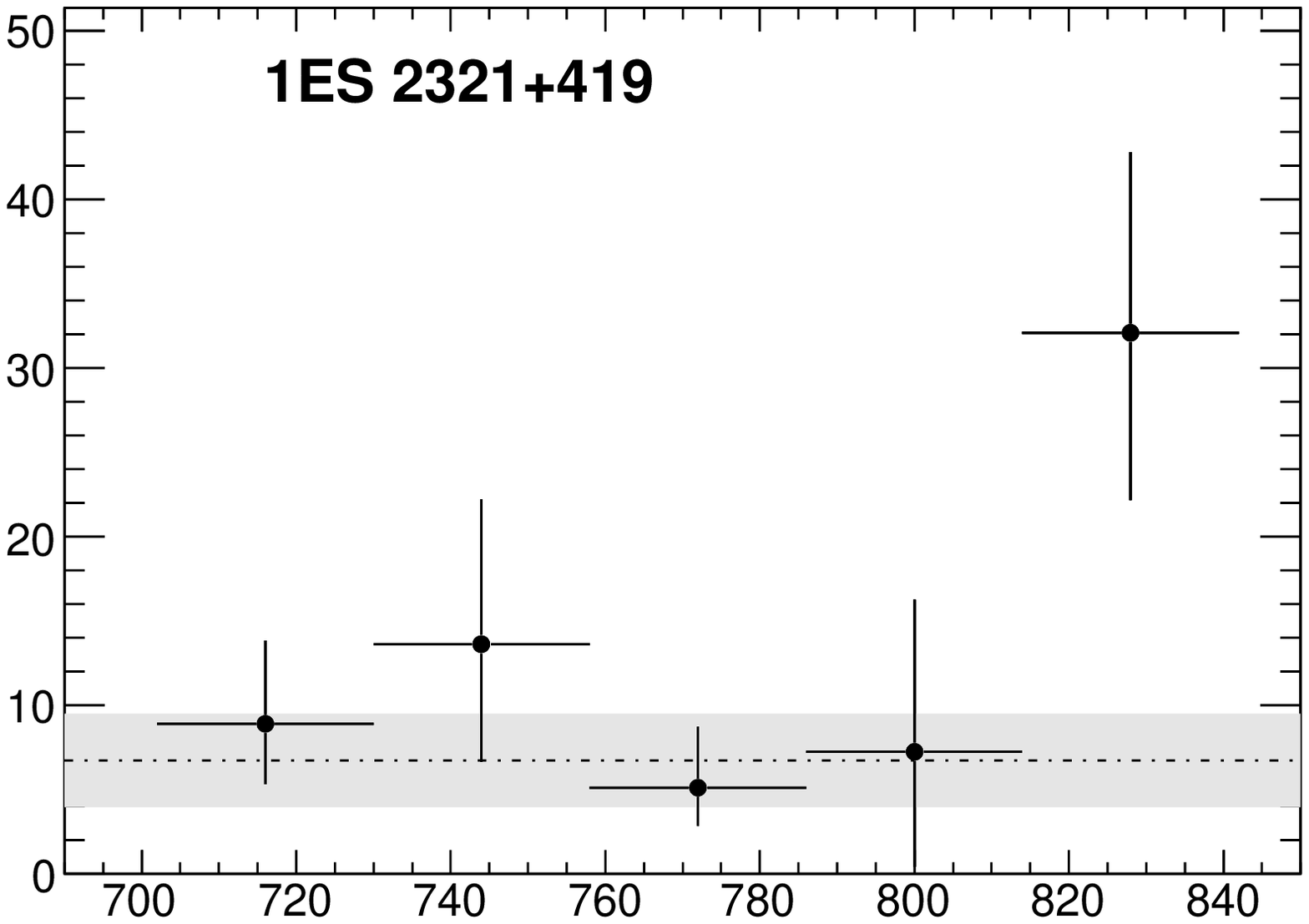}%
\includeLCLune{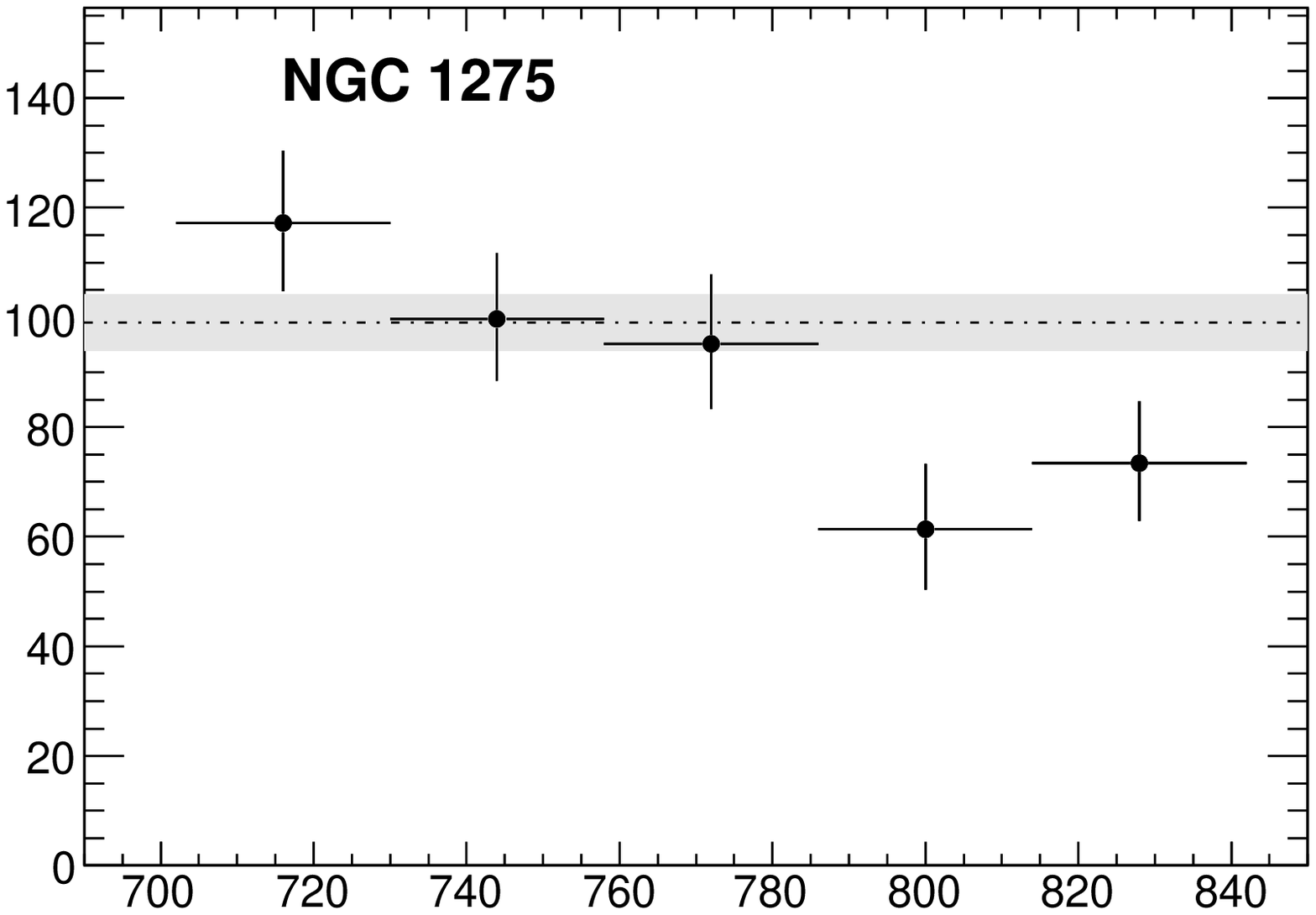}%
\hspace*{\lcLuneWidth}

\end{minipage}

Date -- MJD-54000 [days]

\caption{\label{FIG::LC28_3}Continued}
\end{figure}

\clearpage

\newlength{\lcTenWidth}
\setlength{\lcTenWidth}{0.3\textwidth}
\newcommand{\includeLCTen}[1]{\includegraphics[bb=25 0 510 345,clip,width=\lcLuneWidth]{#1}}


\begin{figure}[p]
\centering
\rotatebox{90}{\makebox[0mm][r]{Flux -- $F(>200\mathrm{MeV})$ [$10^{-9}$\cmsc]}}%
\begin{minipage}[t]{0.91\textwidth}
\

\includeLCTen{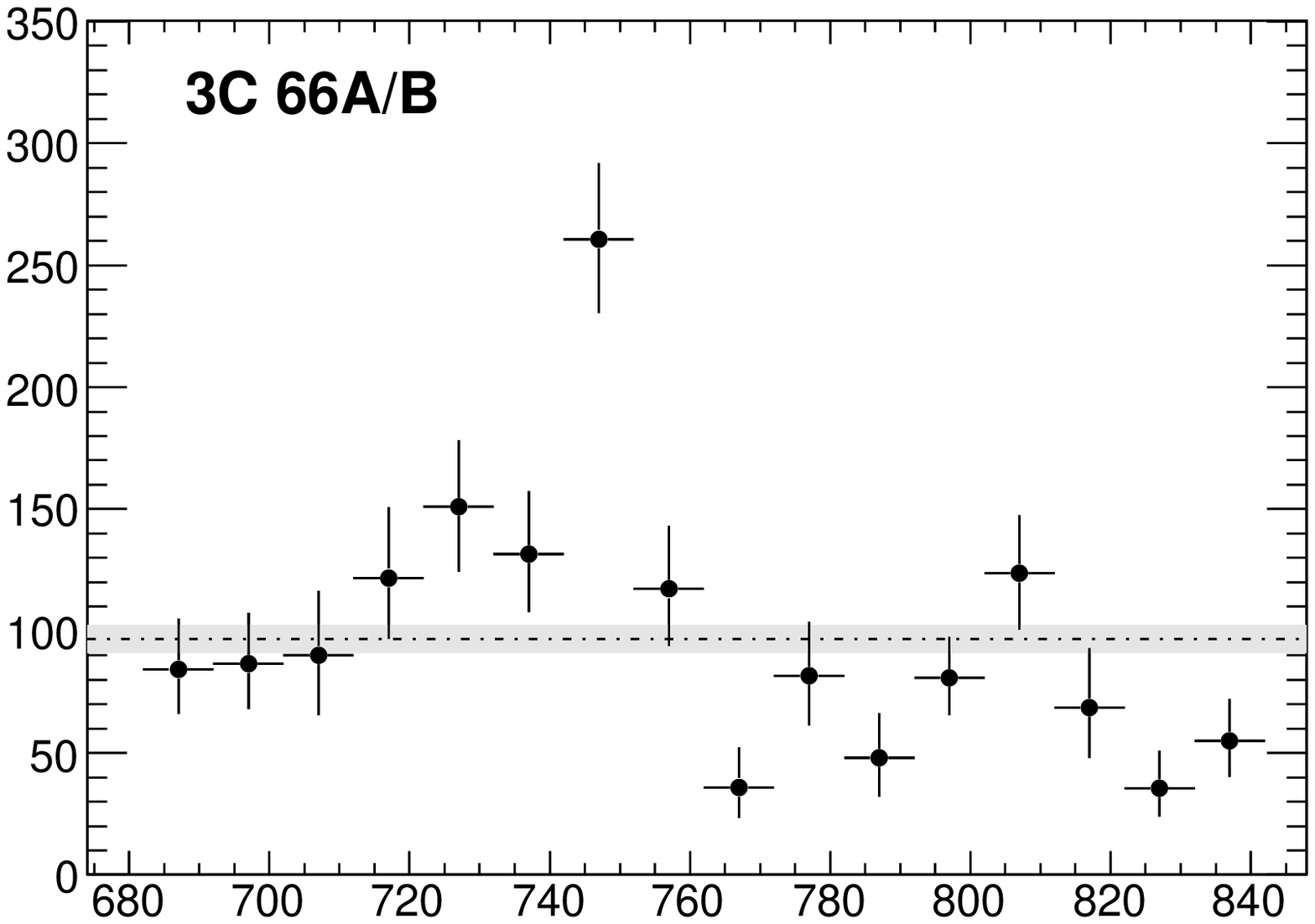}%
\includeLCTen{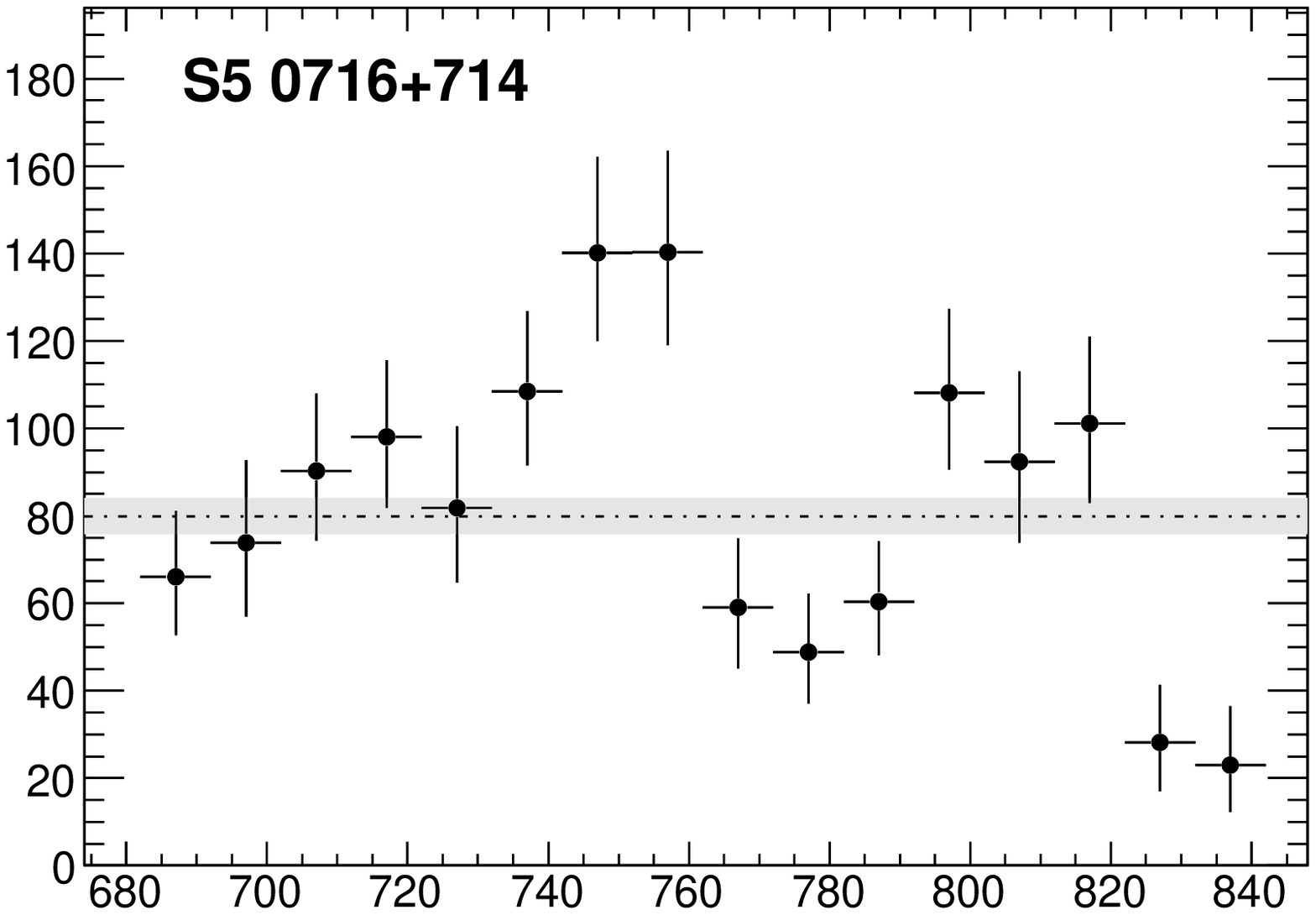}%
\includeLCTen{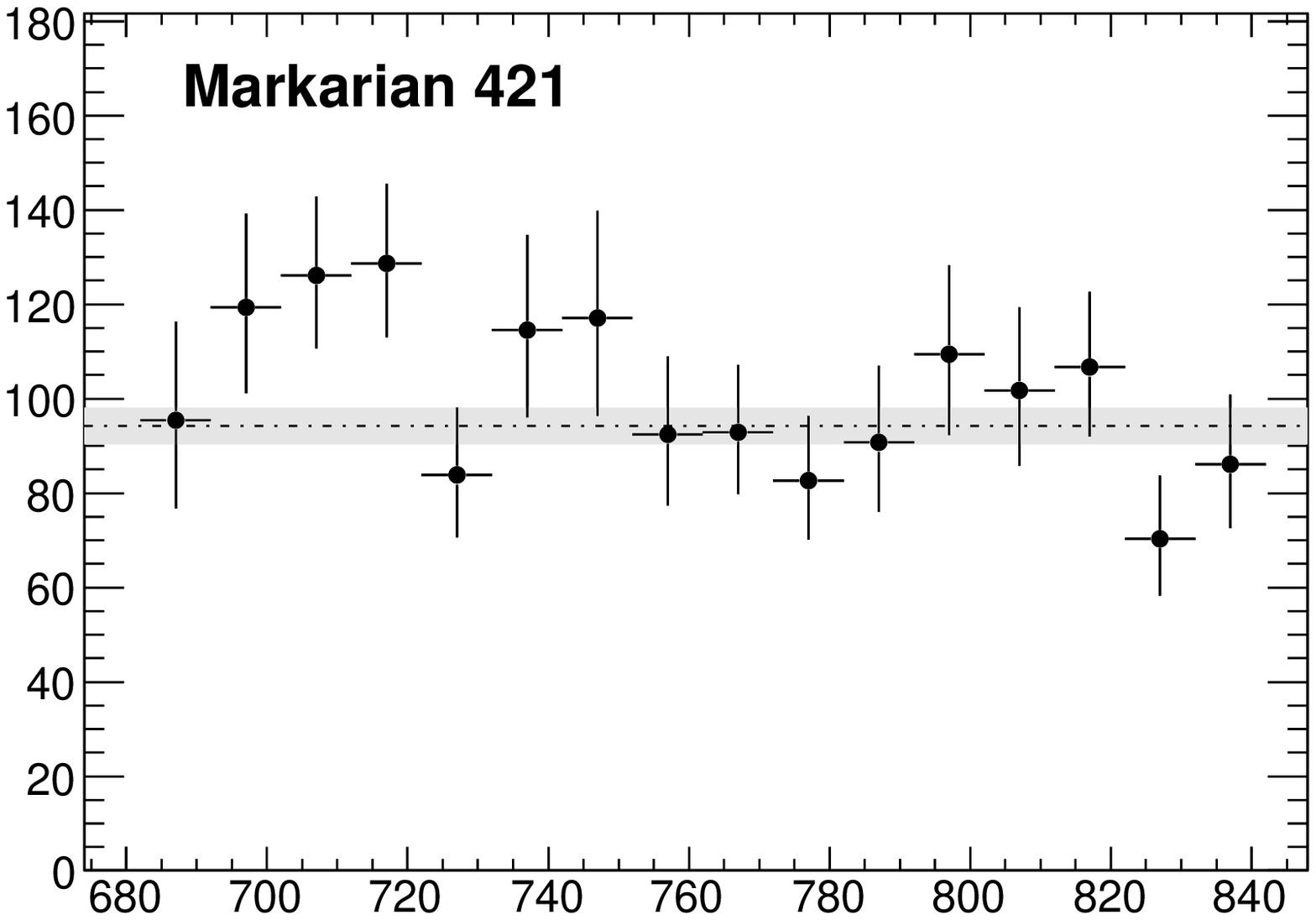}

\includeLCTen{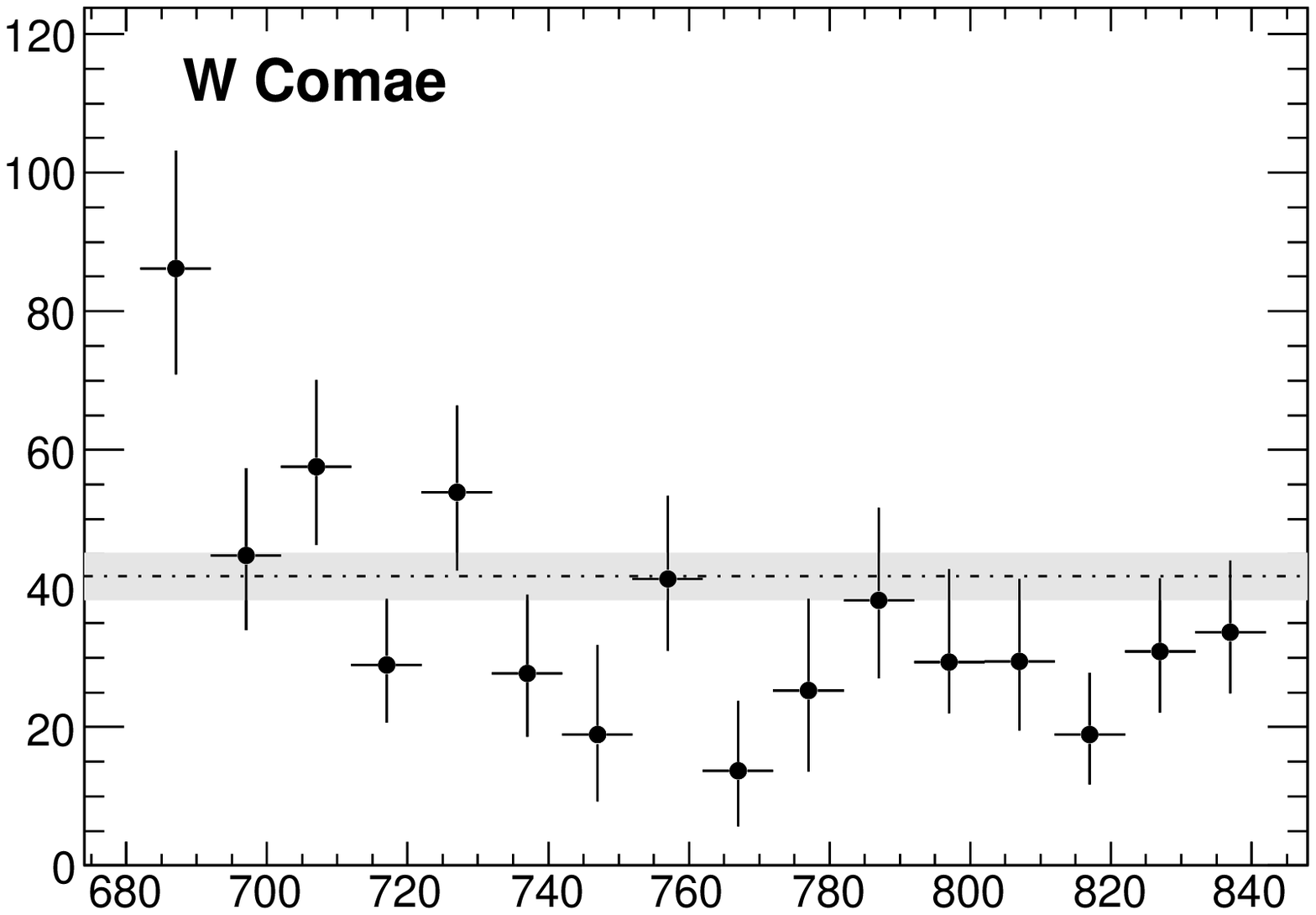}%
\includeLCTen{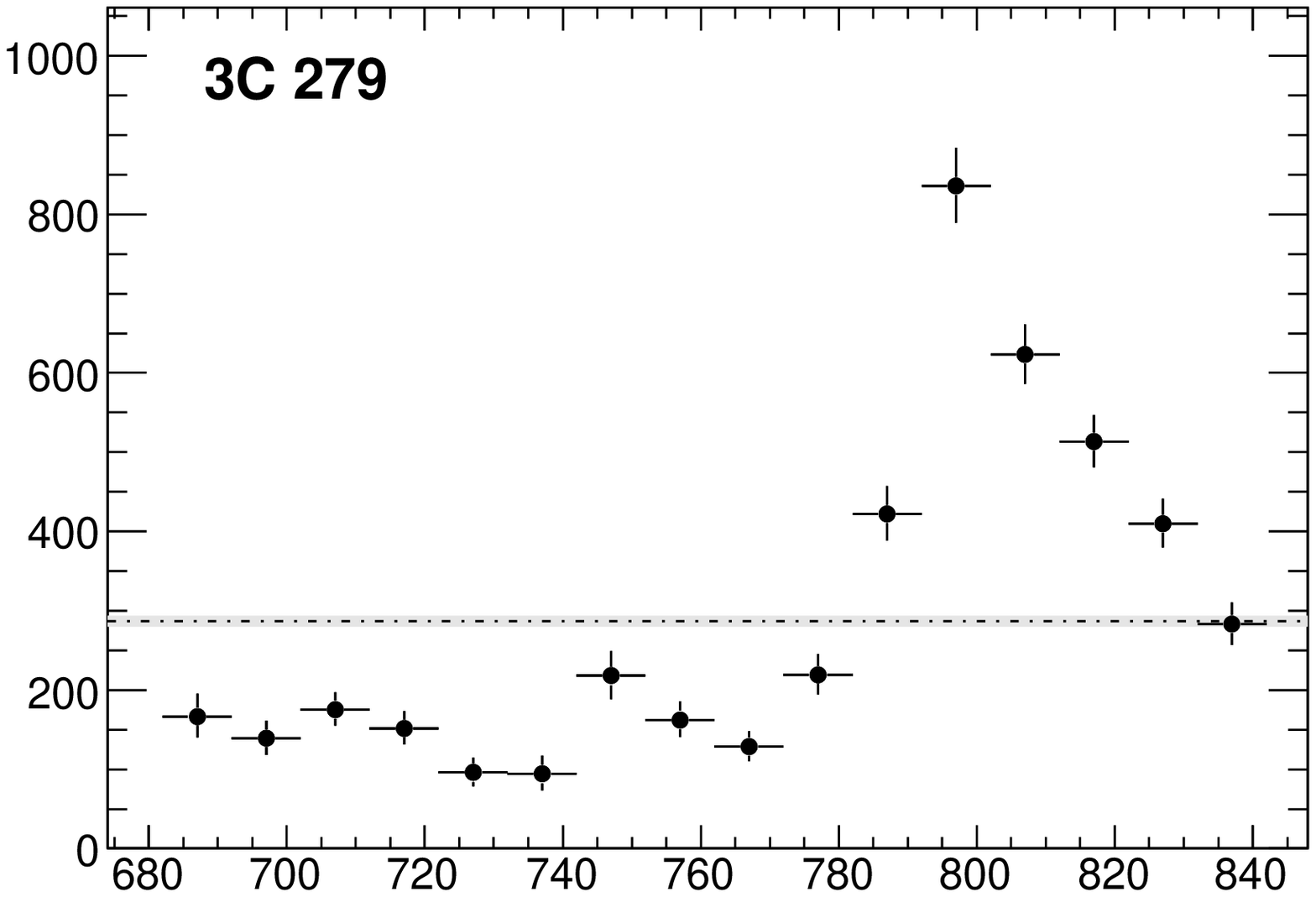}%
\includeLCTen{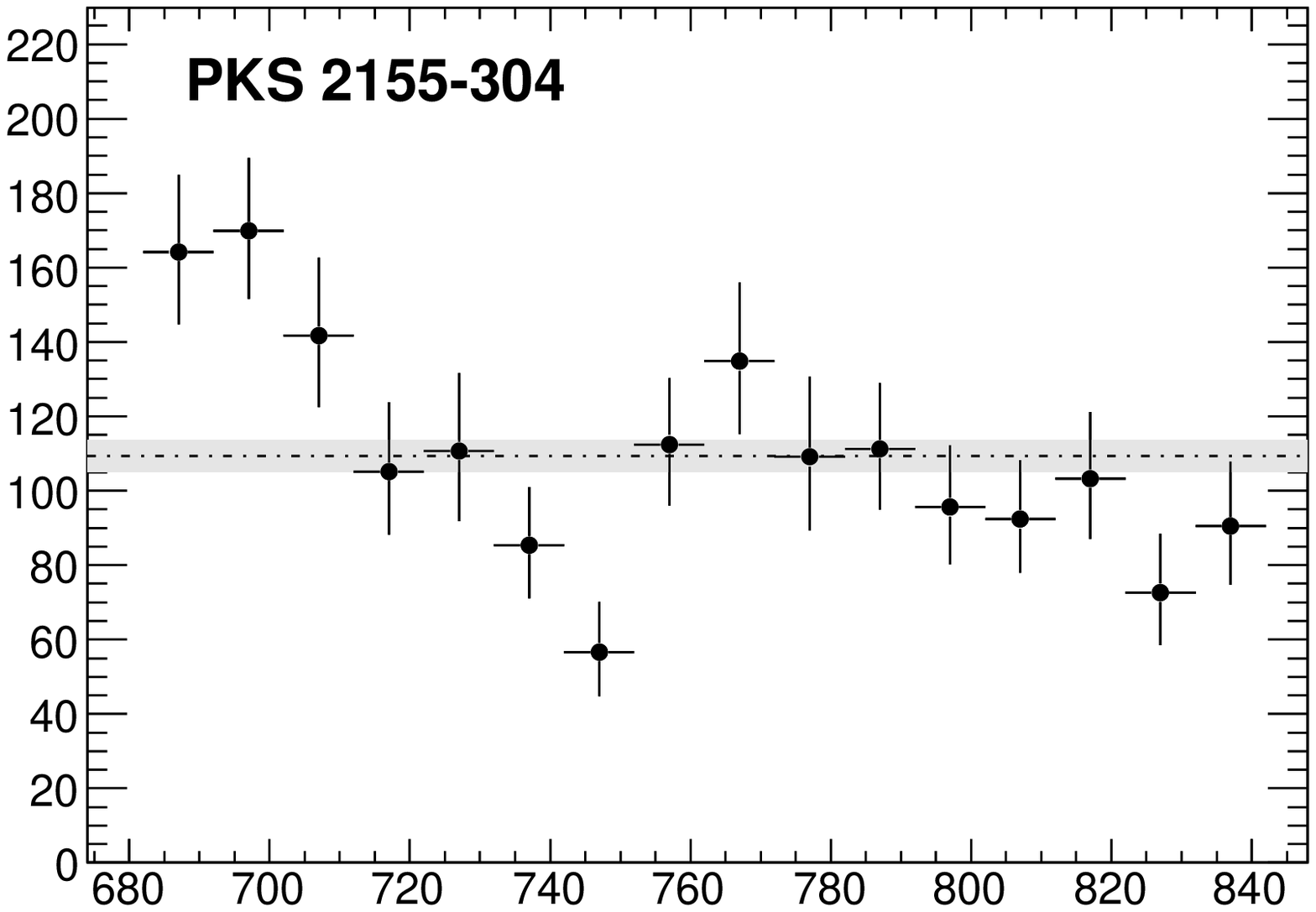}

\includeLCTen{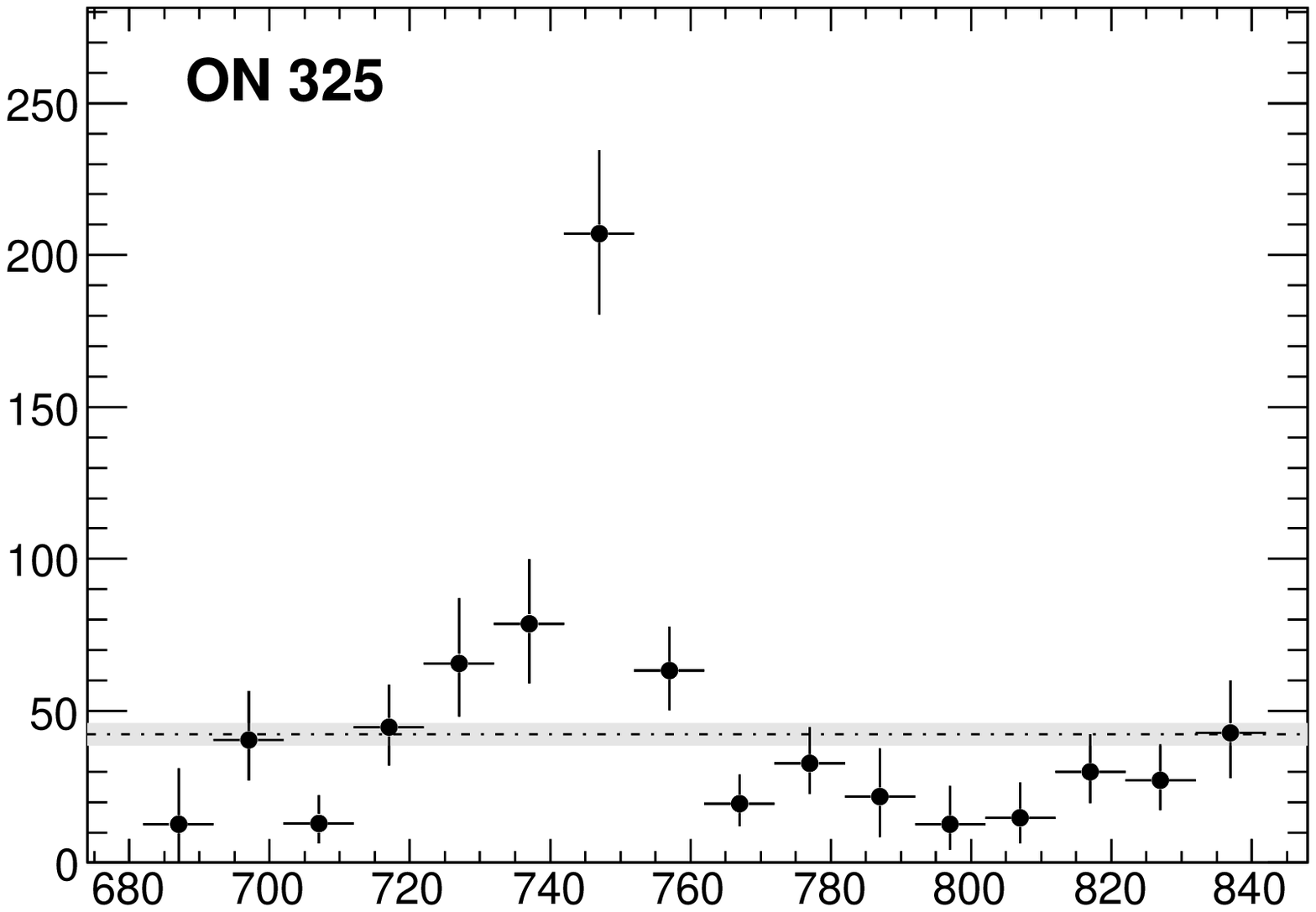}%
\includeLCTen{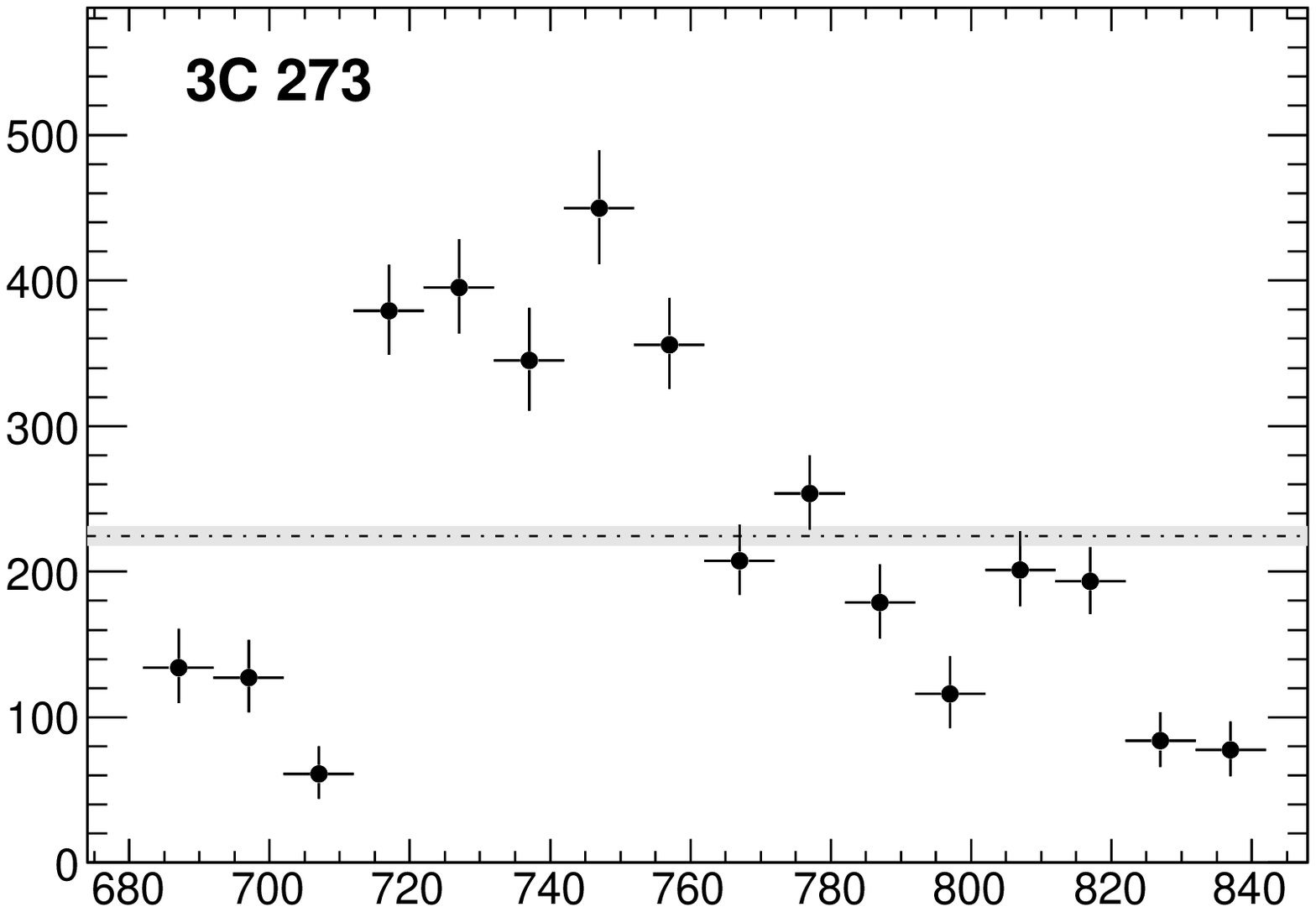}%
\includeLCTen{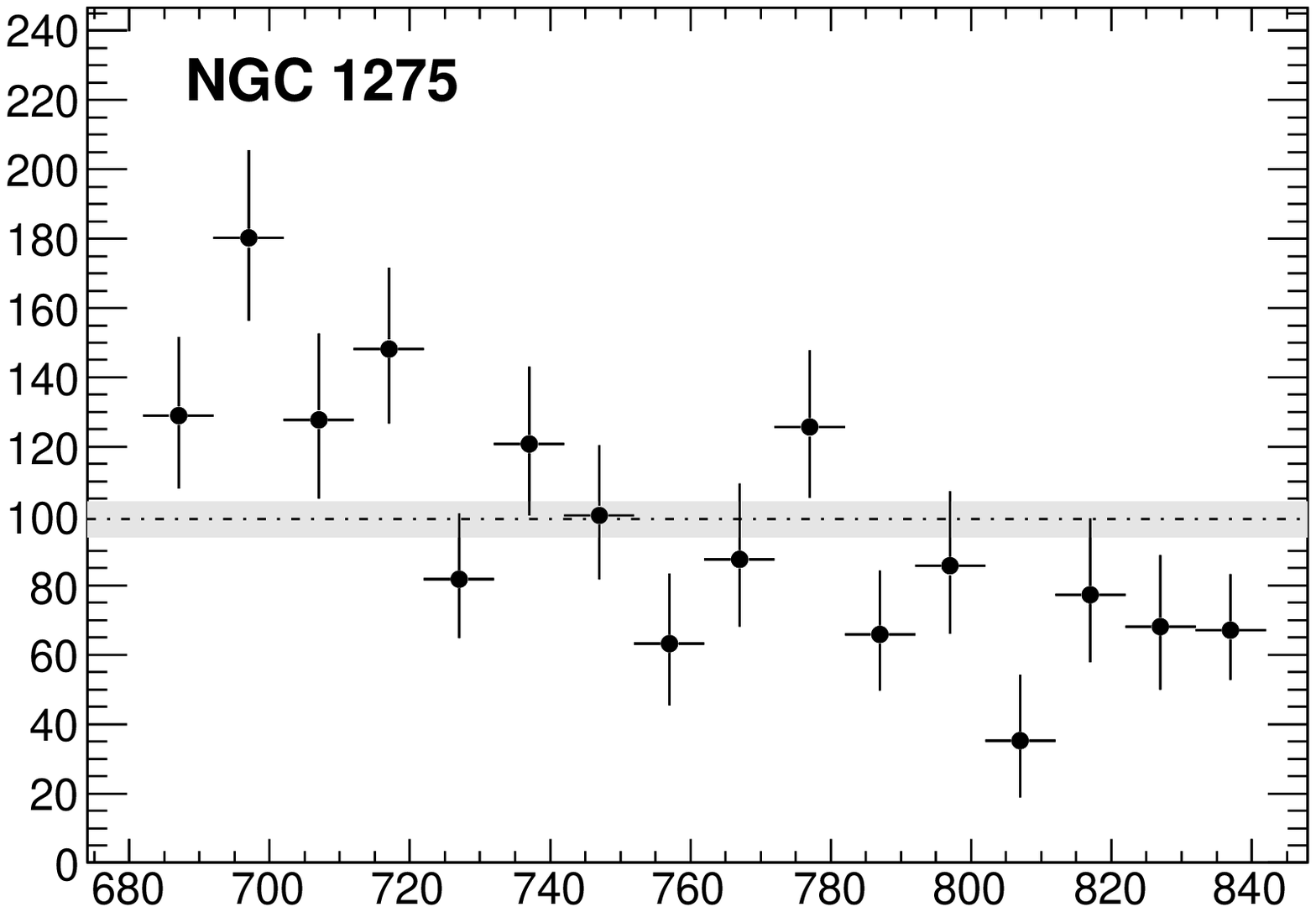}

\end{minipage}

Date -- MJD-54000 [days]

\caption{\label{FIG::LC10} 10-day light curves for selected 
\Fermic-detected sources. See caption of Figure~\ref{FIG::LC28_1} for 
explanation of what is indicated on each panel.}
\end{figure}

\begin{figure}[t]
\centering
\includegraphics[bb=25 0 527 374,clip,width=0.9\textwidth]{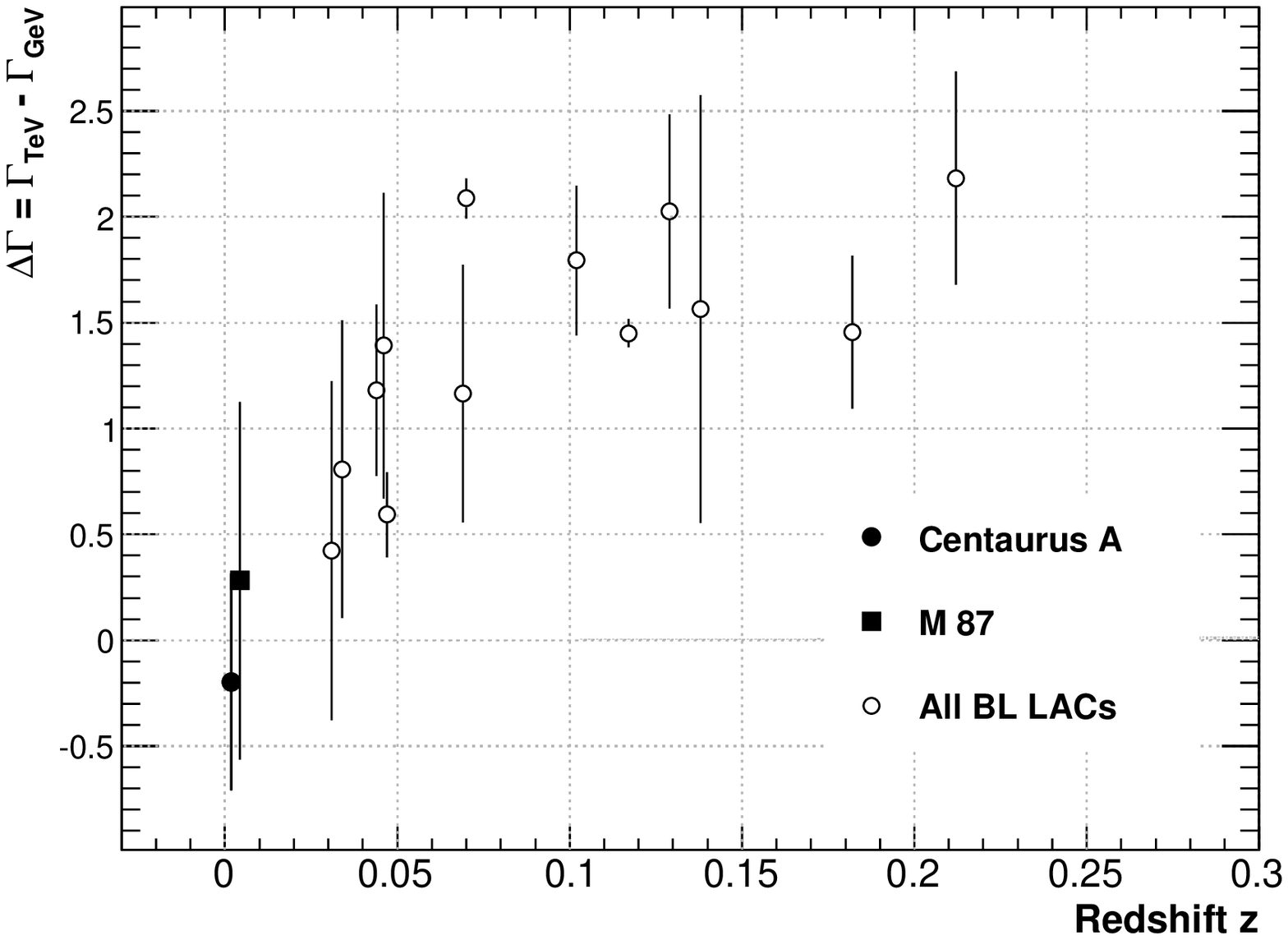}
\caption{\label{FIG::DELTAG_VS_Z}
Difference, $\Delta\Gamma$, between the measured TeV and \Fermi photon
indices as a function of the redshift. Empty circles denote the
BL~Lacs, the filled circle denotes Cen~A and the filled square M~87.}
\end{figure}

\end{document}